\documentclass[journal]{IEEEtran}
\usepackage{graphicx}
\usepackage{subfigure}
\usepackage{algpseudocode}
\usepackage{algorithm}
\usepackage{mathrsfs}
\usepackage{amsmath, bm}
\usepackage{amssymb}
\usepackage{color}
\usepackage{multirow}
\usepackage{epstopdf}
\usepackage{cite}

\newcommand{\mathcalbf}[1]{\bm{\mathcal{#1}}}

\ifCLASSINFOpdf
\else
\fi
\begin{document}
	%
	\title{Non-local tensor completion for multitemporal remotely sensed images inpainting}
	%
	%
	%

	\author{Teng-Yu~Ji,
		Naoto~Yokoya,~\IEEEmembership{Member,~IEEE,}
		Xiao~Xiang~Zhu,~\IEEEmembership{Senior~Member,~IEEE,}
		and~Ting-Zhu~Huang 
		\thanks{The work of T.-Y. Ji and T.-Z. Huang was supported by NSFC (61772003, 61402082). The work of  N. Yokoya was supported by Japan Society for the Promotion of Science (JSPS) KAKENHI 15K20955 and Alexander von Humboldt Fellowship for postdoctoral researchers. The work of X. X. Zhu has received funding from the European Research Council (ERC) under the European Union��s Horizon 2020 research and innovation programme (grant agreement No [ERC-2016-StG-714087]), as well as from Helmholtz Association under the framework of the Young Investigators Group ��SiPEO�� (VH-NG-1018, www.sipeo.bgu.tum.de). ({\em Corresponding authors: Xiao Xiang Zhu and Ting-Zhu Huang.})}
		\thanks{T.-Y. Ji and T.-Z. Huang are with  the School of Mathematical Sciences, University of Electronic Science and Technology of China, Chengdu 610054, China (e-mail: tengyu\_ji@126.com; tingzhuhuang@126.com).}
		\thanks{N. Yokoya is with the Department of Advanced Interdisciplinary Studies, University of Tokyo 153-8904, Japan, the Remote Sensing Technology Institute (IMF), German Aerospace Center (DLR) Wessling 82234, Germany, and Signal Processing in Earth Observation (SiPEO), Technical University of Munich (TUM), Munich 80333, Germany (e-mail: yokoya@sal.rcast.u-tokyo.ac.jp).}
		\thanks{X. X. Zhu is with the Remote Sensing Technology Institute (IMF), German Aerospace Center (DLR), 82234 Wessling, Germany, and also with Signal Processing in Earth Observation (SiPEO), Technical University of Munich (TUM), 80333 Munich, Germany (e-mail: xiao.zhu@dlr.de).}}

\markboth{IEEE Transactions on Geoscience and Remote Sensing, in press}%
{Shell \MakeLowercase{\textit{et al.}}: Bare Demo of IEEEtran.cls for IEEE Transactions on Magnetics Journals}
	\maketitle

	\begin{abstract}
			\textit{This is the pre-acceptance version, to read the final version please go to IEEE Transactions on Geoscience and Remote Sensing on IEEE Xplore.} 
       Remotely sensed images may contain some missing areas because of poor weather conditions and sensor failure. Information of those areas may play an important role in the interpretation of multitemporal remotely sensed data. The paper aims at reconstructing the missing information by a non-local low-rank tensor completion method (NL-LRTC). First, non-local correlations in the spatial domain are taken into account by searching and grouping similar image patches in a large search window. Then low-rankness of the identified 4-order tensor groups is promoted to consider their correlations in spatial, spectral, and temporal domains, while reconstructing the underlying patterns. Experimental results on simulated and real data demonstrate that the proposed method is effective both qualitatively and quantitatively. In addition, the proposed method is computationally efficient compared to other patch based methods such as the recent proposed PM-MTGSR method.
	\end{abstract}


	%
	\IEEEpeerreviewmaketitle

	\section{Introduction}
	%
	%
	%
	%
	\IEEEPARstart{R}{emotely} sensed images are important tools for exploring the properties of our living environment and have been used in many applications, such as hyperspectral unmixing \cite{Iordache2011unHy,Iordache2012unHy,DiasHSIUnmixOverView,Zhao2013TGRS,yokoya2012unmixing,yokoya2014unmixing,bieniarz_joint_2014}, classification \cite{lunga2014classReview,tarabalka2010class,harsanyi1994class,matsuki2015class,ghamisi2016class, mou2017rnn,mou2018cnn}, and target detection \cite{manolakis2003tardetec,willett2014tardetec,nasrabadi2014tardetecReview,yokoya2010detec,yokoya2015tardetec,shahzad2016detec}. However, these applications are largely limited by the missing information that is introduced 
	when acquiring these data by both/either the defective sensor and/or the poor atmospheric conditions. For example, three-quarters of the detectors (in band 6) of the Aqua Moderate Resolution Imaging Spectroradiometer (MODIS) are ineffective \cite{Shen2014CSInpainting,wang2006modis}, the scan line corrector (SLC) of the Landsat enhanced thematic mapper plus (ETM+) sensor has permanently failed \cite{zeng2013etmslc,Shen2015MissingReview}, and the ozone monitoring instrument (OMI) onboard the Aura satellite suffers a row anomaly problem. On the other hand, the clouds cover approximately 35\% of the Earth's surface at any one time \cite{lin2014cloud}. Owing to the above two reasons, missing information is inevitable in optical remotely sensed images, particularly in the multitemporal image analysis. Thus, reconstructing the missing information is highly desirable.

	Recently, many reconstruction methods for remotely sensed images have been proposed, which can be classified into four categories: spatial-based, spectral-based, temporal-based, and hybrid methods. The spatial-based methods take advantage of the relationships between different pixels in the spatial dimension without any other spectral and temporal auxiliary images and include interpolation methods \cite{yu2011kriging}, propagated diffusion methods \cite{maalouf2009bandelet,ballester2001filling}, variation-based methods \cite{bugeau2010comprehensive,He2016TGRSLRMFTV,Cheng2014InpaintingRS,Shen2014CSInpainting,shen2009map,shen2010universalReconstruc}, and exemplar-based methods \cite{criminisi2004region,efros1999texture}. This kind of method cannot reconstruct a large missing area because there is not enough reference information.

	The spectral-based methods borrow the correlative information from other spectral data to reconstruct the missing area. The basic idea of this kind of method is to estimate the relationship of the known areas between the complete and incomplete bands and then use the relationship to reconstruct the missing area. The typical example of this kind of method is Aqua MODIS band 6 inpainting. For example, Wang et al. \cite{wang2006modis}, Rakwatin et al. \cite{rakwatin2009modis}, and Shen et al. \cite{shen2011modis} reconstructed the missing area of band 6 by considering the spectral relationship with band 7 because these two bands are highly correlated. Gladkova et al. \cite{gladkova2012modis} and Li et al. \cite{li2014modis} took the relationships between band 6 and the other six bands into consideration. These methods can reconstruct a large area and get a better result than the spatial-based methods. However, for the most remotely sensed images, all bands contain the same missing areas. For this case, the spectral-based methods fail in getting a promising result.

	The third class of methods is to reconstruct the missing area by making use of other data taken at the same location and different periods. Temporal-based methods have been widely studied for remotely sensed inpainting, especially cloud removal. The clouds vary with time. Thus, the missing areas in different images are diverse. The basic temporal-based method is to replace the missing area with the same area of different periods \cite{zeng2013etmslc,lin2014cloud,zeng2015reconstructing}. Inspired by the temporal filter methods for the one-dimensional signal denoising, many researchers developed temporal filter methods by regarding the temporal fibers as signals \cite{julien2010comparison,zhu2012changingWeight,savitzky1964smoothing}. Recently, temporal learning model-based methods exploit the compressed sensing and regression technologies to reconstruct the missing information \cite{lorenzi2013missing,li2014recovQuantita}. More recently, Wang et al. \cite{wang2016removing} proposed a temporally contiguous robust matrix completion model for cloud removal by making the best use of the temporal correlations: low-rankness in time-space and temporal smoothness. As the method (ALM-IPG) considers the local temporal correlation, temporally contiguous property, it is good at processing the data whose adjacent temporal images are similar.

	The above three classes of methods make use of only one kind of relationship (spatial domain, spectral domain, or temporal domain). In some cases, they are powerful, but sometimes they are not. To get a better result, the hybrid methods were introduced to extract the complementary information from two or three domains. This kind of method includes the joint spatio-temporal methods \cite{zeng2013etmslc,cheng2014spatio-temporal}, joint spatio-spectral methods \cite{benabdelkader2008spatio-spectral}, and joint spectral-temporal methods \cite{Li2015SparseRS}.
	Recently, Li et al. \cite{Li2016PM} proposed the patch matching-based multitemporal group sparse representation (PM-MTGSR) that makes use of the local sparsity in the temporal domain and the non-local similarity in the spatial domain to reconstruct the missing information. Obviously, PM-MTGSR is a joint spatio-temporal method, namely, it also makes use of only two domains relationships.

	The hybrid methods perform better than each of the three classes of methods. This indicates that the results would be better if a method takes advantage of more latent structures information in the observed data.
	In this paper, we present a new methodology that makes full use of spatial, spectral, and temporal relationships for the reconstruction of missing data in multitemporal remotely sensed images. The proposed method is designed to be good at processing not only the temporally contiguous data but also the data that have a large difference between the adjacent temporal images. Low-rank tensor-based methods characterize the global correlations for each dimension. Inspired by this, the paper introduces a non-convex low-rank approximation for tensor rank to make the best use of the global correlations in the spatial, spectral, and temporal domains. Similar concept has been applied for time series analysis of radar data \cite{Kang2018}. To take advantage of the three domains similarities, we group similar patches and consider their low-rankness. Experimental results on both cloud removal and destriping experiments show that our low-rank approach is capable of achieving more accurate reconstruction than other state-of-the-art approaches.


	This paper is organized as follows. Some notations are introduced in Section \ref{Sec:Preli}. Section \ref{Sec:Method} describes the proposed algorithm for the multitemporal remotely sensed image reconstruction. Section \ref{Sec:Exp} presents the experimental results and discussion, and the conclusion is given in Section \ref{Sec:Conclusion}.

	\section{Preliminary}\label{Sec:Preli}
	In this paper, we use non-bold low-case letters for scalars, e.g., $x$, bold low-case letter for vectors, e.g., $\bm{x}$, bold upper-case letters for matrices, e.g., $\bm{X}$, and bold calligraphic letters for tensors, e.g., $\mathcalbf{X}$. Moreover, we also use bold norm up-case letters for clusters of some variables, e.g., $\textbf{M}=(\mathcalbf{M}_1, \cdots, \mathcalbf{M}_N)$. An $N$-order tensor is defined as $\mathcalbf{X}\in\mathbb{R}^{J_1\times\cdots\times J_N}$, and $x_{j_1, \cdots, j_N}$ is its $(j_1, \cdots, j_N)$-th component.

\begin{figure}[h]
		\centering
		\includegraphics[width=0.155\textwidth]{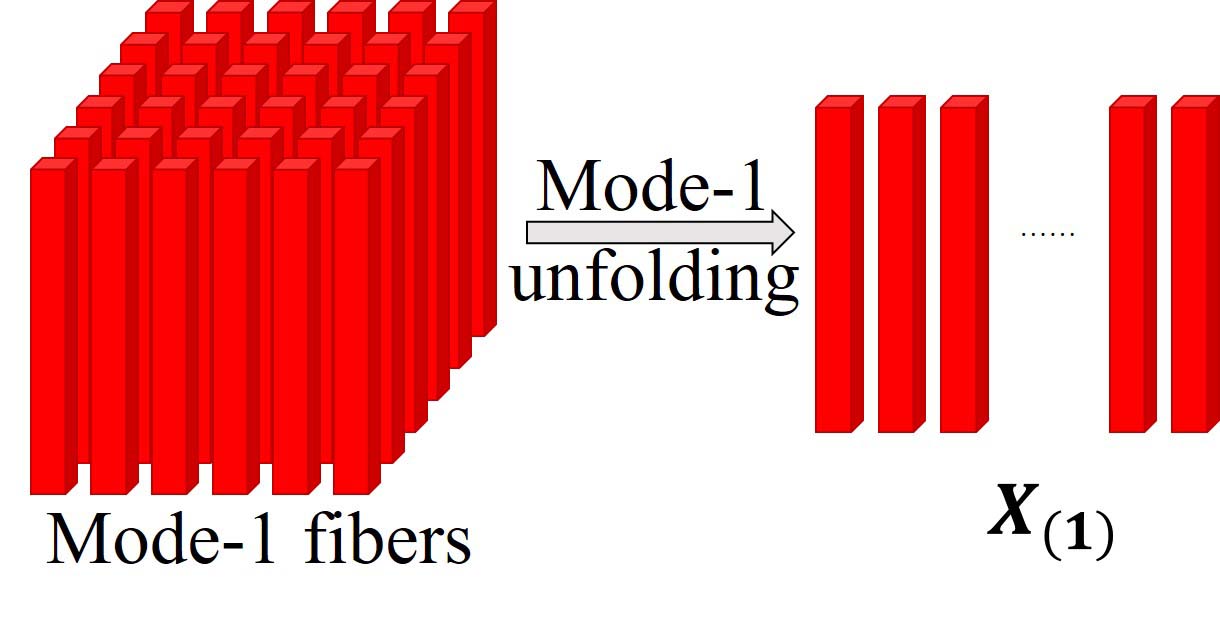}
		\includegraphics[width=0.155\textwidth]{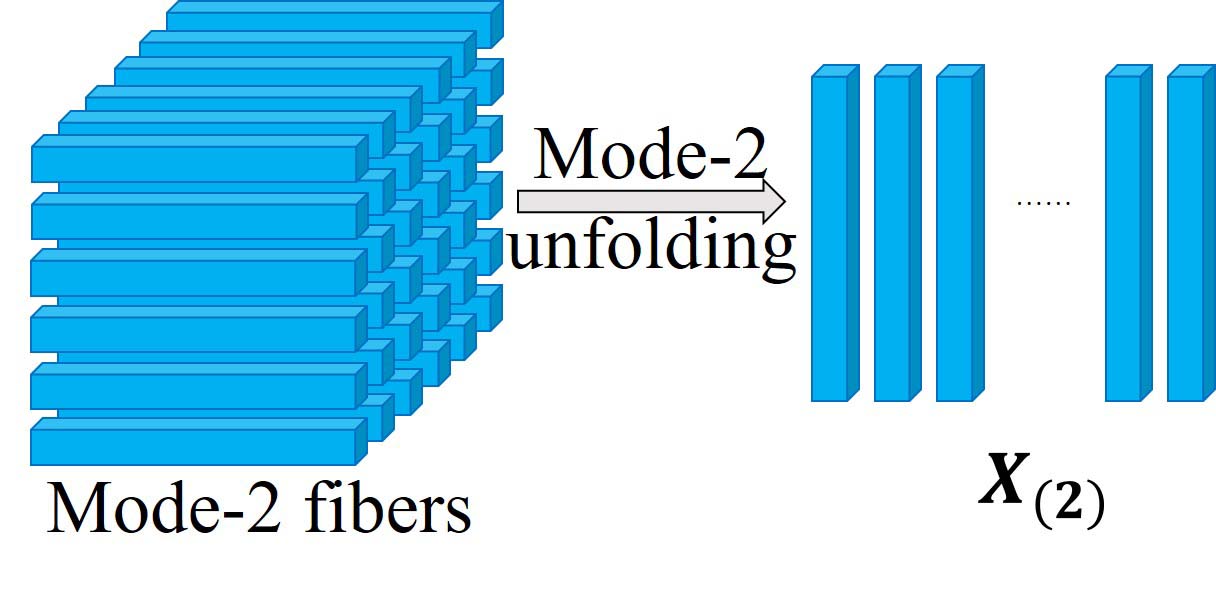}
		\includegraphics[width=0.155\textwidth]{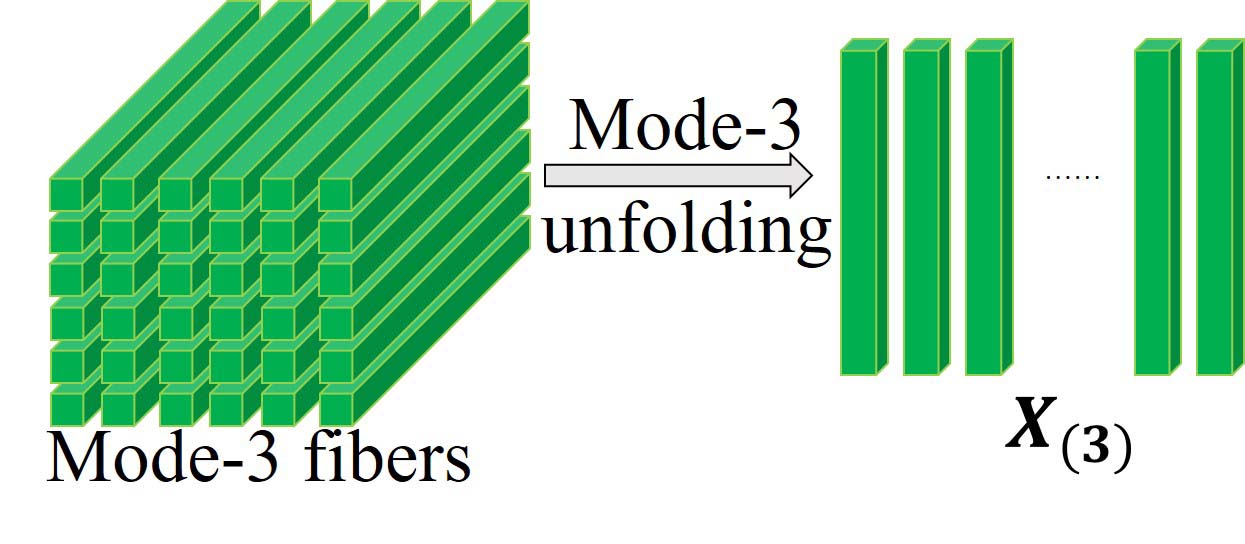}
		\caption{\footnotesize Mode-$n$ fibers and corresponding unfoldings of a 3-order tensor.}
		\label{Fig:unfolding}
	\end{figure}

	\textbf{Fibers} are the higher-order analogue of matrix rows and columns. A fiber is defined by fixing every index but one. For example, $\bm{x}_{: j_2 \cdots j_N} = (x_{1, j_2,\cdots,j_N}, \cdots, x_{J_1, j_2,\cdots,j_N})$ is one of mode-1 fibers of $N$-order tensor $\mathcalbf{X}\in\mathbb{R}^{J_1\times\cdots\times J_N}$. The mode-$n$ \textbf{unfolding} of a tensor $\mathcalbf{X}$ is denoted as $\bm{X}_{(n)}\in\mathbb{R}^{J_n\times\prod_{j\neq n}J_j}$, whose columns are the mode-$n$ fibers of $\mathcalbf{X}$ in the lexicographical order. Fig. \ref{Fig:unfolding} shows the mode-$n$ $(n=1,2,3)$ fibers and unfoldings for a 3-order tensor. The inverse operator of unfolding is denoted as ``fold'', i.e., $\mathcalbf{X}=\text{fold}_n(\bm{X}_{(n)})$. Then $n$\textbf{-rank} of an $N$-order tensor $\mathcalbf{X}$, denoted as $\text{rank}_n(\mathcalbf{X})$, is the rank of $\bm{X}_n$, and the rank of $\mathcalbf{X}$ based on $n$-rank is defined as an array: $\text{rank}(\mathcalbf{X})=(\text{rank}_1(\mathcalbf{X}), \cdots, \text{rank}_N(\mathcalbf{X}))$. The tensor $\mathcalbf{X}$ is low-rank, if $\bm{X}_{(n)}$ is low-rank for all $n$. Please refer to \cite{Kolda2009SIAMreview} for its extensive overview.

	\section{Methodology}\label{Sec:Method}
	Missing information is inevitable in the observation process for remotely sensed images. The existing methods that characterize the correlation are mostly interpolation \cite{yu2011kriging}, sparse \cite{Li2016PM}, smooth \cite{wang2016removing}, and low-rank technologies \cite{wang2016removing}, no matter which of the four methods (spatial, spectral, temporal, or hybrid) is adopted. For example, PM-MTGSR \cite{Li2016PM} characterizes the local relationships in the temporal domain using the group sparse technology, and ALM-IPG \cite{wang2016removing} characterizes the local and global correlations in the temporal domain using the smooth and low-rank technology, respectively. Although these two methods take advantage of spatial and temporal relationships, they prefer the relationships in the temporal domain. Recently, low-rank tensor based methods have attracted much attention regarding the completion of high-dimensional images because the tensors rank can characterize the correlations in different domains \cite{Liu2013PAMItensor,JiTVINS}. Combined with the definition of $n$-rank that is a vector composed of ranks of mode-$ n $ unfoldings, it can be seen from the Fig. \ref{Fig:unfolding} that the rank of mode-$n$ unfolding describes the correlations of mode-$n$ fibers. To present the motivation in detail, we analyze the low-rankness of some 4-order tensor groups stacked by the 3-order similar patches that are extracted from the 4-temporal cloud-free Landsat-8 data (``Image 3'' in Fig. \ref{Fig:SimulateMUdata}, Section IV); see Tab. \ref{Tab:MotivateLowrank}. For example, the first group is of size $4\times 4\times 3\times 708$, where the four dimensions correspond to the numbers of pixels, observations, spectral channels, and patches, and it has 8496 mode-1 fibers, 8496 mode-2 fibers, 11328 mode-3 fibers, and 48 mode-4 fibers. The dimensions of the spaces (DimSpac) generated by mode-1, -2, -3, and -4 fibers are 2, 3, 2, and 13, respectively. That means the mode-$n$ $(n=1,\cdots,4)$ fibers are highly correlated. i.e., it is possible to reconstruct the missing area using tensor low-rank technology. Inspired by this, we introduce the tensors rank to characterize the global correlations in the spatial, spectral, and temporal domains to reconstruct the missing information of remotely sensed images after grouping the similar patches. It should be noted that missing areas are detected before their reconstruction.

	\begin{table}[!t]
		\caption{\footnotesize Analysis of low-rankness of 4-order groups.  }
		\footnotesize
		\begin{center}
			\begin{tabular}{c | c | c c c c }
				\hline
				Group                         &                    & mode-1 & mode-2 & mode-3 & mode-4 \\ \hline
				\multirow{2}{*}{1} & \# of fibers & 8496      & 8496     & 11328     & 48 \\
				& DimSpac    & 2           & 3            & 2            & 13 \\ \hline
				\multirow{2}{*}{2} & \# of fibers & 360       & 360        & 480        & 48 \\
				& DimSpac   & 2           & 3            & 2            & 9 \\ \hline
				\multirow{2}{*}{3} & \# of fibers & 5424      & 5424     & 7232     & 48 \\
				& DimSpac    & 2           & 3            & 2            & 12 \\ \hline
				\multirow{2}{*}{4} & \# of fibers & 1956      & 1956     & 2608     & 48 \\
				& DimSpac    & 2           & 2            & 2            & 12 \\ \hline
				\multirow{2}{*}{5} & \# of fibers & 3444      & 3444     & 4592     & 48 \\
				& DimSpac    & 2           & 3            & 2            & 14 \\ \hline
				\multirow{2}{*}{6} & \# of fibers & 11484      & 11484     & 15312     & 48 \\
				& DimSpac    & 2           & 2            & 2            & 15 \\ \hline
			\end{tabular}
		\end{center}
		\label{Tab:MotivateLowrank}
	\end{table}

\begin{figure*}[ht]
		\begin{center}
			\includegraphics[width=1\textwidth]{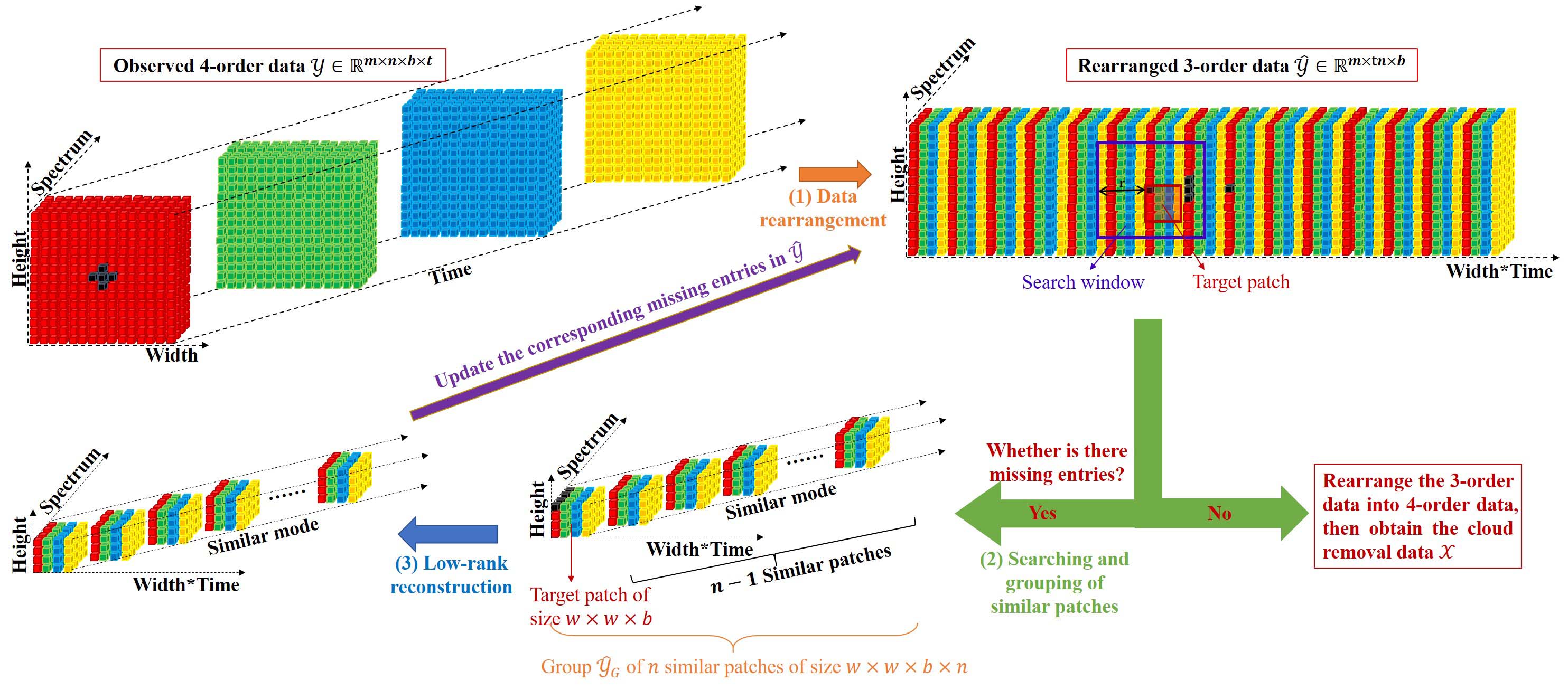}
		\end{center}
		\caption{\footnotesize Illustration for the proposed NL-LRTC method. ``Height'' denotes one of the spatial mode, ``Width'' denotes another mode of spatial domain, ``Width*Time'' means this mode contains the information of both spatial (Width) and temporal mode. The proposed method comprises three steps: (1) reshape the 4-order observed data into 3-order data; (2) search and group similar patches; (3) reconstruct each group using the low-rank tensor completion method.}
		\label{Fig:flowchart}
	\end{figure*}

	The flowchart of the proposed NL-LRTC is shown in Fig. \ref{Fig:flowchart}. The proposed NL-LRTC method consists of three parts. The method first reshapes the observed 4-order tensor into a 3-order tensor so that the pixels at the different periods but the same location become adjoining. Next, we search and group similar patches in a searching window. Last, the missing information of every group is reconstructed using the low-rank tensor completion method.

	\subsection{Data Rearrangement}
	The observed multitemporal remotely sensed data set $\mathcalbf{Y}\in\mathbb{R}^{m\times n\times b\times t}$  is a 4-order tensor, where $m\times n$ denotes the number of pixels of remotely sensed images, $b$ denotes the number of spectral channels of remote sensors, and $t$ is the number of time series. The indices set $\bm{\Omega}\in\mathbb{R}^{m\times n\times b\times t}$ is also a 4-order tensor, where the position $(i,j,k,l)\in \mathbb{Z}^m\times\mathbb{Z}^n\times\mathbb{Z}^b\times\mathbb{Z}^t$ is covered by cloud if $\bm{\Omega}_{i,j,k,l}=0$ and is cloud free if $\bm{\Omega}_{i,j,k,l}=1$. 

	To make the best use of correlations between different periods and find more accurate similar patches, it is necessary to reshape the observed data. The reshaping process is illustrated in the first step of Fig. \ref{Fig:flowchart}. 
	PM-MTGSR also reshaped the data before searching the similar patches. The difference between PM-MTGSR and our method is that PM-MTGSR reshapes the 4-order tensor into a matrix, while the result of our reshaping is a 3-order tensor. The difference leads to another difference between these two methods: the similar patches in our method are 3-order tensors but matrices in PM-MTGSR.


	As the description above indicates, the aim is to reshape the observed data into a 3-order tensor to take advantage of the temporal correlations. Detailedly, we stack the mode-1 slices at the same locations and different periods one by one, i.e., the observed 4-order tensor $\mathcalbf{Y}$ is reshaped into a 3-order tensor $\hat{\mathcalbf{Y}}\in\mathbb{R}^{m\times tn\times b}$ which is defined by $\hat{\mathcalbf{Y}}_{u,v,w}=\mathcalbf{Y}_{i,j,k,l}$ when $u=i$, $v=(j-1)n+l$, and $w=k$. Similarly, we reshape the indices tensor $\bm{\Omega}\in\mathbb{R}^{m\times n\times b\times t}$ into a 3-order tensor $\hat{\bm{\Omega}}\in\mathbb{R}^{m\times tn\times b}$ which is defined by $\hat{\bm{\Omega}}_{u,v,w}=\bm{\Omega}_{i,j,k,l}$ when $u=i$, $v=(j-1)n+l$, and $w=k$. Let $\mathcalbf{X}$ be the recovery data we are seeking and $\hat{\mathcalbf{X}}\in\mathbb{R}^{m\times tn\times b}$ be the corresponding reshaped 3-order tensor.

	\subsection{Grouping of Similar Patches}
	This section is to search and group the similar patches for missing area pixels of the reshaped data $\hat{\mathcalbf{Y}}$. The second step of Fig. \ref{Fig:flowchart} shows the process of similar patch searching after reshaping. The red square denotes the target patch $\hat{\mathcalbf{Y}}_{i,j}=\hat{\mathcalbf{Y}}(i:(i+w-1),j:(j+w-1),:)\in\mathbb{R}^{w\times w\times b}$, where $(i,j)$ denotes the coordinate of the target patch, and  $w\times w\times b$ denotes the patch size. When the target patch is given, similar patches are searched for in the surrounding window with a radius of $r$ in the reshaped data $\hat{\mathcalbf{Y}}$. According to the reshaping procedure only the similar information in the square window of size $(2(r/t)+1)\times(2(r/t)+1)$ in the original data $\mathcalbf{Y}$ is used.
	Given a metric of the similarity indicator between the target patch and the other patches and an indicator threshold $\gamma_2$, a similar patch can be found when the indicator reaches the given condition. There are many indicators for measuring similarity between two vectors \cite{Deza2009distance,Li2016PM}, such as the Euclidean distance, the mean relative error, normalized cross-correlation, cosine coefficients, and so on. This work adopts the normalized cross-correlation defined as:
		\begin{equation*}
		Q= \frac{\sum\limits_{j_1 \cdots j_N}(x_{j_1, \cdots, j_N}-\mu_{\mathcalbf{X}})(y_{j_1, \cdots, j_N}-\mu_{\mathcalbf{Y}})}{\sqrt{\sum\limits_{j_1 \cdots j_N}(x_{j_1, \cdots, j_N}-\mu_{\mathcalbf{X}})^2}\sqrt{\sum\limits_{j_1 \cdots j_N}(y_{j_1, \cdots, j_N}-\mu_{\mathcalbf{Y}})^2}},
		\end{equation*}
		where $\mathcalbf{X}, \mathcalbf{Y}\in \mathbb{R}^{J_1\times\cdots\times J_N}$, $\mu_{\mathcalbf{X}}, \mu_{\mathcalbf{Y}}$ are the mean values of $\mathcalbf{X}$ and $\mathcalbf{Y}$, respectively.  The mean value of an $N$-order tensor $\mathcalbf{X}$ is defined as $\mu_{\mathcalbf{X}}:=\frac{1}{N}\sum_{j_1, \cdots, j_N}x_{j_1, \cdots, j_N}$.

	After completing a search for similar patches, these 3-order-tensor patches are stacked into a 4-order tensor $\hat{\mathcalbf{Y}}_G\in\mathbb{R}^{w\times w\times b\times n}$, where $n$ is the number of similar patches. The corresponding indices set $\hat{\bm{\Omega}}_G$ can be obtained according to the coordinates of patches in $\hat{\mathcalbf{Y}}_G$.

	\subsection{Low-rank Reconstruction}
\begin{figure*}[ht]
		\centering
		\includegraphics[width=1\textwidth]{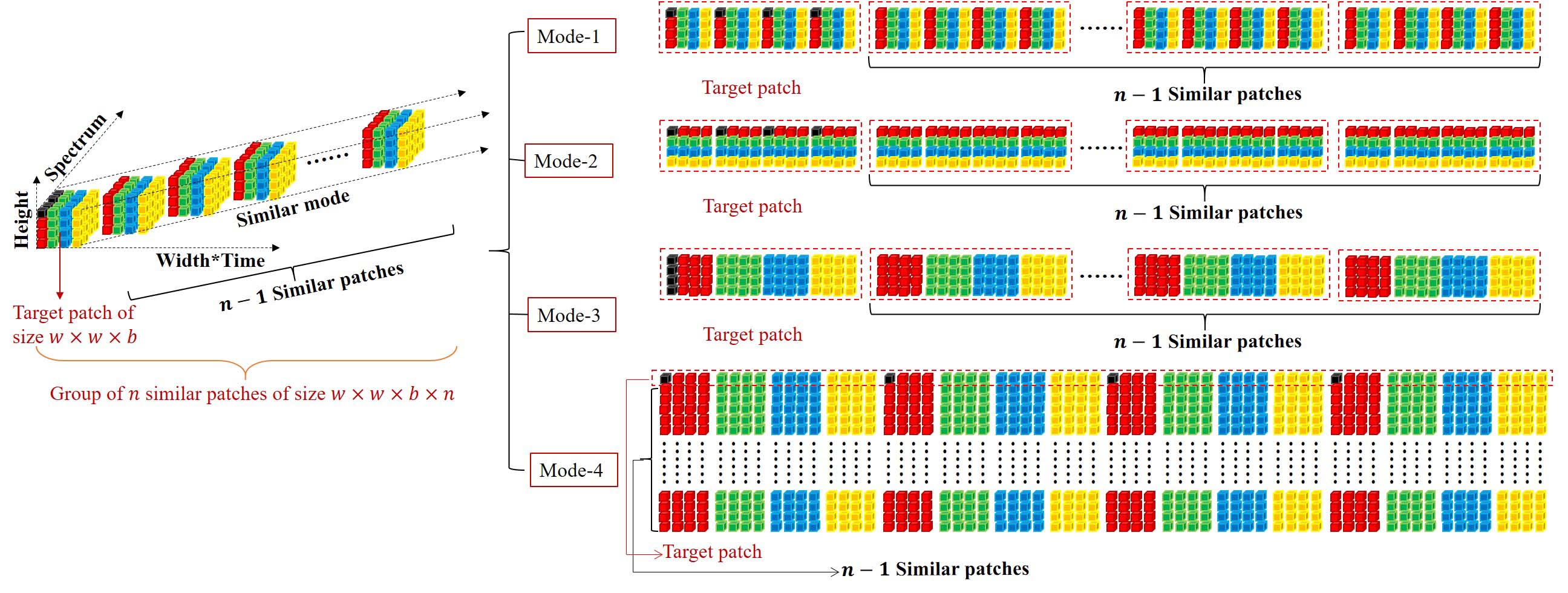}
		\caption{\footnotesize Illustration for how to exploit the low-rankness of the group of similar patches using four unfolding matrices. The mode-1 unfolding is of size $w\times wbn$, the mode-2 unfolding is of size $w\times wbn$, the mode-3 unfolding is of size $b\times w^2n$, and the mode-4 unfolding is of size $n\times w^2b$. }
		\label{Fig:GroupAnalysis}
	\end{figure*}

	In this section, we propose a low-rank method to reconstruct the missing information in the 4-order tensor $\hat{\mathcalbf{Y}}_G$ obtained in the last subsection. Different from the low-rank matrix methods which consider only one mode  correlation, e.g., \cite{wang2016removing}, NL-LRTC studies the low-rankness of $\hat{\mathcalbf{Y}}_G$ from four aspects, i.e., NL-LRTC considers the correlations of mode-$i (i=1,2,3,4)$ using $\text{rank}(\hat{\mathcalbf{Y}}_G)$. In fact, the four dimensions of $\hat{\mathcalbf{Y}}_G$ denote spatial, temporal, spectral, and patch similarity, respectively. As mentioned in the description about tensor rank previously with Fig.  \ref{Fig:unfolding}, $\text{rank}(\hat{\mathcalbf{Y}}_G)$ takes advantage of all of the spatial, spectral, and temporal relationships. This can be found in Fig. \ref{Fig:GroupAnalysis} where NL-LRTC considers the low-rankness of the group of similar patches by analyzing the low-rankness of four unfolding matrices that can describe the correlations of the spatial, temporal, and spectral domains. In contrast, the group of PM-MTGSR only describes the spatial and temporal domains relationships (seen Fig. 3 of \cite{Li2016PM}). This is another difference between NL-LRTC and PM-MTGSR. Tensor nuclear norm and the corresponding algorithm (HaLRTC) were developed in order to make it possible to minimize the tensor rank \cite{Liu2013PAMItensor}.
		However, the tensor nuclear norm cannot treat the different singular values accurately according to their different importance. For the proposed non-convex surrogate of tensor rank,  the larger singular values that are associated with the major projection orientations and are more important can be shrunk less to preserve the major data components \cite{Ji2017AMM,Gu2014CVPRWNNM}. This is one of the differences between HaLRTC and NL-LRTC. Another difference is that HaLRTC is without the patch strategy.

	In the last section, $\hat{\mathcalbf{X}}_G$ and $\hat{\bm{\Omega}}_G$ have been obtained. Next, we reconstruct the missing areas in $\hat{\mathcalbf{X}}_G$ group by group using the following model:
	\begin{equation}
	\label{Eq:RecoverModel}
	\begin{aligned}
	\underset{\hat{\mathcalbf{X}}_G}{\text{min}}\ & \text{logDet}(\hat{\mathcalbf{X}}_G, \bm{\varepsilon})=\sum_{i=1}^{4}\alpha_i L(\hat{\bm{X}}_{G_{(i)}}) \\
	\text{s.t.} \ &\hat{\mathcalbf{X}}_{G_{\hat{\bm{\Omega}}_G}}=\hat{\mathcalbf{Y}}_{G_{\hat{\bm{\Omega}}_G}},
	\end{aligned}
	\end{equation}
	where $L(\hat{\bm{X}}_{G_{(i)}}) = \text{log det}( (\hat{\bm{X}}_{G_{(i)}}\hat{\bm{X}}_{G_{(i)}}^{T})^{1/2} + \varepsilon_{i} \bm{I}_{i})$, $\bm{I}_i$ is the Identity matrix, $\alpha_{i}$s are constants satisfying $\alpha_{i}\geq 0$ and $\sum_{i=1}^{N}\alpha_i=1$, $\bm{\varepsilon}=(\varepsilon_1, \cdots, \varepsilon_N)^T>0$, and $\hat{\bm{X}}_{G_{(i)}}$ is the mode-$i$ unfolding of $\hat{\mathcalbf{X}}_G$.

	The $L(\hat{\bm{X}}_{G_{(i)}})$ can be approximated by using the first-order Taylor expansion:
	\begin{equation}
	\label{Eq:TaylorApproximation}
	\begin{aligned}
	L(\hat{\bm{X}}_{G_{(i)}}) \approx  & \sum_{j=1}^{J_i} \frac{\sigma_j(\hat{\bm{X}}_{G_{(i)}})}{\sigma_{j}^{k}(\hat{\bm{X}}_{G_{(i)}})+\varepsilon_{i}} + \text{constant}\\
	= & (\bm{\omega}^{k})^{T} \bm{\sigma}+ \text{constant}\overset{\vartriangle}{=}L_{\bm{\omega}^{k}}(\hat{\bm{X}}_{G_{(i)}}),
	\end{aligned}
	\end{equation}
	where $\sigma_{j}^{k}(\hat{\bm{X}}_{G_{(i)}})$s are the solutions obtained in the $k$-th iteration,  $\bm{\omega}^{k}=(1/(\sigma_1^k(\hat{\bm{X}}_{G_{(i)}})+\varepsilon_{i}),\cdots,1/(\sigma_{J_i}^k(\hat{\bm{X}}_{G_{(i)}})+\varepsilon_{i}))^{T}$, $\bm{\sigma}=(\sigma_1(\hat{\bm{X}}_{G_{(i)}}),\cdots,\sigma_{J_i}(\hat{\bm{X}}_{G_{(i)}}))^{T}$, and $(J_1, J_2, J_3, J_4) = (w, w, b, n)$. From the approximate function (\ref{Eq:TaylorApproximation}), we can see that the proposed function logDet indeed shrinks the larger singular values less.

	Next, we present a computationally efficient algorithm that is based on the alternating direction method of multipliers (ADMM) \cite{lin2010admm,He2012Admm,zhao2014SISconvex} to solve the problem (\ref{Eq:RecoverModel}) by replacing $L(\hat{\bm{X}}_{G_{(i)}})$ with $L_{\bm{\omega}^{k}}(\hat{\bm{X}}_{G_{(i)}})$ and introducing some auxiliary values. Thus, the problem (\ref{Eq:RecoverModel}) can be rewritten as:
	\begin{equation}
	\label{Eq:ModelADMM}
	\begin{aligned}
	\underset{\hat{\mathcalbf{X}}_G,\mathcalbf{M}_1, \ldots, \mathcalbf{M}_4}{\text{min}} &\bm{1}_{\hat{\mathcalbf{Y}}_{G}}^{\hat{\bm{\Omega}}_G}(\hat{\mathcalbf{X}}_G) + \sum_{i=1}^{4}\alpha_{i} L_{\bm{\omega}^{k}}(\bm{M}_{i,(i)}) \\
	\text{s.t.} \quad\ \ & \mathcalbf{M}_1 = \hat{\mathcalbf{X}}_G, \cdots, \mathcalbf{M}_4 = \hat{\mathcalbf{X}}_G,
	\end{aligned}
	\end{equation}
	where $\bm{1}_{\hat{\mathcalbf{Y}}_{G}}^{\hat{\bm{\Omega}}_G}(\hat{\mathcalbf{X}}_G)=0$ if $\hat{\mathcalbf{X}}_{G_{\hat{\bm{\Omega}}_G}}=\hat{\mathcalbf{Y}}_{G_{\hat{\bm{\Omega}}_G}}$, otherwise $\bm{1}_{\hat{\mathcalbf{Y}}_{G}}^{\hat{\bm{\Omega}}_G}(\hat{\mathcalbf{X}}_G)=\infty$.

	By attaching the Lagrangian multiplier  $\{\Lambda_i\}_{i=1}^{4}$ that have the same size with $\hat{\mathcalbf{X}}_G$ to the linear constraint, the augmented Lagrangian function of (\ref{Eq:ModelADMM}) is given by:
	\begin{equation}
	\label{Eq:ALF}
	\begin{aligned}
	&\mathcal{L}(\hat{\mathcalbf{X}}_G, \mathcalbf{M}_1, \ldots, \mathcalbf{M}_4, \Lambda_1, \ldots \Lambda_4)
	=\bm{1}_{\hat{\mathcalbf{Y}}_{G}}^{\hat{\bm{\Omega}}_G}(\hat{\mathcalbf{X}}_G)+\\
	&\sum_{i=1}^{4}\left( \alpha_{i} L_{\bm{\omega}^{k}}(\bm{M}_{i,(i)})
	+ \frac{\beta}{2} \| \hat{\mathcalbf{X}}_G-\mathcalbf{M}_{i} + \frac{\Lambda_i}{\beta}\|_{F}^{2}\right),
	\end{aligned}
	\end{equation}
	where $\beta$ is the penalty parameter for the violation of the linear constraints, and $\langle \cdot, \cdot\rangle$ is the sum of the elements of the Hadamard product. Thus, $\hat{\mathcalbf{X}}_G$ and $\{\mathcalbf{M}_i\}_{i=1}^{4}$ can be obtained by minimizing the augmented Lagrangian function (\ref{Eq:ALF}), alternately. The Lagrangian multipliers are updated as	$\Lambda_i^{k+1} = \Lambda_i^{k} + \beta(\hat{\mathcalbf{X}}_G^{k+1}-\mathcalbf{M}_i^{k+1})$ for $i=1,\ldots,4$.

	First, $\hat{\mathcalbf{X}}_G$ is obtained by solving the following optimization subproblem:
	\begin{equation}
	\label{Eq:subX}
	\begin{aligned}
	\min_{\hat{\mathcalbf{X}}_G} \left\{\bm{1}_{\hat{\mathcalbf{Y}}_{G}}^{\hat{\bm{\Omega}}_G}(\hat{\mathcalbf{X}}_G)
	+  \sum_{i=1}^{4}
	\frac{\beta}{2} \| \hat{\mathcalbf{X}}_G-\mathcalbf{M}_{i}^k +\frac{\Lambda_{i}^k}{\beta}\|_{F}^{2}\right\}.
	\end{aligned}
	\end{equation}
	It is obvious that the objective function of (\ref{Eq:subX}) is differentiable, thus $\hat{\mathcalbf{X}}_G^{k+1}$ has a closed form solution:
	\begin{equation}
	\label{Eq:Xsol}
	\hat{\mathcalbf{X}}_G^{k+1} = \frac{1}{4\beta}\left(\sum_{i=1}^{4}(\beta \mathcalbf{M}_{i}^{k}-\Lambda_{i}^{k})\right)_{\hat{\bm{\Omega}}_G^{c}} + \hat{\mathcalbf{Y}}_{G_{\hat{\bm{\Omega}}_G}},
	\end{equation}
	where $\hat{\bm{\Omega}}_G^{c}$ is the complementary set of the indices set $\hat{\bm{\Omega}}_G$.

	Next, $\{\mathcalbf{M}_i\}_{i=1}^{4}$-subproblems are solved. Note that $\mathcalbf{M}_i$-subproblems are independent, and thus we can solve them separately. Without loss of generality, the typical variable $\mathcalbf{M}_i$ is solved through the following problem:
	\begin{equation}
	\label{Eq:subM2}
	\begin{aligned}
	\min_{M_{i,(i)}} \frac{\alpha_{i}}{\beta} L_{\bm{\omega}^{k}}(\bm{M}_{i,(i)})
	+ \frac{1}{2} \| \bm{M}_{i,(i)} -\hat{\bm{X}}_{G_{(i)}}^{k+1}- \frac{\Lambda_{i, (i)}^k}{\beta} \|_{F}^{2},
	\end{aligned}
	\end{equation}
	where $\bm{\omega}^{k}=(1/(\sigma_1(M_{i,(i)}^{k})+\varepsilon_{i}),\cdots,1/(\sigma_{J_i}(M_{i,(i)}^{k})+\varepsilon_{i}))^{T}$, $(J_1, J_2, J_3, J_4)=(w,w,b,n)$. $\bm{M}_{i,(i)}^{k+1}$ can be obtained using a thresholding operator \cite{Ji2017AMM,Dong2014logdet},
	\begin{equation}
	\label{Eq:Ysol}
	\begin{aligned}
	\bm{M}_{i,(i)}^{k+1}
	= \bm{U}^{k}(\bm{\Sigma}^{k} - \tau \text{diag}(\bm{\omega}^{k}))_{+}(\bm{V}^{k})^{T},
	\end{aligned}
	\end{equation}
	where $\bm{U}^{k}\bm{\Sigma}^{k} (\bm{V}^{k})^{T}$ is the SVD of $\hat{\bm{X}}_{G_{(i)}}^{k+1} + \frac{1}{\beta}\Lambda_{i,(i)}^{k}$ and $(\bm{X})_{+}=\max\{\bm{X},0\}$. Thus, $\mathcalbf{M}_i^{k+1}=\text{fold}_{i}(\bm{M}_{i,(i)}^{k+1})$.

	Based on the previous derivation, we develop the low-rank method to reconstruct missing information in multitemporal remotely sensed images, as outlined in \textbf{Algorithm \ref{Alg:recovery}}. Then the proposed NL-LRTC method is outlined in \textbf{Algorithm \ref{Alg:flowchart}}.


	\begin{algorithm}
		\caption{NL-LRTC for multitemporal remotely sensed images inpainting.}
		\label{Alg:flowchart}
		\begin{algorithmic}[1]
			\Require Data $\mathcalbf{Y}$ and  index set $\bm{\Omega}$, radius of searching window $r$, patch size $w$, and indicator threshold $\gamma_2$, parameters $\beta$, $\alpha$, and $\varepsilon$.
			\State Obtain the 3D tensors $\hat{\mathcalbf{Y}}$ and $\hat{\bm{\Omega}}$ by rearranging the data $\mathcalbf{Y}$ and $\bm{\Omega}$, respectively;
			\While{$\hat{\bm{\Omega}}^c\neq 0$}  
			\State Find $(i,j)$ subject to $\hat{\bm{\Omega}}_{i,j}=0$, that means the pixel $(i,j)$ is covered by clouds;
			\State Search the similar patches for patch $\hat{\mathcalbf{Y}}_{i,j}$ in the searching window;
			\State Stack these similar patches as a group $\hat{\mathcalbf{Y}}_G$, and obtain the corresponding index set $\hat{\bm{\Omega}}_G$;
			\State Estimate the missing pixel values in $\hat{\mathcalbf{Y}}_G$ using Algorithm \ref{Alg:recovery} and set $\hat{\bm{\Omega}}_G=1$;
			\State Replace the corresponding entries in $\hat{\Omega}$ and $\hat{\mathcalbf{Y}}$ with $\hat{\bm{\Omega}}_G$ and $\hat{\mathcalbf{X}}_G$, respectively.
			\EndWhile  
			\Ensure Recovered data $\mathcalbf{X}$.
		\end{algorithmic}
	\end{algorithm}

	\begin{algorithm}
		\caption{Low-rank reconstruction of missing information via ADMM.}
		\label{Alg:recovery}
		\begin{algorithmic}[1]
			\Require Group $\hat{\mathcalbf{Y}}_G$ and  index set $\hat{\bm{\Omega}}_G$, parameters $\beta$, $\alpha$, and $\varepsilon$.
			\State \textbf{Initialize:} $\{\mathcalbf{M}_i\}_{i=1}^{4}=0$, $\{\Lambda_i\}_{i=1}^{4}=0$, $\text{tol}=10^{-5}$, and $K=100$.
			\For{$k=1$ to $K$}  
			\State Update $\hat{\mathcalbf{X}}_G^{k+1}$ by (\ref{Eq:Xsol});
			\For{$i = 1$ to 4}
			\State Update $\mathcalbf{M}_i$ by (\ref{Eq:Ysol});
			\State Update $\Lambda_{i}^{k+1}$ by: $\Lambda_i^{k+1} = \Lambda_i^{k} + \beta(\hat{\mathcalbf{X}}_G^{k+1}-\mathcalbf{M}_i^{k+1})$;
			\EndFor
			\State If
			$\left\| \hat{\mathcalbf{X}}_G^{k+1} - \hat{\mathcalbf{X}}_G^{k}\right\|_F/\left\| \hat{\mathcalbf{X}}_G^{k}\right\|_F<\text{tol}$, stop iteration.
			\EndFor  
			\Ensure Recovered data $\hat{\mathcalbf{X}}_G$ for group $\hat{\mathcalbf{Y}}_G$.
		\end{algorithmic}
	\end{algorithm}

	\section{Experiments and Discussion}\label{Sec:Exp}
	\subsection{Test Data}
\begin{figure}[h]
		\centering
		\subfigure[M102014]{\includegraphics[width=0.15\textwidth]{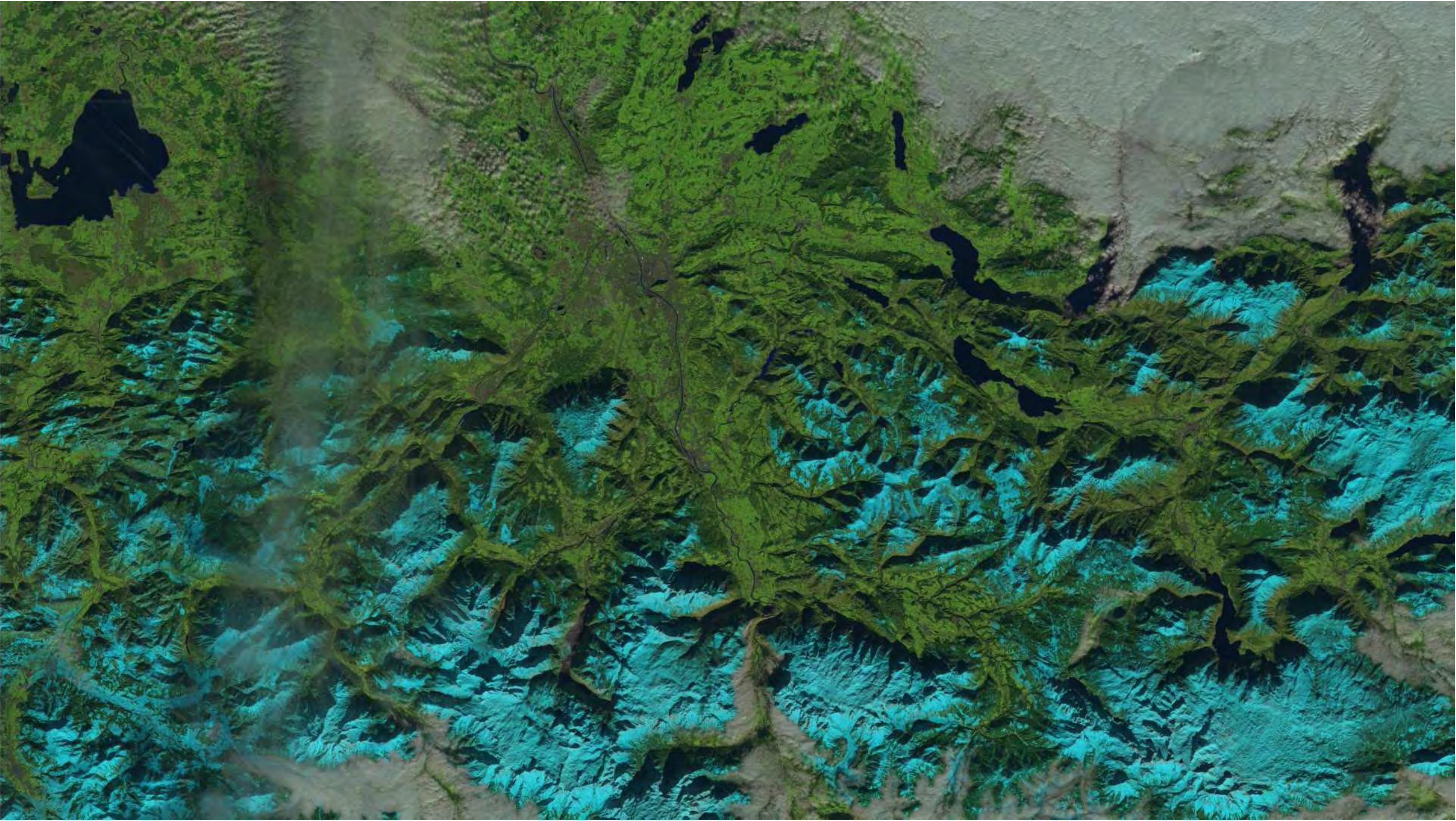}}
		\subfigure[M012015]{\includegraphics[width=0.15\textwidth]{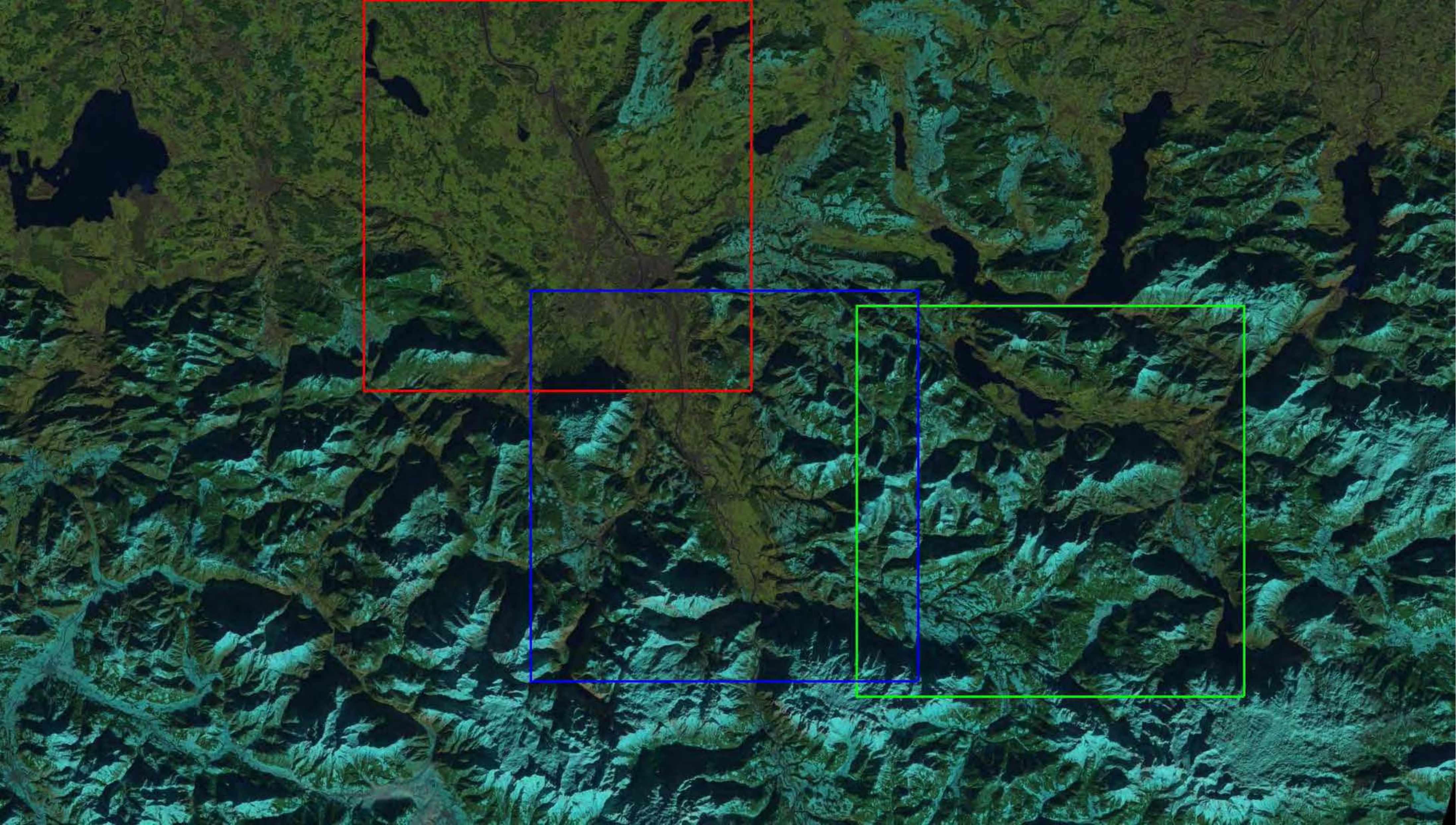}}
		\subfigure[M022015]{\includegraphics[width=0.15\textwidth]{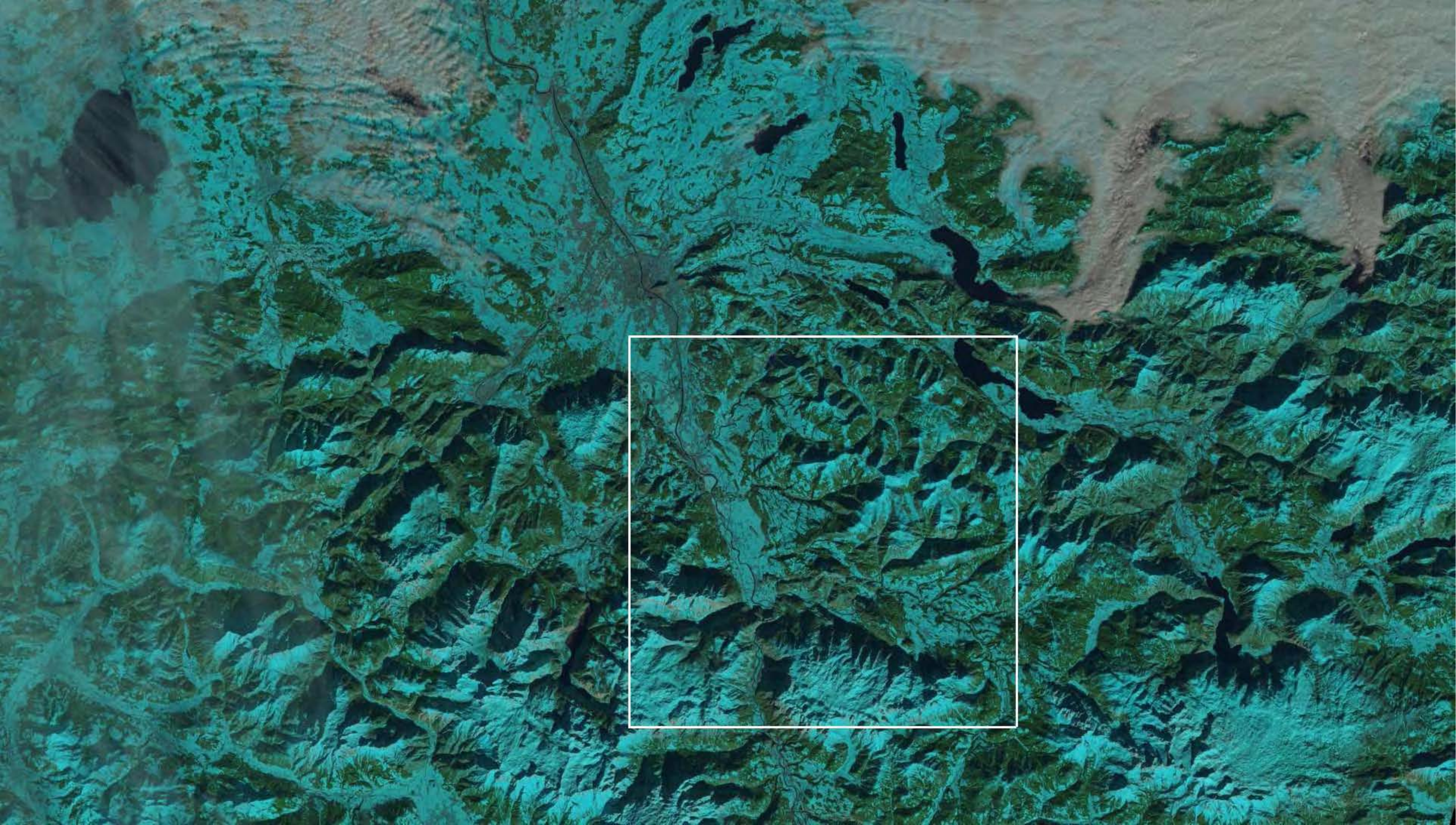}}
		\subfigure[M032015]{\includegraphics[width=0.15\textwidth]{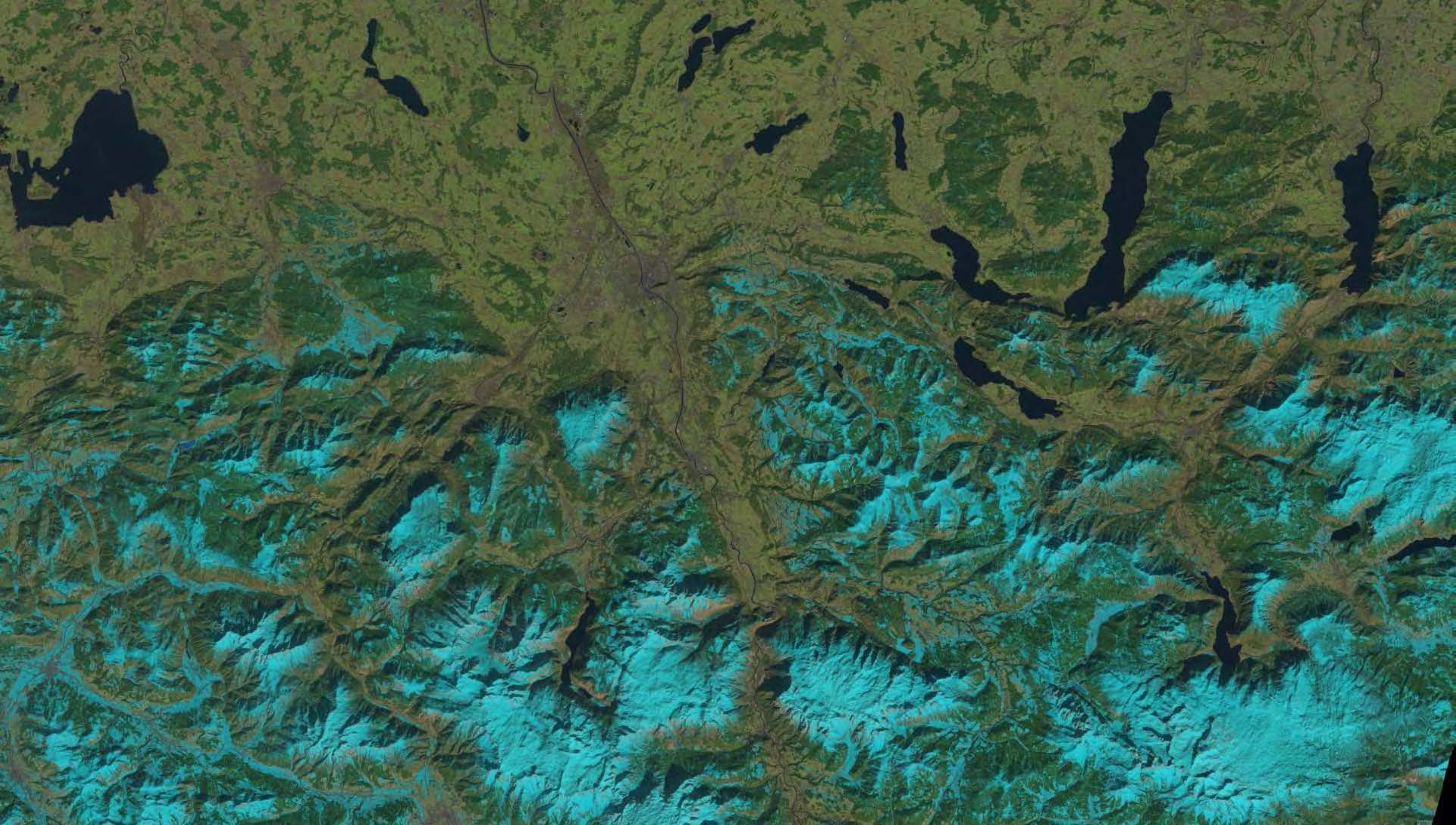}}
		\subfigure[M042015]{\includegraphics[width=0.15\textwidth]{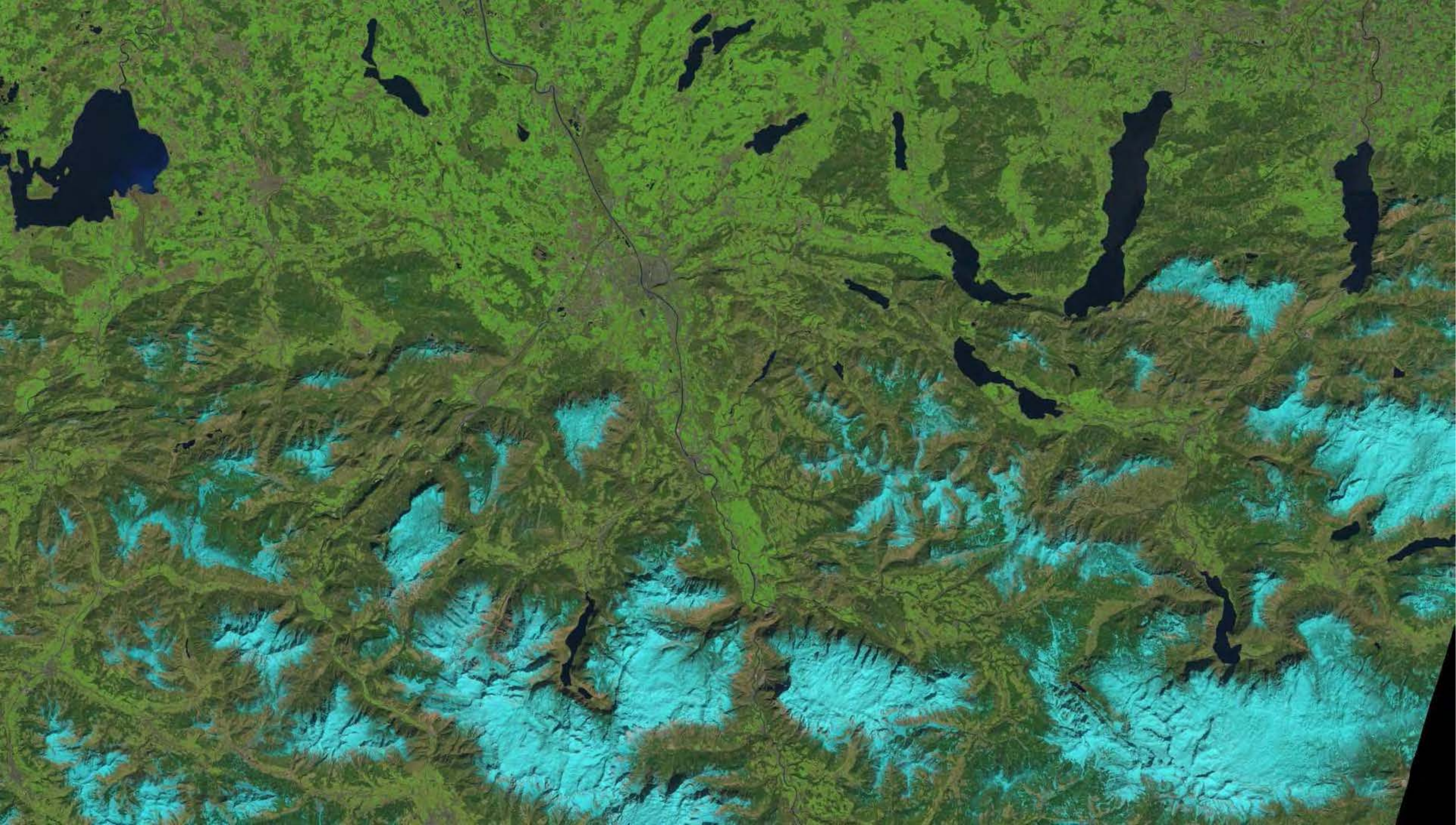}}
		\subfigure[M082015]{\includegraphics[width=0.15\textwidth]{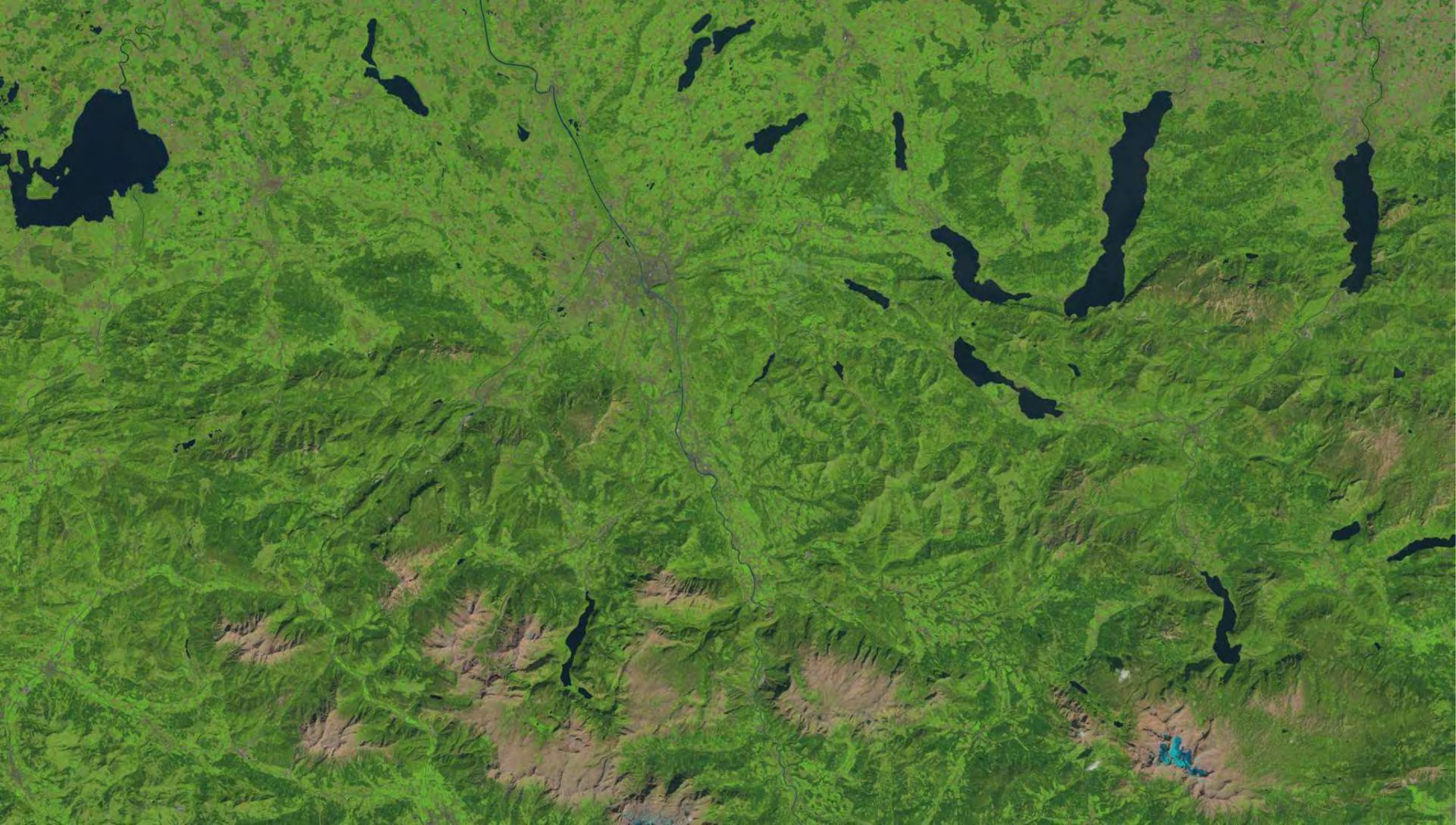}}
		\caption{\footnotesize Data set taken by Landsat-8. ``MXXYYYY'' means the image is taken over Munich in XX-th month YYYY-year. }
		\label{Fig:MUdata}
	\end{figure}

\begin{figure}[h]
		\centering
		\subfigure[BJ122015]{\includegraphics[width=0.09\textwidth]{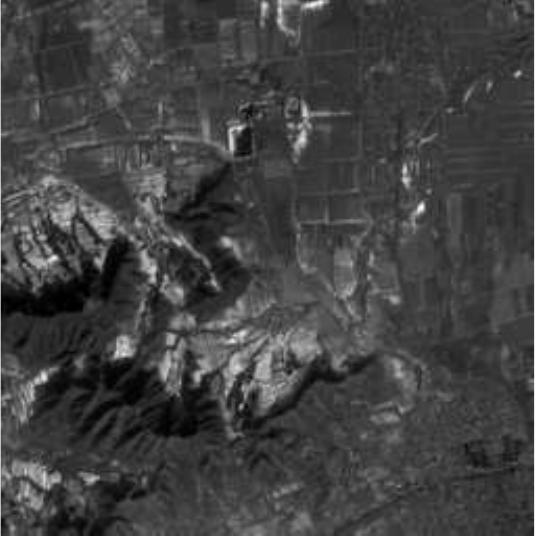}}
		\subfigure[BJ032016]{\includegraphics[width=0.09\textwidth]{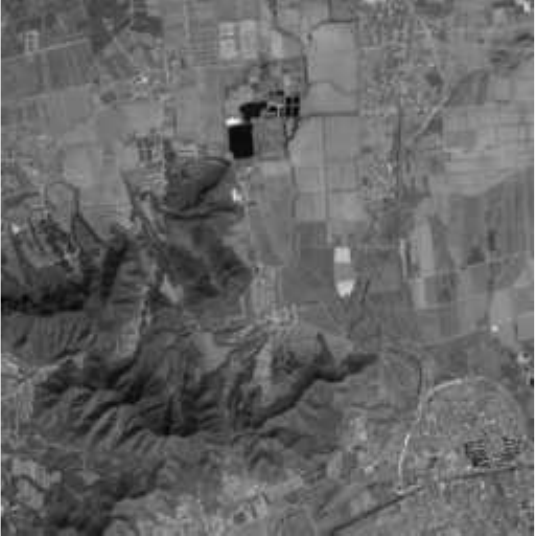}}
		\subfigure[BJ072016]{\includegraphics[width=0.09\textwidth]{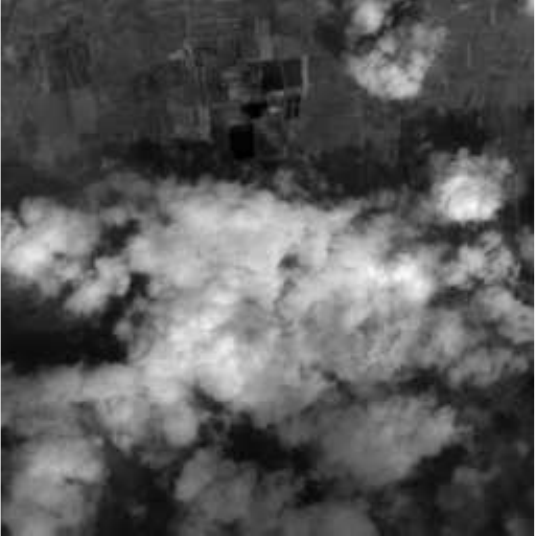}}
		\subfigure[BJ082016]{\includegraphics[width=0.09\textwidth]{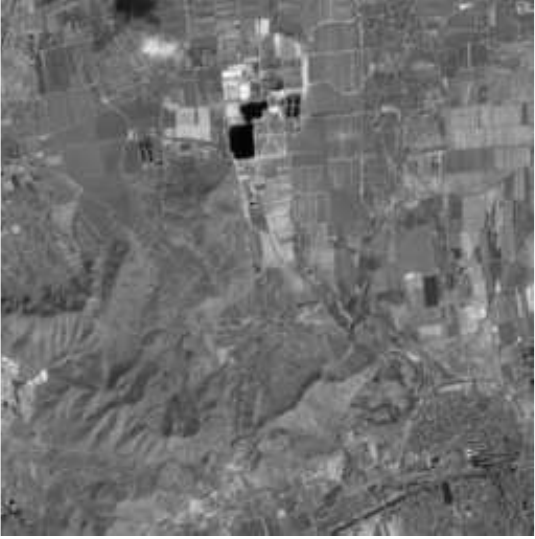}}
		\subfigure[BJ092016]{\includegraphics[width=0.09\textwidth]{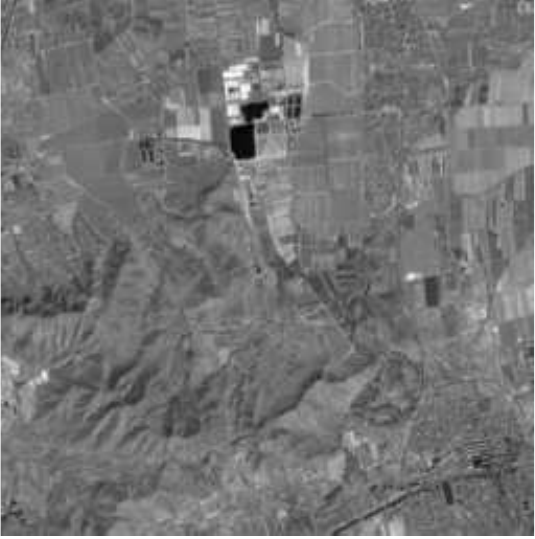}}
		\caption{\footnotesize Band 6 of Beijing data. ``BJXXYYYY'' means the image is taken over Beijing in XX-th month YYYY-year.}
		\label{Fig:BJdata}
	\end{figure}

\begin{figure}[h]
		\centering
		\subfigure[EU102016]{\includegraphics[width=0.23\linewidth]{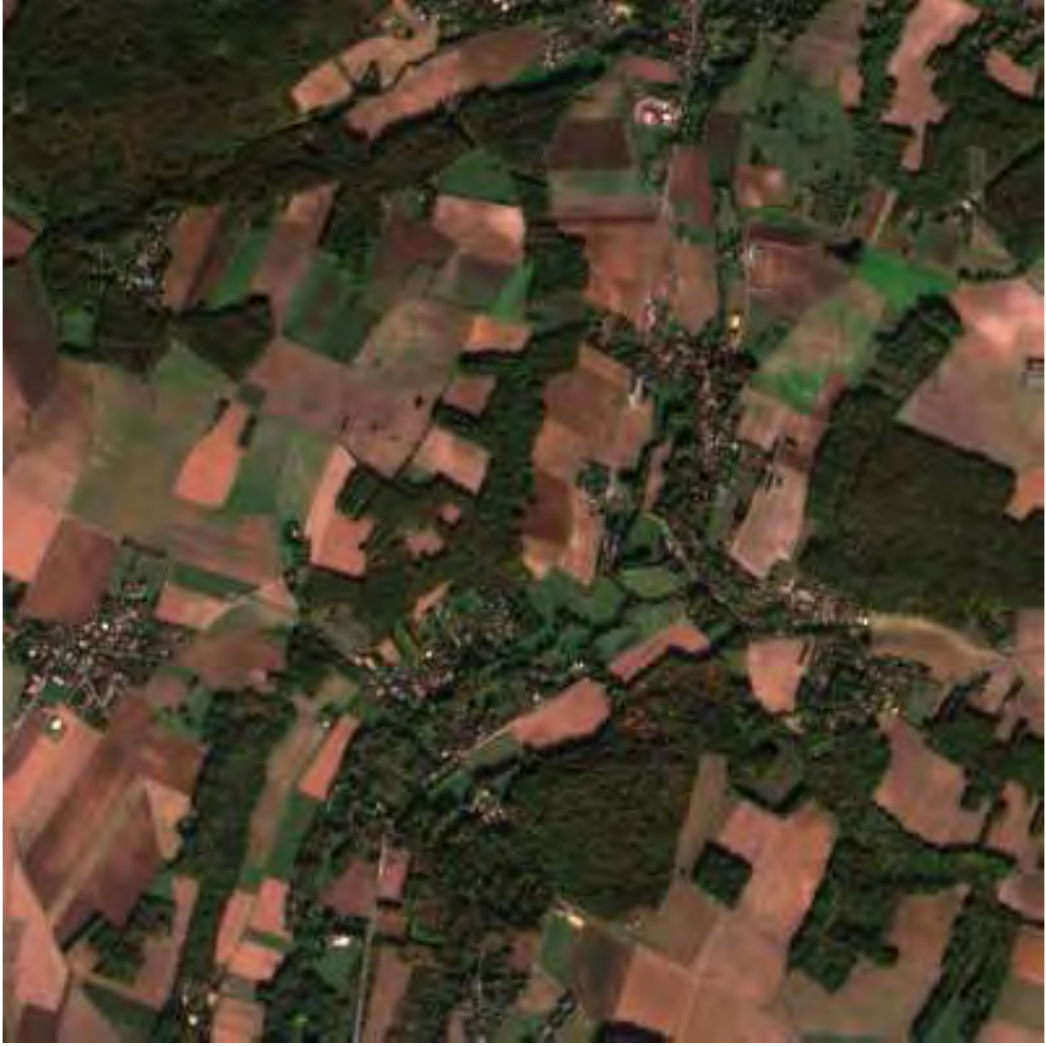}}
		\subfigure[EU012017]{\includegraphics[width=0.23\linewidth]{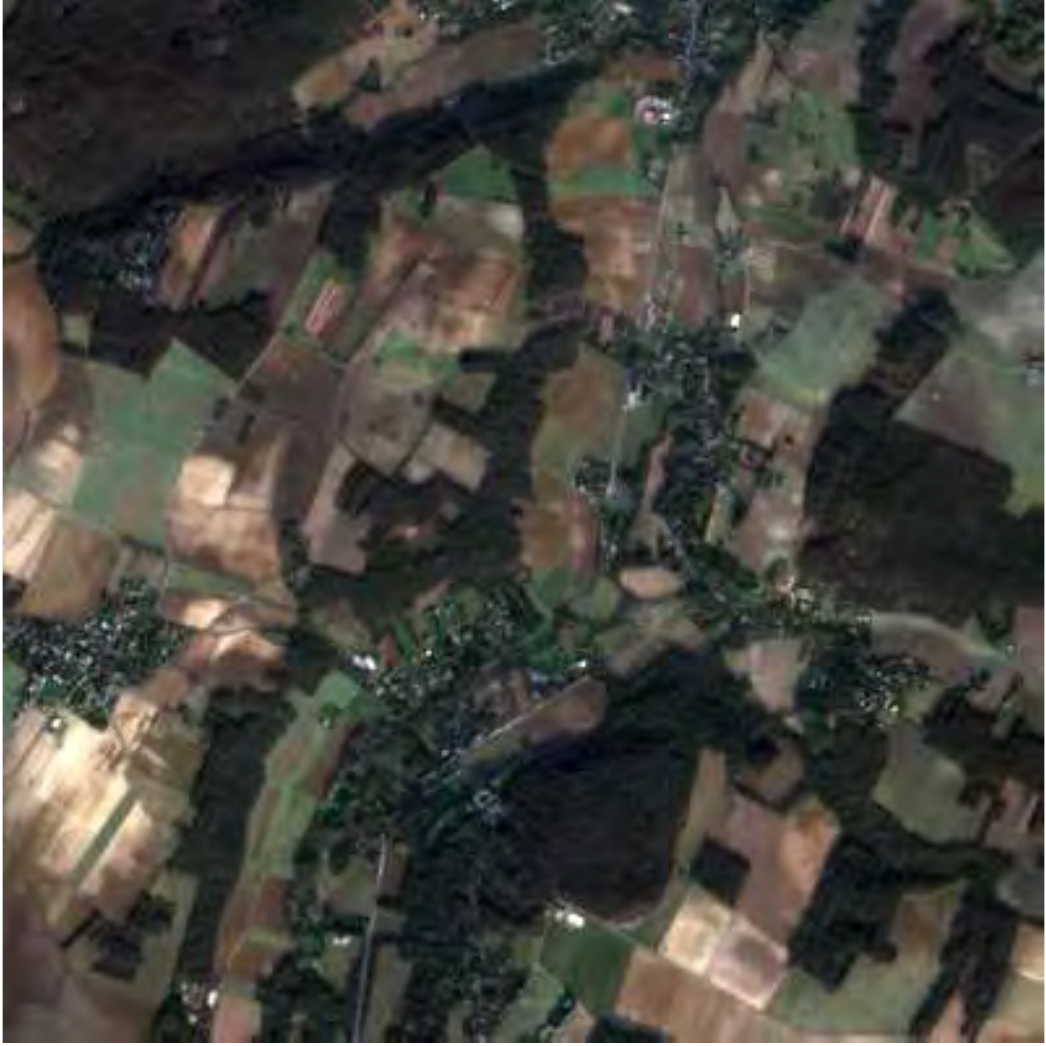}}
		\subfigure[EU052017]{\includegraphics[width=0.23\linewidth]{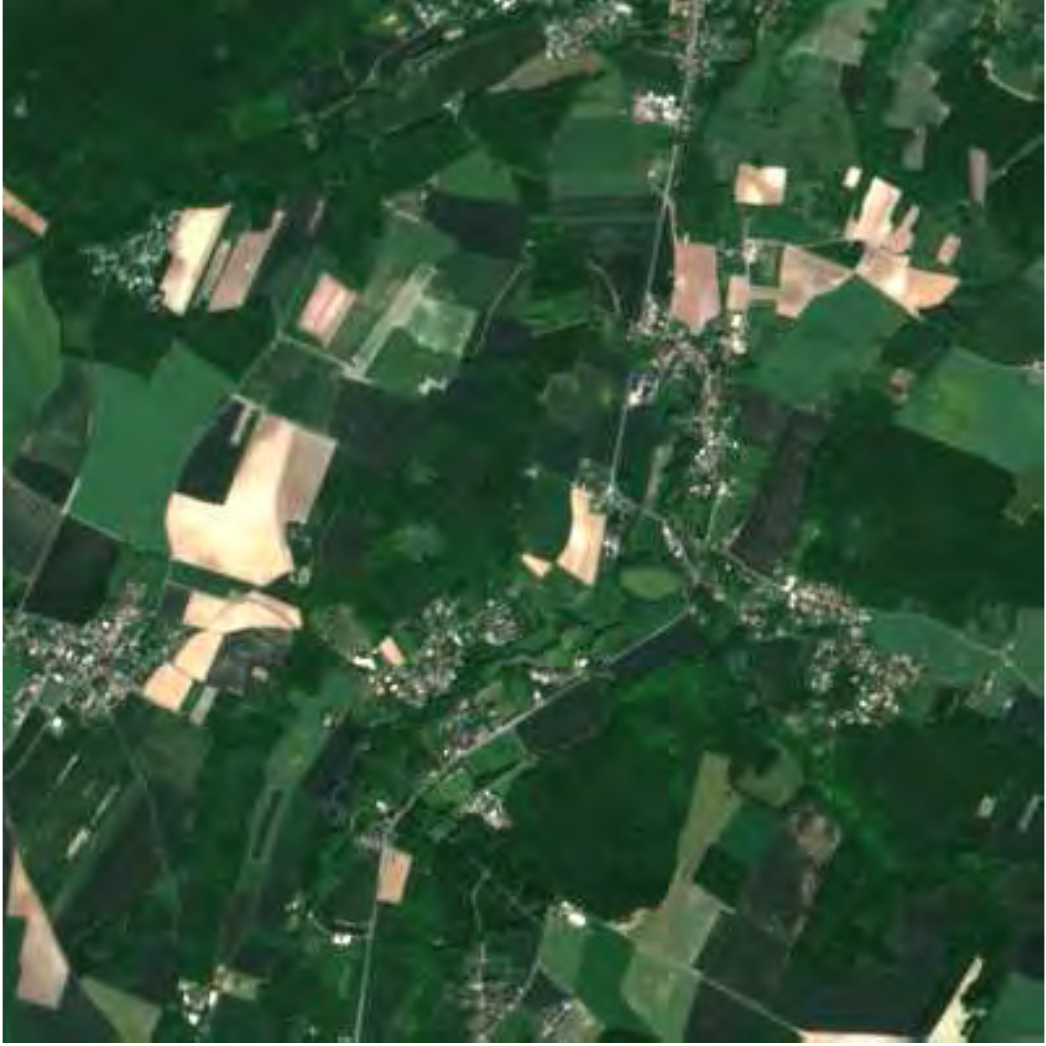}}
		\subfigure[EU082017]{\includegraphics[width=0.23\linewidth]{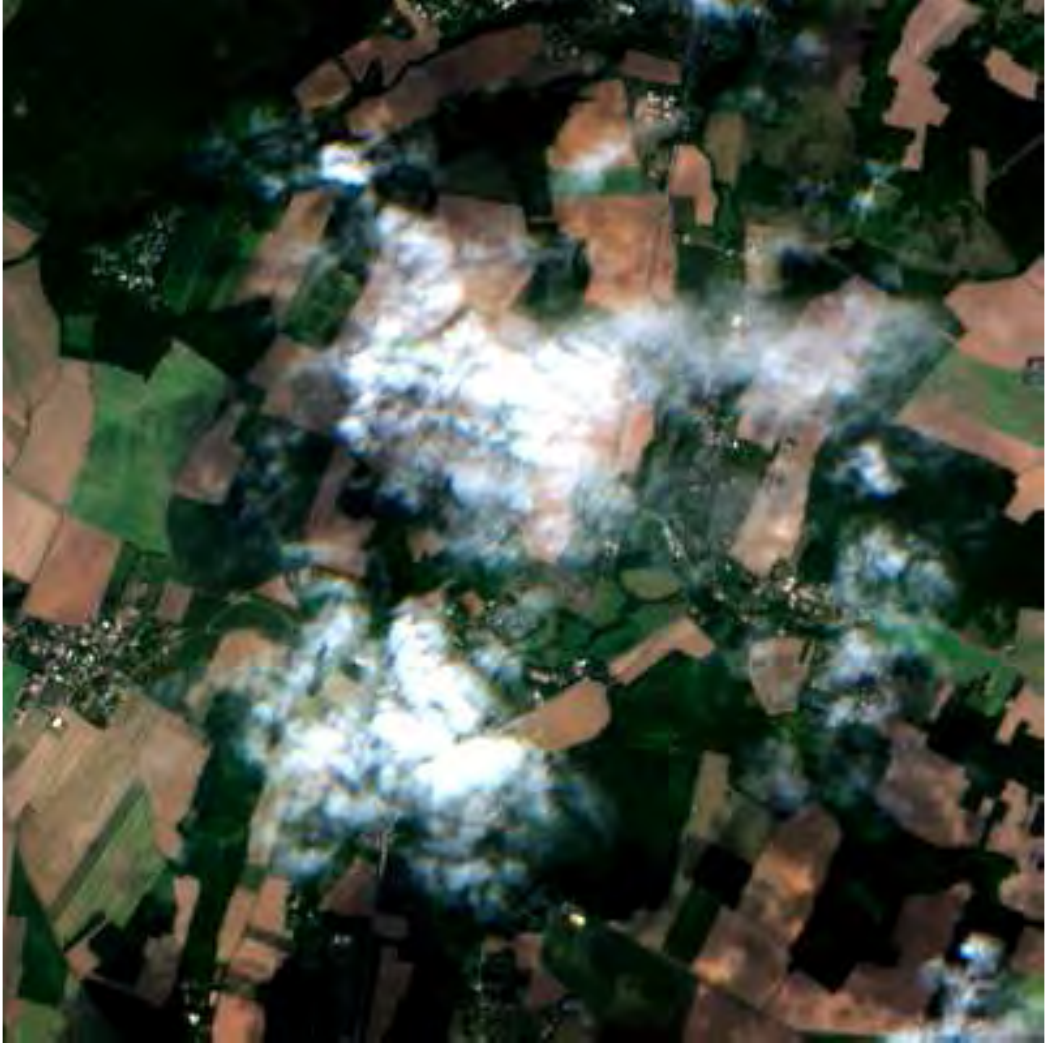}}
		\caption{\footnotesize RGB bands (bands 4, 3, and 2) of Eure data. ``EUXXYYYY'' denotes the image taken over Eure in XX-th month YYYY-year. }
		\label{Fig:EUdata}
	\end{figure}

	The proposed reconstruction method, NL-LRTC, for multitemporal remotely sensed images is applied to three data sets for simulated and real-data experiments. The first data set was taken over Munich, Germany,  by Landsat-8. The data set has nine bands, and three bands with 30-m resolution (red, green, blue) are used. The data set was acquired over the Munich suburbs (which consist of forests, mountains, hills, etc.) and includes six temporal images denoted as ``M102014'', ``M012015'', ``M022015'', ``M032015'', ``M042015'', and ``M082015'', where ``MXXYYYY'' means the data is taken over Munich in XX-th month YYYY-year; see Fig. \ref{Fig:MUdata}. 
	The second data set was taken over Beijing, China, by Sentinel-2 with six spectral bands at a ground sampling distance of 20 meters (bands 5, 6, 7, 8A, 11, 12). The data set was acquired over the Beijing suburbs (which consist of villages, mountains, etc.) and includes five temporal images denoted as ``BJ122015'', ``BJ032016'', ``BJ072016'', ``BJ082016'', and ``BJ092016'', where ``BJXXYYYY'' means the data is taken over Beijing in XX-th month YYYY-year; see Fig. \ref{Fig:BJdata}. 
	The third data set was taken over Eure, France, by Sentinel-2 and atmospheric correction has been processed by MAYA \cite{Lonjou2016Data}. The data set includes four temporal images and four spectral bands at a ground sampling distance of 10 meters (bands 2, 3, 4, and 8), see Fig. \ref{Fig:EUdata}.
	In our experiments, the observed multitemporal remotely sensed data contain four different temporal data, namely, the observed tensor $\mathcalbf{Y}$ is of size $m\times n\times b\times 4$. For the first data set, ``M032015'', ``M042015'', and ``M082015'' are the reference data; four subimages of ``M012015'' and ``M022015'' are used for the simulated experiments in that the size of the tested images is $512\times 512$ in the spatial domain; ``M102014'' is used for the real experiment in that the size of the tested images is $1080\times 1920$ in the spatial domain. For the Munich area, the surface reflectance is changed with time due to snow, seasonal change of vegetation, etc. Thus, ``M012015'', ``M022015'', and ``M102014'' are greatly different from the other reference data. We also study how NL-LRTC performs when the temporal difference is not so great with the second data set. For this data set, ``BJ092016'' and ``BJ072016'' are used for simulated and real experiment, respectively. The other temporal data are as reference data. The size of tested Beijing images is $256\times 256$ in the spatial domain. We test another real experiment with ``EU082017'' in the third data set whose size is $400\times 400$ in the spatial domain.

	\subsection{Performance Evaluation}
	In the simulated experiments, the performance of multitemporal remotely sensed images reconstruction is quantitatively evaluated by peak signal-to-noise ratio (PSNR) \cite{Shen2014CSInpainting}, structural similarity (SSIM) index \cite{Wang2004SSIM}, metric Q \cite{Zhu2010Q}, average gradient (AG) \cite{Li2005AG}, and blind image quality assessment (BIQA) \cite{Gabarda2007BIQA}. The PSNR and SSIM assess the recovered image by comparing it with the original image from the gray-level fidelity and the structure-level fidelity aspects, respectively. The metric Q, AG, and BIQA assess the recovered image without the reference image based on the human vision system. Given a reference image $\tilde{\bm{X}}\in\mathbb{R}^{m\times n}$, the PSNR of a reconstructed image $\bm{X}\in\mathbb{R}^{m\times n}$ is computed by the standard formula
	\begin{equation}
	\label{Eq:psnr}
	\text{PSNR}(\bm{X}, \tilde{\bm{X}}) = 10\log_{10}\frac{N\tilde{\bm{X}}_\text{max}^{2}}{\|\tilde{\bm{X}}-\bm{X}\|_{F}^{2}},
	\end{equation}
	where $N=mn$ denotes the number of the pixels in the image, and $\tilde{\bm{X}}_\text{max}$ is the maximum pixel value of the original image. The SSIM of the estimated image $\bm{X}$ is defined by
	\begin{equation}
	\label{Eq:ssim}
	\text{SSIM}(\bm{X}, \tilde{\bm{X}}) = \frac{(2\mu_{\bm{X}}\mu_{\tilde{\bm{X}}}+c_{1})(2\sigma_{\bm{X}\tilde{\bm{X}}}+c_{2})}{(\mu_{\bm{X}}^{2}+\mu_{\tilde{\bm{X}}}^{2}+c_{1})(\sigma_{\bm{X}}^{2}+\sigma_{\tilde{\bm{X}}}^{2}+c_{2})},
	\end{equation}
	where $\mu_{\bm{X}}$ and $\mu_{\tilde{\bm{X}}}$ represent the average gray values of the recovered image $\bm{X}$ and the original clear image $\bm{\tilde{\bm{X}}}$, respectively. $\sigma_{\bm{X}}$ and $\sigma_{\tilde{\bm{X}}}$ represent the standard deviation of $\bm{X}$ and $\bm{\tilde{\bm{X}}}$, respectively. $\sigma_{\bm{X}\tilde{\bm{X}}}$ represents the covariance between $\bm{X}$ and $\bm{\tilde{\bm{X}}}$. The metric Q of an image is defined by
	\begin{equation}
	\text{Q}(\bm{X})=s_1\frac{s_1-s_2}{s_1+s_2},
	\end{equation}
	where $s_1$ and $s_2$ are two singular values of the gradient matrix of the image $\bm{X}$. The AG is computed by
	\begin{equation}
	\text{AG}(\bm{X})=\frac{1}{(m-1)(n-1)}\sum_{i=1}^{m-1}\sum_{j=1}^{n-1}\sqrt{(\Delta_1 x_{i,j}^2+\Delta_2 x_{i,j}^2)/2},
	\end{equation}
	where $\Delta_1 x_{i,j}$ and $\Delta_2 x_{i,j}$ are the first differences along both directions, respectively. The BIQA can be calculated according to \cite{Gabarda2007BIQA} and its code is available online\footnote{https://cn.mathworks.com/matlabcentral/fileexchange/30800-blind-image-quality-assessment-through-anisotropy}. In the real experiments, the performance is quantitatively evaluated by the metric Q, AG, and BIQA. In our experiments, the PSNR, SSIM, Q, AG, and BIQA values of a multispectral image are the average values of those for all bands. For all the five indicators, the larger the value, the better the results.


	Without any special instructions, the parameters are set as following: the number of time series $t=4$, patch size $w=4$ (patch size $w$ should be multiples of $t$ due to the rearrangement procedure), indicator thresholding value $\gamma_2$ is 0.91, radius of searching windows $r=100$ ($r/t>20$ is recommended to make the search region large enough), searching step is 2 (half of patch size $w$), penalty parameter $\beta = 1 \text{ or } 10$, and $\varepsilon=10^{-4} \text{, } 10^{-2}$ for the images whose value ranges are $[0, 255]$ and $[0, 1]$ respectively.  Three of the most advanced missing information reconstruction methods, HaLRTC \cite{Liu2013PAMItensor}, ALM-IPG \cite{wang2016removing}, and PM-MTGSR \cite{Li2016PM}, are compared with NL-LRTC.
		The parameters for these three compared methods are tuned to maximize the reconstruction PSNR value for each data set.
	All the experiments are performed under Windows 10 and Matlab Version 9.0.0.341360 (R2016a) running on a desktop with an Inter(R) Core(TM) i7-6700 CPU at 3.40 GHz and 16 GB of memory.

	\subsection{Simulated Experiments}
\begin{figure}[!ht]
		\centering
		\rotatebox{90}{\footnotesize \textbf{Image 1}}
		\includegraphics[width=0.110\textwidth]{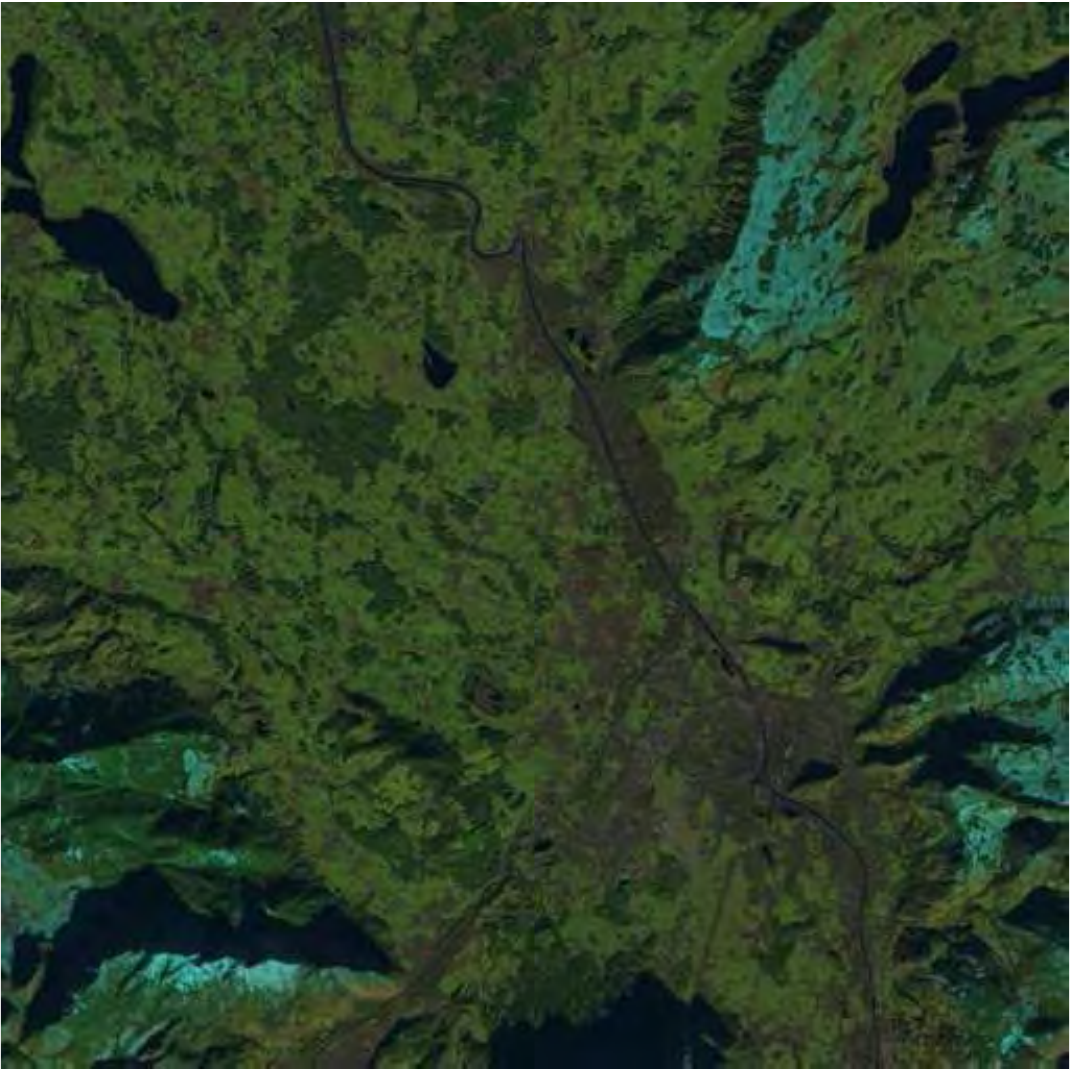}
		\includegraphics[width=0.110\textwidth]{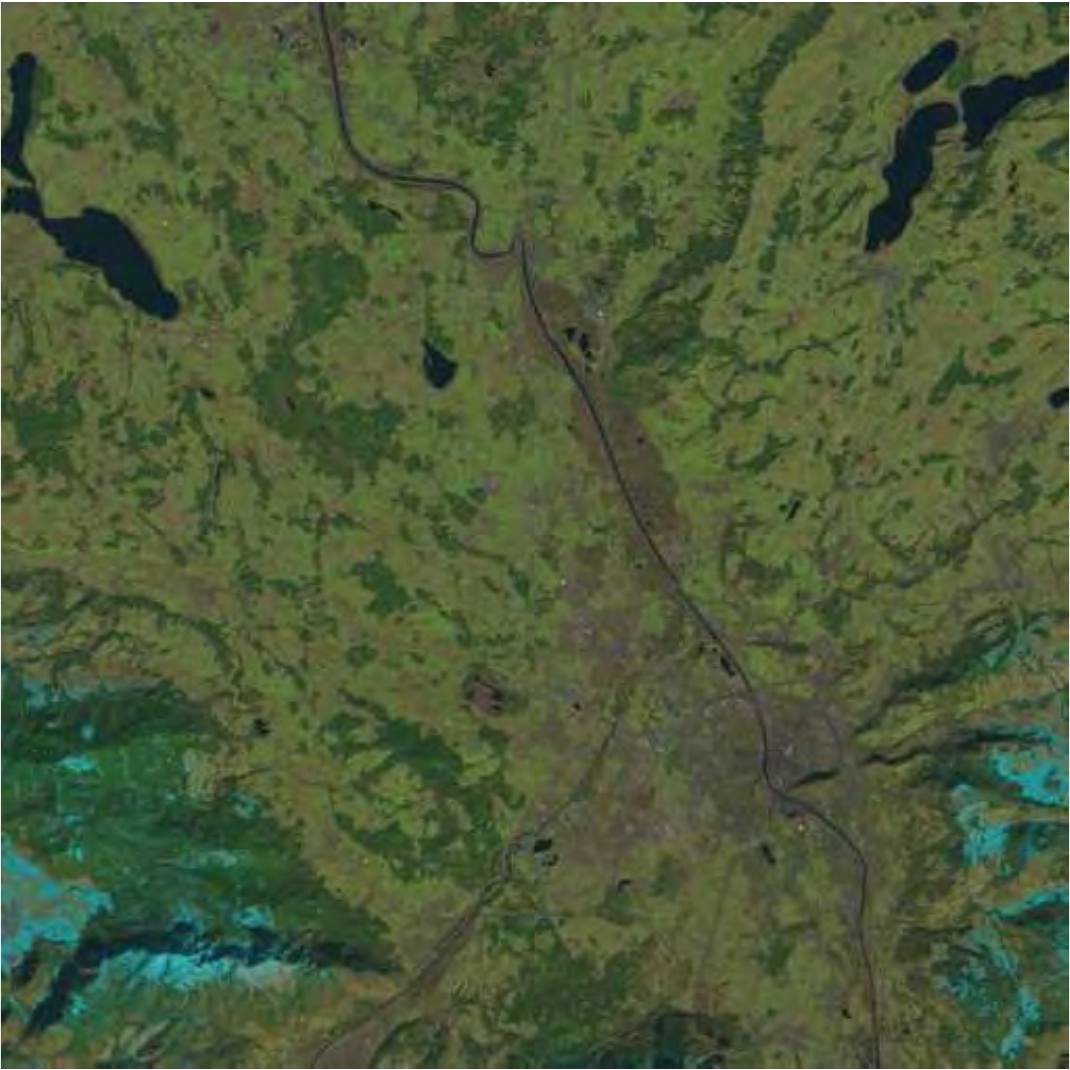}
		\includegraphics[width=0.110\textwidth]{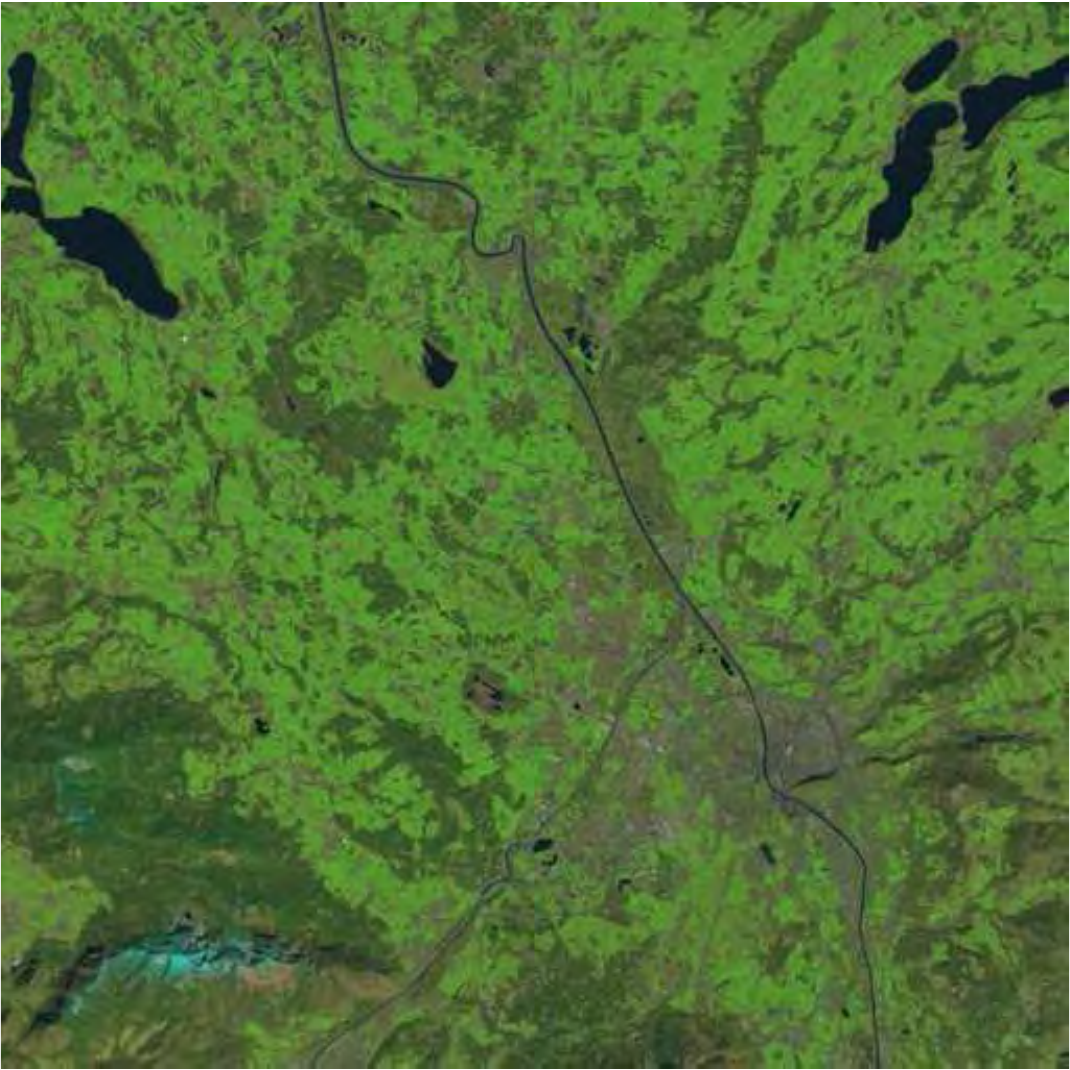}
		\includegraphics[width=0.110\textwidth]{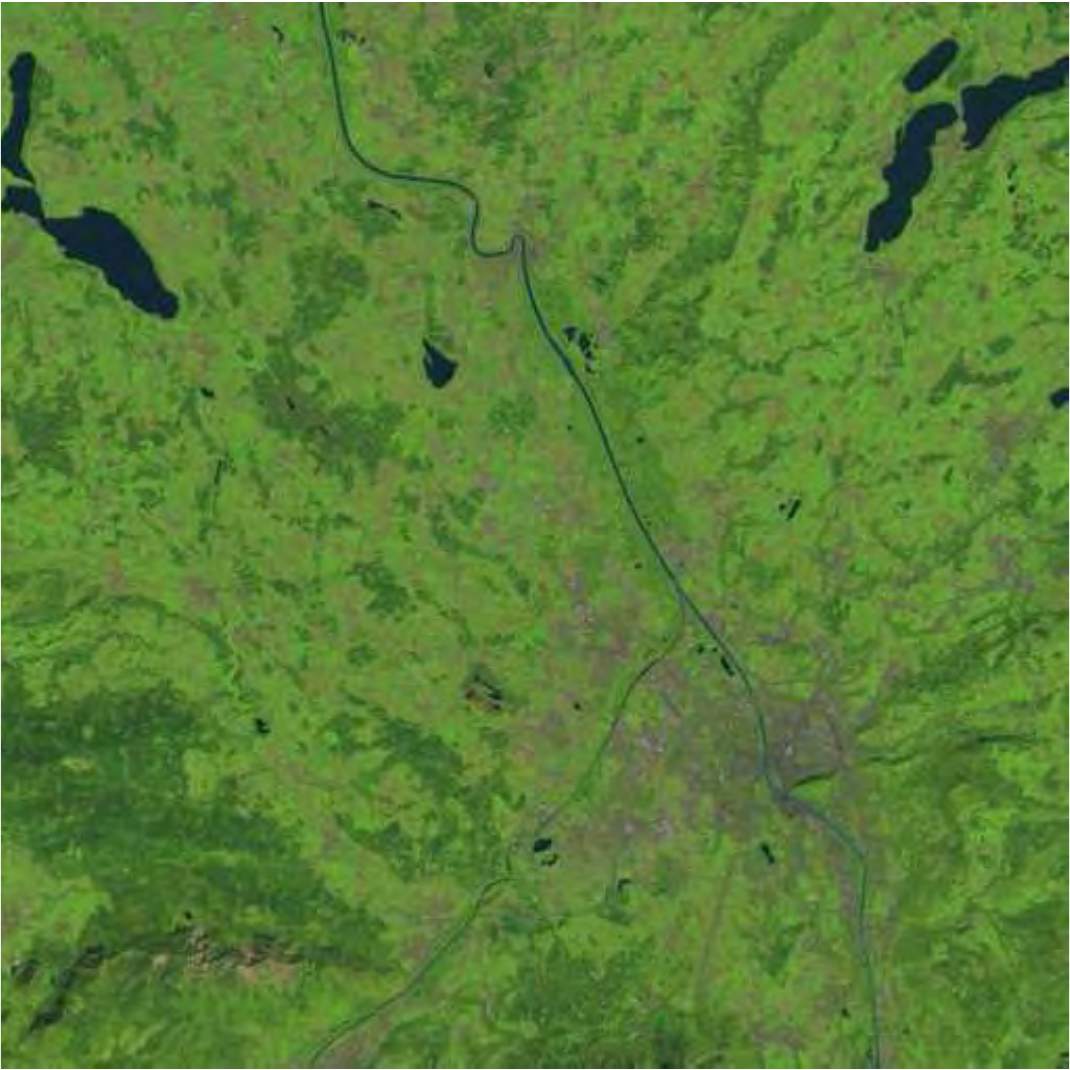}	\\ [0.05cm]
		\rotatebox{90}{\footnotesize \textbf{Image 2}}
		\includegraphics[width=0.110\textwidth]{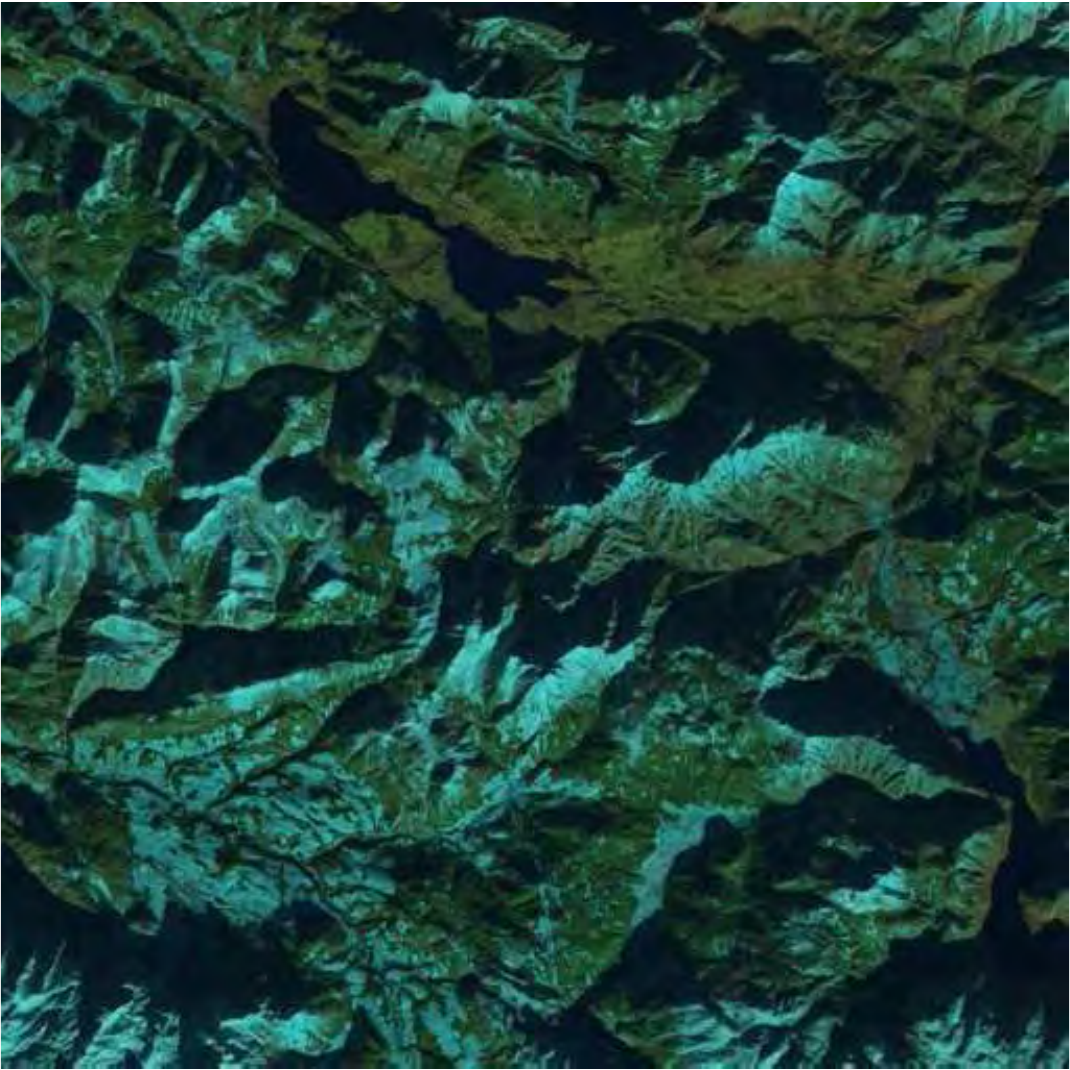}
		\includegraphics[width=0.110\textwidth]{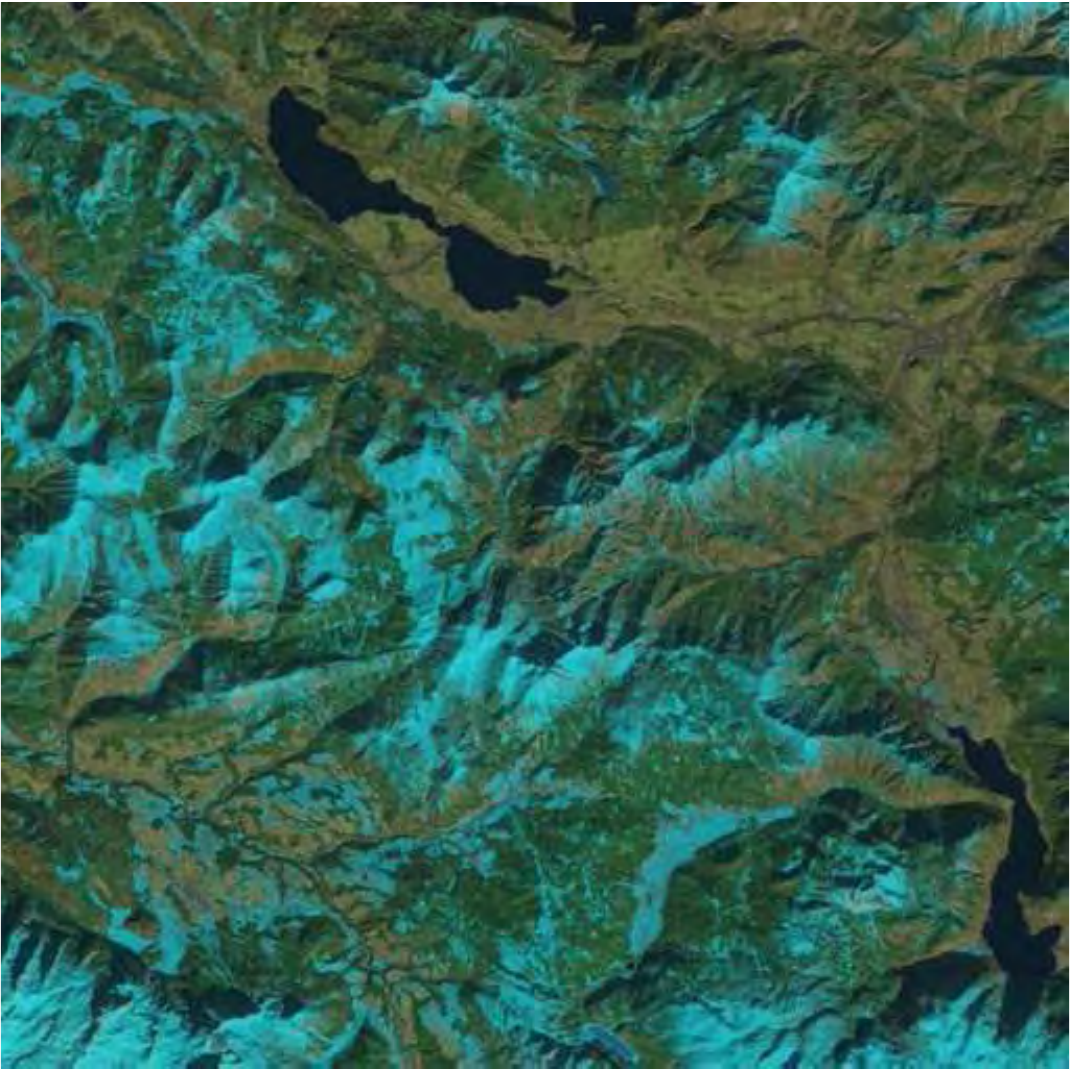}
		\includegraphics[width=0.110\textwidth]{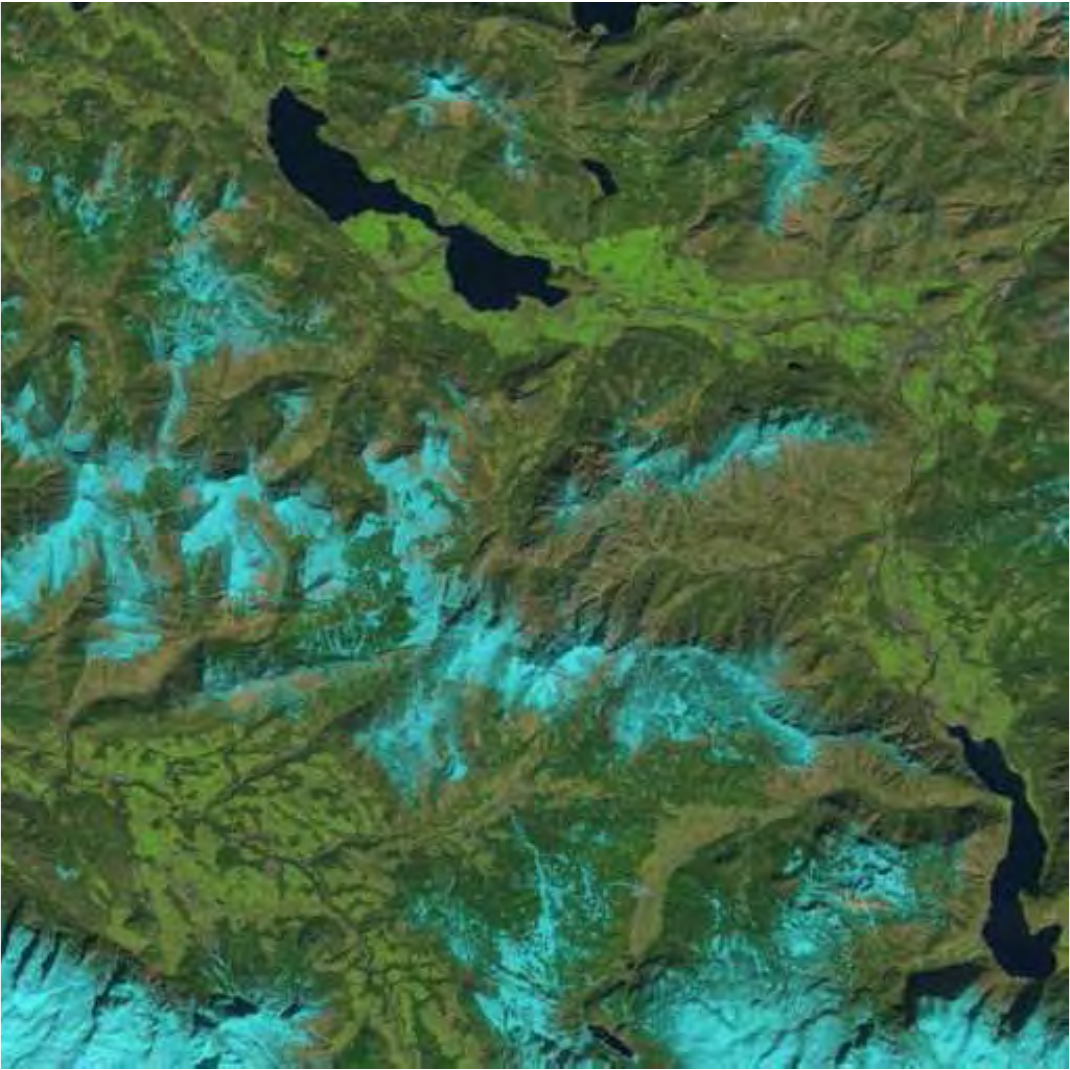}
		\includegraphics[width=0.110\textwidth]{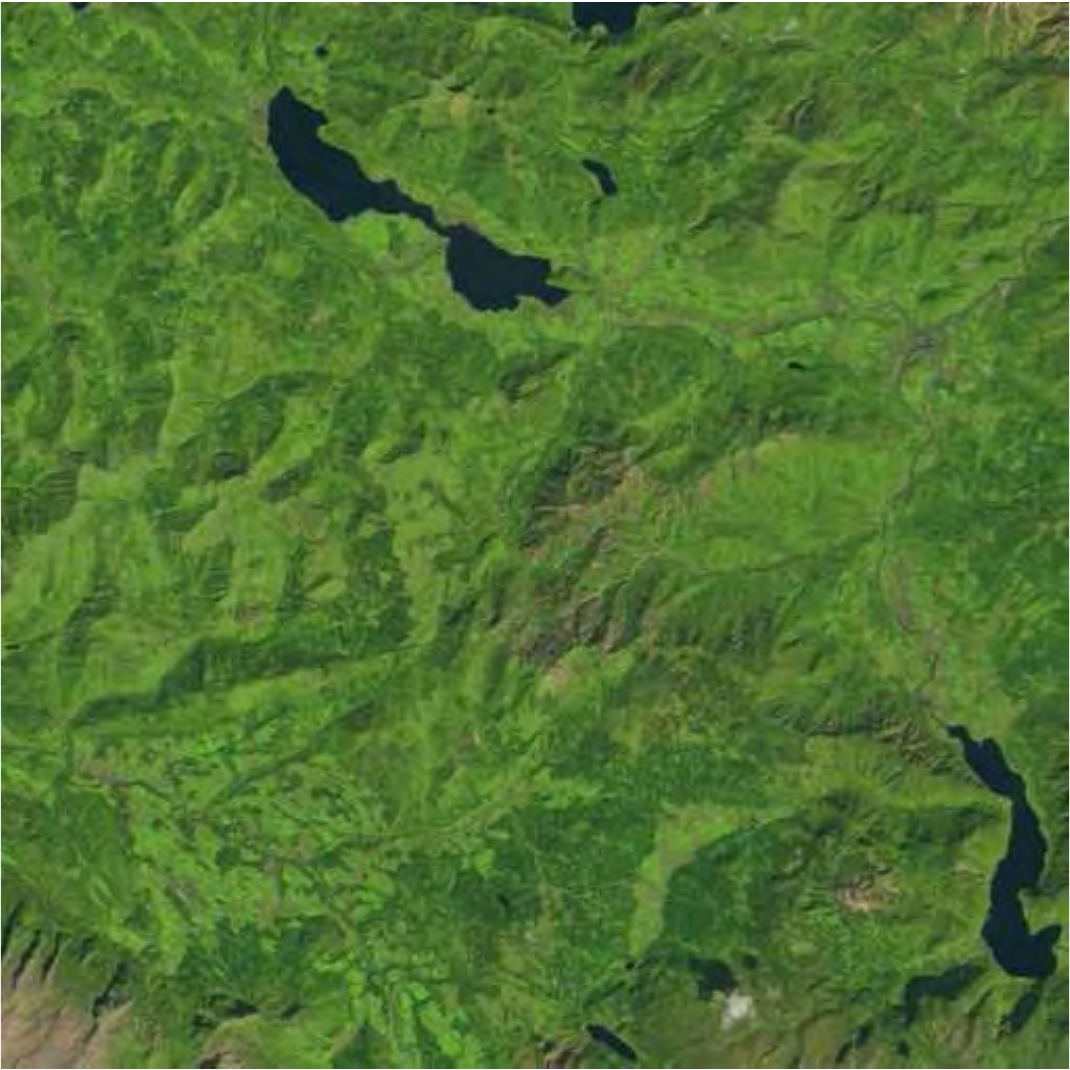}\\ [0.05cm]
		\rotatebox{90}{\footnotesize \textbf{Image 3}}
		\includegraphics[width=0.110\textwidth]{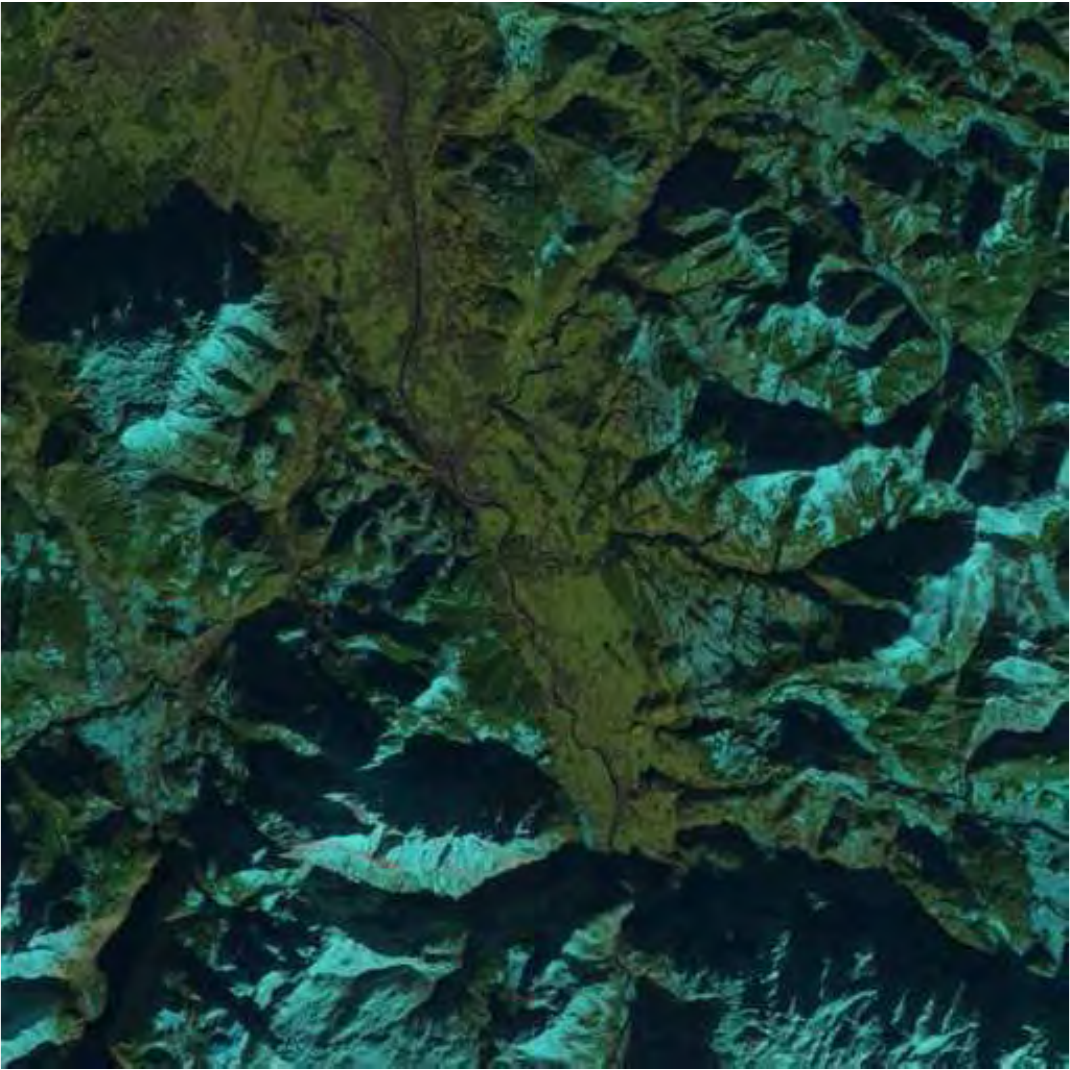}
		\includegraphics[width=0.110\textwidth]{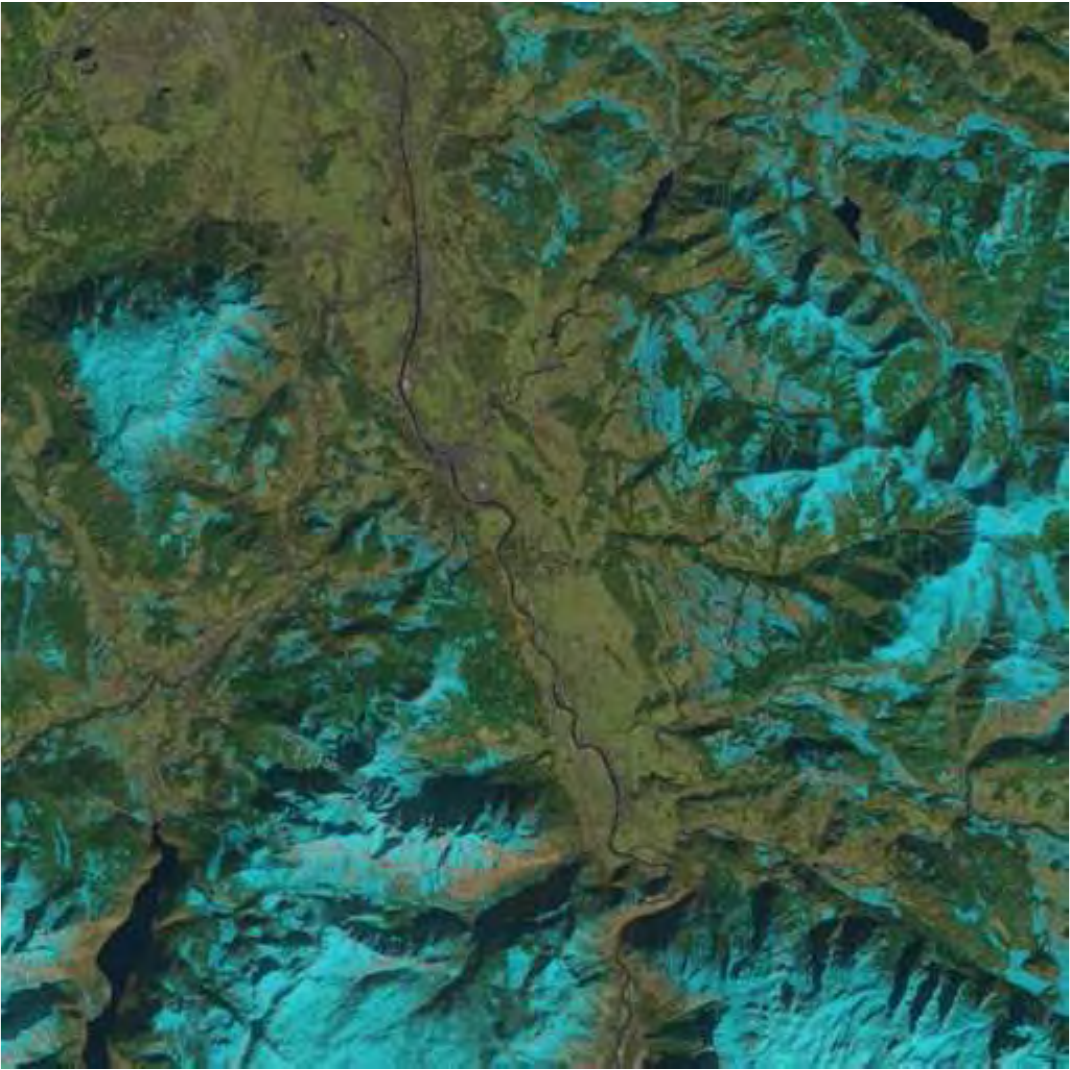}
		\includegraphics[width=0.110\textwidth]{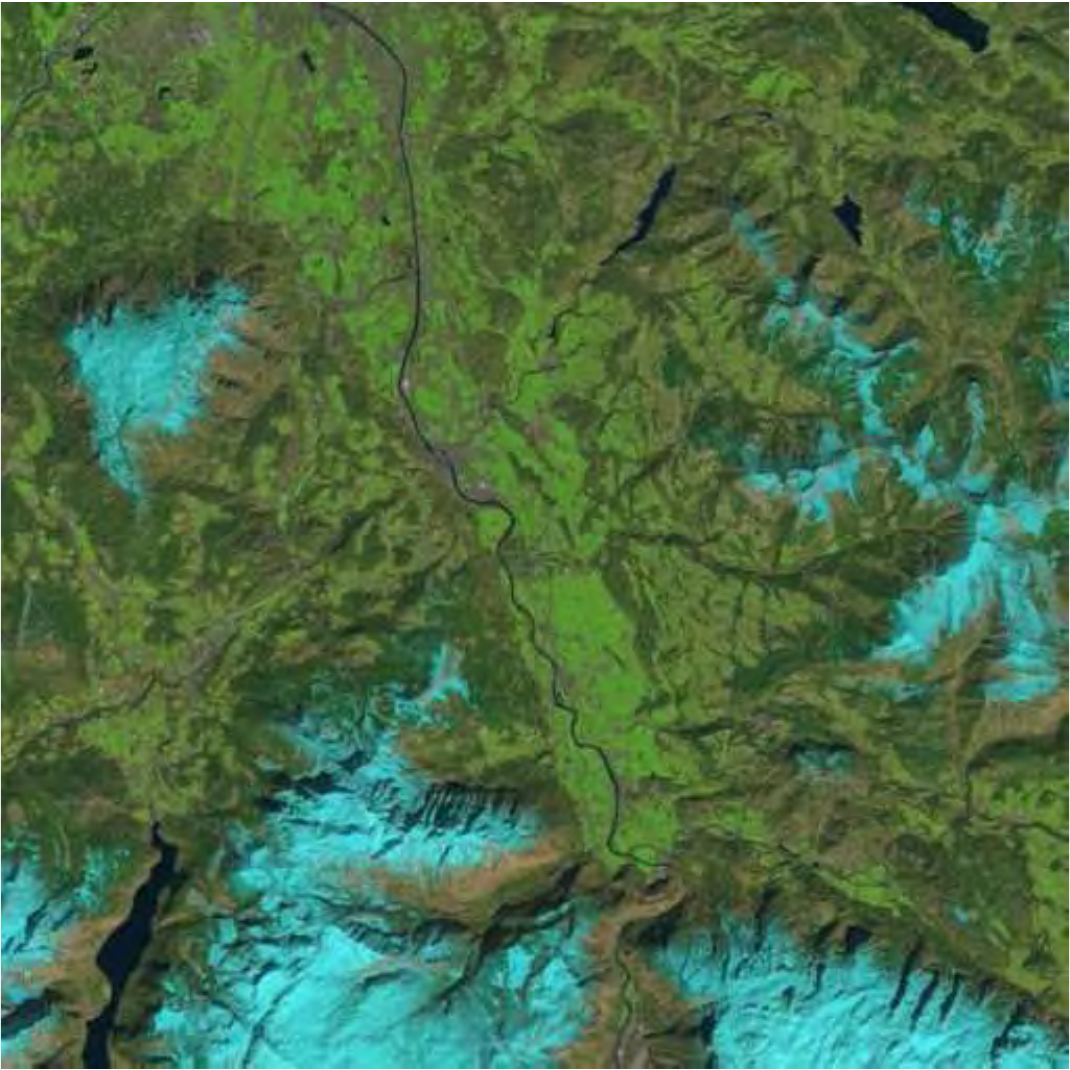}
		\includegraphics[width=0.110\textwidth]{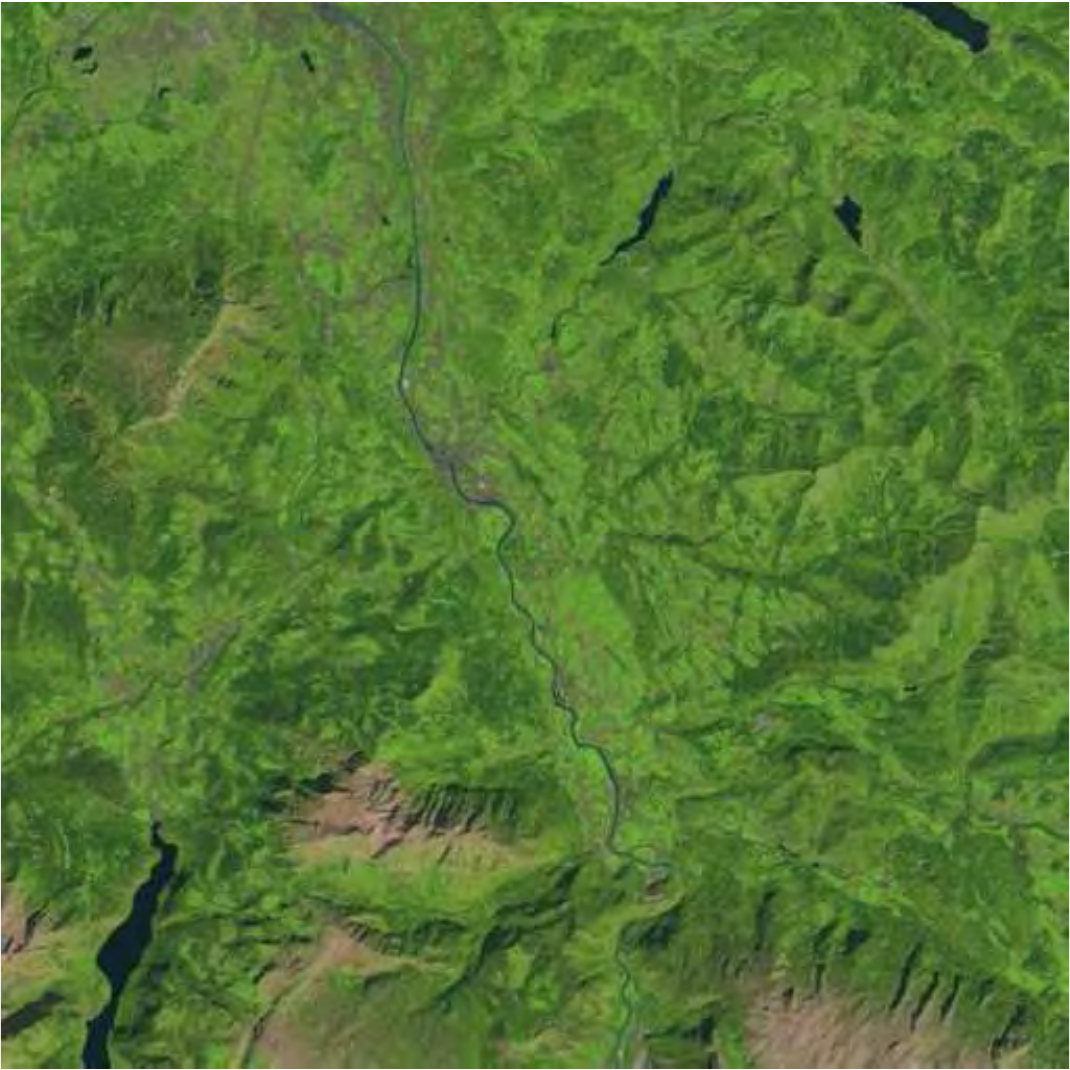}\\ [0.05cm]
		\rotatebox{90}{\footnotesize \textbf{Image 4}}
		\includegraphics[width=0.110\textwidth]{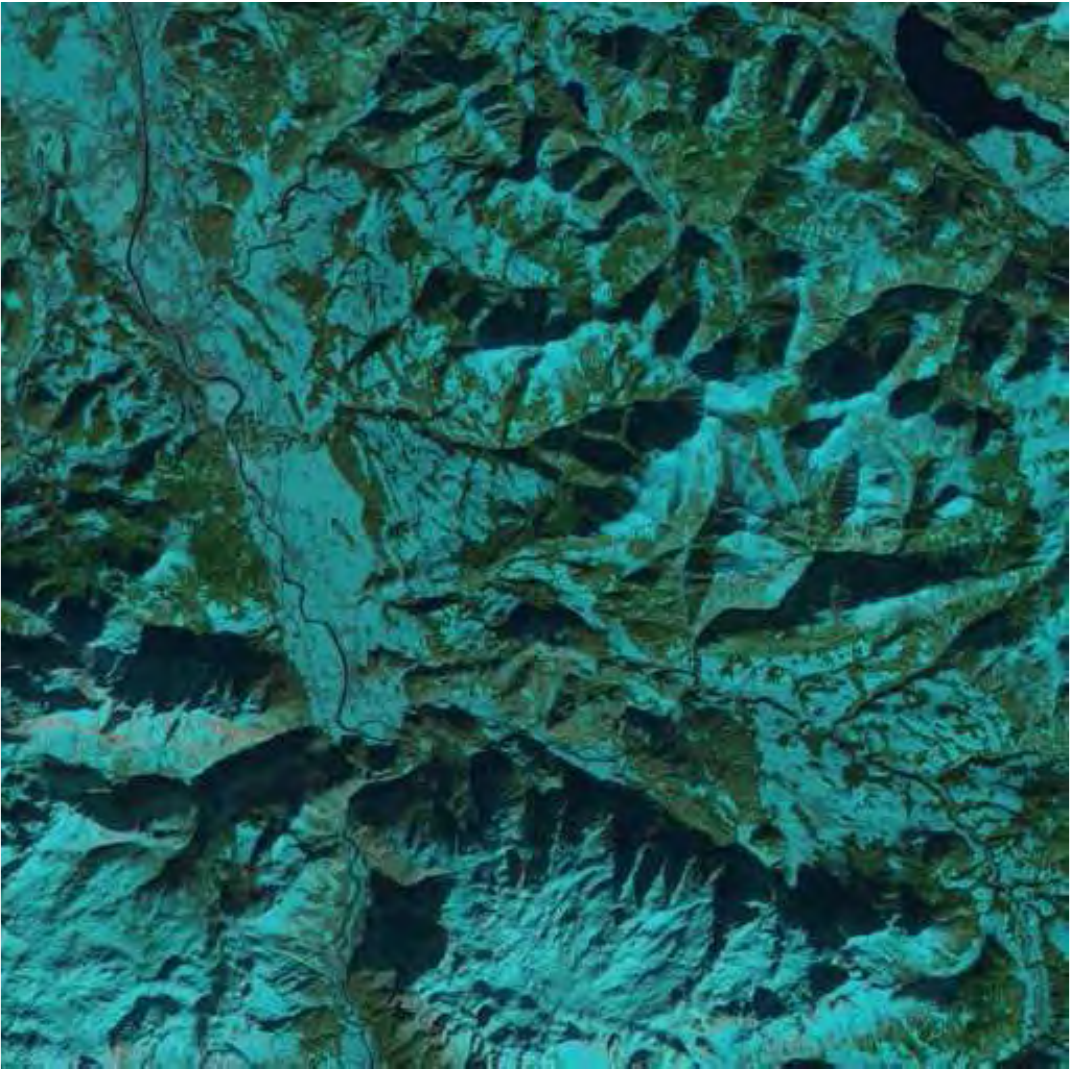}
		\includegraphics[width=0.110\textwidth]{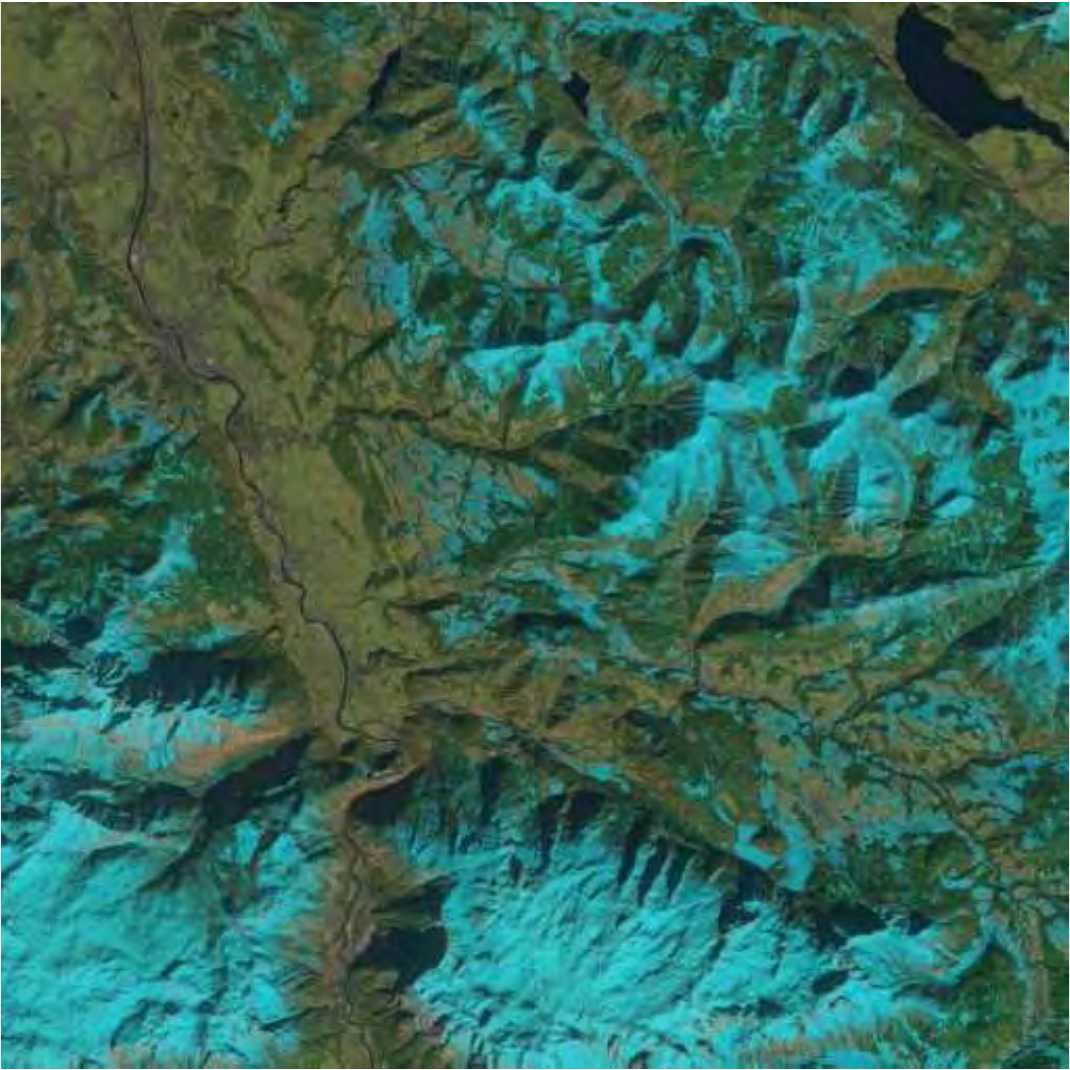}
		\includegraphics[width=0.110\textwidth]{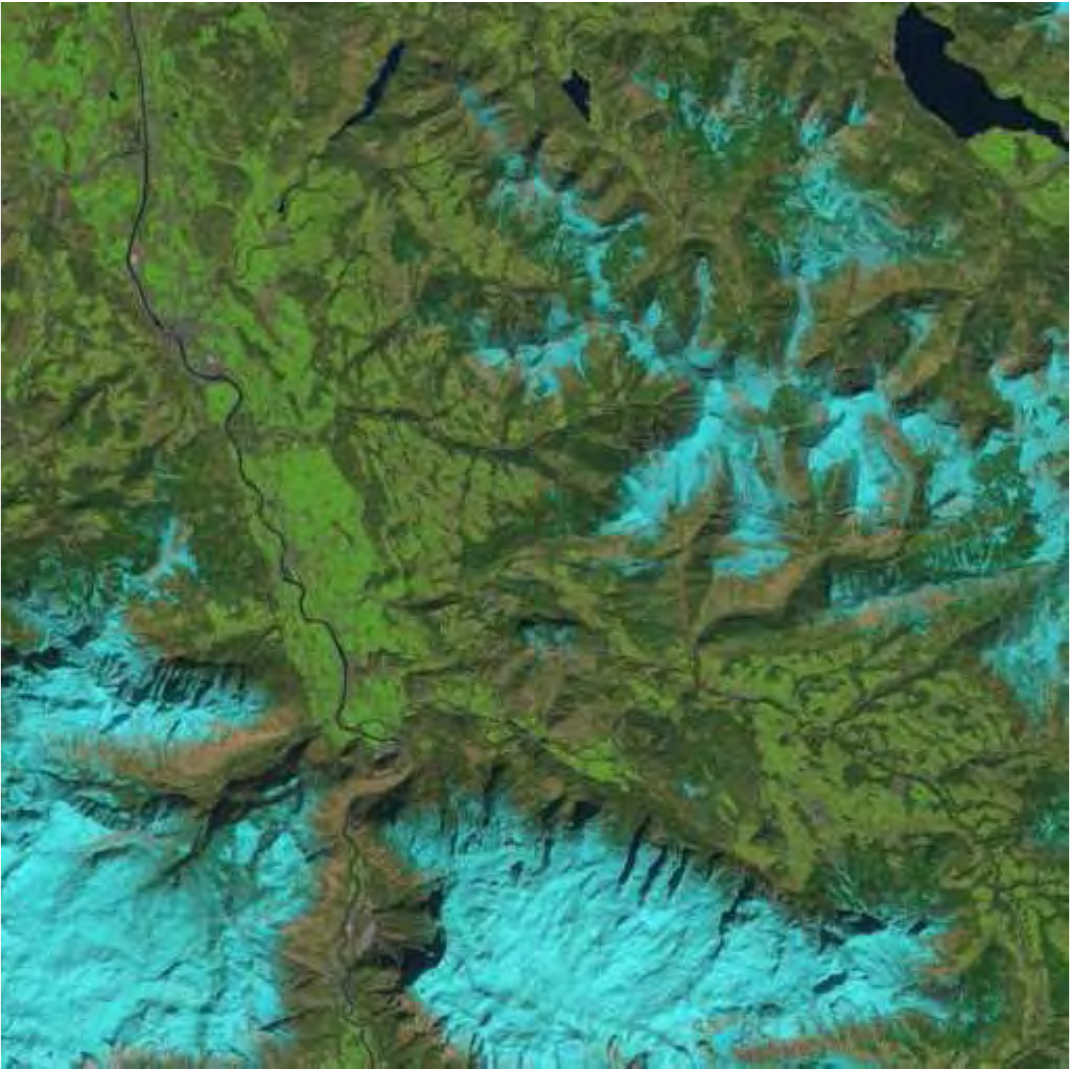}
		\includegraphics[width=0.110\textwidth]{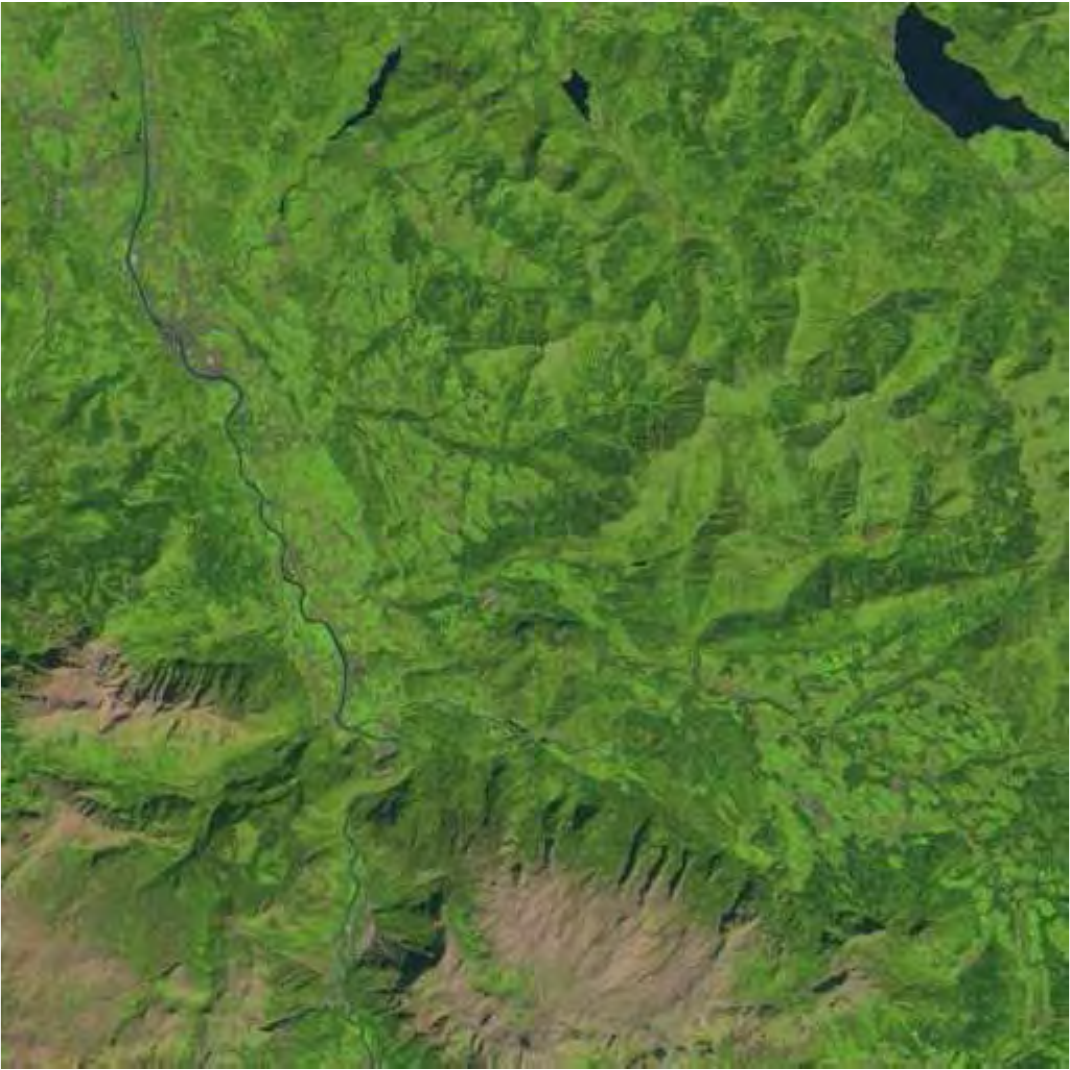}
		\caption{\footnotesize Simulated images cut from Fig. \ref{Fig:MUdata}. From top to bottom, data set denoted as ``Image 1'', ``Image 2'', ``Image 3'', and ``Image 4'', respectively. }
		\label{Fig:SimulateMUdata}
	\end{figure}

\begin{figure*}[!ht]
		\centering
		\rotatebox{90}{\footnotesize \textbf{Exp. 1}}
		\includegraphics[width=0.18\textwidth]{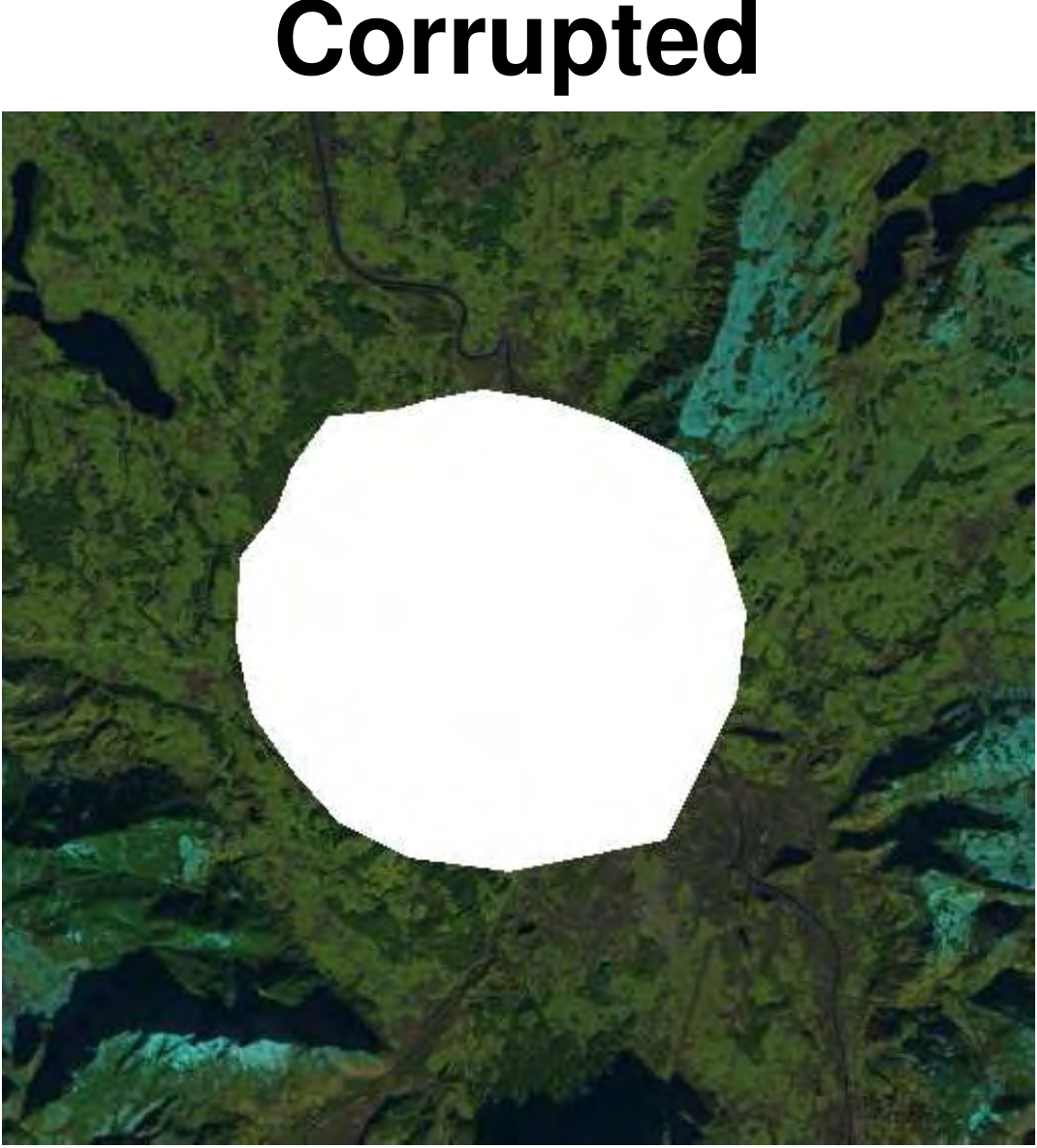}
		\includegraphics[width=0.18\textwidth]{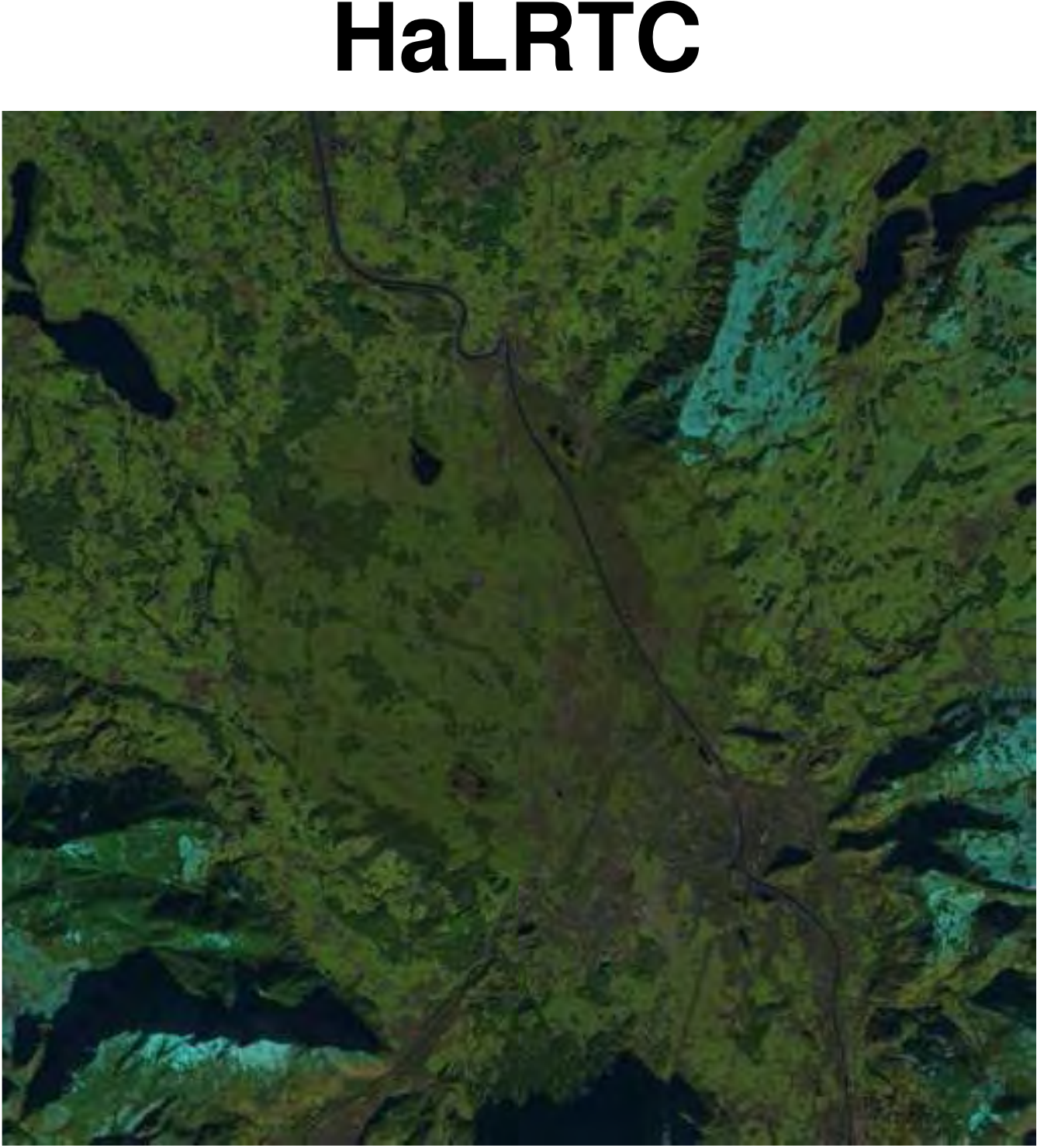}
		\includegraphics[width=0.18\textwidth]{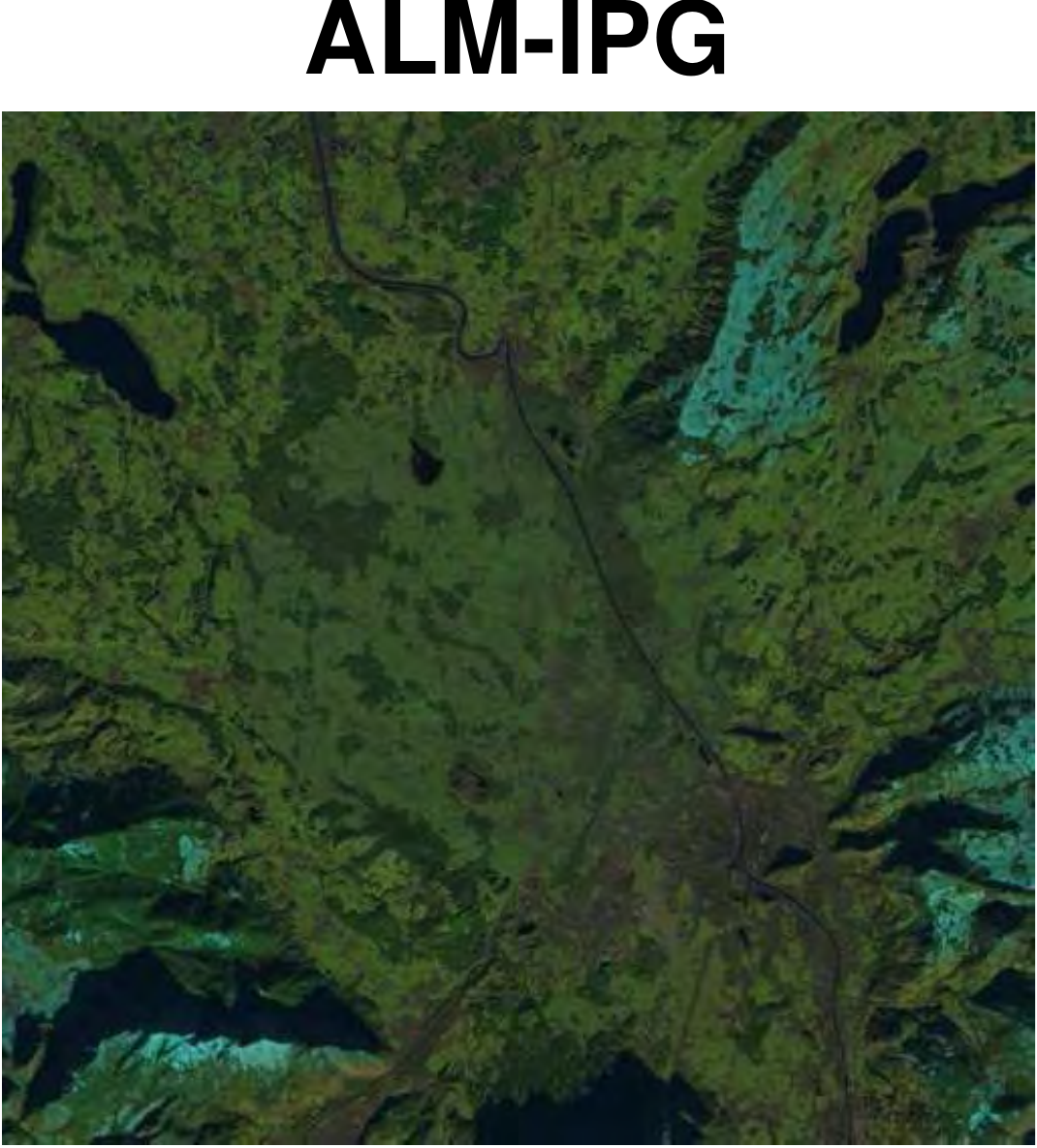}
		\includegraphics[width=0.18\textwidth]{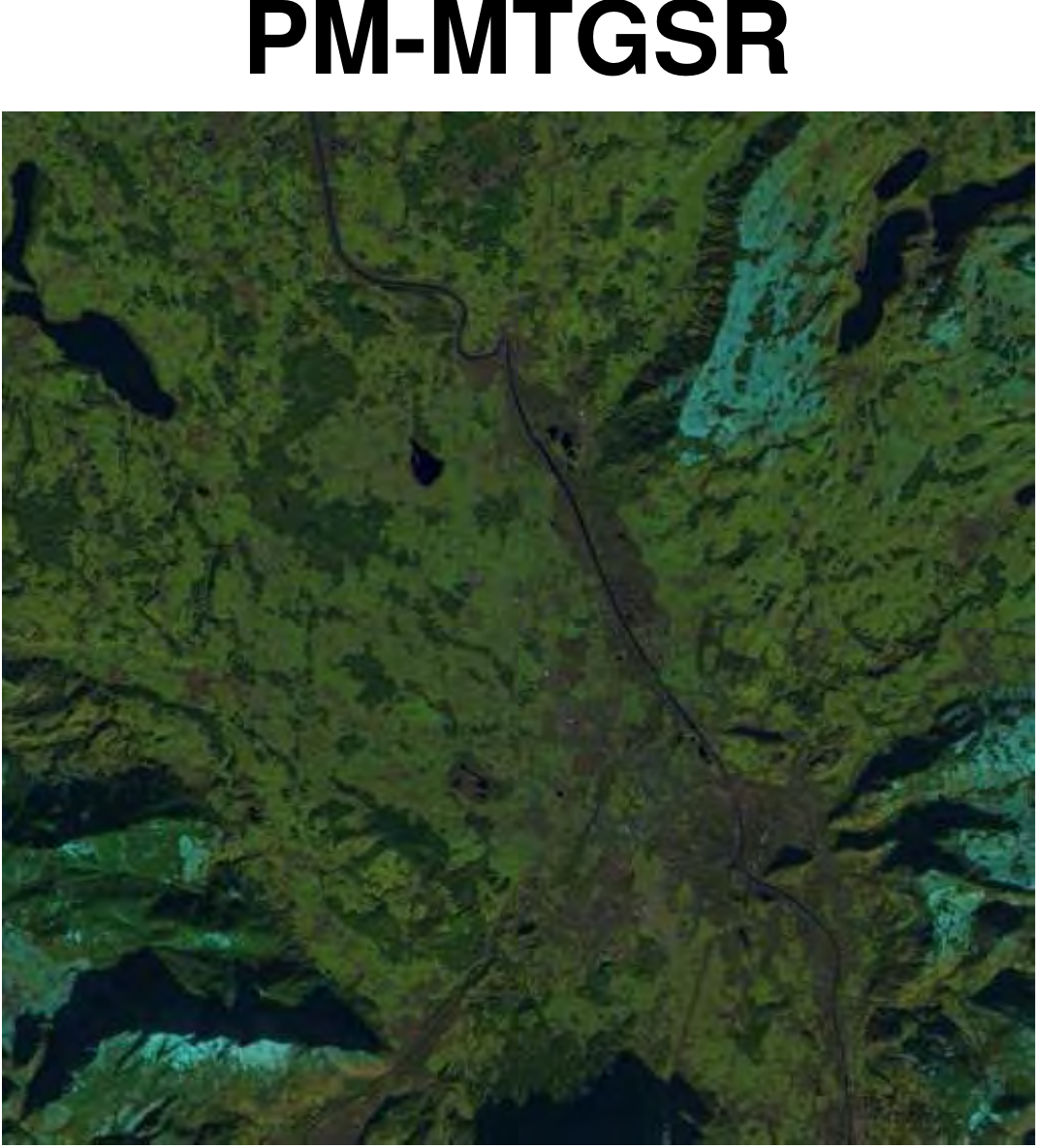}
		\includegraphics[width=0.18\textwidth]{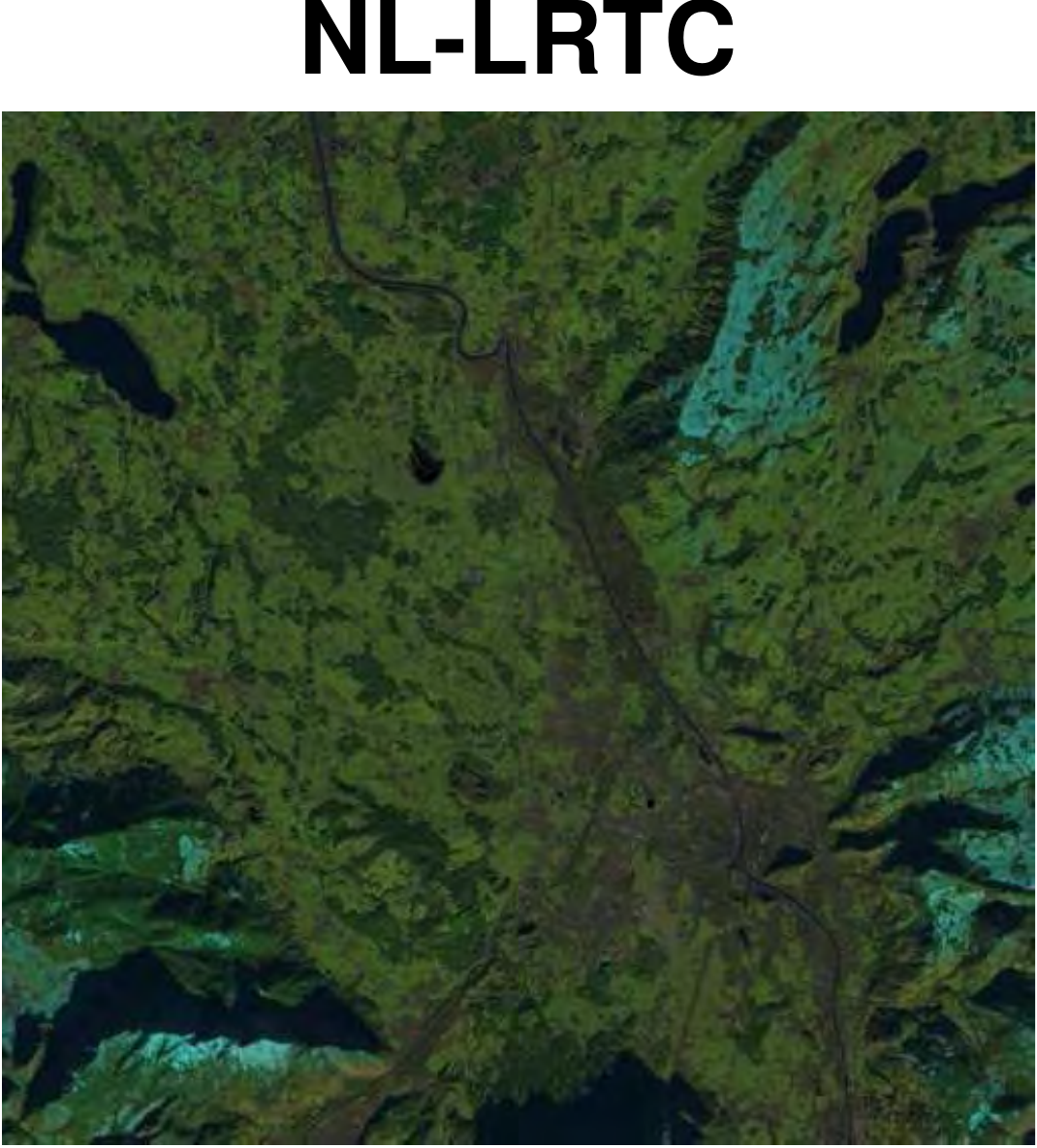} \\[0.05cm]
		\rotatebox{90}{\footnotesize \textbf{Exp. 2}}
		\includegraphics[width=0.18\textwidth]{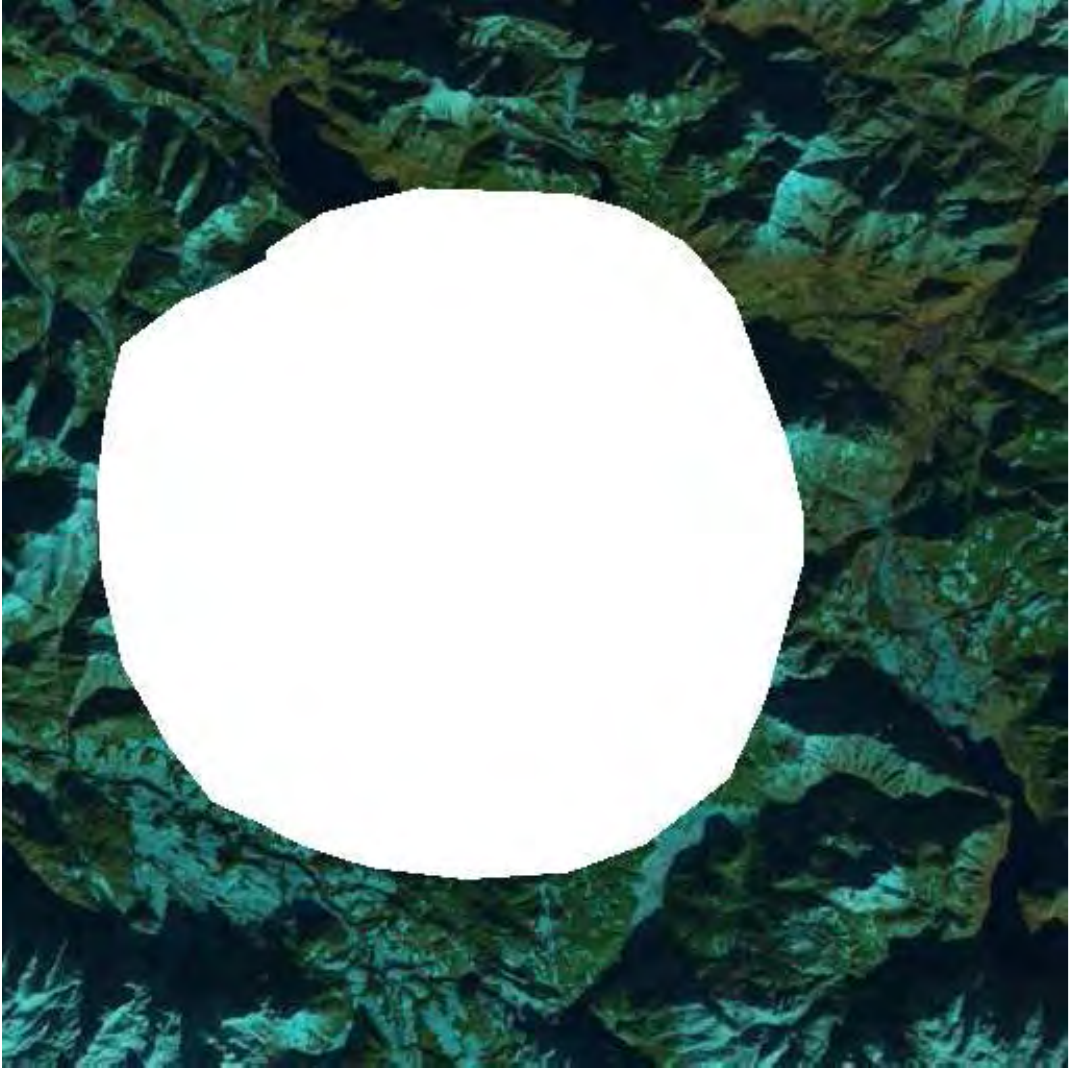}
		\includegraphics[width=0.18\textwidth]{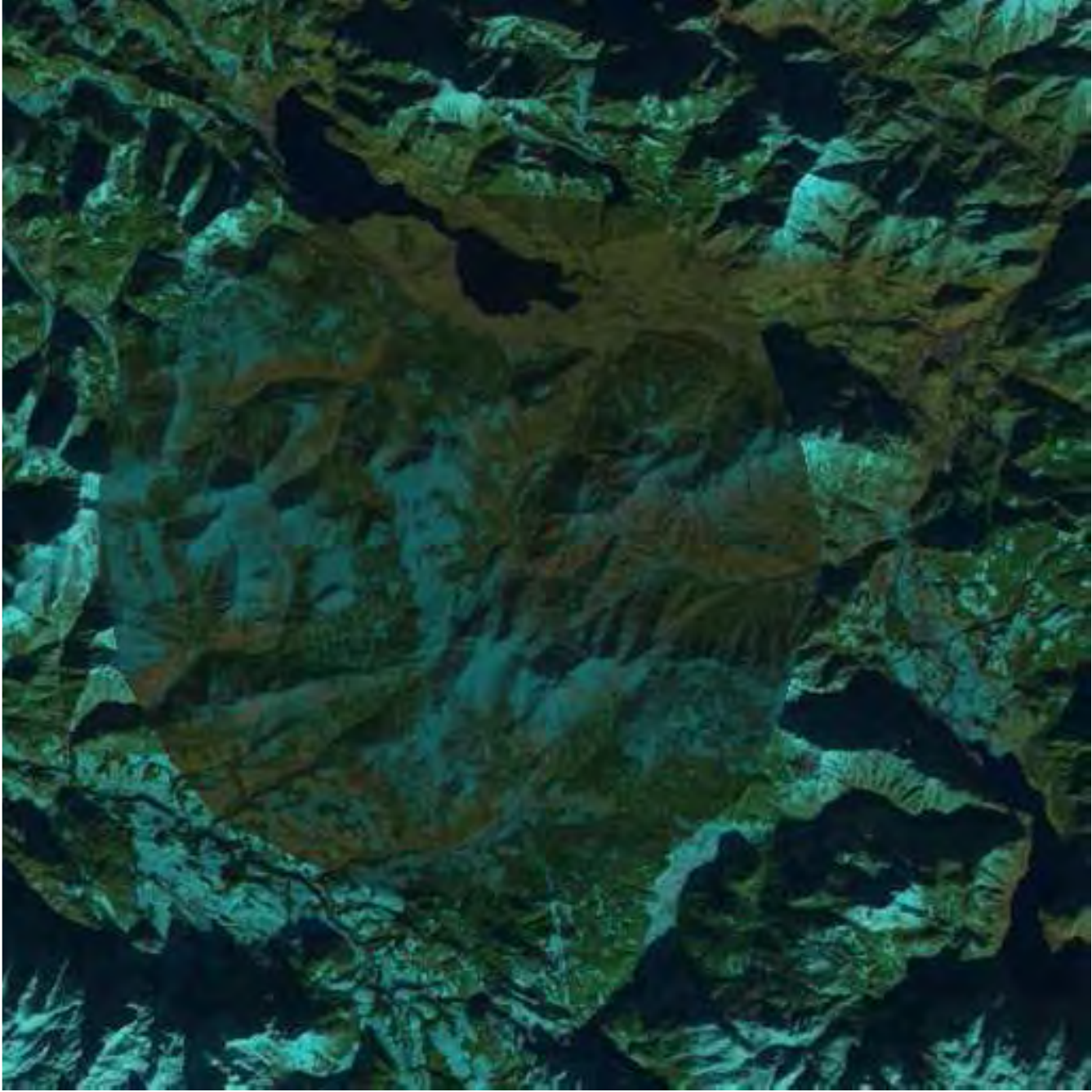}
		\includegraphics[width=0.18\textwidth]{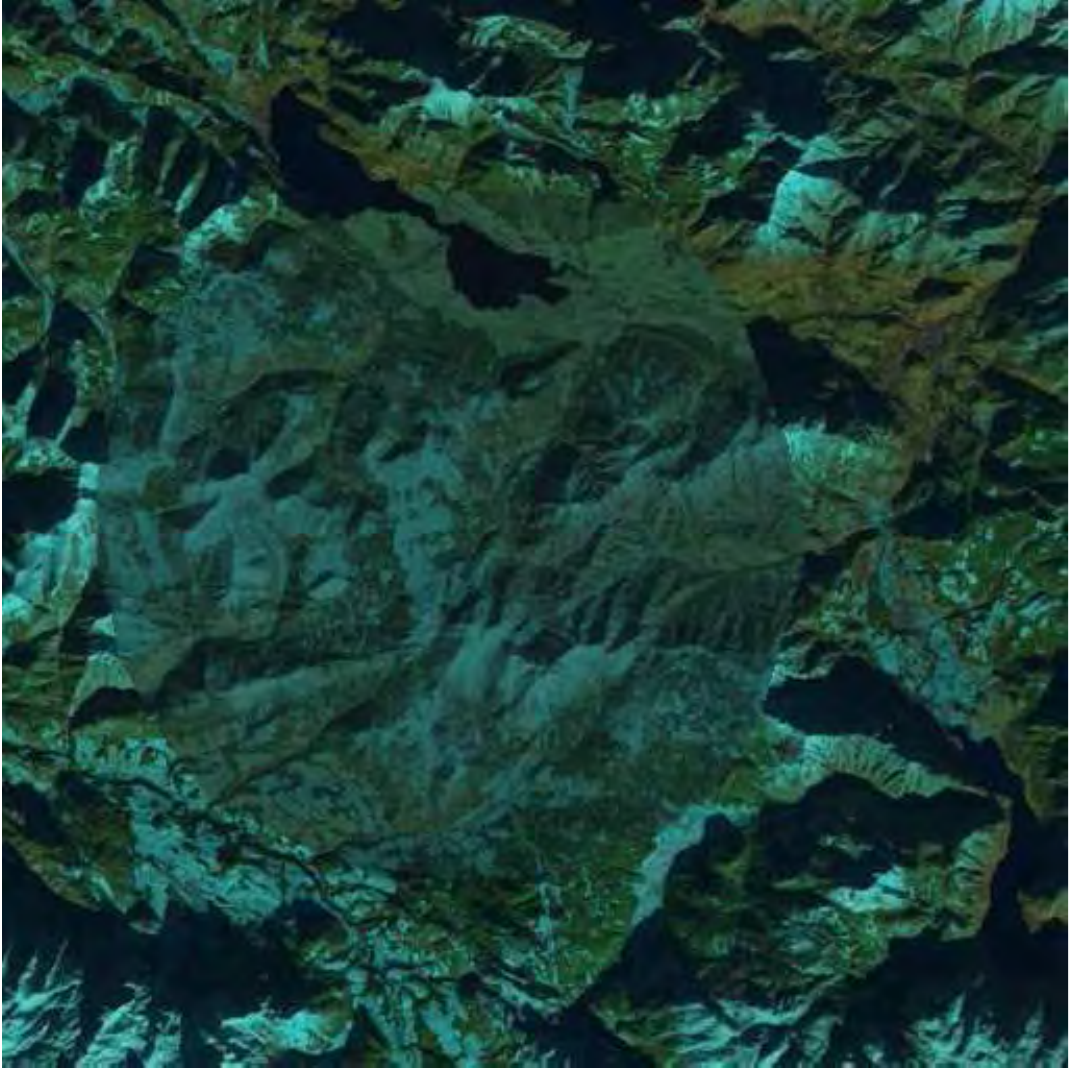}
		\includegraphics[width=0.18\textwidth]{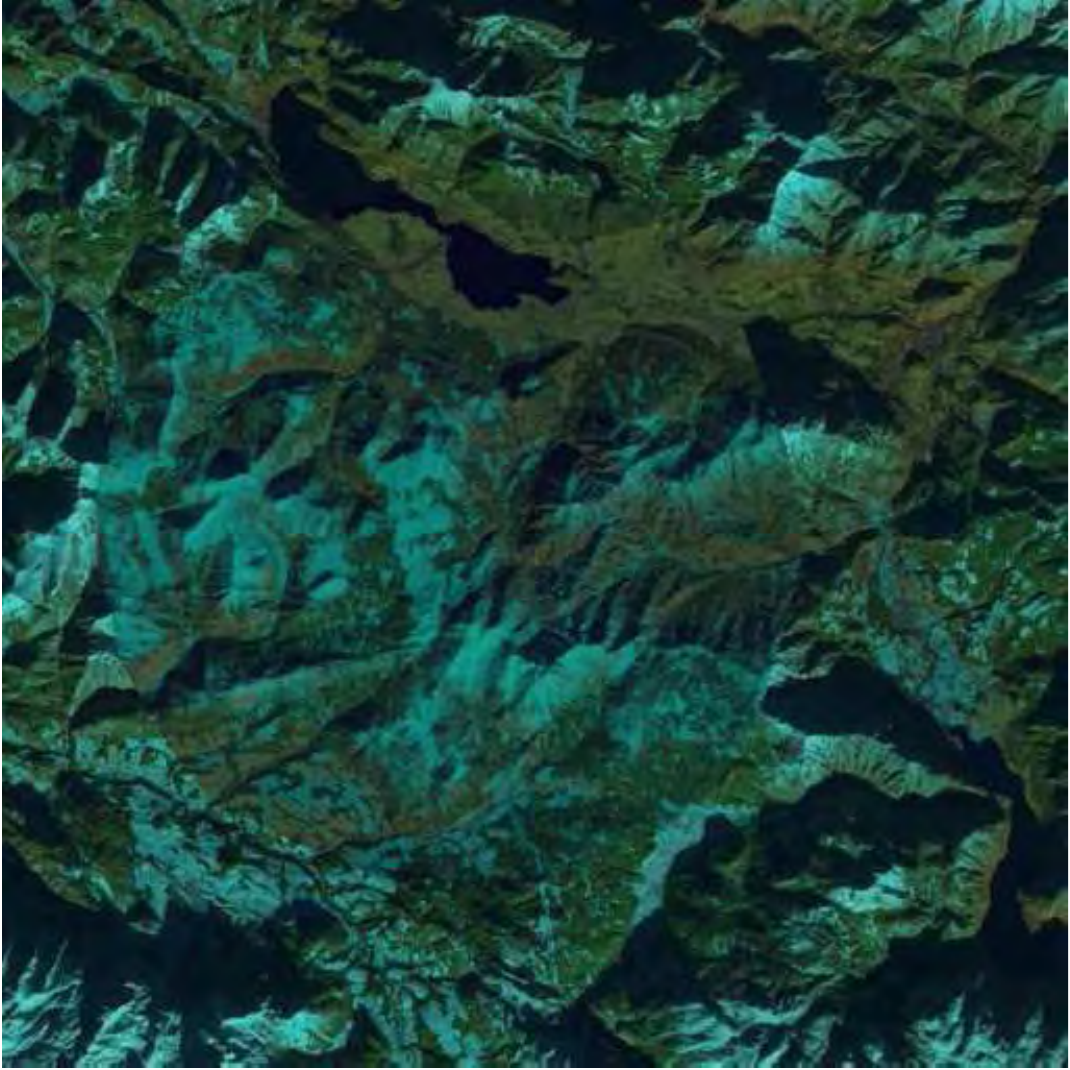}
		\includegraphics[width=0.18\textwidth]{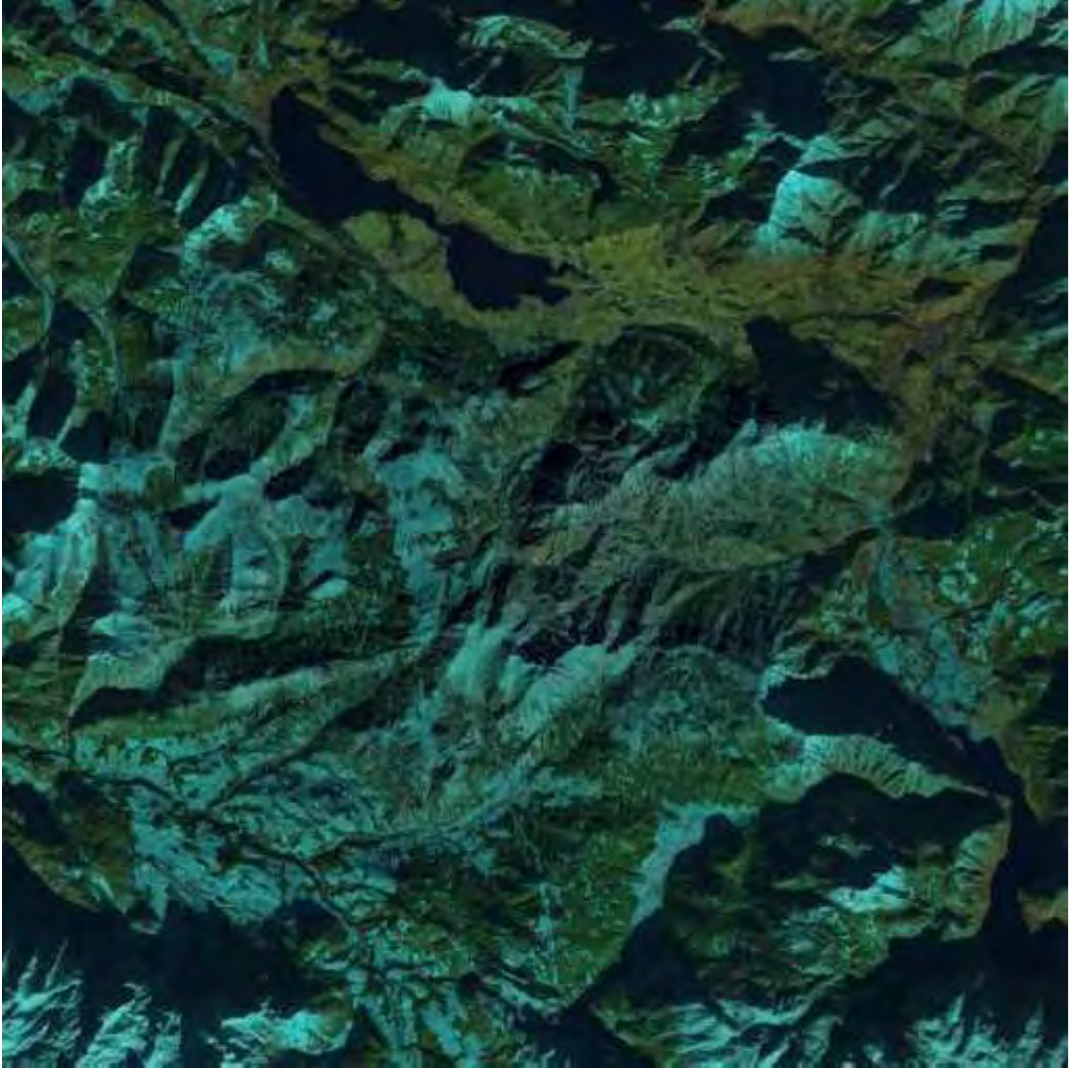} \\[0.05cm]
		\rotatebox{90}{\footnotesize \textbf{Exp. 3}}
		\includegraphics[width=0.18\textwidth]{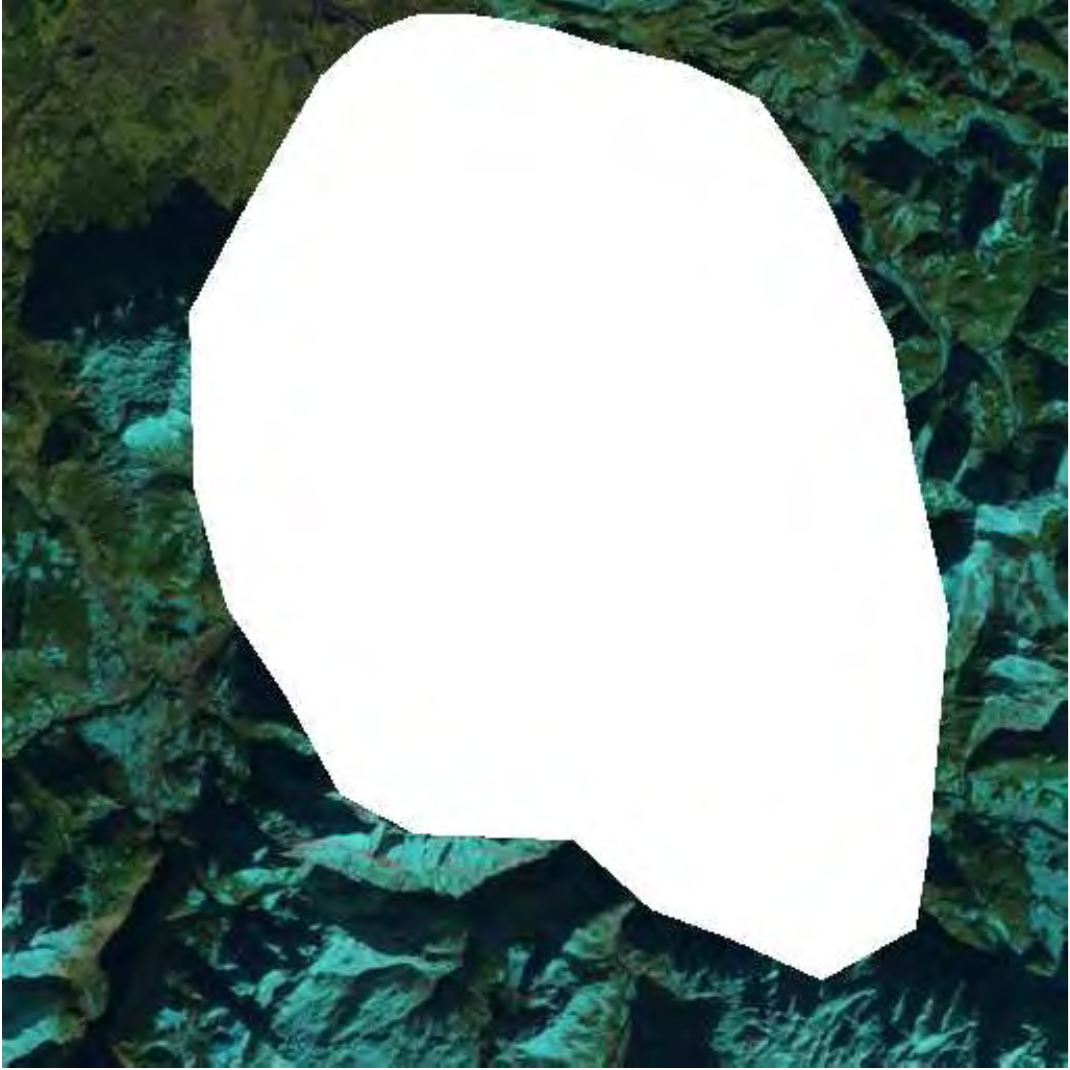}
		\includegraphics[width=0.18\textwidth]{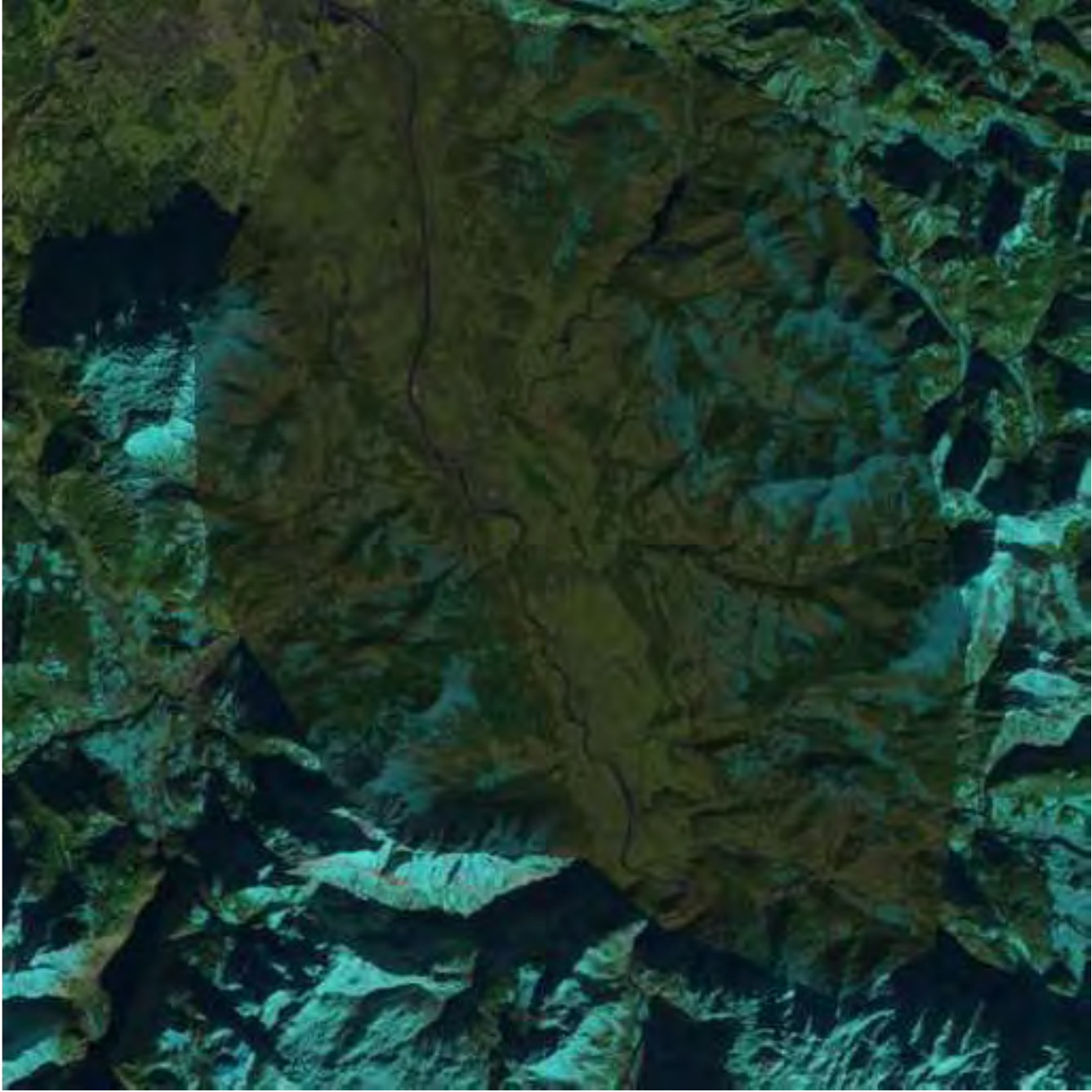}
		\includegraphics[width=0.18\textwidth]{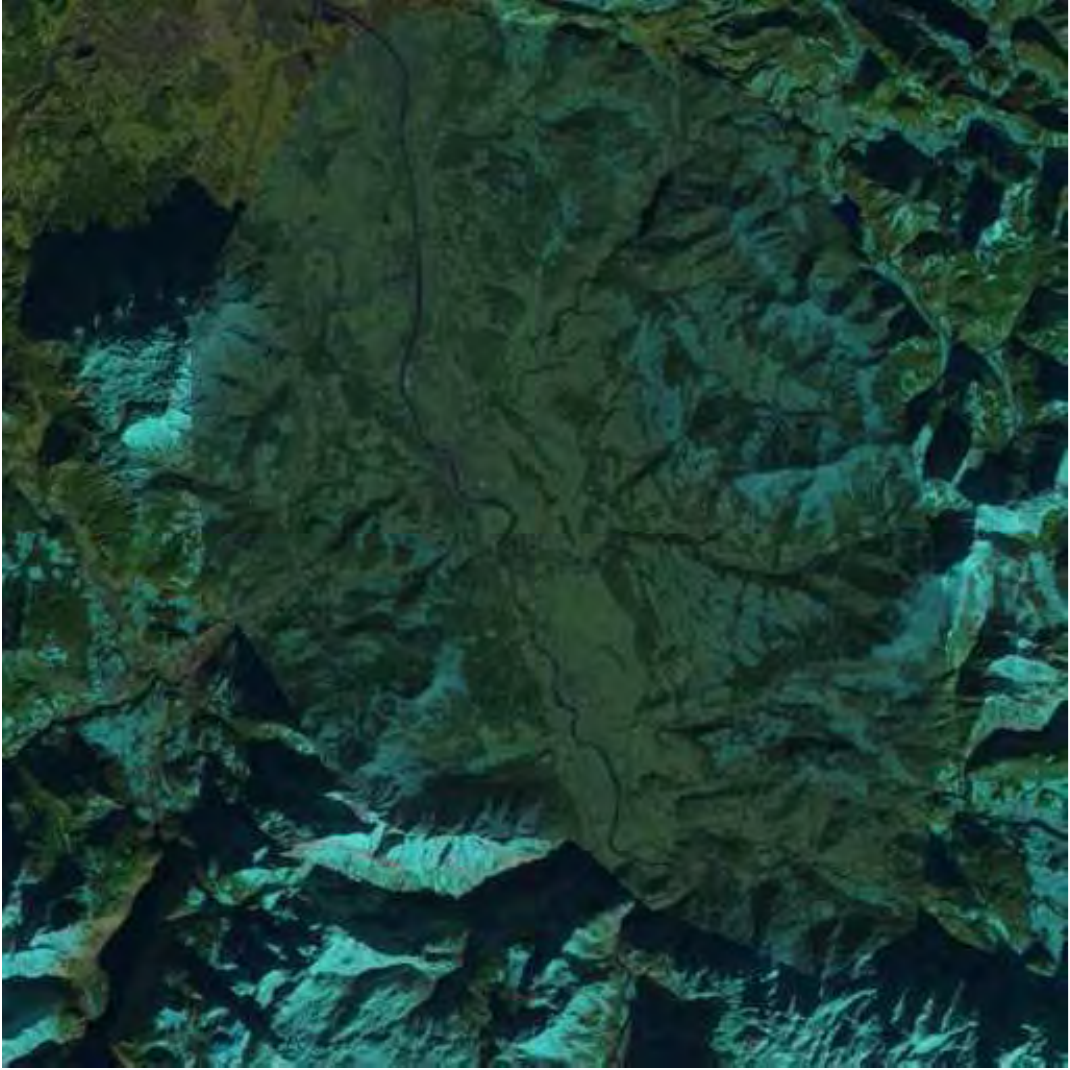}
		\includegraphics[width=0.18\textwidth]{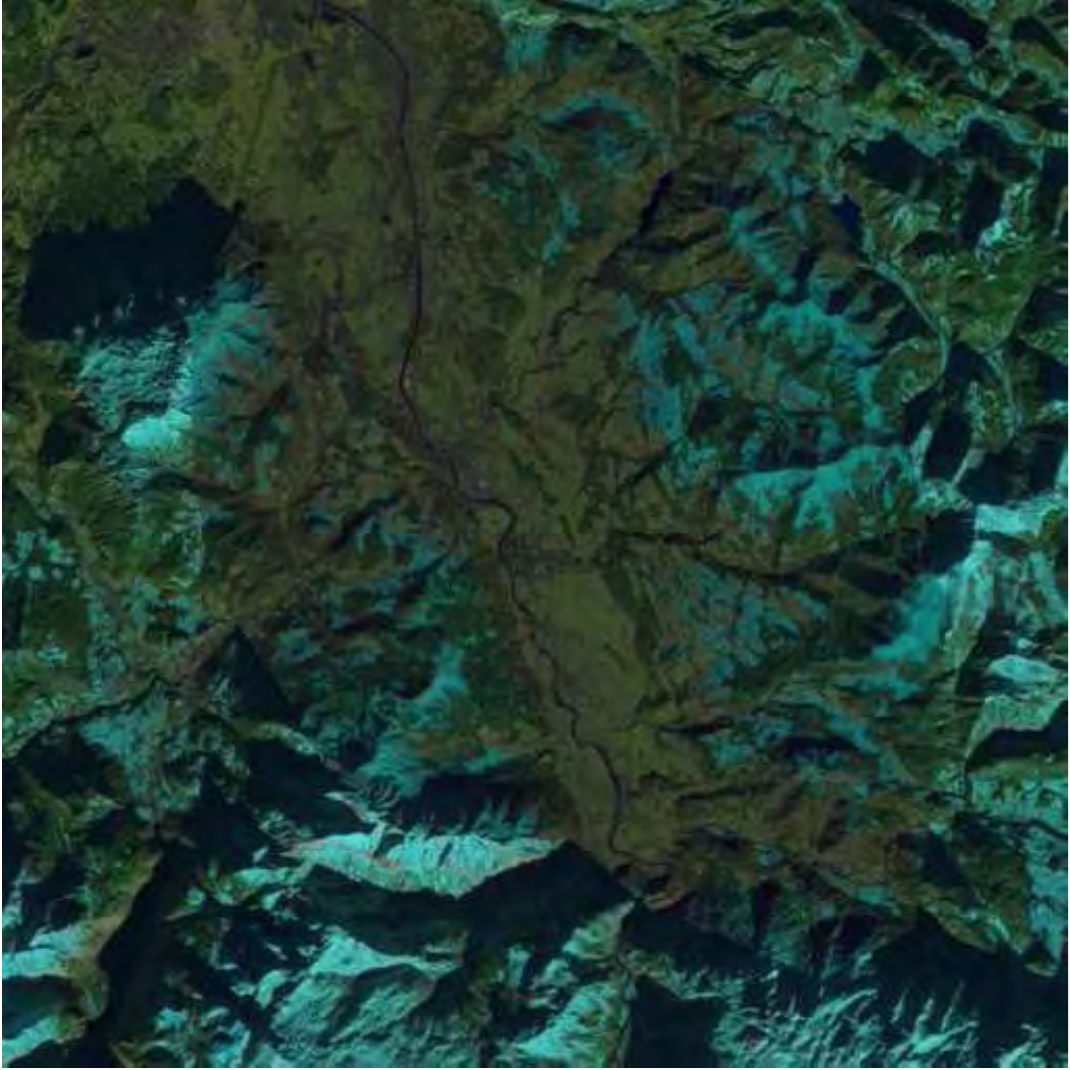}
		\includegraphics[width=0.18\textwidth]{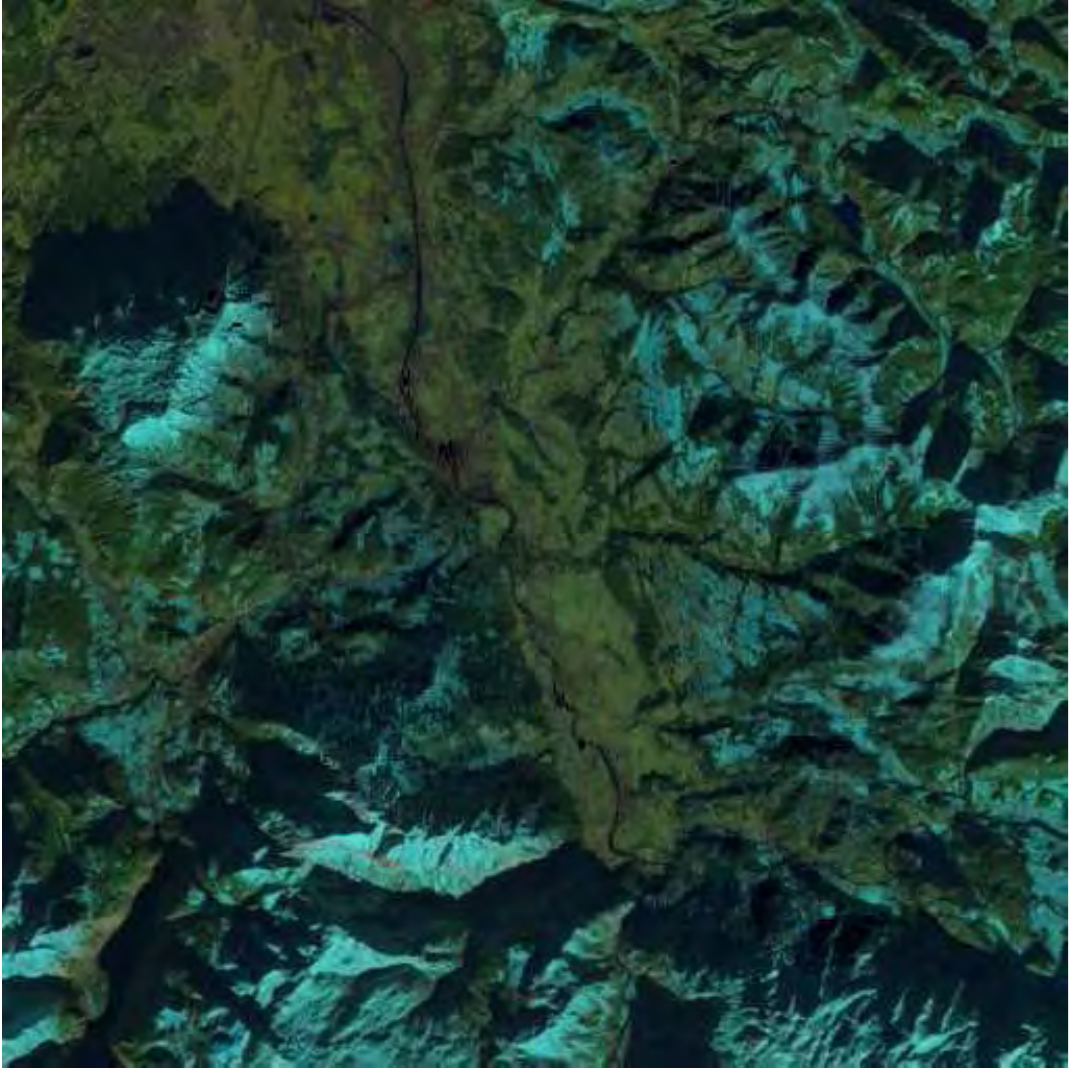}\\[0.05cm]
		\rotatebox{90}{\footnotesize \textbf{Exp. 4}}
		\includegraphics[width=0.18\textwidth]{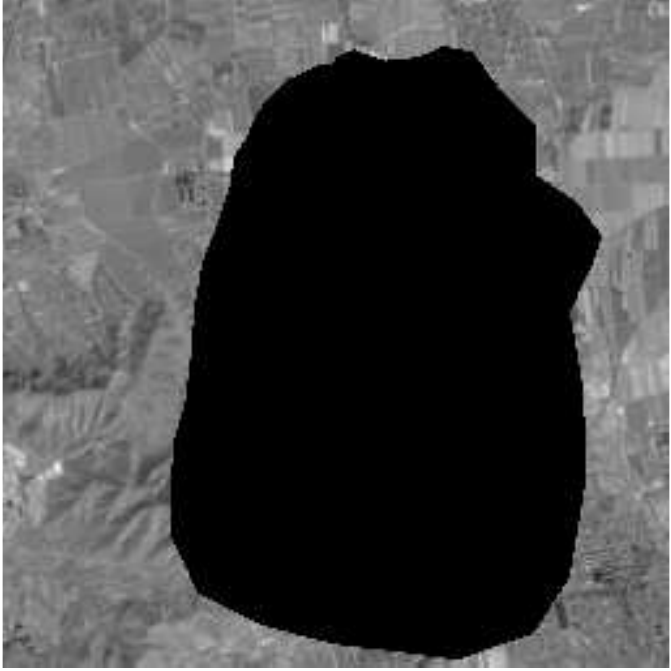}
		\includegraphics[width=0.18\textwidth]{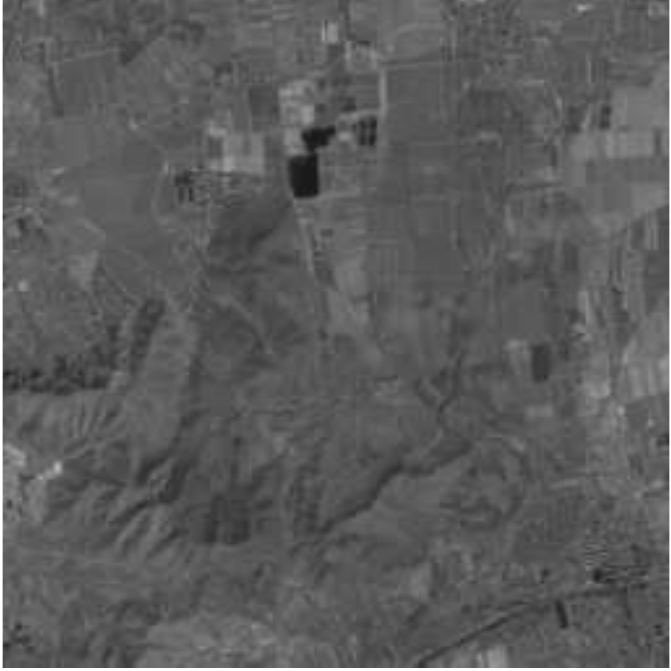}
		\includegraphics[width=0.18\textwidth]{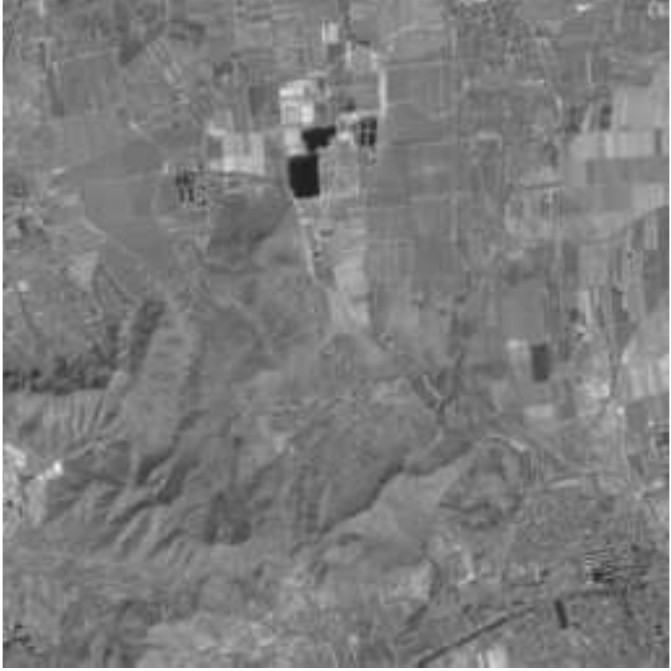}
		\includegraphics[width=0.18\textwidth]{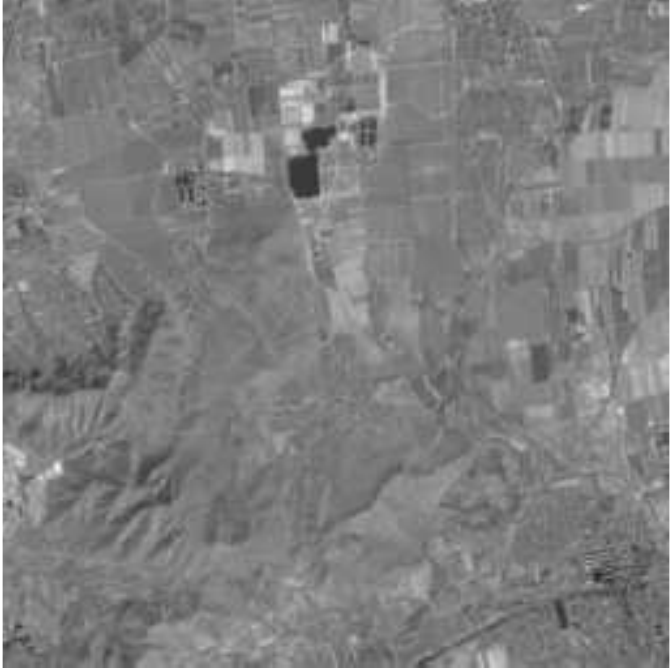}
		\includegraphics[width=0.18\textwidth]{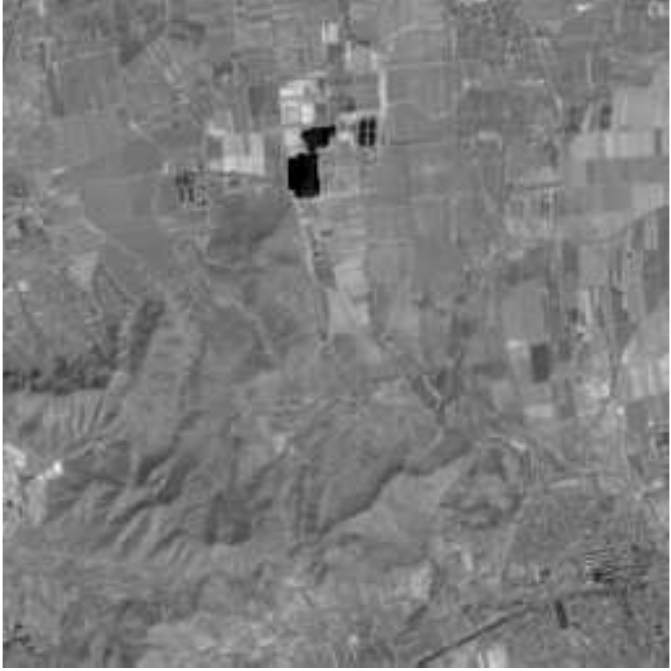}\\[0.05cm]
		\rotatebox{90}{\footnotesize \textbf{Exp. 5}}
		\includegraphics[width=0.18\textwidth]{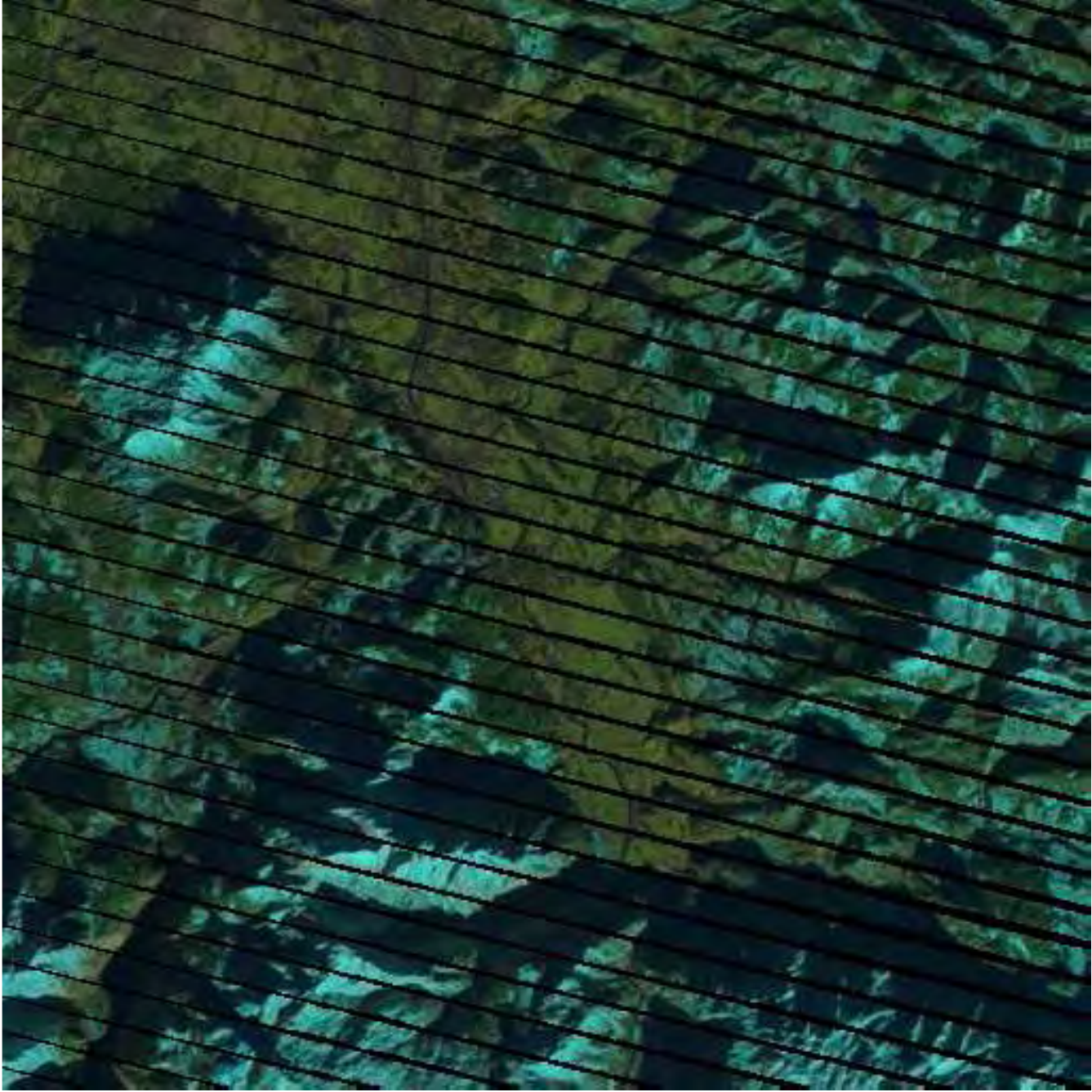}
		\includegraphics[width=0.18\textwidth]{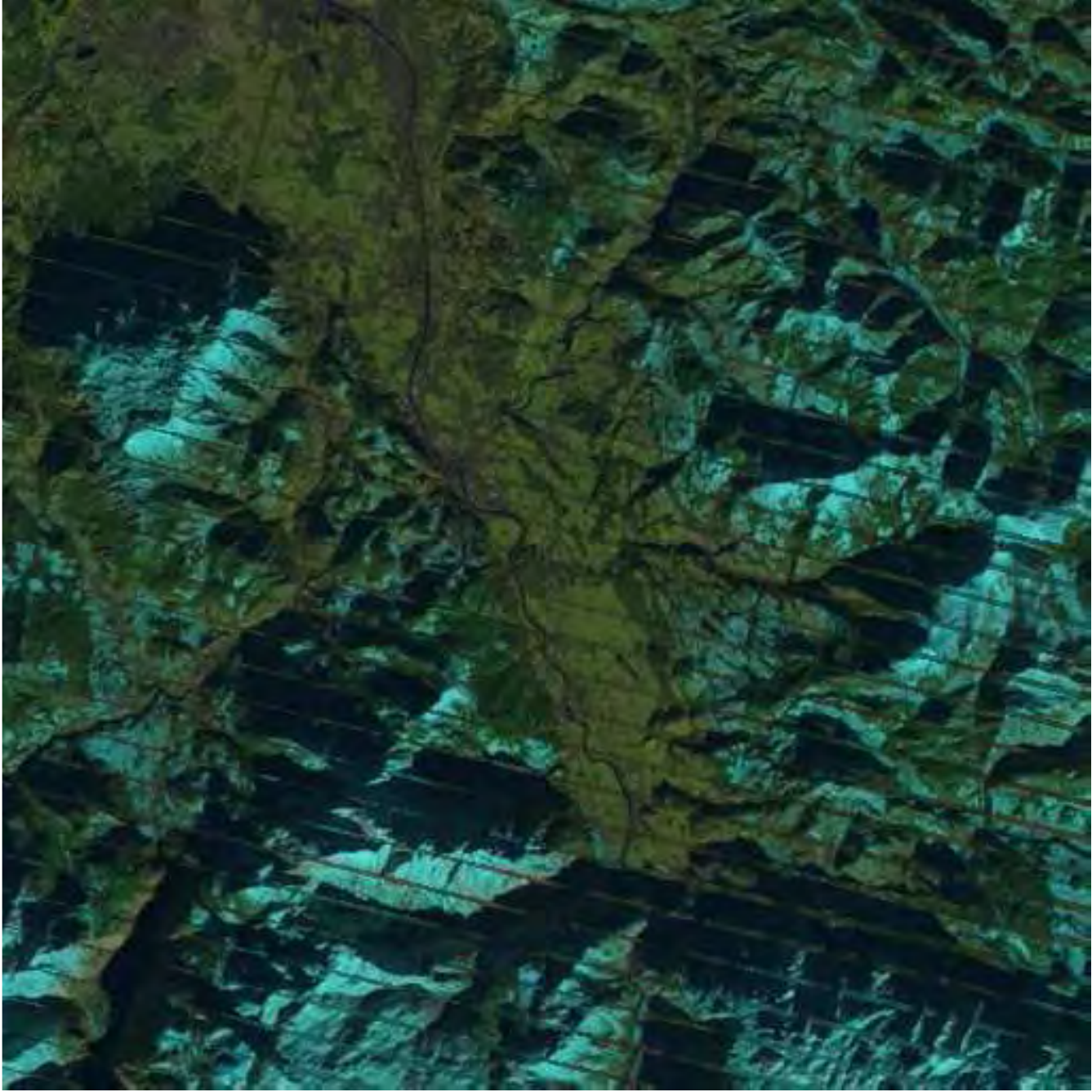}
		\includegraphics[width=0.18\textwidth]{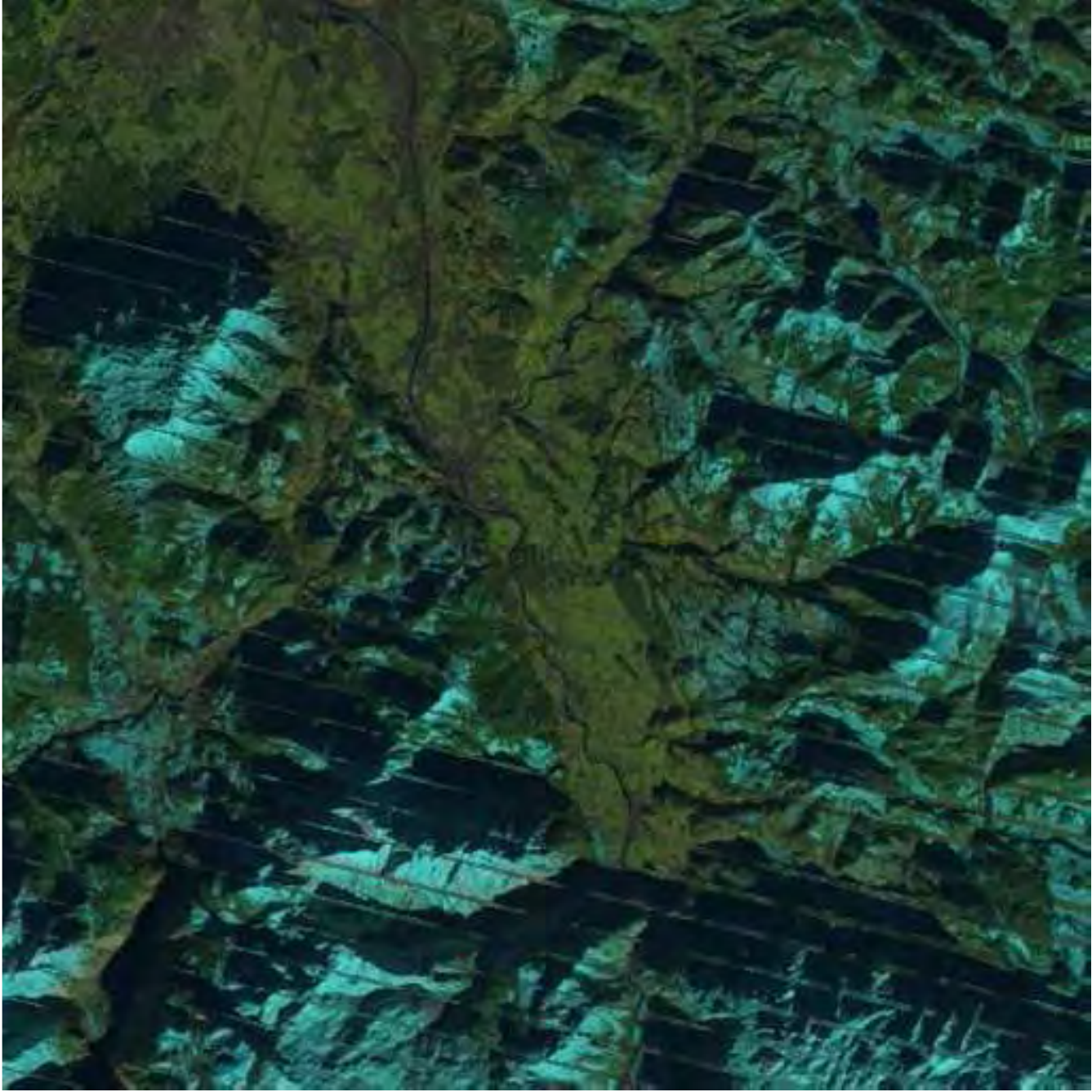}
		\includegraphics[width=0.18\textwidth]{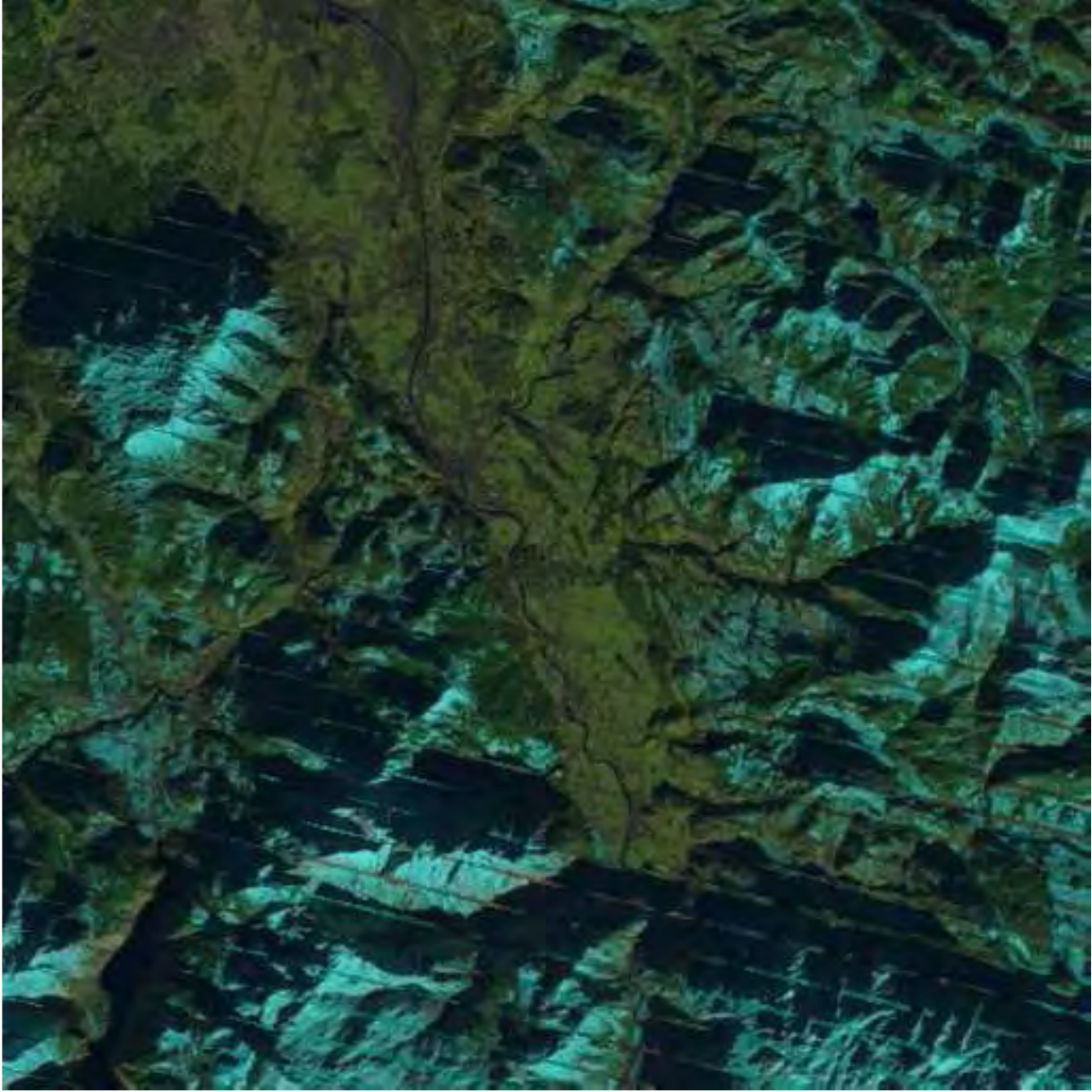}
		\includegraphics[width=0.18\textwidth]{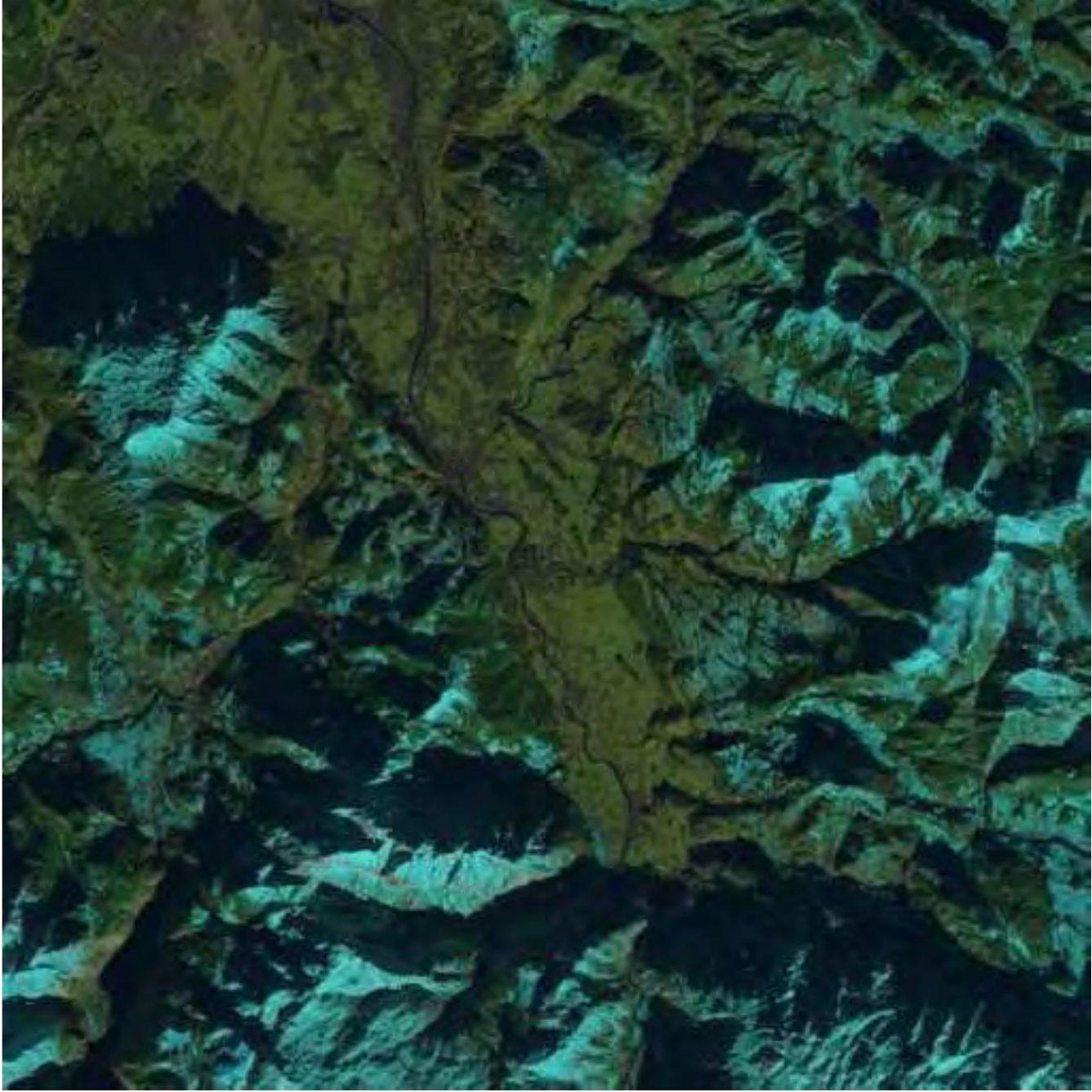}  \\[0.05cm]
		\rotatebox{90}{\footnotesize \textbf{Exp. 6}}
		\includegraphics[width=0.18\textwidth]{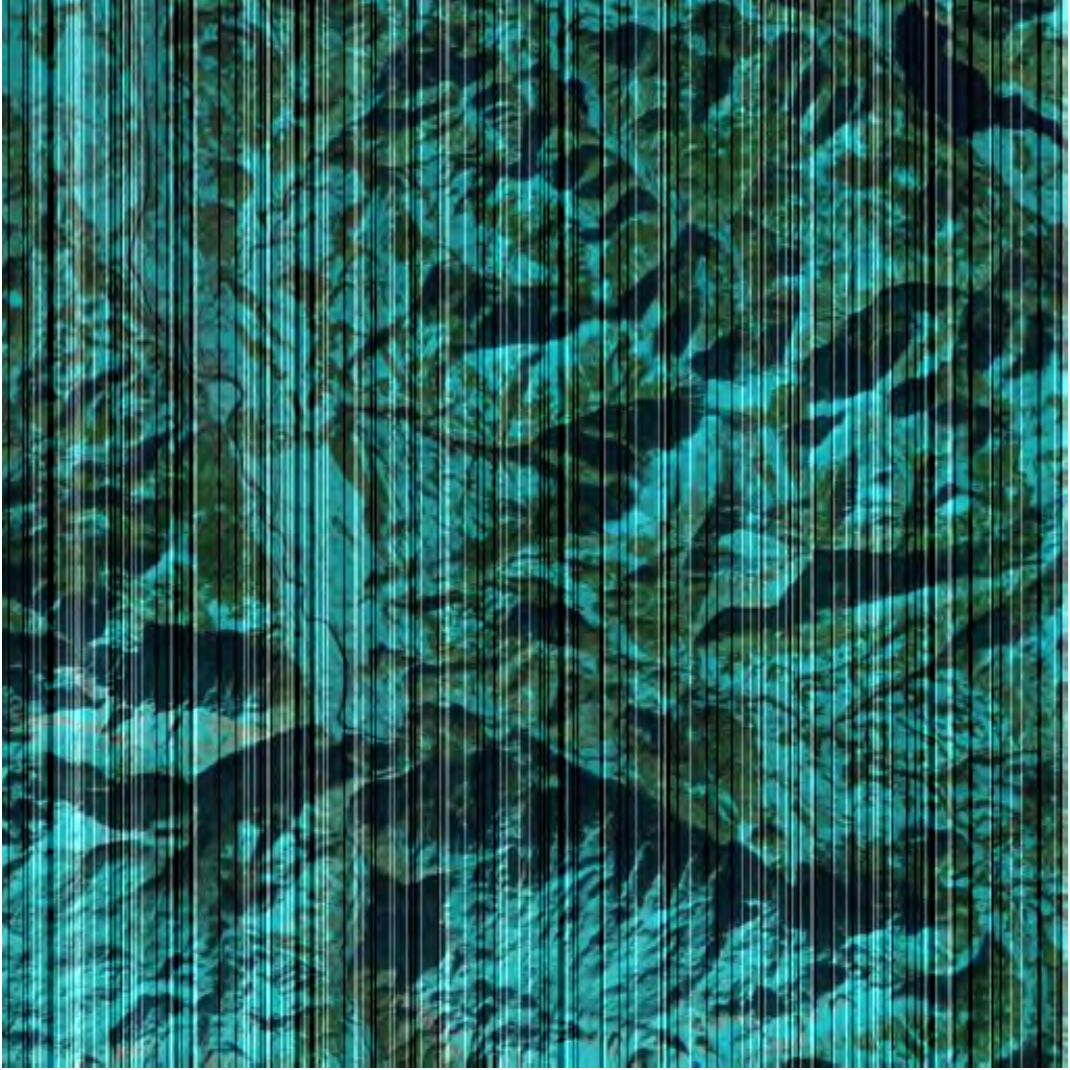}
		\includegraphics[width=0.18\textwidth]{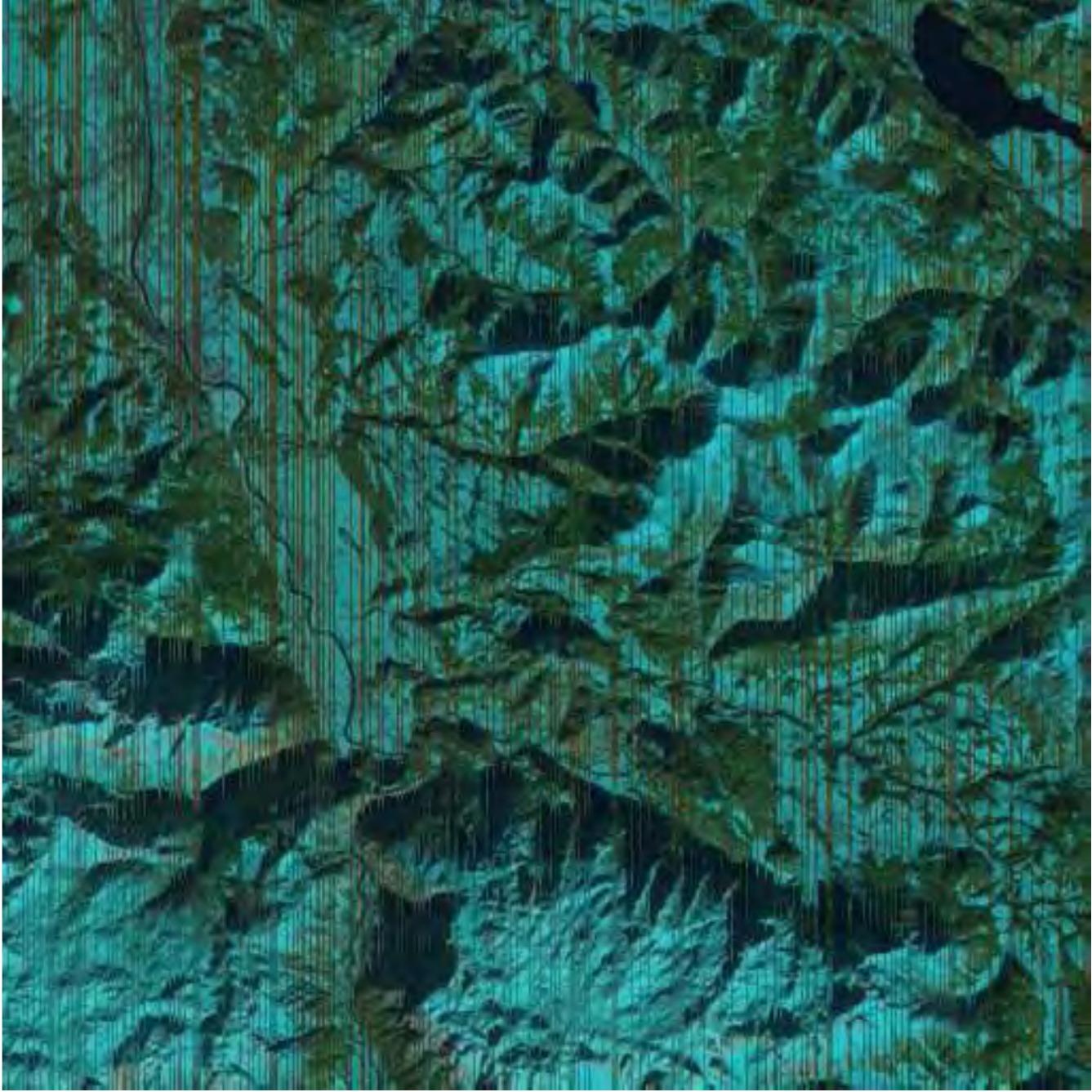}
		\includegraphics[width=0.18\textwidth]{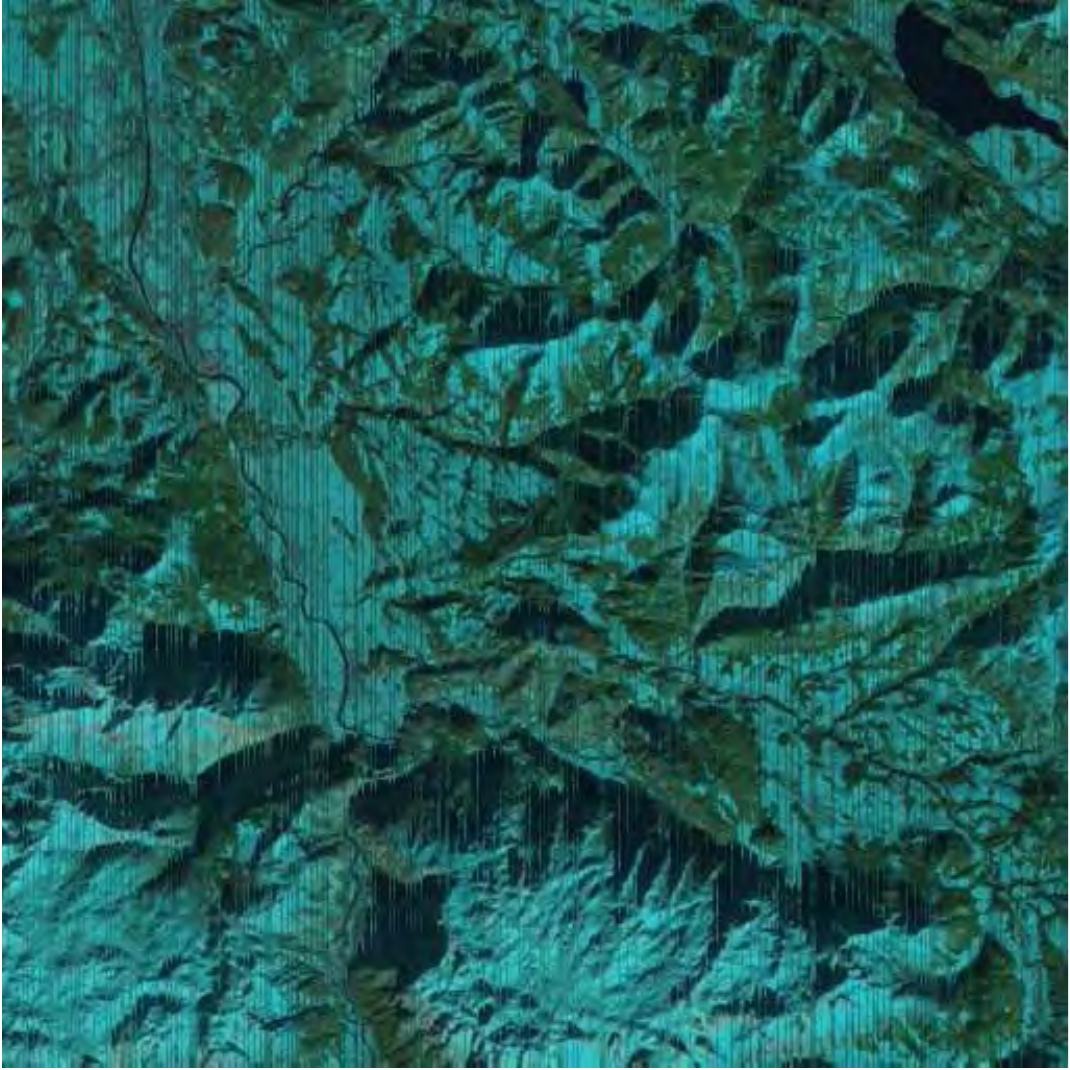}
		\includegraphics[width=0.18\textwidth]{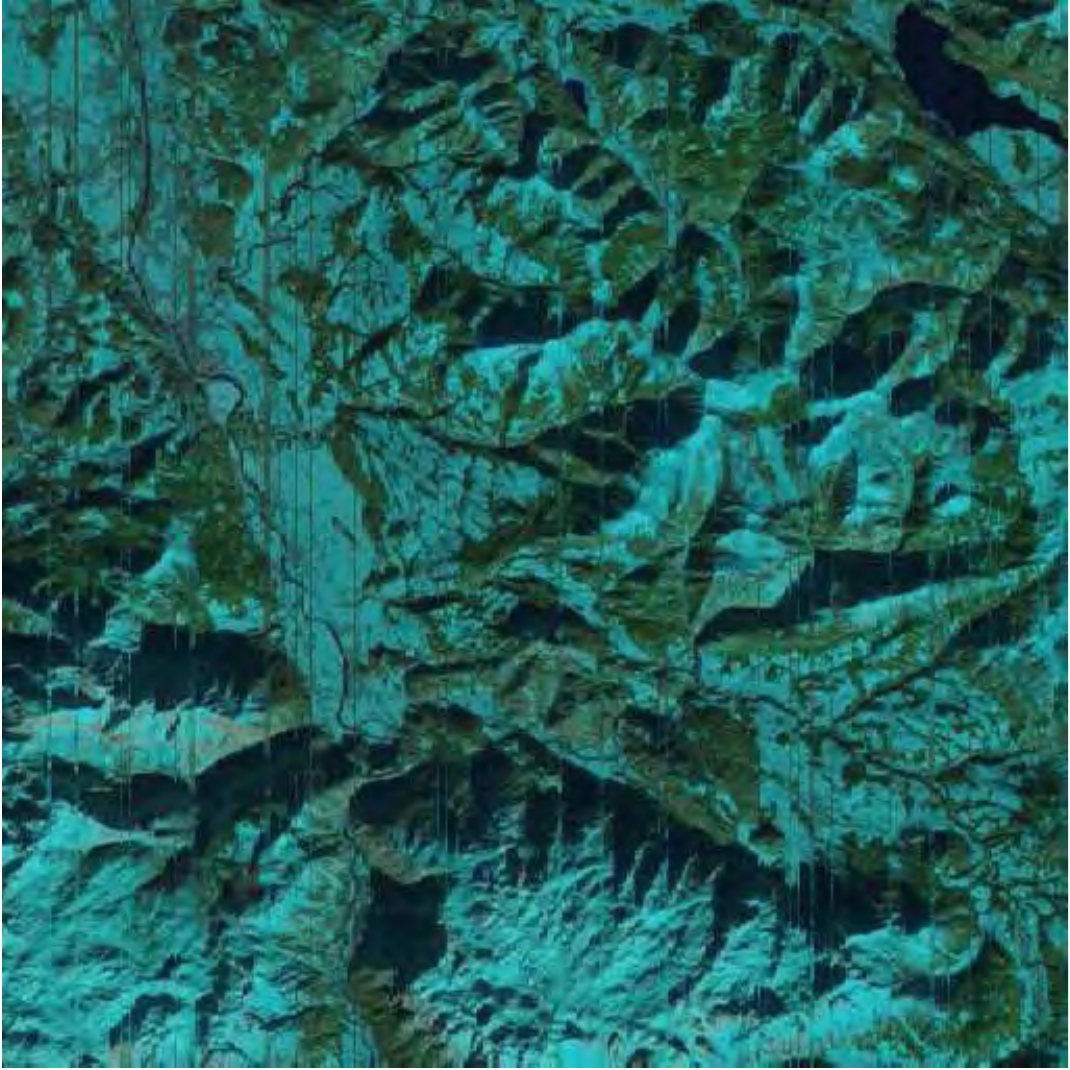}
		\includegraphics[width=0.18\textwidth]{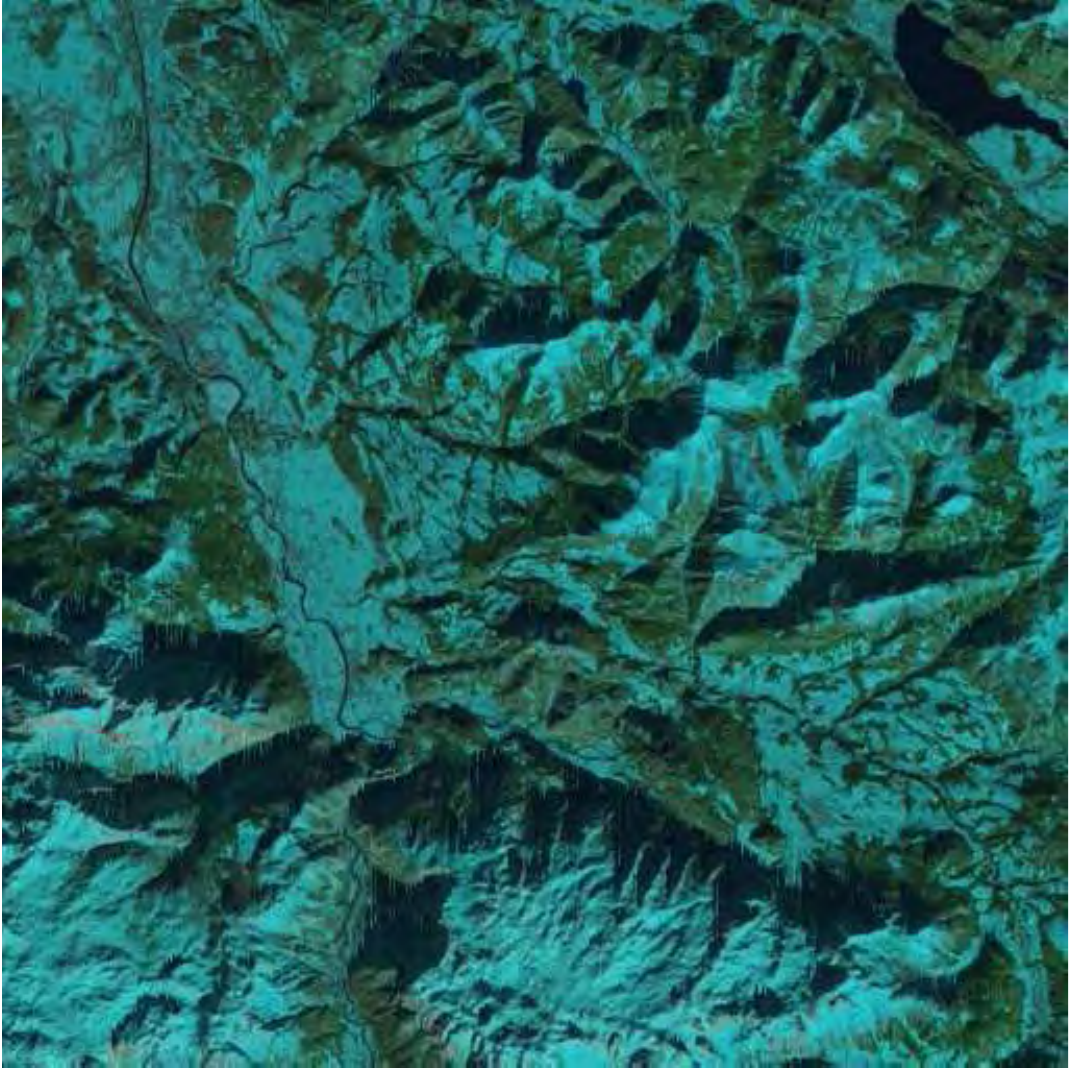}
		\caption{\footnotesize Simulated experiments, Exps. 1--6, for clouds and stripes removal. Exps. 1--3 are cloud removal for ``Images 1--3'', respectively; Exp. 4 is cloud removal for Beijing data (results for band 6 are shown in this figure); Exps. 5 and 6 are stripes removal for ``Image 3'' and ``Image 4'', respectively.  }
		\label{Fig:SimulateCloudResult}
	\end{figure*}

	In this section, the simulated experiments are presented to test NL-LRTC. The test data of Munich are ``M012015'' and ``M022015''. To assess the performance of NL-LRTC fully and efficiently, three subimages of ``M012015'' and one subimage of ``M022015'' shown in Fig. \ref{Fig:SimulateMUdata} are tested in the simulated experiments. These subimages are of size $512\times 512\times 3$. The structure and details of these subimages are different: the data set ``Image 1'' is cut from Fig. \ref{Fig:MUdata}(b) (red square) and corresponding areas of Fig. \ref{Fig:MUdata}(d)-(f) and mostly contains vegetation areas with relatively low contrast due to flat terrains; ``Image 2'' is cut from Fig. \ref{Fig:MUdata}(b) (green square) and corresponding areas of Fig. \ref{Fig:MUdata}(d)-(f) and mainly contains mountains, hills, and rivers; and ``Image 3'' and ``Image 4'' are extracted from Fig. \ref{Fig:MUdata}(b) (blue square) and (c) (white square), respectively, and the corresponding areas of Fig. \ref{Fig:MUdata}(d)-(f). The data sets ``Image 3'' and ``Image 4'' contain both characteristics of ``Image 1'' and ``Image 2''. The simulated clouds and stripes removal results are shown in Fig. \ref{Fig:SimulateCloudResult}, where Exps. 1--4 are for clouds removal and Exps. 5 and 6 are for stripes removal.

\begin{figure}[h]
		\begin{center}
			\includegraphics[width=0.45\textwidth]{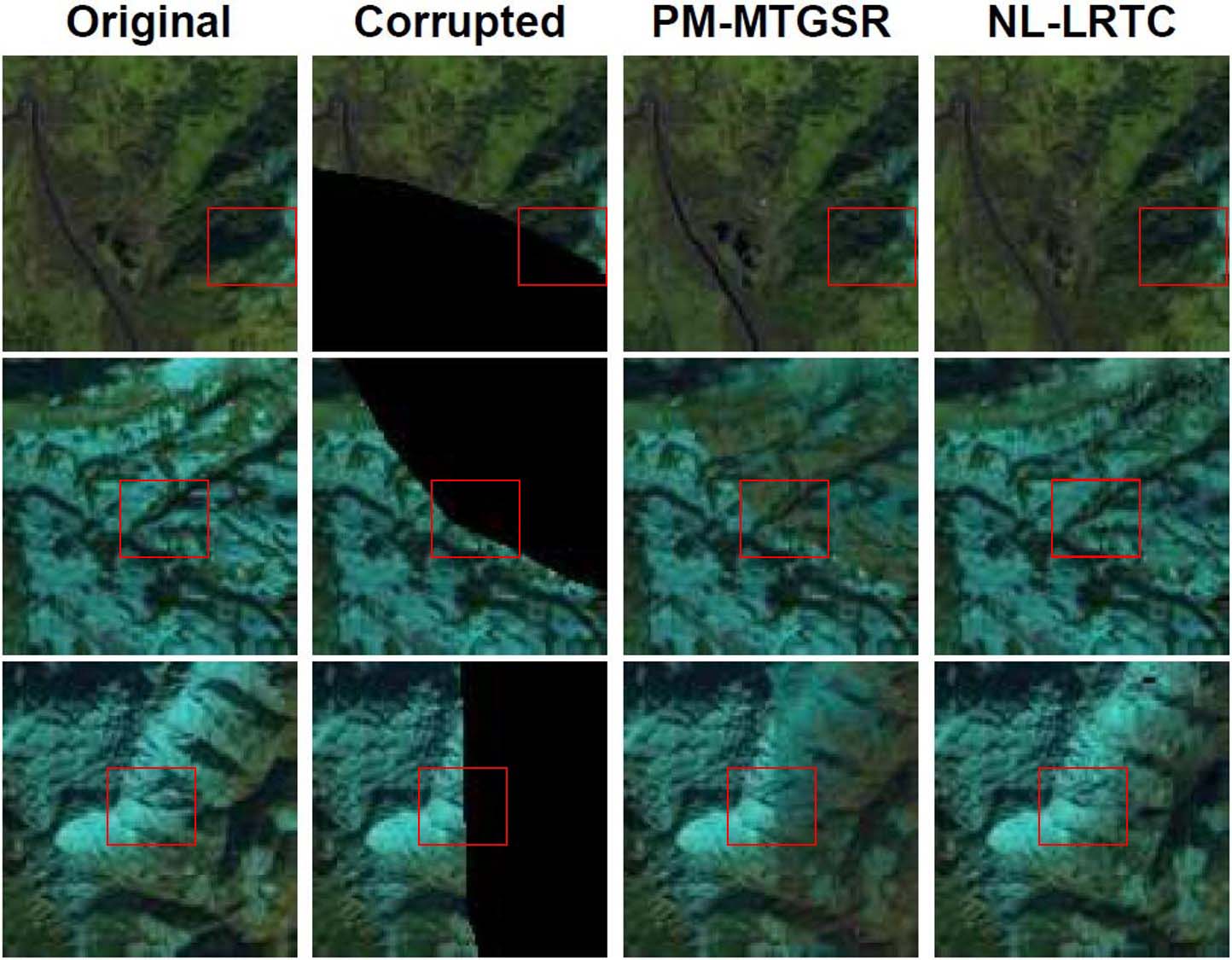}
		\end{center}
		\caption{\footnotesize Zoom results of PM-MTGSR and NL-LRTC for Exps. 1--3. From left to right: original data, corrupted data, zoom part of results reconstructed by PM-MTGSR and NL-LRTC, respectively. From top to bottom: zoom results for Exps. 1, 2, 3, respectively. For the corrupted data, the black areas are missing. }
		\label{Fig:SimulateCloudZoomResult}
	\end{figure}

	First, the cloud removal experiments (Exps. 1--3) are undertaken using ``Image 1'', ``Image 2'', and ``Image 3'' shown in Fig. \ref{Fig:SimulateMUdata}: the first column is the simulated corrupted data, and the other three columns are the supplementary data. The simulated cloud masks are generated by Matlab manually. The recovery results compared with HaLRTC, ALM-IPG, and PM-MTGSR are shown in Fig. \ref{Fig:SimulateCloudResult}. Exps. 1--3 indicate that PM-MTGSR and NL-LRTC succeed in estimating the missing entries. However, the results of HaLRTC and ALM-IPG have a noticeable difference between the known and the estimated areas. HaLRTC utilizes only the low-rankness of the observed 4-order multitemporal data, in another word, it does not adopt patch strategy. ALM-IPG mainly studies the smooth relationships between the continuous temporal images, while the temporal series of the given data is not continuous, i.e., the image acquired in the adjacent time is different largely. To compare the results of PM-MTGSR and NL-LRTC in detail, a zoomed region is displayed in Fig. \ref{Fig:SimulateCloudZoomResult}. From this figure, we can see that there are obvious edges between the estimated and known areas for the PM-MTGSR results, especially for Exps. 2 and 3. In contrast, the estimated areas of NL-LRTC have the similar tint with the known areas, i.e., the estimated areas of NL-LRTC are in harmony with the known areas.

\begin{figure*}[!ht]
		\centering
		\rotatebox{90}{\footnotesize \textbf{Exp. 1}}
		\includegraphics[width=0.23\textwidth]{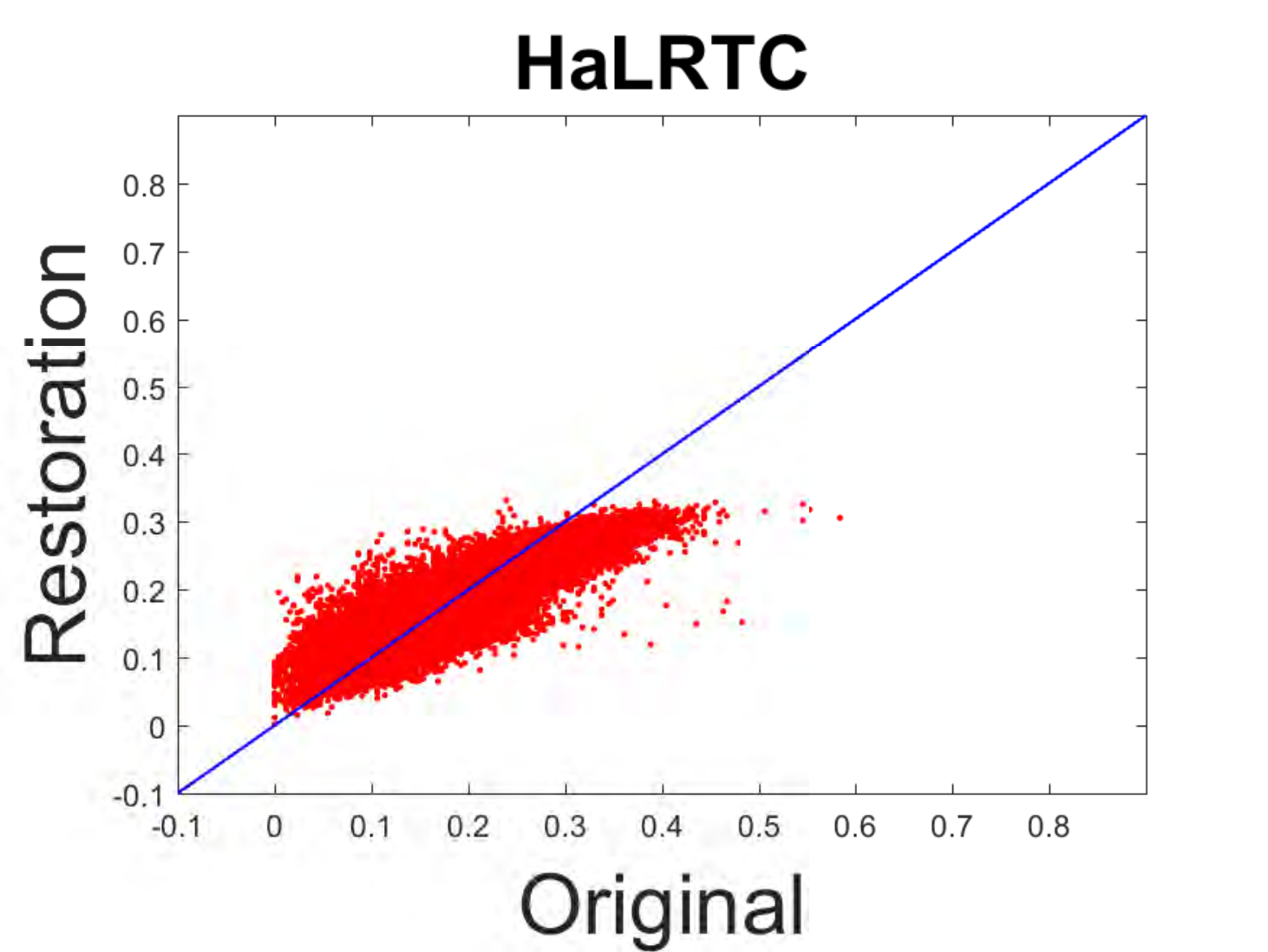}
		\includegraphics[width=0.23\textwidth]{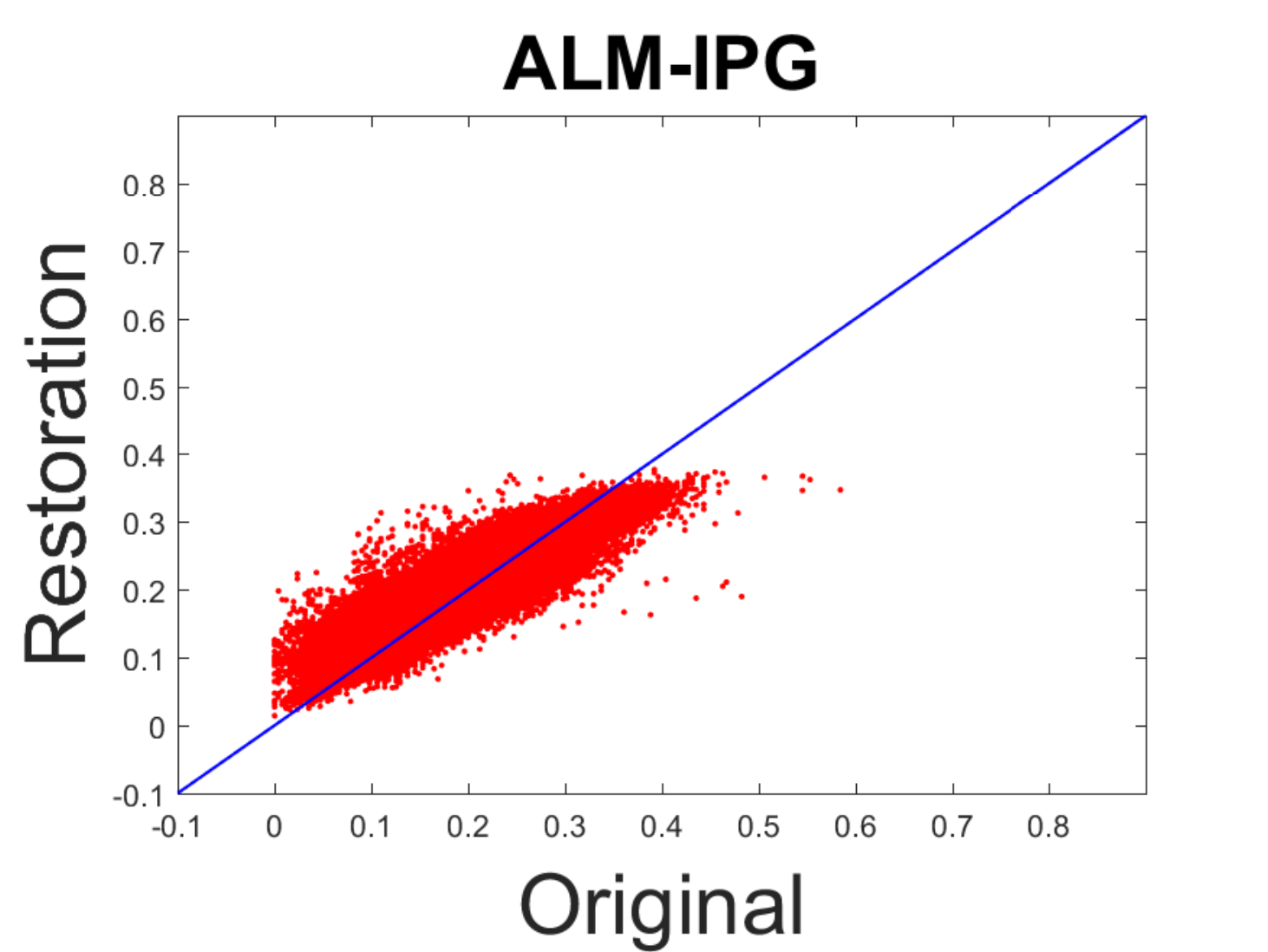}
		\includegraphics[width=0.23\textwidth]{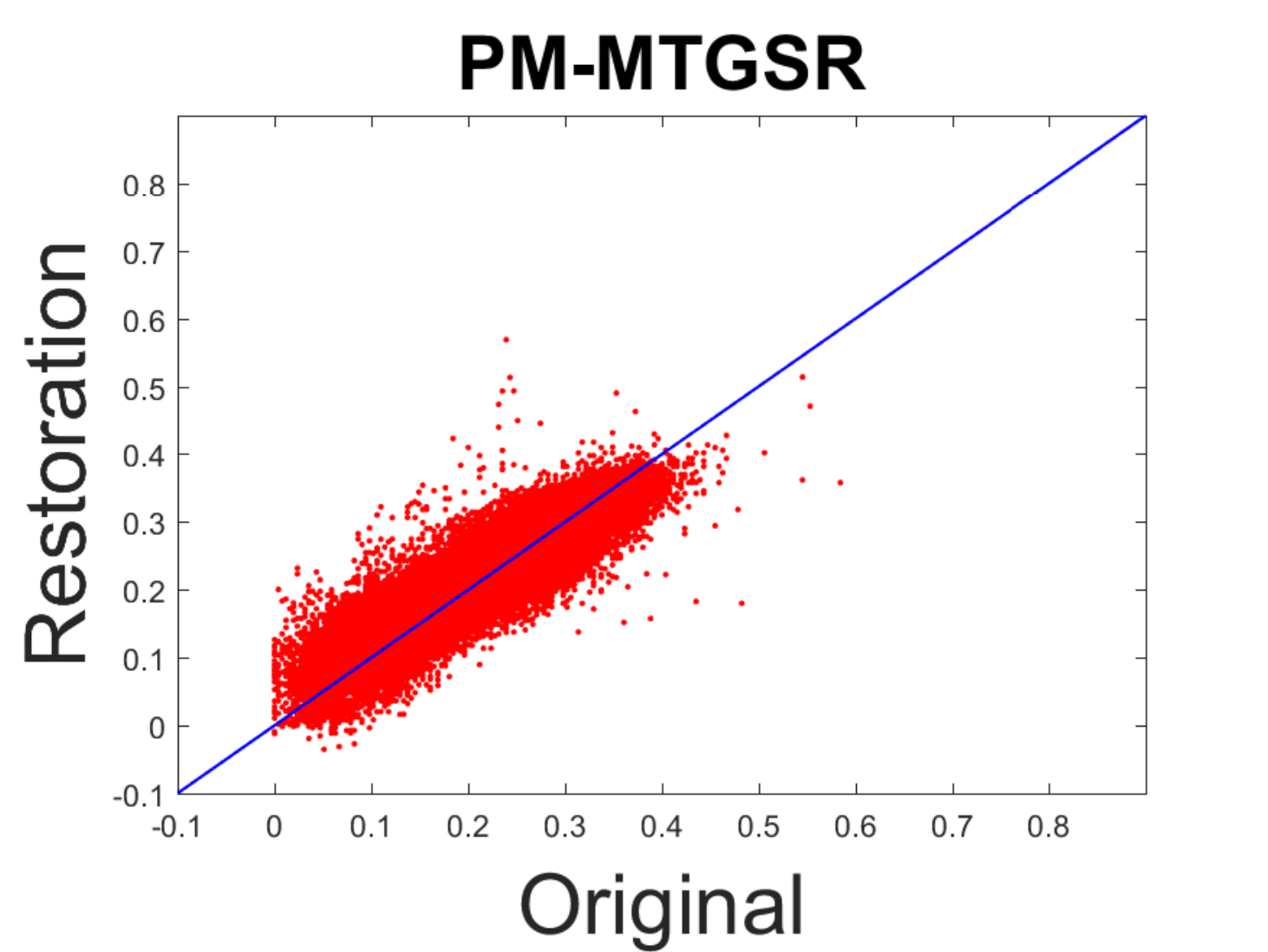}
		\includegraphics[width=0.23\textwidth]{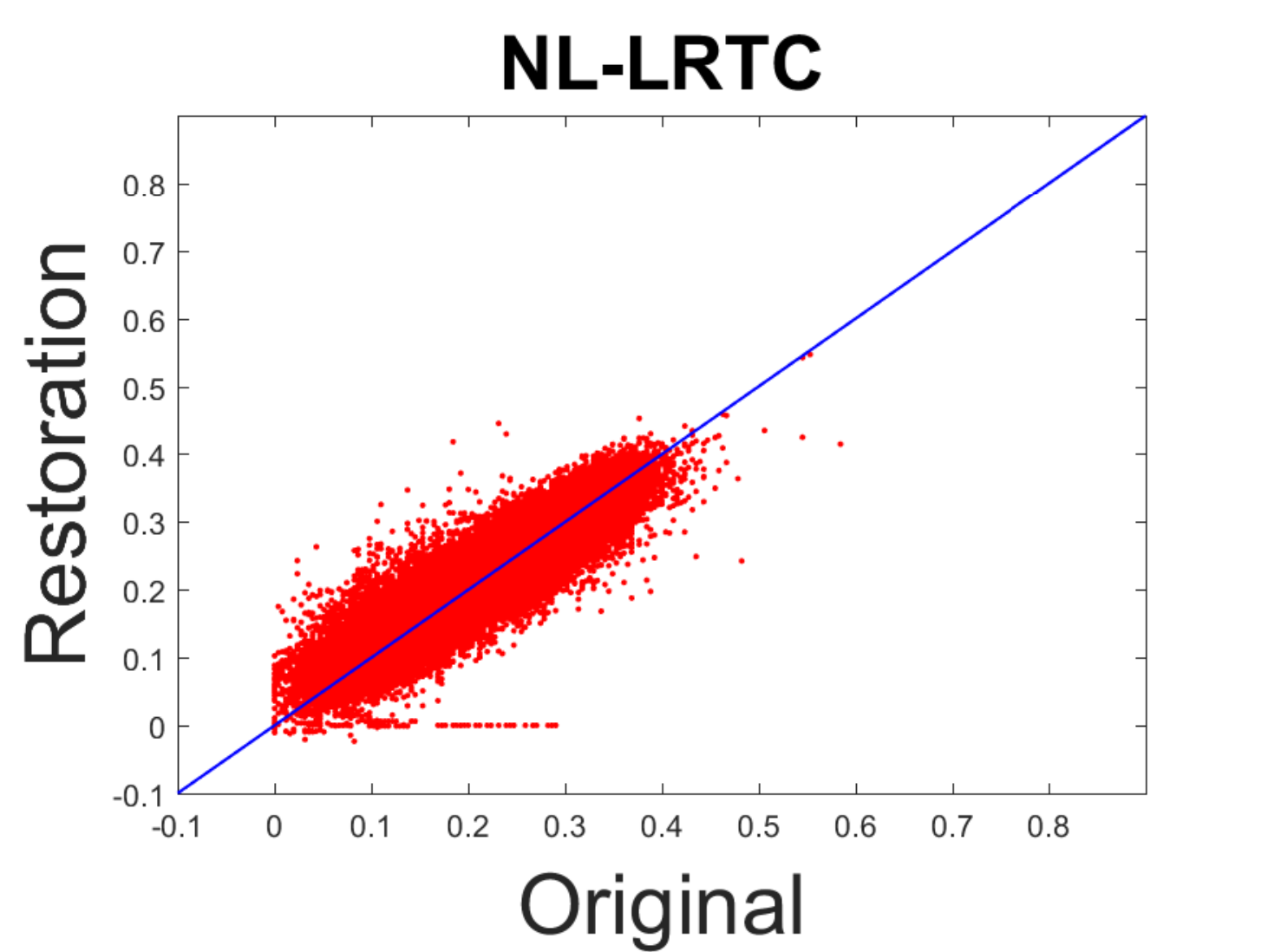}

		\rotatebox{90}{\footnotesize \textbf{Exp. 2}}
		\includegraphics[width=0.23\textwidth]{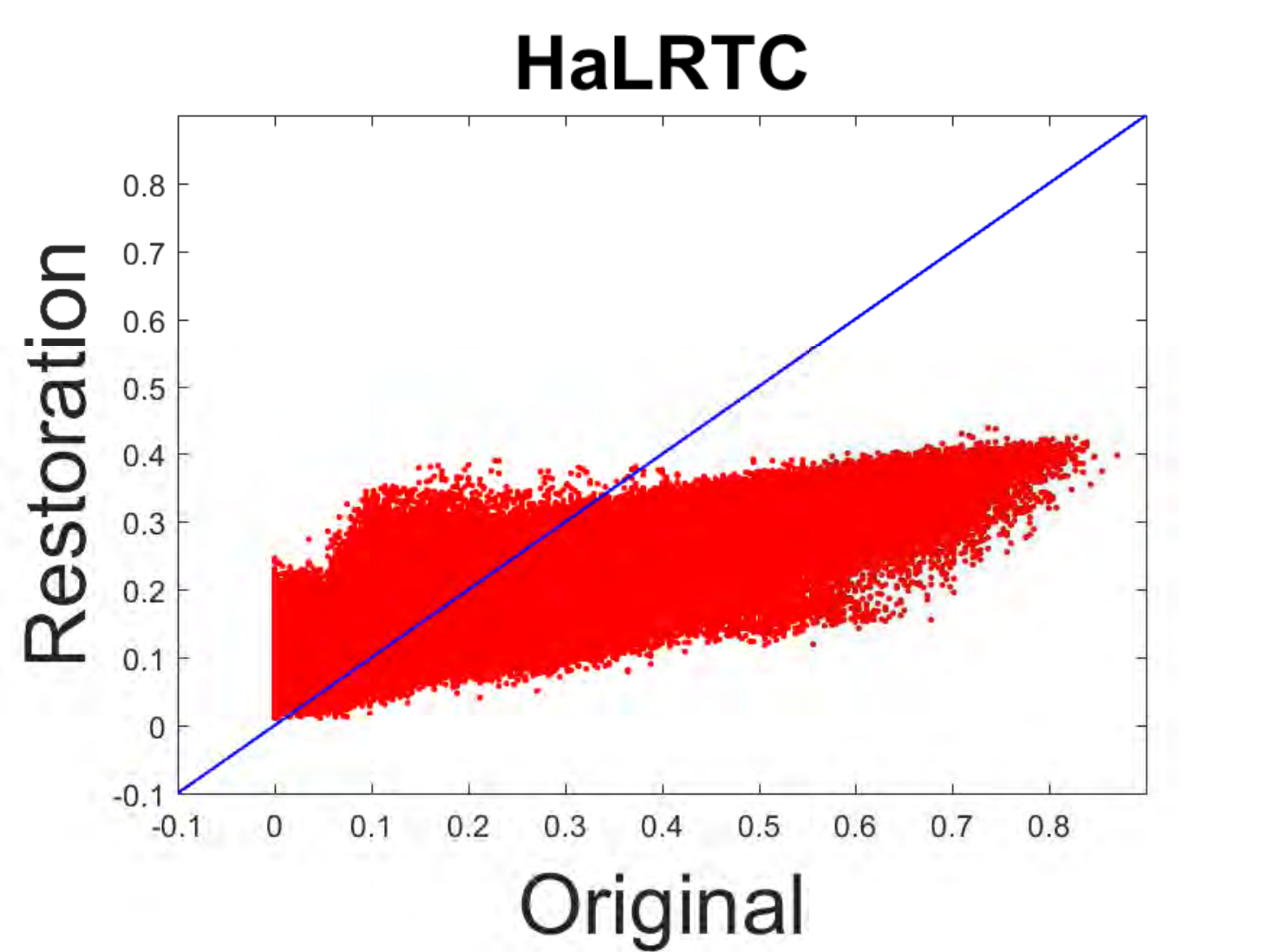}
		\includegraphics[width=0.23\textwidth]{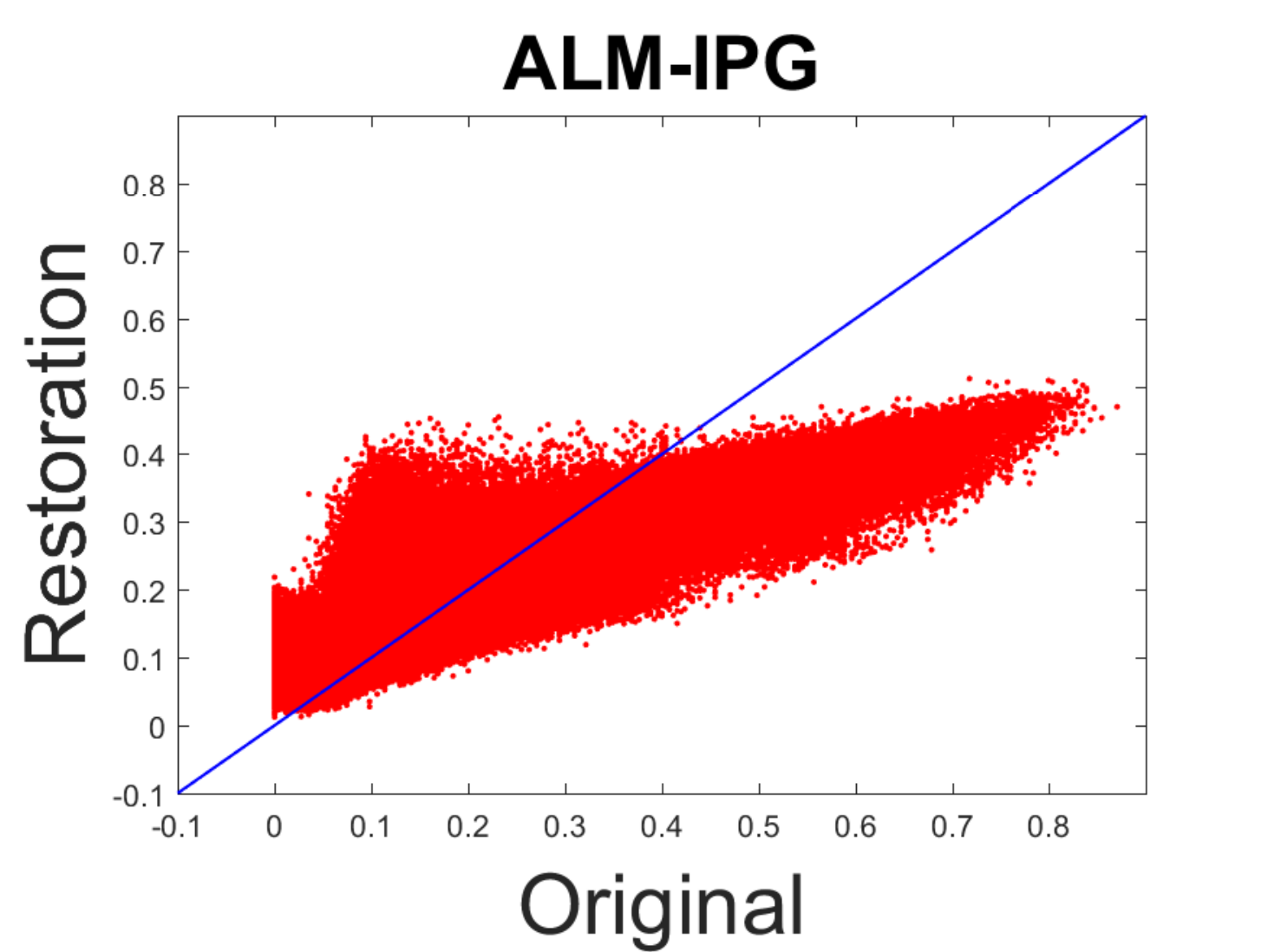}
		\includegraphics[width=0.23\textwidth]{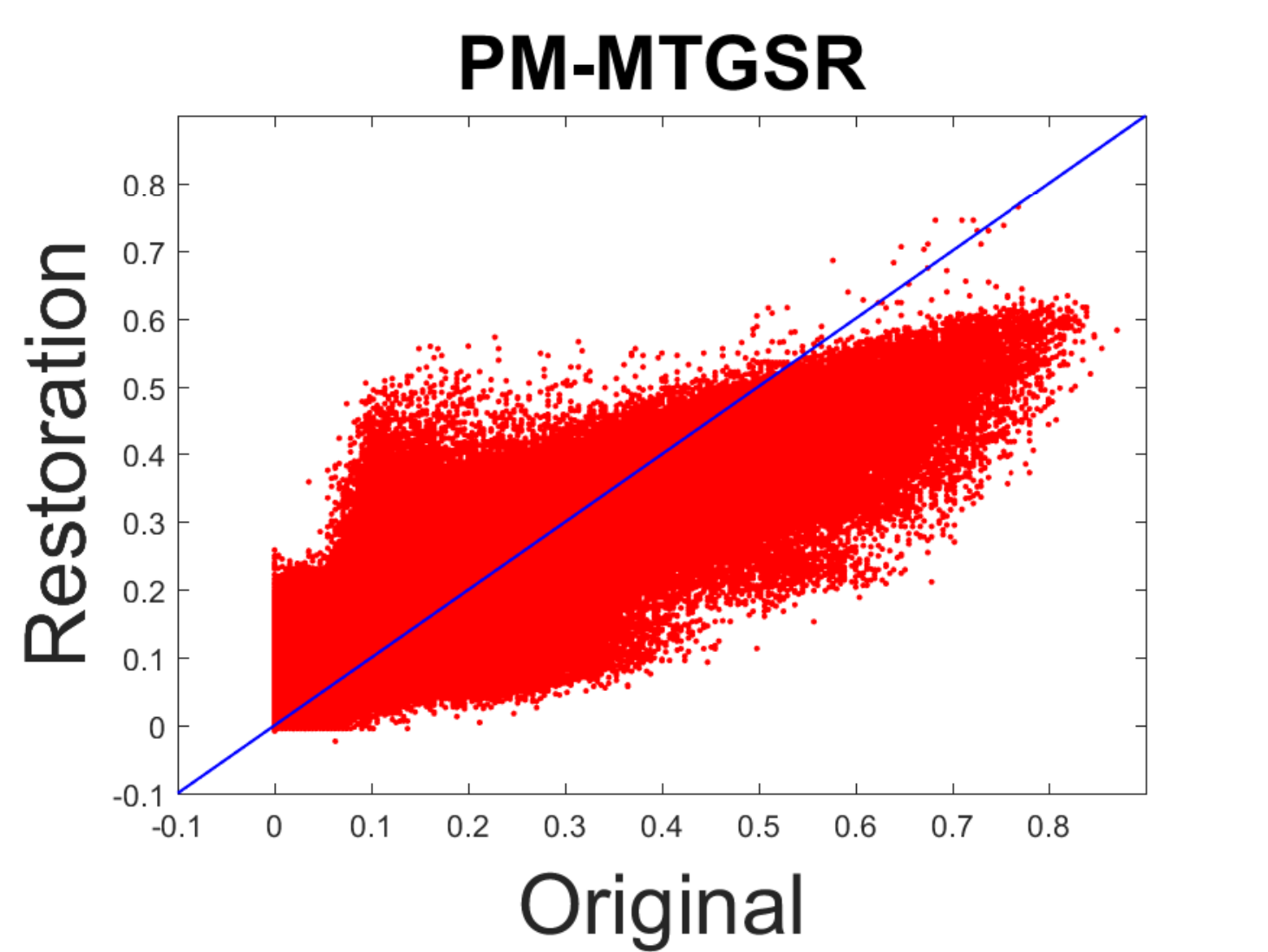}
		\includegraphics[width=0.23\textwidth]{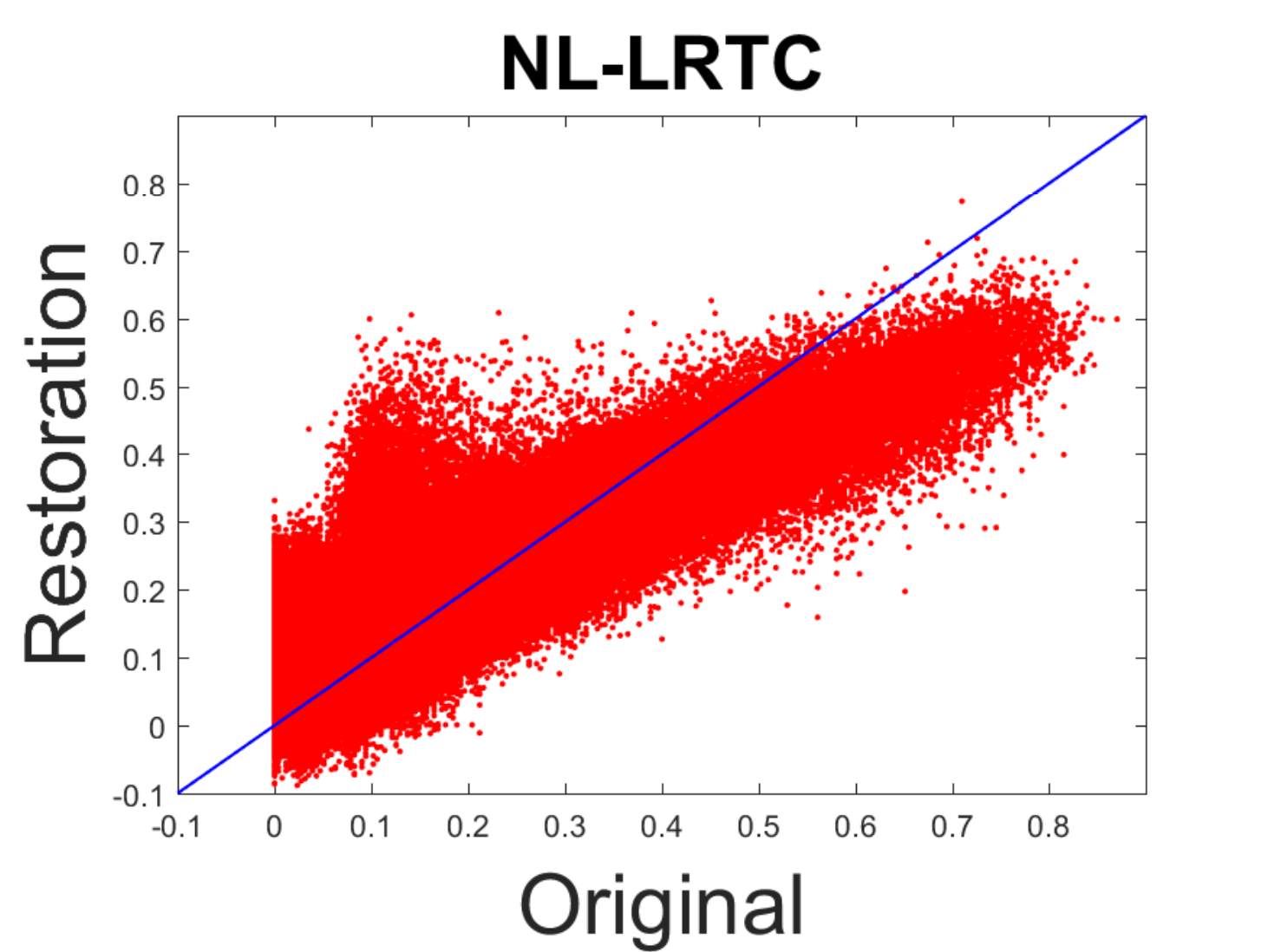}

		\rotatebox{90}{\footnotesize \textbf{Exp. 3}}
		\includegraphics[width=0.23\textwidth]{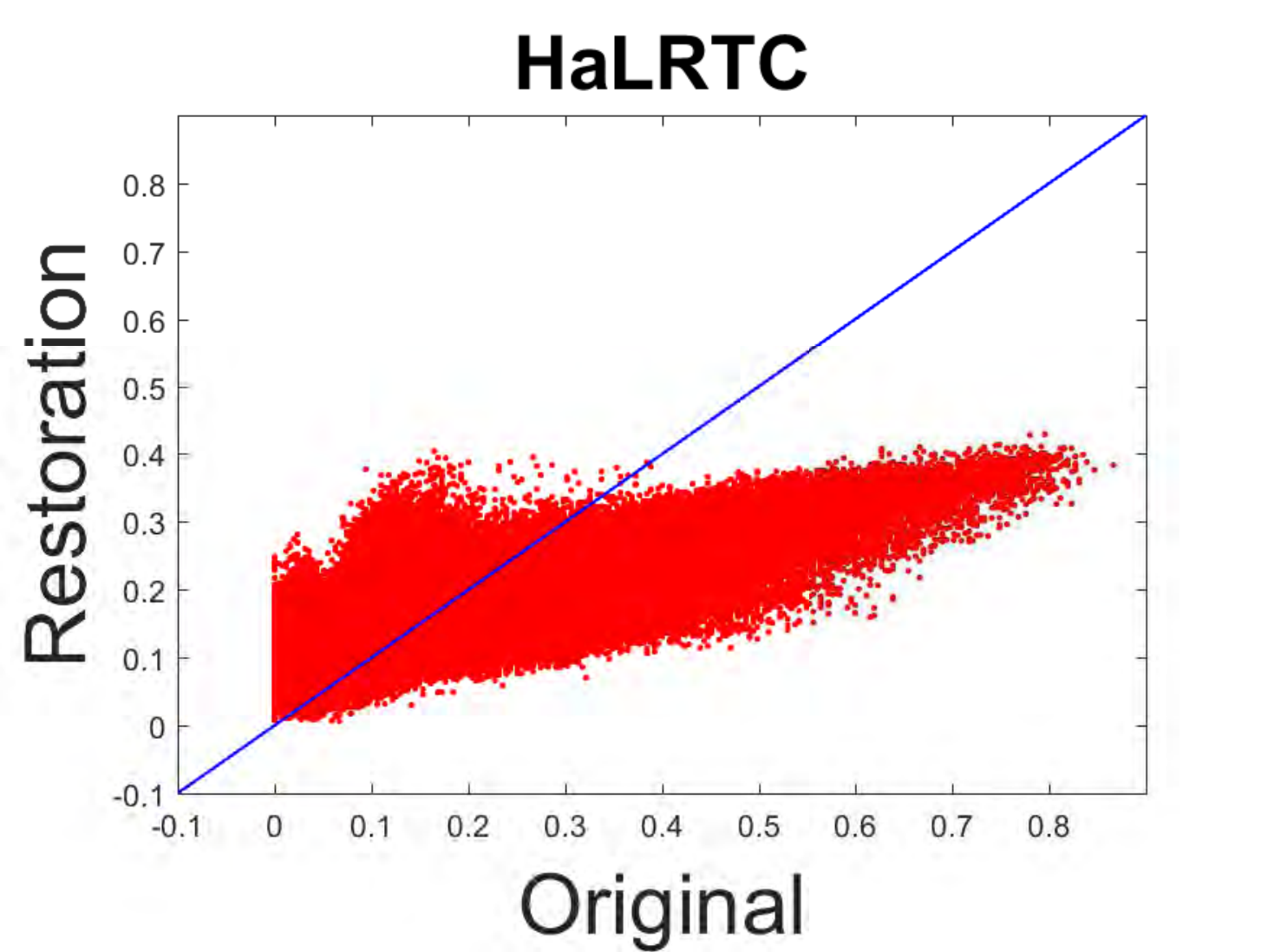}
		\includegraphics[width=0.23\textwidth]{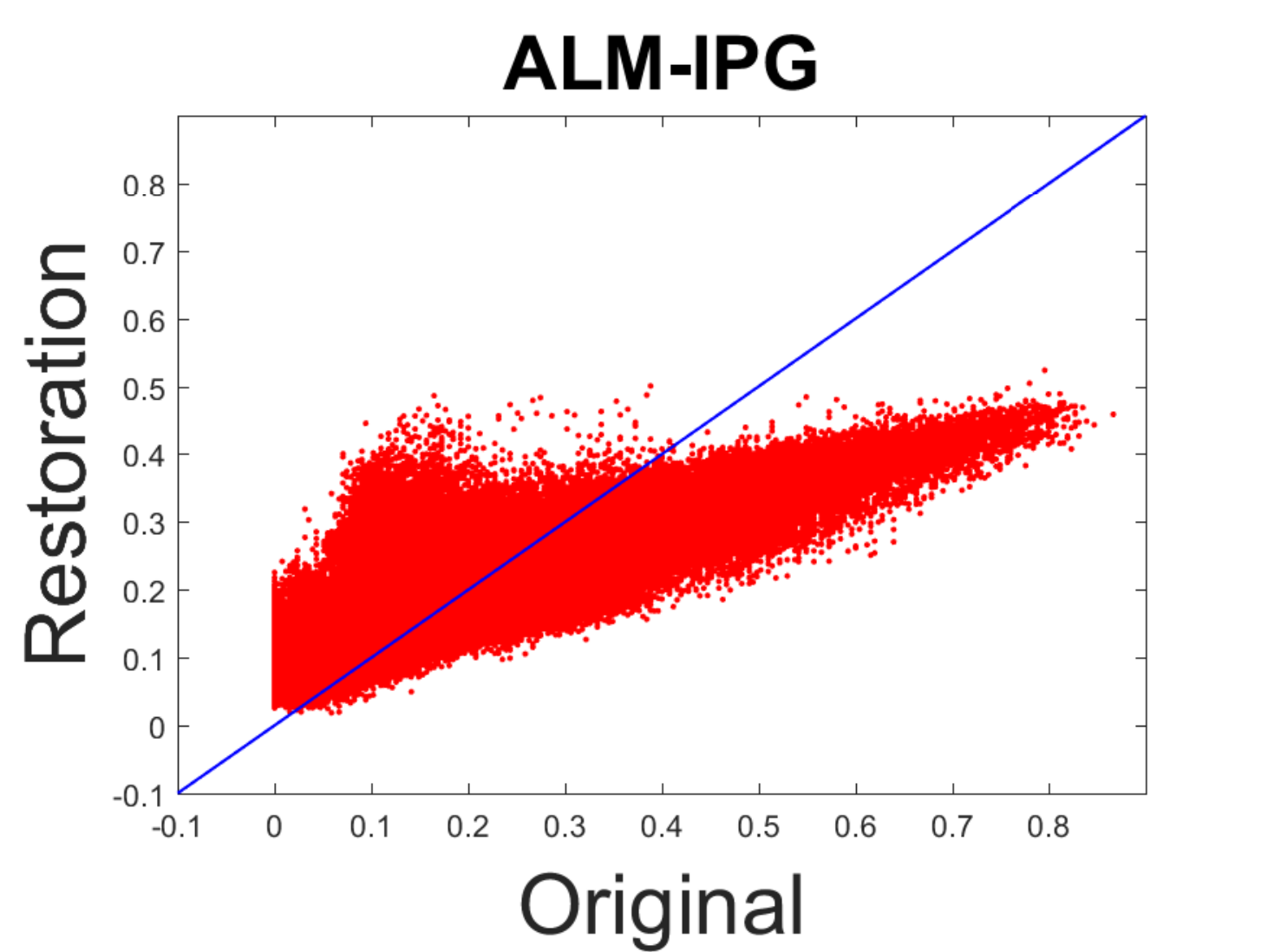}
		\includegraphics[width=0.23\textwidth]{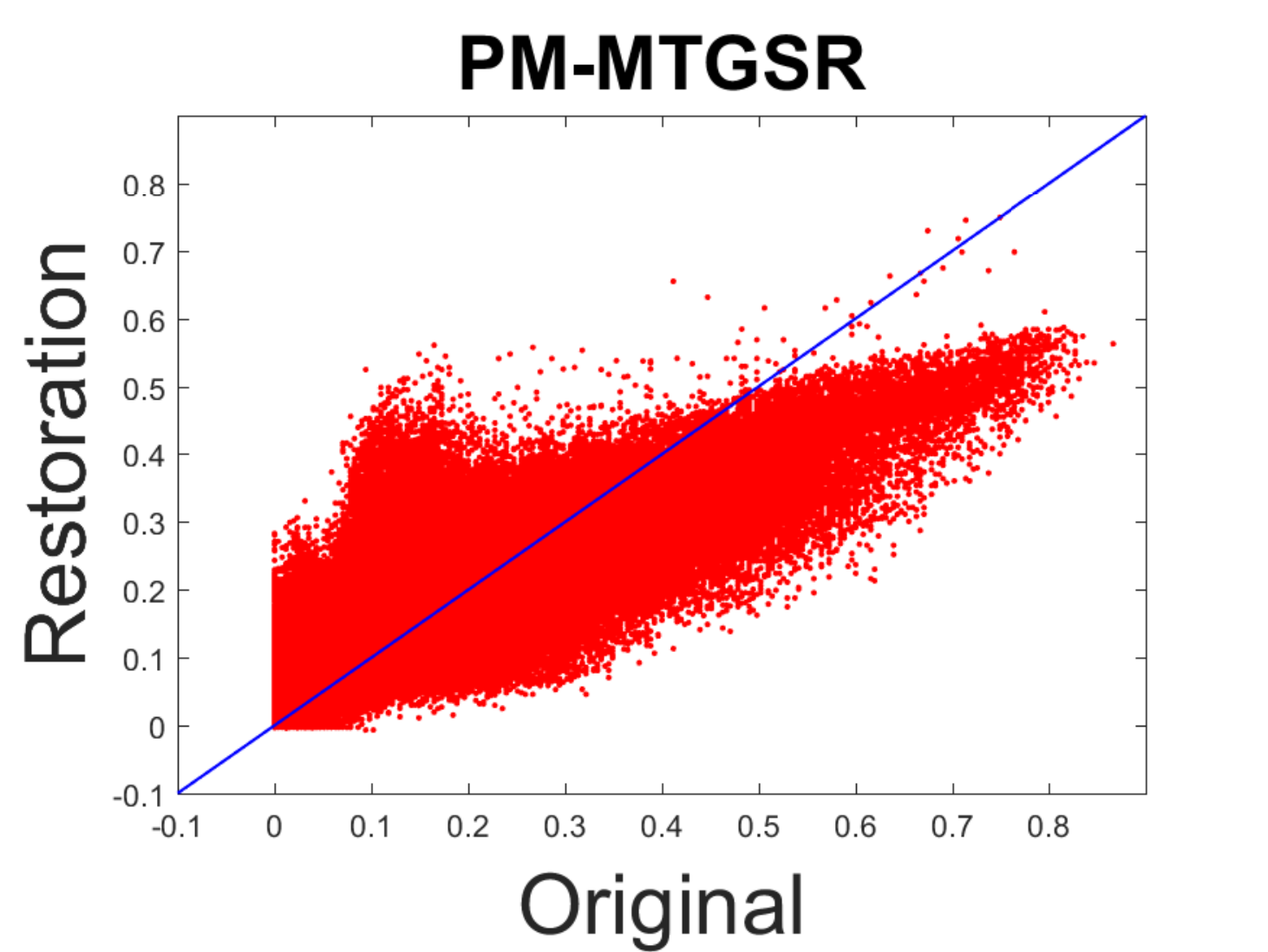}
		\includegraphics[width=0.23\textwidth]{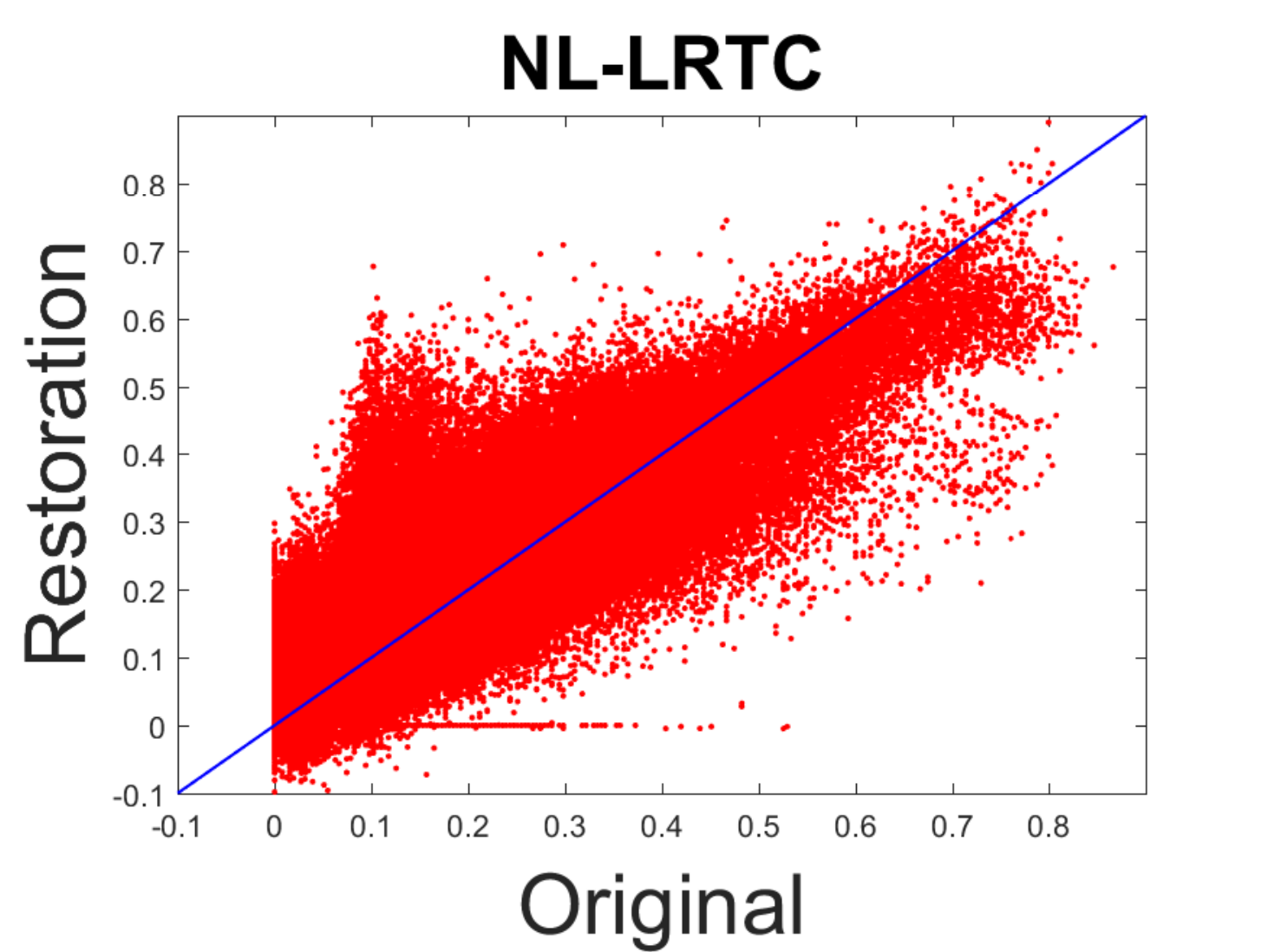}

		\rotatebox{90}{\footnotesize \textbf{Exp. 4}}
		\includegraphics[width=0.23\textwidth]{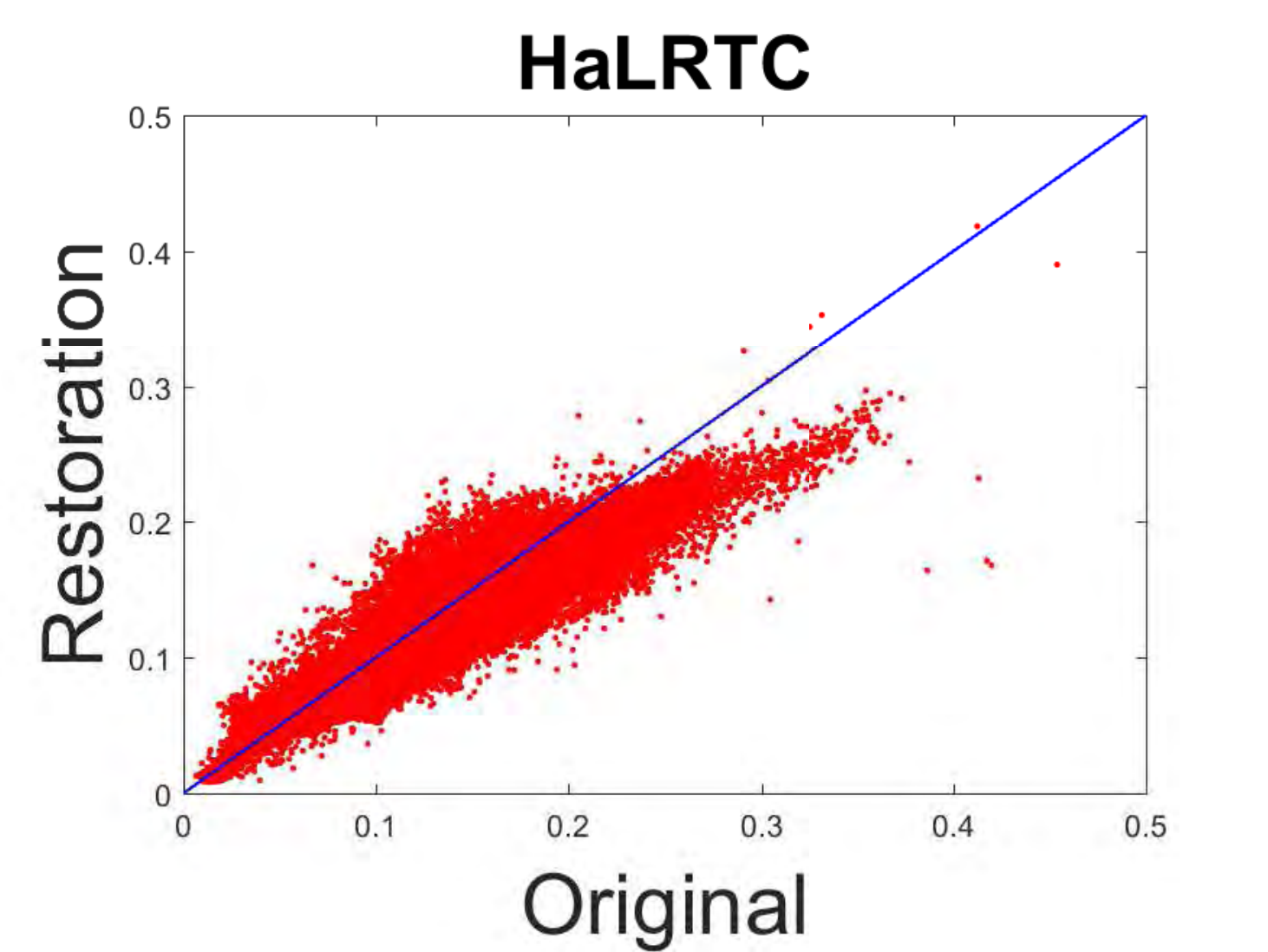}
		\includegraphics[width=0.23\textwidth]{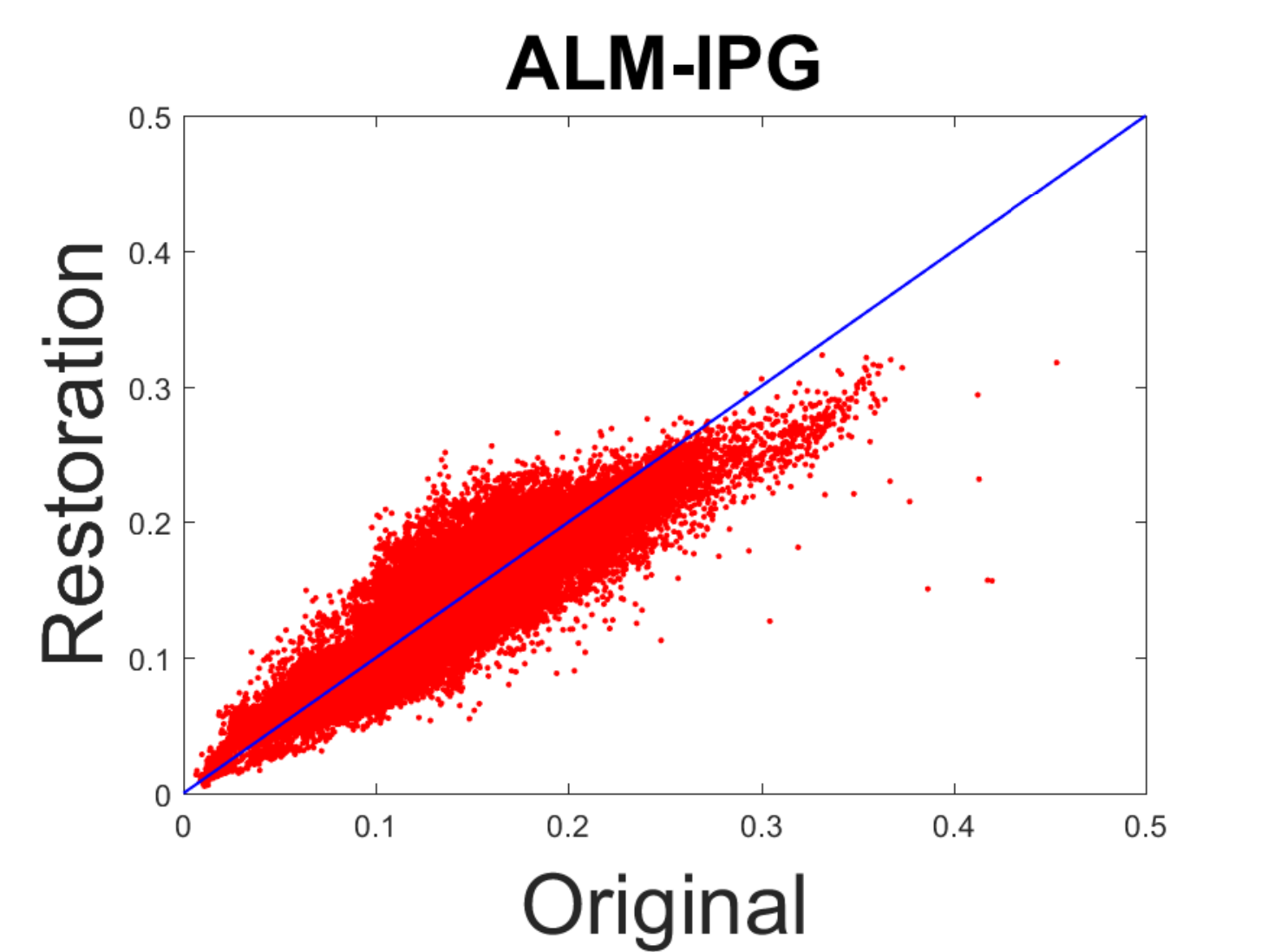}
		\includegraphics[width=0.23\textwidth]{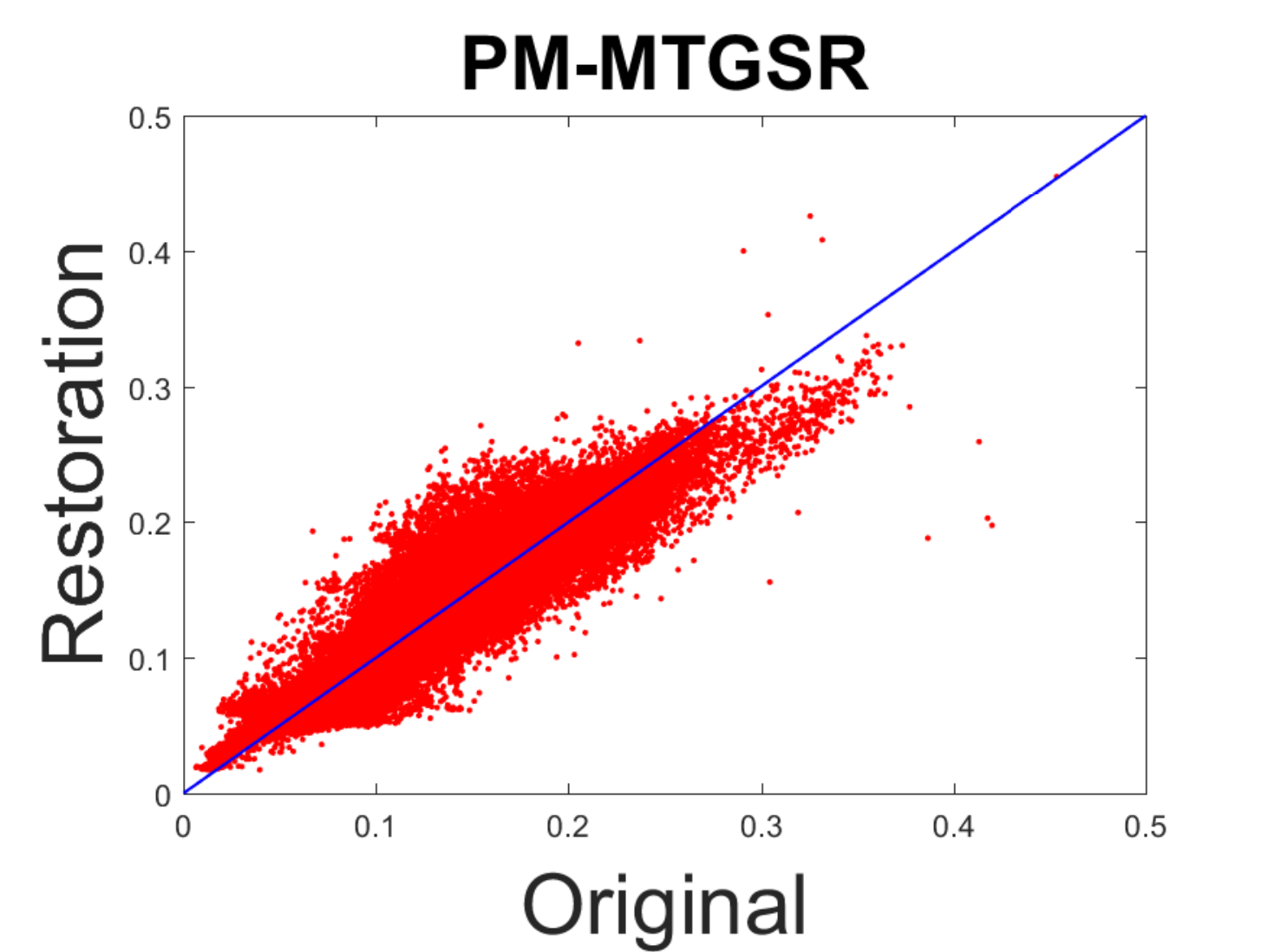}
		\includegraphics[width=0.23\textwidth]{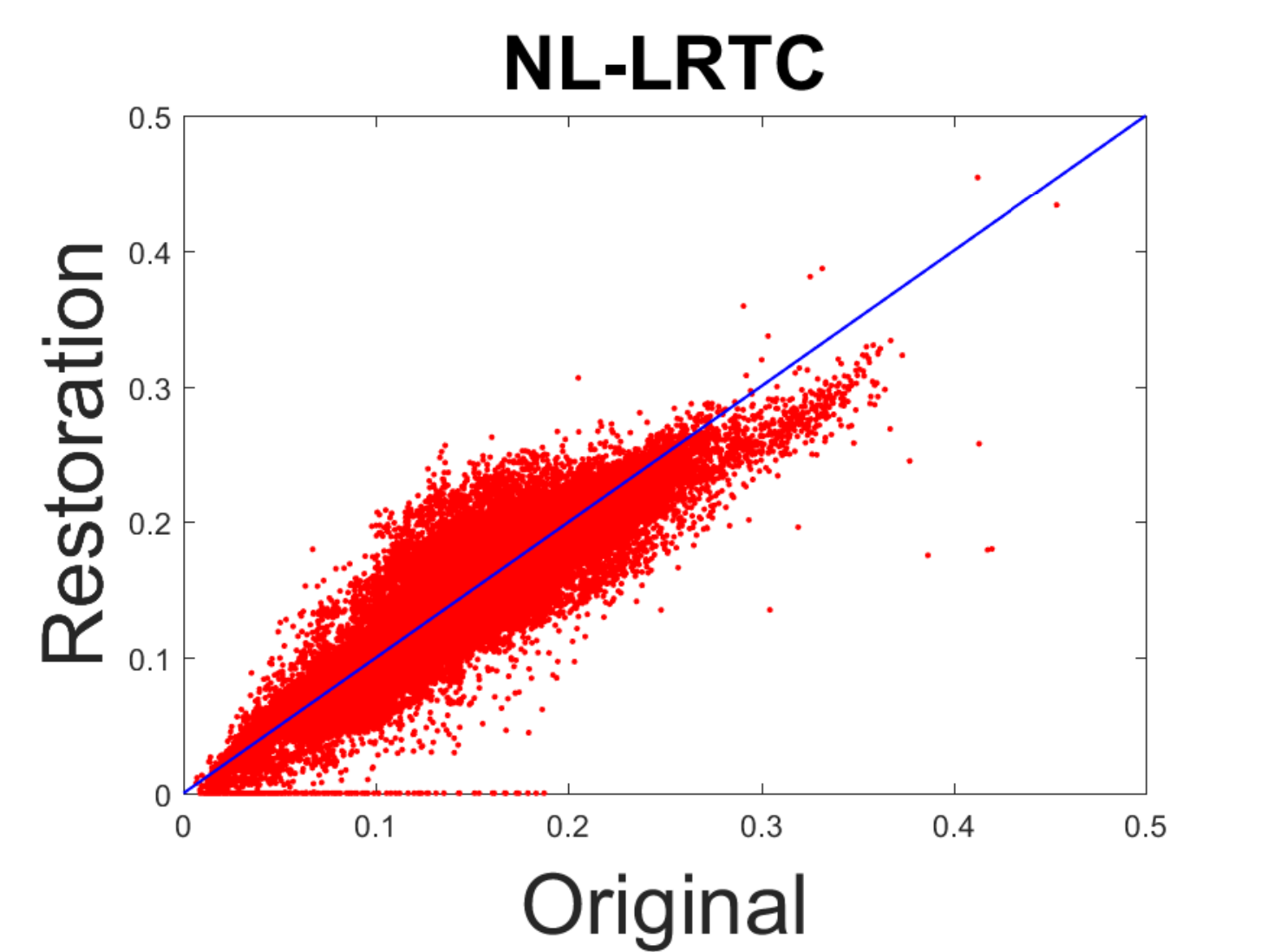}

		\rotatebox{90}{\footnotesize \textbf{Exp. 5}}
		\includegraphics[width=0.23\textwidth]{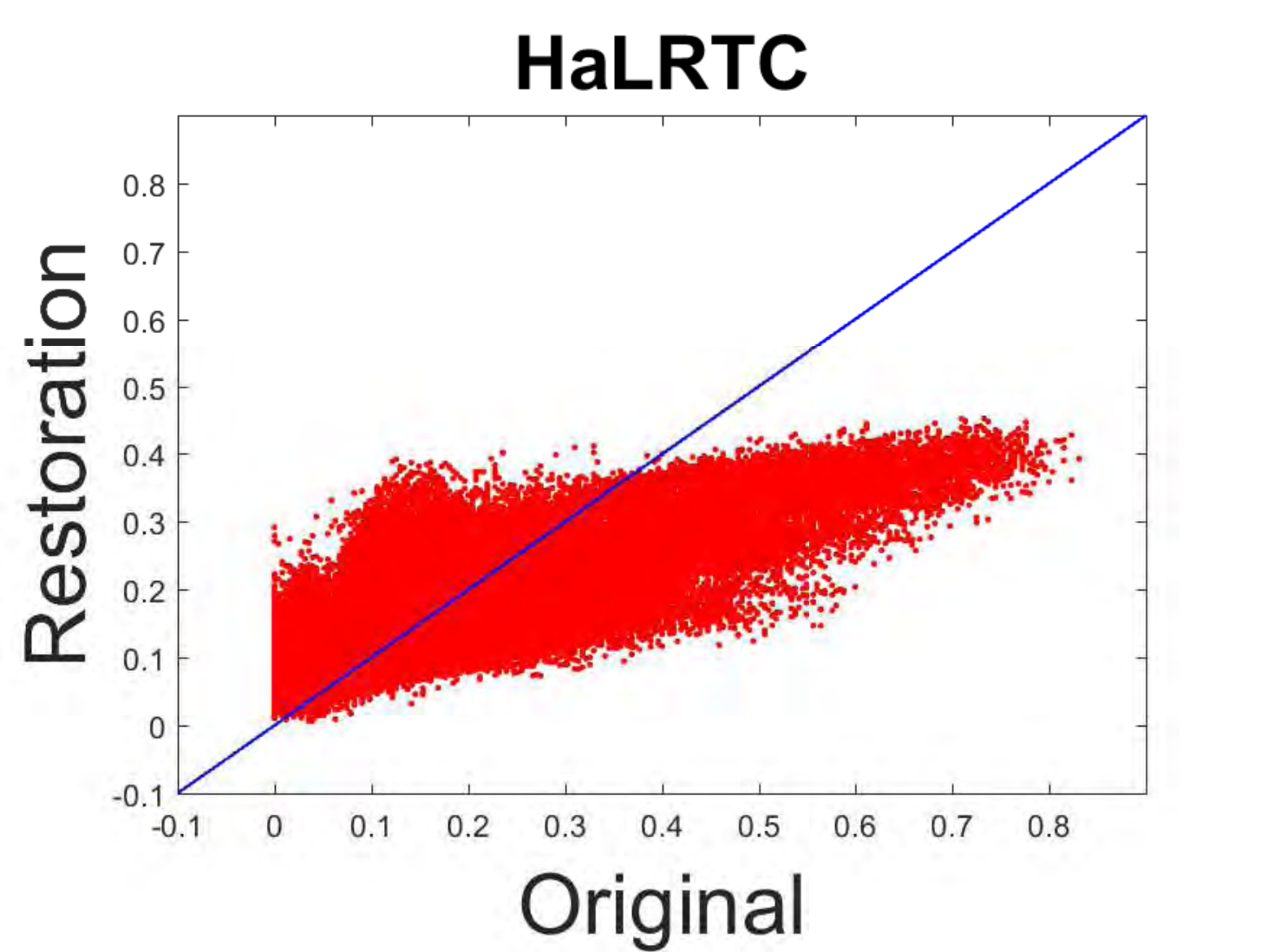}
		\includegraphics[width=0.23\textwidth]{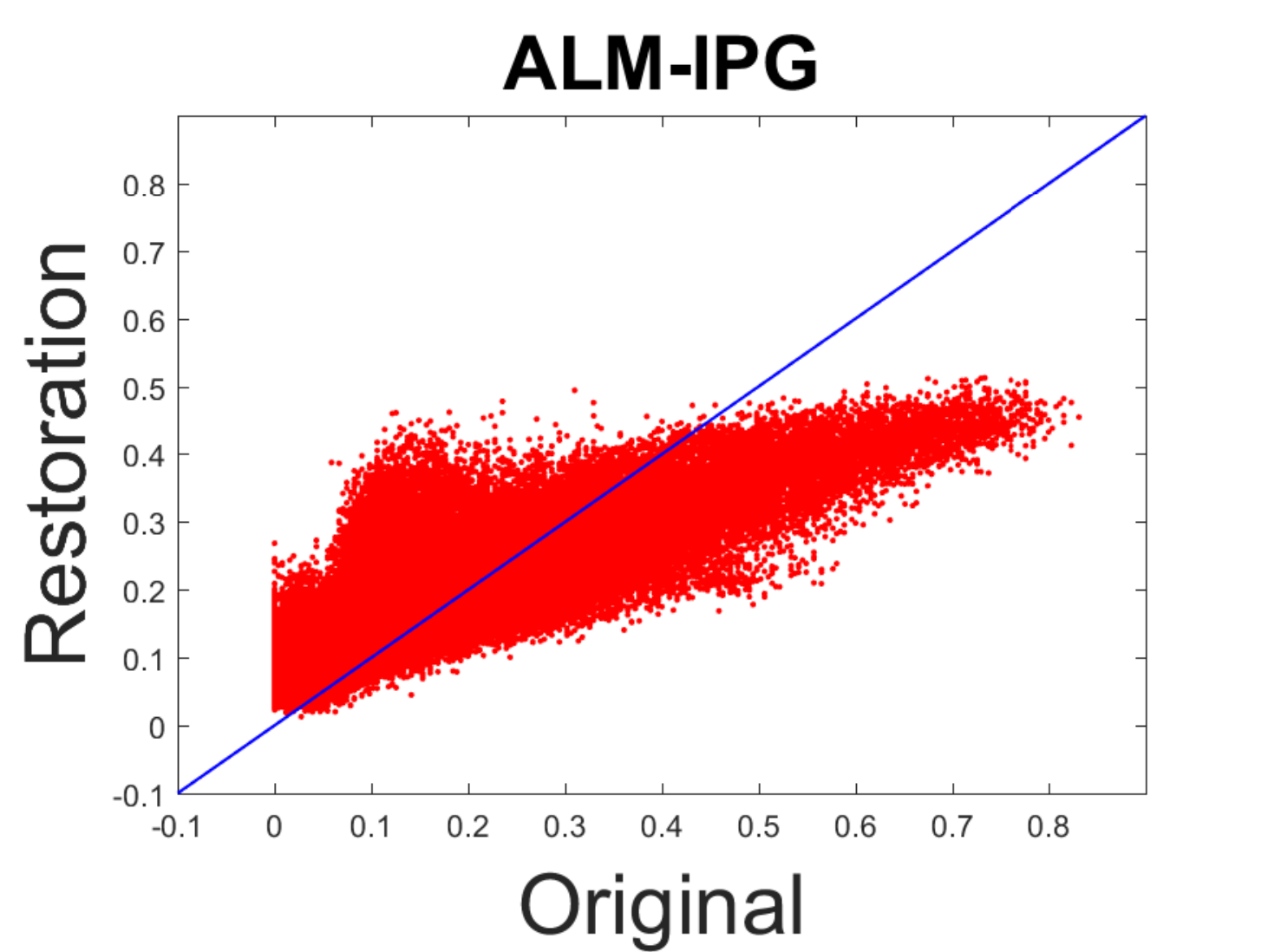}
		\includegraphics[width=0.23\textwidth]{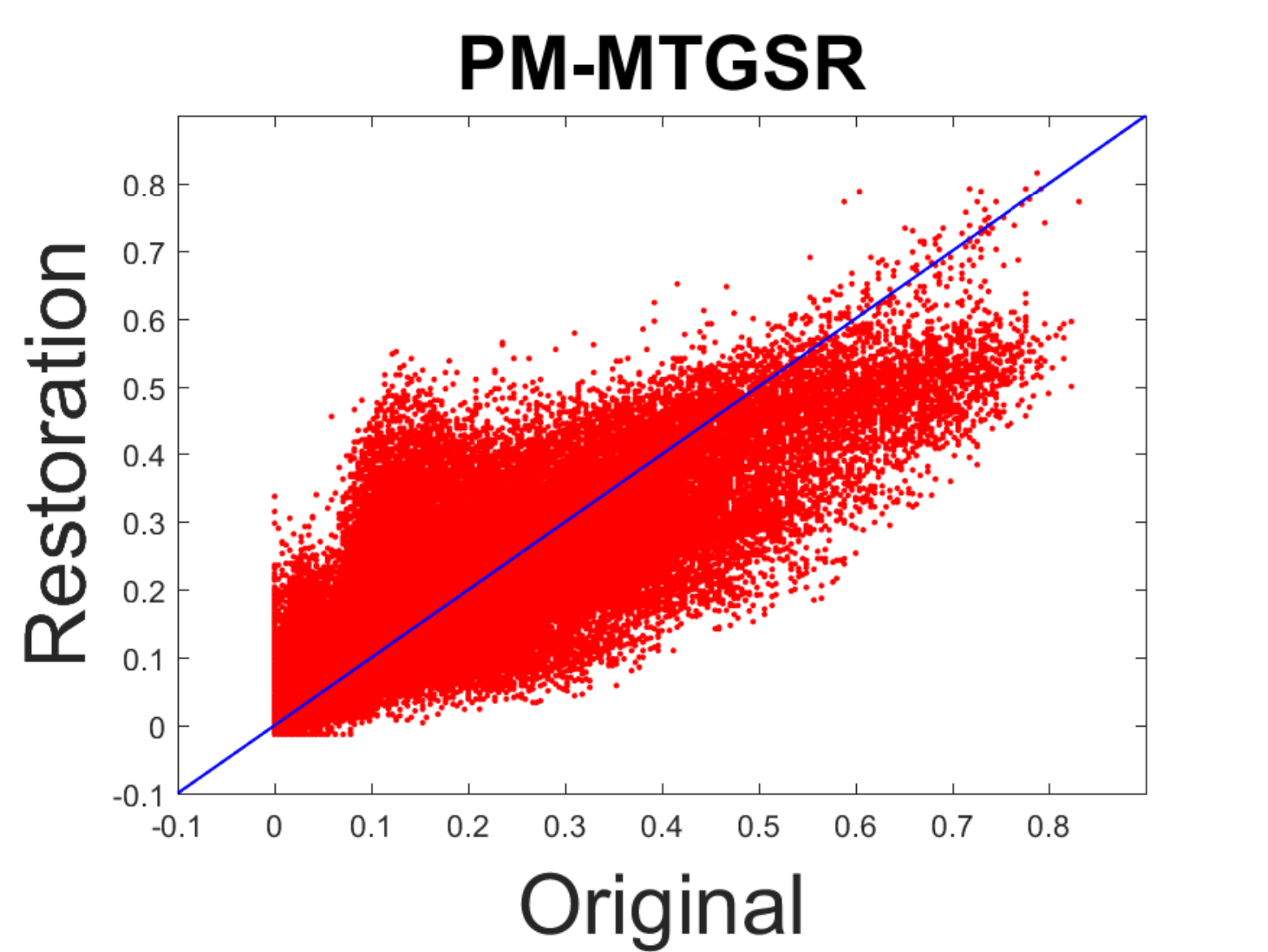}
		\includegraphics[width=0.23\textwidth]{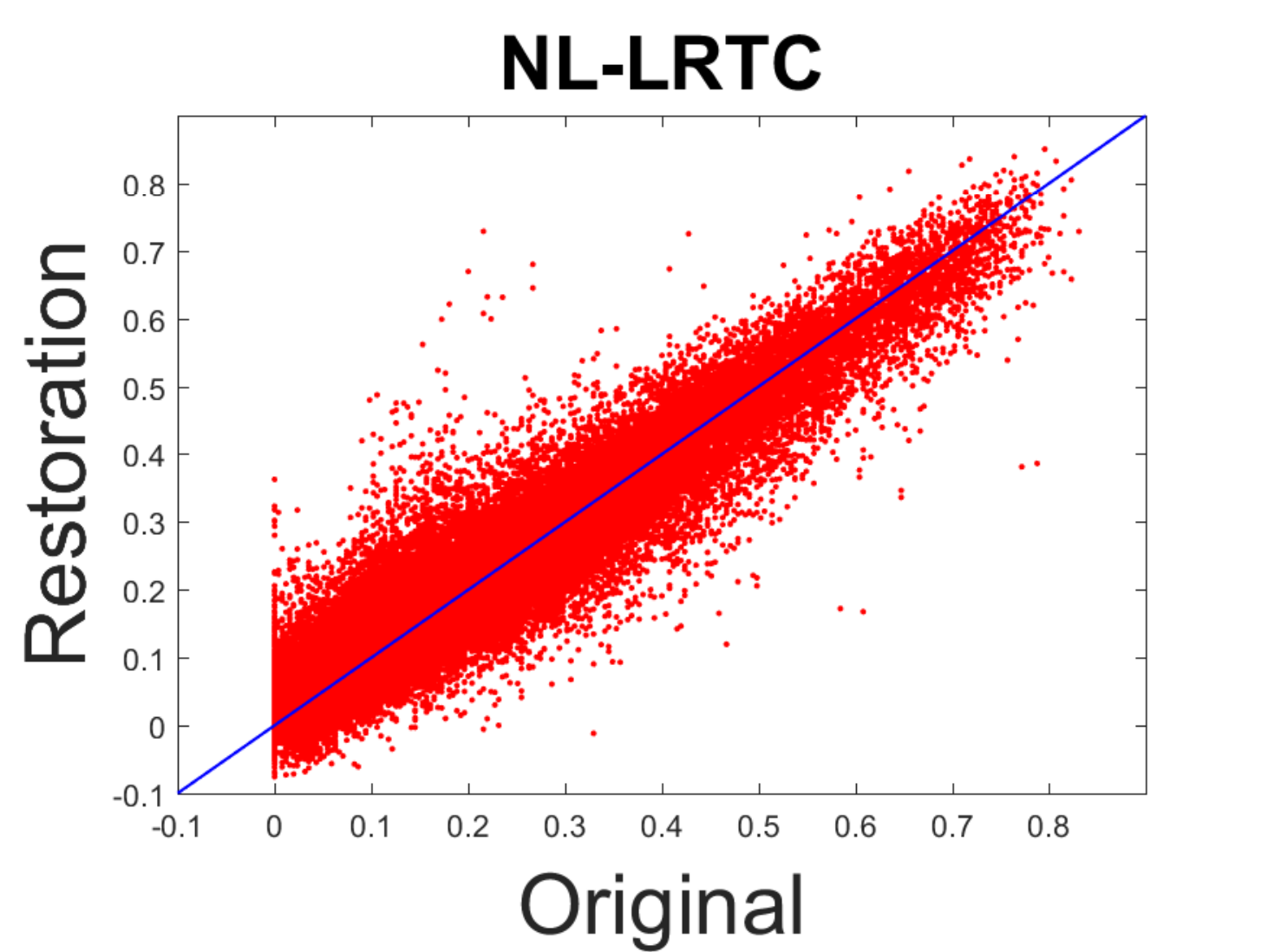}

		\rotatebox{90}{\footnotesize \textbf{Exp. 6}}
		\includegraphics[width=0.23\textwidth]{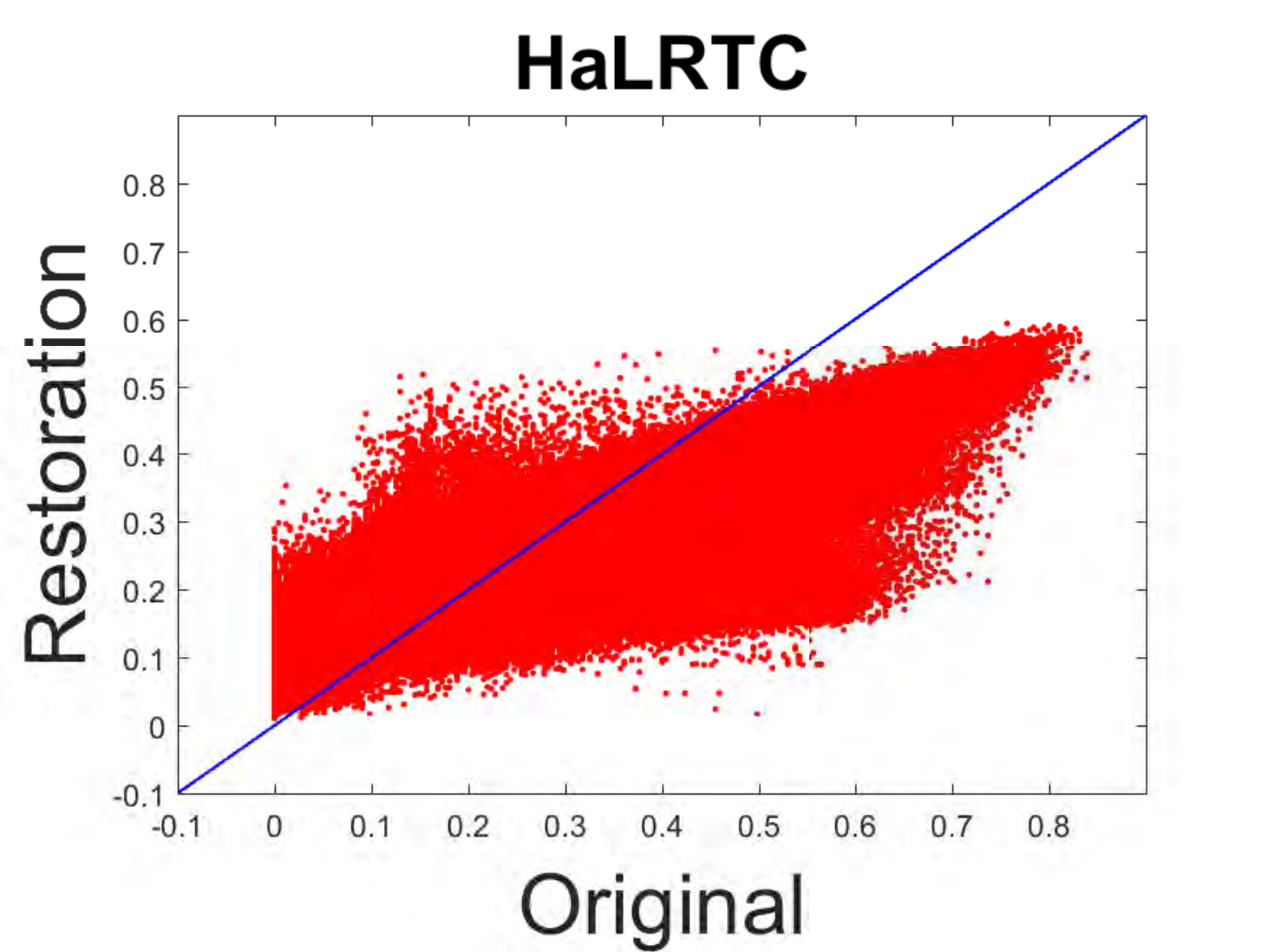}
		\includegraphics[width=0.23\textwidth]{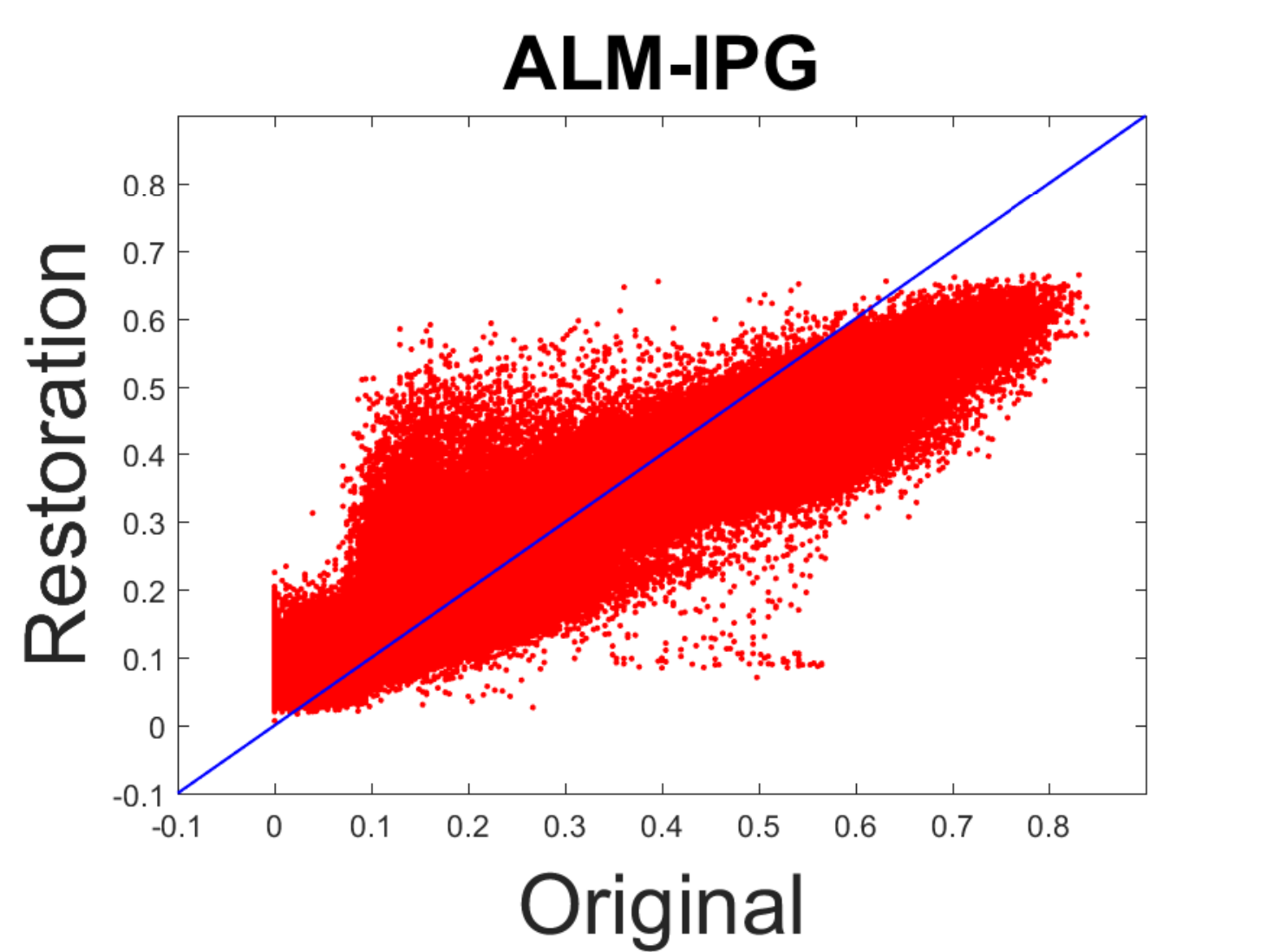}
		\includegraphics[width=0.23\textwidth]{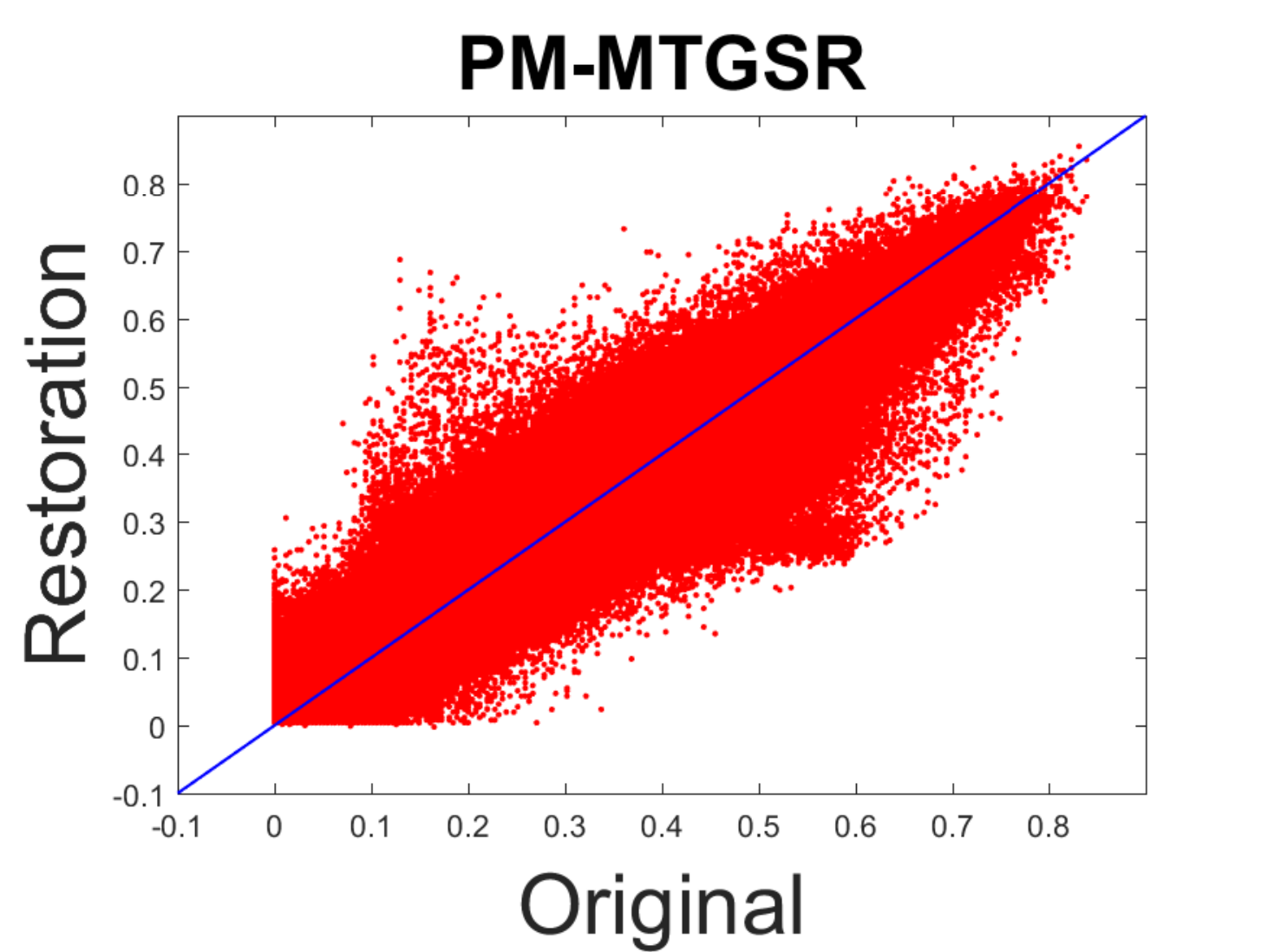}
		\includegraphics[width=0.23\textwidth]{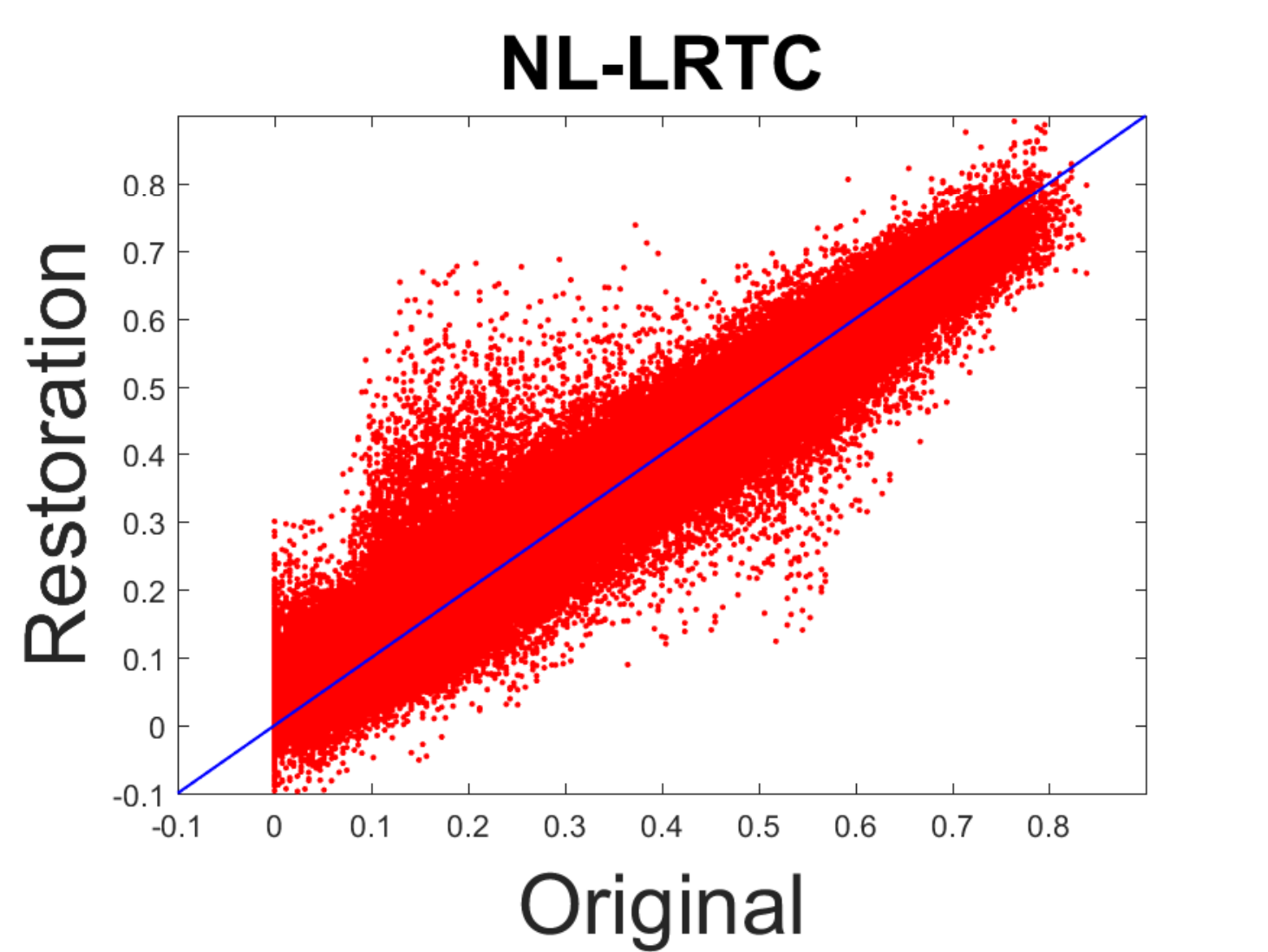}
		\caption{\footnotesize  Scatter plots between the original and restoration pixels in the missing areas for Exps. 1--6. }
		\label{Fig:ScatterplotsCloud}
	\end{figure*}

	To further contrast the original and reconstructed pixel values, scatter plots between the original and restoration pixels in the missing areas for Exps. 1--3 are shown in Fig. \ref{Fig:ScatterplotsCloud}.  The scatter plots results of Exp. 1 are visually better than those of Exp. 2 and Exp. 3 because the difference between the target image and reference images in Exp. 1 are slighter than those of Exp. 2 and Exp. 3. The points on the scatter plot of NL-LRTC show a more compact distribution than those of HaLRTC, ALM-IPG, and PM-MTGSR, but the difference between PM-MTGSR and NL-LRTC is not obvious for Exp. 1. For Exps. 2 and 3, the scatter plots of HaLRTC, ALM-IPG, and PM-MTGSR are obviously worse than those of ours, especially for the larger pixel values. This is consistent with the visual results of Exps. 2 and 3 shown in Figs. \ref{Fig:SimulateCloudResult} and \ref{Fig:SimulateCloudZoomResult} where the reconstructed areas by HaLRTC, ALM-IPG, and PM-MTGSR are noticeably different from the known areas.

	Furthermore, one more simulated Beijing data shown in Fig. \ref{Fig:BJdata} is tested to demonstrate the effectiveness of NL-LRTC (denoted as Exp. 4). In this experiment, we test how the four methods perform when the temporal difference between cloud-contained data and other temporal cloud-free data is not too large. ``BJ092016'' is adopted as the cloud-contained data. The results of band 6 reconstructed by the four methods for Exp. 4 are shown in Fig. \ref{Fig:SimulateCloudResult}. The Exp. 4 shows that the results by all the four methods are almost visually similar. One can get a similar conclusion from the scatter plots shown in Fig. \ref{Fig:ScatterplotsCloud}. The points on the scatter plots of all the four methods are mostly distributed surrounding the blue diagonal, but there are also a few points deviating from the diagonal line.  In this experiment, we can see that the results obtained by ALM-IPG are improved compared to Exp. 1--3 because ALM-IPG mainly studies the smoothness of adjacent temporal data. In conclusion, when the temporal difference is not large, all the four methods perform in a similarly effective manner.

	Besides the cloud removal experiments, the destriping experiments (Exps. 5 and 6) are also conducted using ``Image 3'' and ``Image 4'' shown in Fig. \ref{Fig:SimulateMUdata}: the first column is the simulated corrupted data, and the other three columns are the supplementary data. Some regular diagonal and random vertical stripes are manually added into ``Image 3'' and ``Image 4'', respectively, as shown in Fig. \ref{Fig:SimulateCloudResult}. It is obvious that the results by NL-LRTC are the best visually. The scatter plots comparing the original and reconstructed pixel values in the missing areas for Exps. 5 and 6 are shown in Fig. \ref{Fig:ScatterplotsCloud}. The points on the scatter plot of HaLRTC and ALM-IPG results obviously deviate from the diagonal. The points for PM-MTGSR and NL-LRTC are better, but the points of PM-MTGSR distributed in the direction orthogonal to the diagonal line are wider than those of NL-LRTC. That means the scatter plots of our method are the best. Overall, the proposed method obtains the best results for the removal of stripes.

	The quantitative comparison is shown in Tab. \ref{Tab:Simulate}. The table shows that all the four methods can recover better results compared to the corrupted image itself. PSNR and SSIM evaluate the recovered image by comparing with the ground truth image. By analyzing the PSNR and SSIM results, HaLRTC obtains the worst results for all the experiments. This is because HaLRTC only takes advantage of the low-rankness of the observed data. ALM-IPG obtains better results than HaLRTC, because it considers the low-rankness and temporal continuous property simultaneously. However, it assumes the smoothness of adjacent temporal images, the results depend on the similarity of the adjacent temporal images. This algorithm is suited to process high-temporal-resolution images, such as videos. PM-MTGSR obtains better results than HaLRTC and ALM-IPG because it makes use of the patch similarity. NL-LRTC also takes the patch similarity into consideration and makes the best use of low-rankness of the three different dimensions. Thus NL-LRTC obtains the best results. For Exp. 4, since the difference between the cloud-contained and reference data is not great, the result of ALM-IPG is better than those of Exps. 1--3. Although the PSNR value of PM-MTGSR is higher than that of NL-LRTC, the difference is slight. Moreover, the SSIM value of NL-LRTC is higher than that of PM-MTGSR. The Q, AG, and BIQA results also show NL-LRTC obtains the best results for Exps. 1--4. The elapsed times of PM-MTGSR and NL-LRTC are longer than those of HaLRTC and ALM-IPG because they are patch-based methods in which searching similar patches costs much more time. NL-LRTC is much faster than PM-MTGSR because PM-MTGSR processes multispectral images band by band, while NL-LRTC reconstructs the missing areas of all band at the same time. In conclusion, the proposed method performs the best for cloud and stripe removal when the temporal difference is large and can also obtain promising results when the temporal difference is slight.

	\begin{table*}[!t]
		\caption{ Quantitative results comparison for Exps. 1--6.  }
		\begin{center}
			\begin{tabular}{l | c c c c c c | c c c c c c}
				\hline\hline
				\multirow{2}{*}{Methods} &\multicolumn{6}{c}{Exp. 1}  & \multicolumn{6}{|c}{Exp. 2} \\ \cline{2-7}  \cline{8-13}
				&   SSIM  & PSNR & Q & AG & BIQA  & Time(s)        &  SSIM   & PSNR  & Q & AG & BIQA    & Time(s)\\ \hline
				Corrupted   & 0.8107 & 9.35   &  - &  -   &  -   &  -                    & 0.6237 & 6.54    & - & - & - &  - \\
				HaLRTC      & 0.9660 & 36.78 & 0.0230 & 0.0250 & 0.0036  & 39.53             &  0.8357 & 24.36 & 0.0342 & 0.0303 & 0.0075 & 44.45\\
				ALM-IPG      & 0.9708 & 38.05  & 0.0232 & 0.0252& 0.0036  & 49.94             & 0.8575 & 25.81  & 0.0345 & 0.0318 & 0.0074 & 41.34 \\
				PM-MTGSR  & 0.9722 & 38.40  & 0.0230 & 0.0260 & 0.0036 & 4028.05  & 0.8594 & 25.85  &0.0373 & 0.0330 & 0.0080  &  4032.23 \\
				NL-LRTC & \textbf{0.9769} & \textbf{38.95}  & \textbf{0.0245} & \textbf{0.0266} & \textbf{0.0038} & 478.73      & \textbf{0.8975} & \textbf{27.32}  & \textbf{0.0443} & \textbf{0.0375} & \textbf{0.0090} &  1216.38\\ \hline

				\multirow{2}{*}{Methods} &\multicolumn{6}{c}{Exp. 3}  & \multicolumn{6}{|c}{Exp. 4} \\ \cline{2-7}  \cline{8-13}
				&   SSIM  & PSNR & Q & AG & BIQA  & Time(s)         &  SSIM   & PSNR  & Q & AG & BIQA    & Time(s)\\ \hline
				Corrupted   & 0.5002 & 4.97    & - & - & -  &  -                   & 0.4851 & 22.02  & - & - & -    &  -\\
				HaLRTC     & 0.8117 & 25.39 & 0.0235 & 0.0271 & 0.0052 & 41.16       &  0.9664 & 39.09  & 0.0092 & 0.0080 & 0.0014  & 20.05\\
				ALM-IPG      & 0.8341 & 26.26  & 0.0260 & 0.0291 &  0.0051 & 41.00  & 0.9677 & 40.41  & \textbf{0.0095} & 0.0080 & 0.0013 & 18.81  \\
				PM-MTGSR  & 0.8419 & 26.67  & 0.0286 & 0.0305 & 0.0058 & 4056.01  & 0.9704 & \textbf{41.17}  & 0.0086 & 0.0082 & 0.0013 &  2391.16\\
				NL-LRTC & \textbf{0.8629} & \textbf{27.18} & \textbf{0.0403} & \textbf{0.0351} & \textbf{0.0068} & 1274.66       & \textbf{0.9707} & 41.13 & 0.0085 & \textbf{0.0085} & \textbf{0.0017}  &  345.75\\ \hline

				\multirow{2}{*}{Methods} &\multicolumn{6}{c}{Exp. 5}  & \multicolumn{6}{|c}{Exp. 6} \\ \cline{2-7}  \cline{8-13}
				&   SSIM  & PSNR & Q & AG & BIQA  & Time(s)         &  SSIM   & PSNR  & Q & AG & BIQA    & Time(s)\\ \hline
				Corrupted   & 0.6848 & 21.70    & - & - & -  &  -                & 0.4920 & 15.67   & & &   &  - \\
				HaLRTC      & 0.9308 & 29.90 & 0.0399 & 0.0363 & \textbf{0.0110}  & 37.83           & 0.8749  & 24.23 & \textbf{0.2298} & \textbf{0.0586} & \textbf{0.0151} &43.35\\
				ALM-IPG      & 0.9341 & 30.75  & \textbf{0.0407} & 0.0363 & {0.0106}  & 41.97             & 0.9285 & 27.78  & {0.0873} & {0.0505} & {0.0081} & 41.66 \\
				PM-MTGSR  & 0.9407 & 30.69  & 0.0356 & \textbf{0.0367} & {0.0106}  &  4148.10     & 0.9381 & 28.73  & 0.0694 & 0.0466 & 0.0068  &  4169.75 \\
				NL-LRTC & \textbf{0.9846} & \textbf{36.43}  & 0.0299 & 0.0355 & 0.0073  &  1428.18     & \textbf{0.9679} & \textbf{31.71} & 0.0523 & 0.0483 & 0.0064  &  1338.02\\ \hline
			\end{tabular}
		\end{center}
		\label{Tab:Simulate}
	\end{table*}

\begin{figure*}[ht]
		\begin{center}
			\rotatebox{90}{\footnotesize \textbf{Exp. 7}}
			\includegraphics[width=0.18\textwidth]{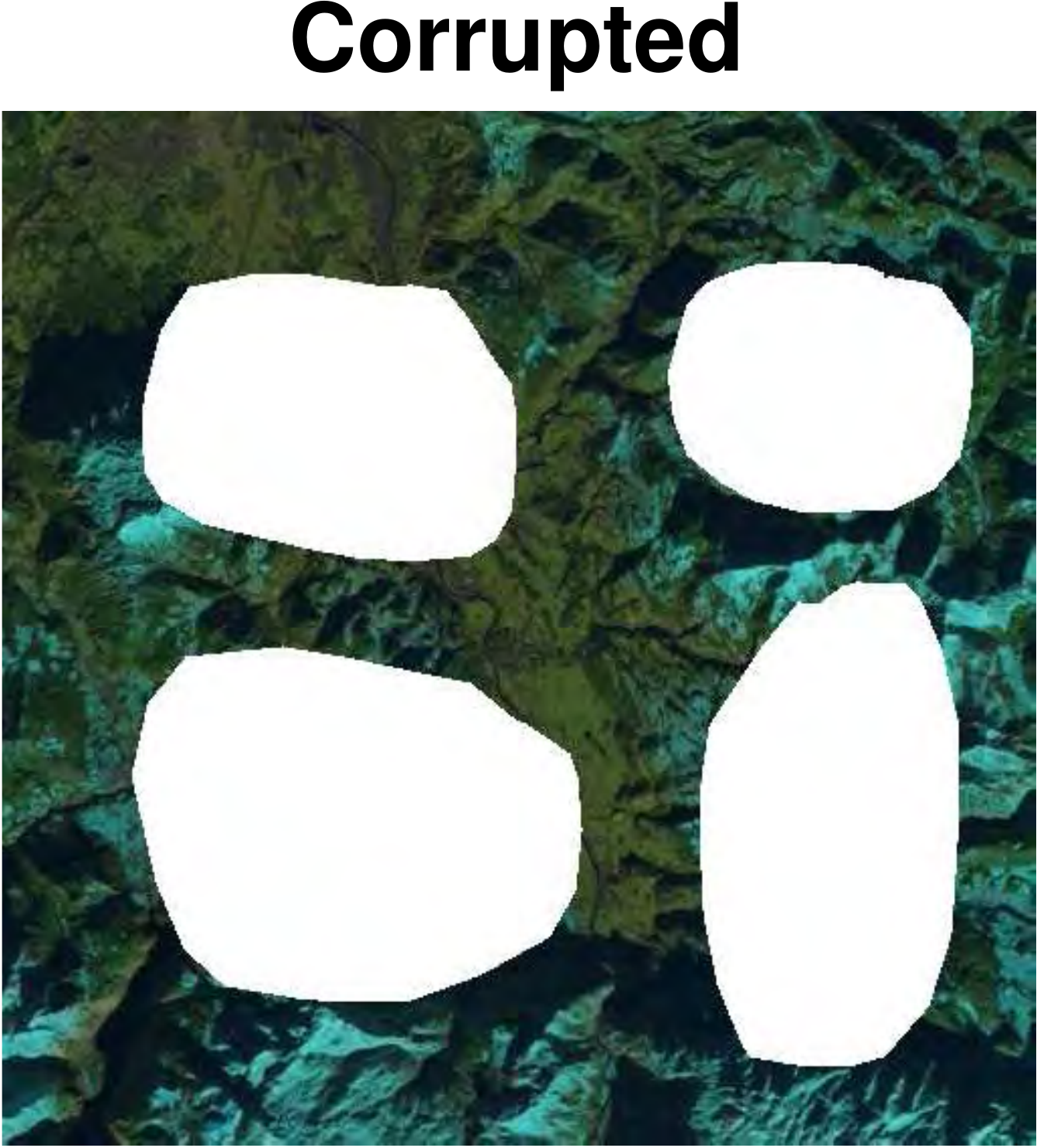}
			\includegraphics[width=0.18\textwidth]{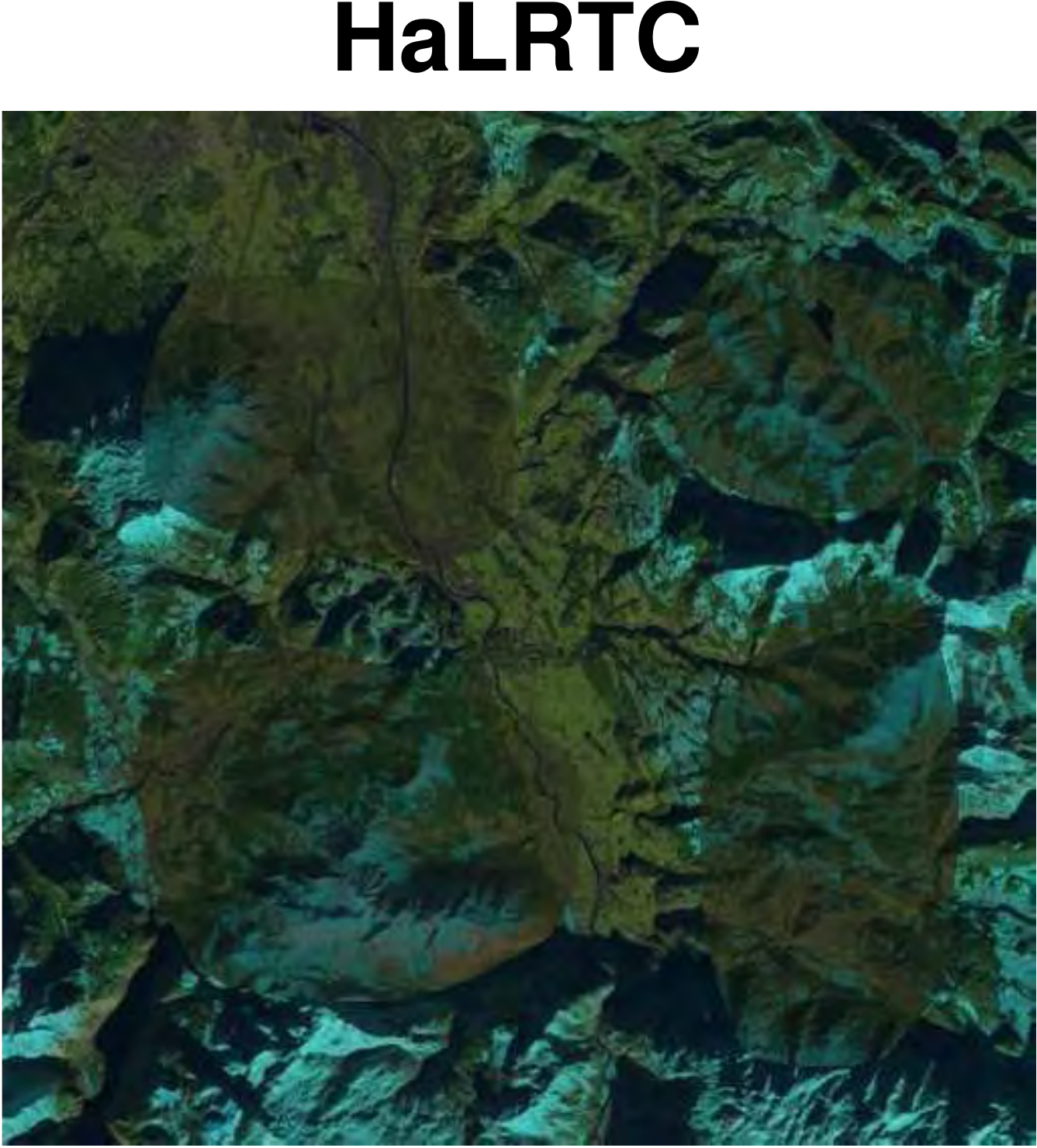}
			\includegraphics[width=0.18\textwidth]{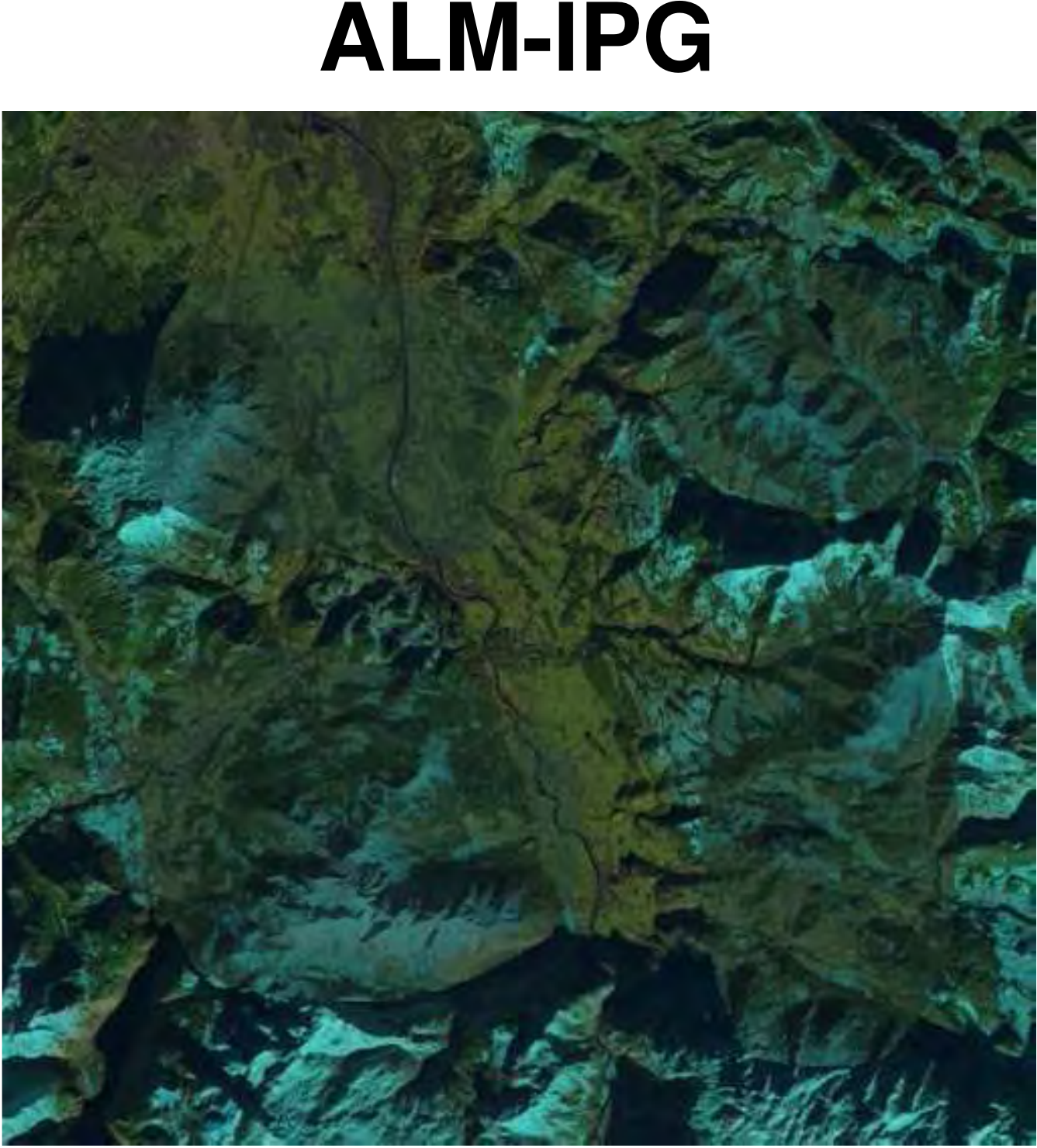}
			\includegraphics[width=0.18\textwidth]{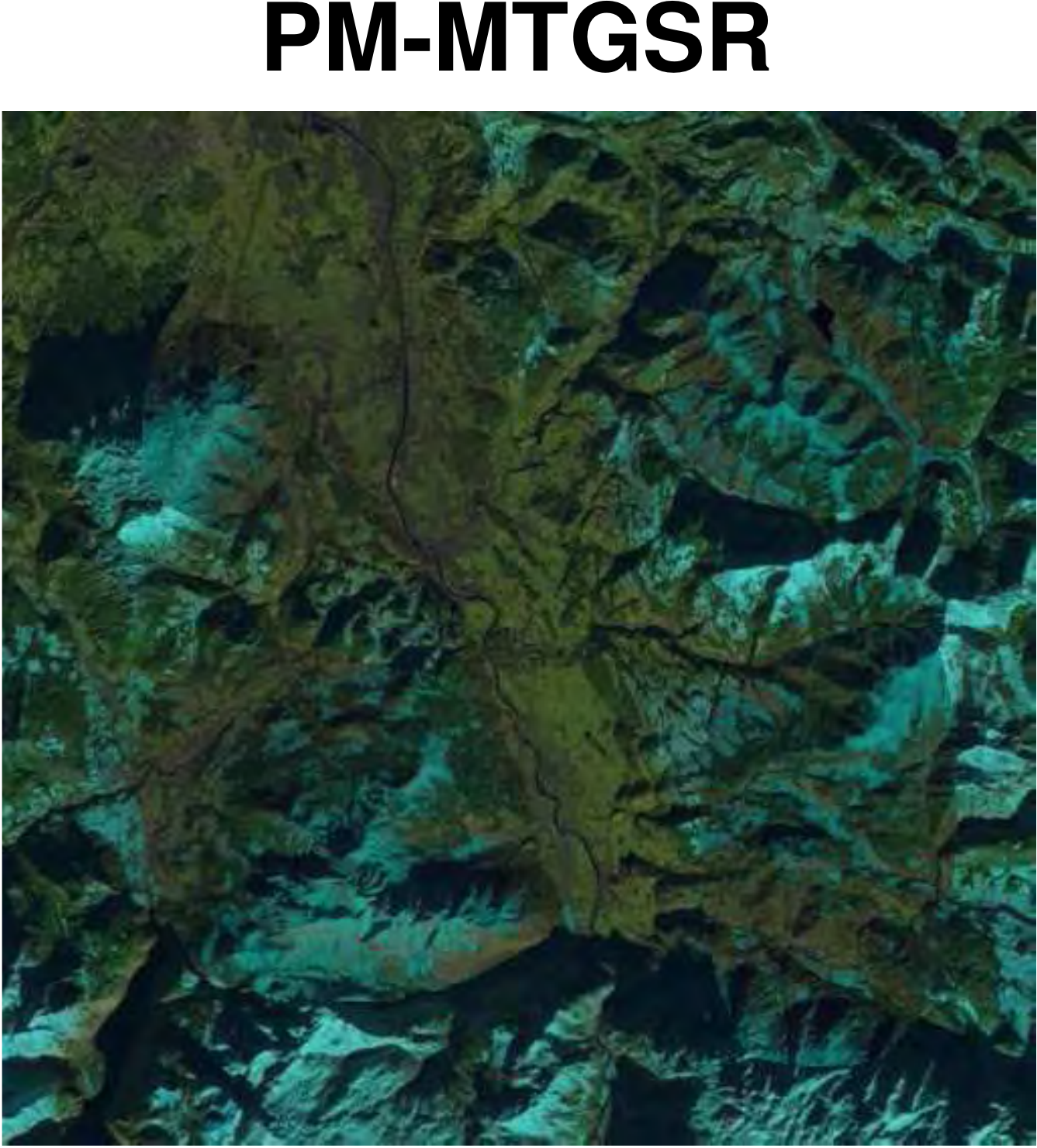}
			\includegraphics[width=0.18\textwidth]{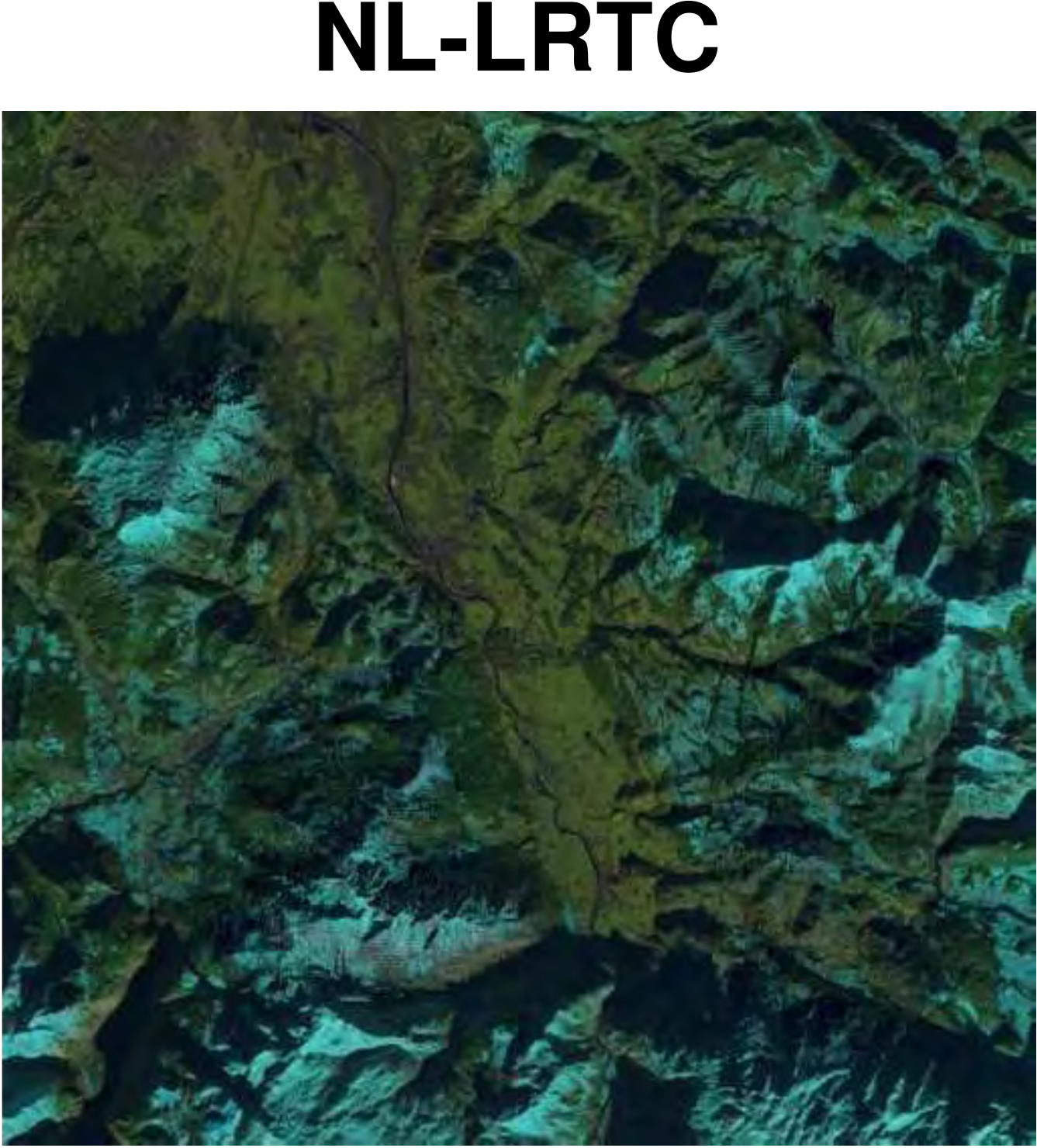}\\[0.05cm]
			\rotatebox{90}{\footnotesize \textbf{Exp. 8}}
			\includegraphics[width=0.18\textwidth]{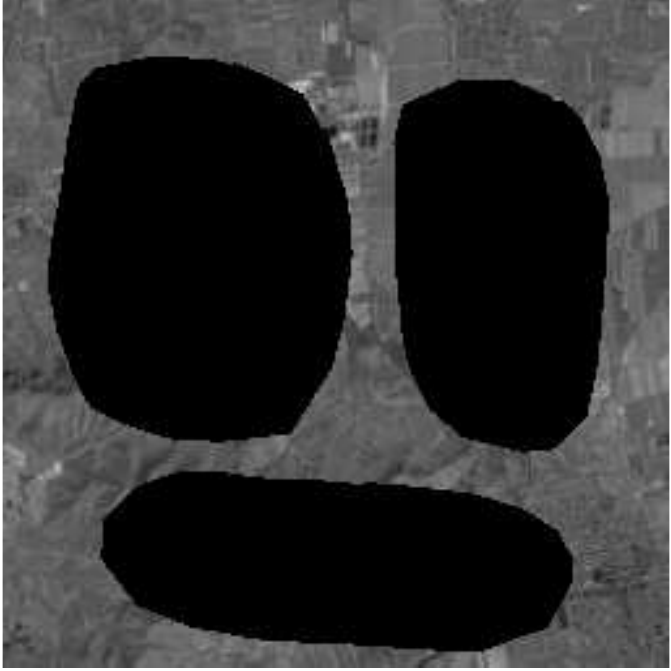}
			\includegraphics[width=0.18\textwidth]{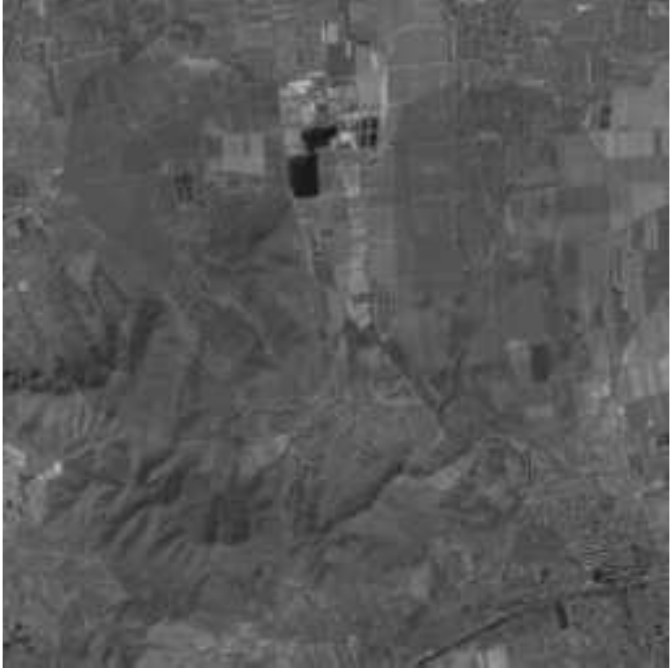}
			\includegraphics[width=0.18\textwidth]{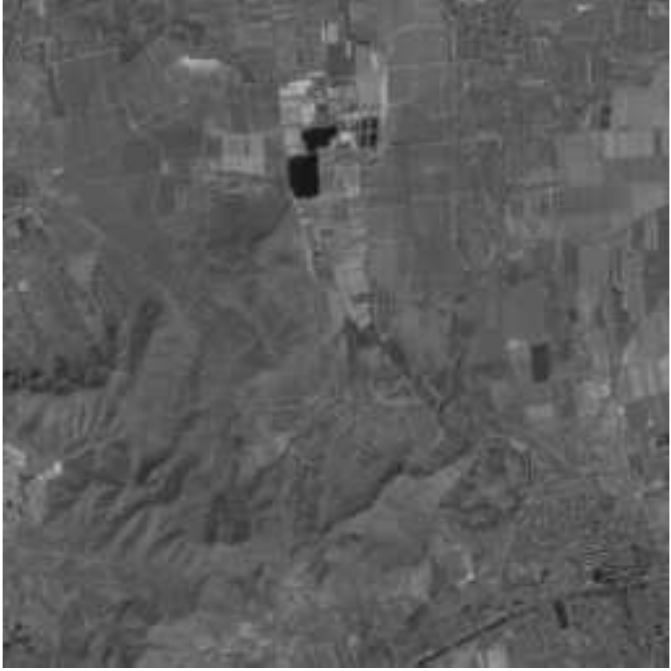}
			\includegraphics[width=0.18\textwidth]{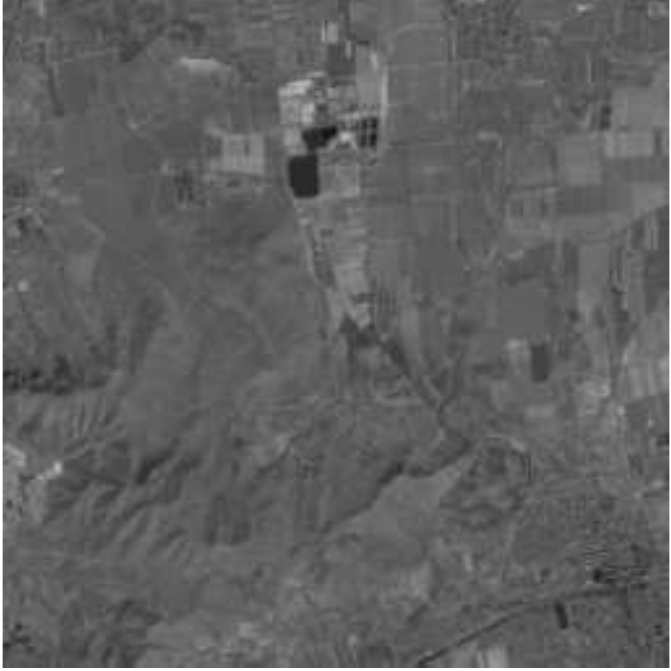}
			\includegraphics[width=0.18\textwidth]{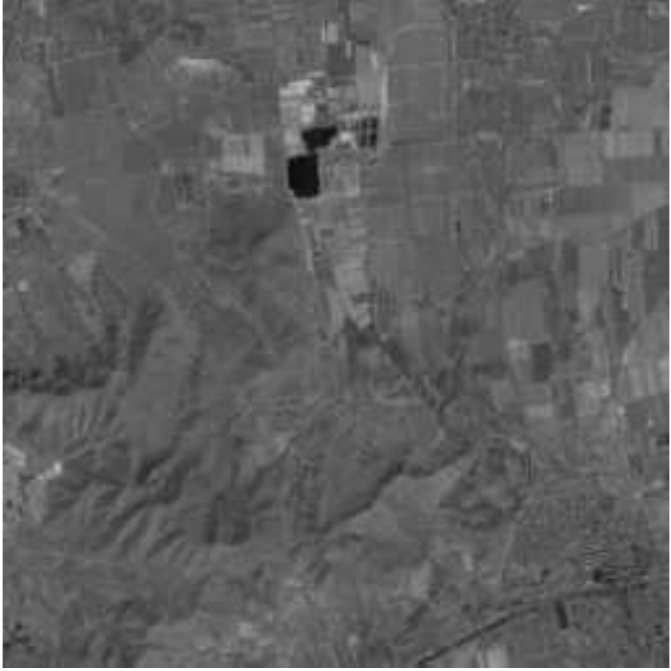}
		\end{center}
		\caption{\footnotesize Cloud removal results for ``Image 3'' and Beijing data containing more than one piece of cloud. Exp. 7 is for ``Image 3'' and Exp. 8 is for Beijing data. }
		\label{Fig:SimulateMultiCloud}
	\end{figure*}

	Next, two simulated data containing more than one piece of cloud are tested. The recovered results are shown in Fig. \ref{Fig:SimulateMultiCloud}, which are for the ``Image 3'' taken by Landsat-8 (Exp. 7) and Beijing data taken by Sentinel-2 (Exp. 8). Exp. 7 and Exp. 8 perform the similar results with Exps. 1--3. and Exp. 4, respectively.  The results recovered by PM-MTGSR and NL-LRTC in Exp. 7 are visually similar, but are visually better than those reconstructed by HaLRTC and ALM-IPG. Exp. 8 shows all the four methods obtain the visually similar results. Tab. \ref{Tab:SimulateMulti} shows that, for both Exps. 7 and 8, NL-LRTC obtains the best quantitative results.

	\begin{table}[!t]
		\caption{Quantitative results comparison for Exps. 7 and 8.  }
		\begin{center}
			\begin{tabular}{c c | c c c c}
				\hline\hline
				& & HaLRTC & ALM-IPG	 & PM-MTGSR & NL-LRTC \\ \hline 
				\multicolumn{1}{c|}{\multirow{2}{*}{Exp. 7}} & SSIM &0.8323 &0.8439 &0.8538 &\textbf{0.8862} \\
				\multicolumn{1}{c|}{} & PSNR &25.32 &26.26 &26.29 &\textbf{27.59} \\ \hline\hline
				\multicolumn{1}{c|}{\multirow{2}{*}{Exp. 8}} & SSIM &0.9618 &0.9645 &0.9692 &\textbf{0.9711} \\
				\multicolumn{1}{c|}{} & PSNR &38.25 &39.81 &40.96 & \textbf{41.61}\\ \hline
			\end{tabular}
		\end{center}
		\label{Tab:SimulateMulti}
	\end{table}

\begin{figure}[ht]
		\centering
		\includegraphics[width=0.24\textwidth]{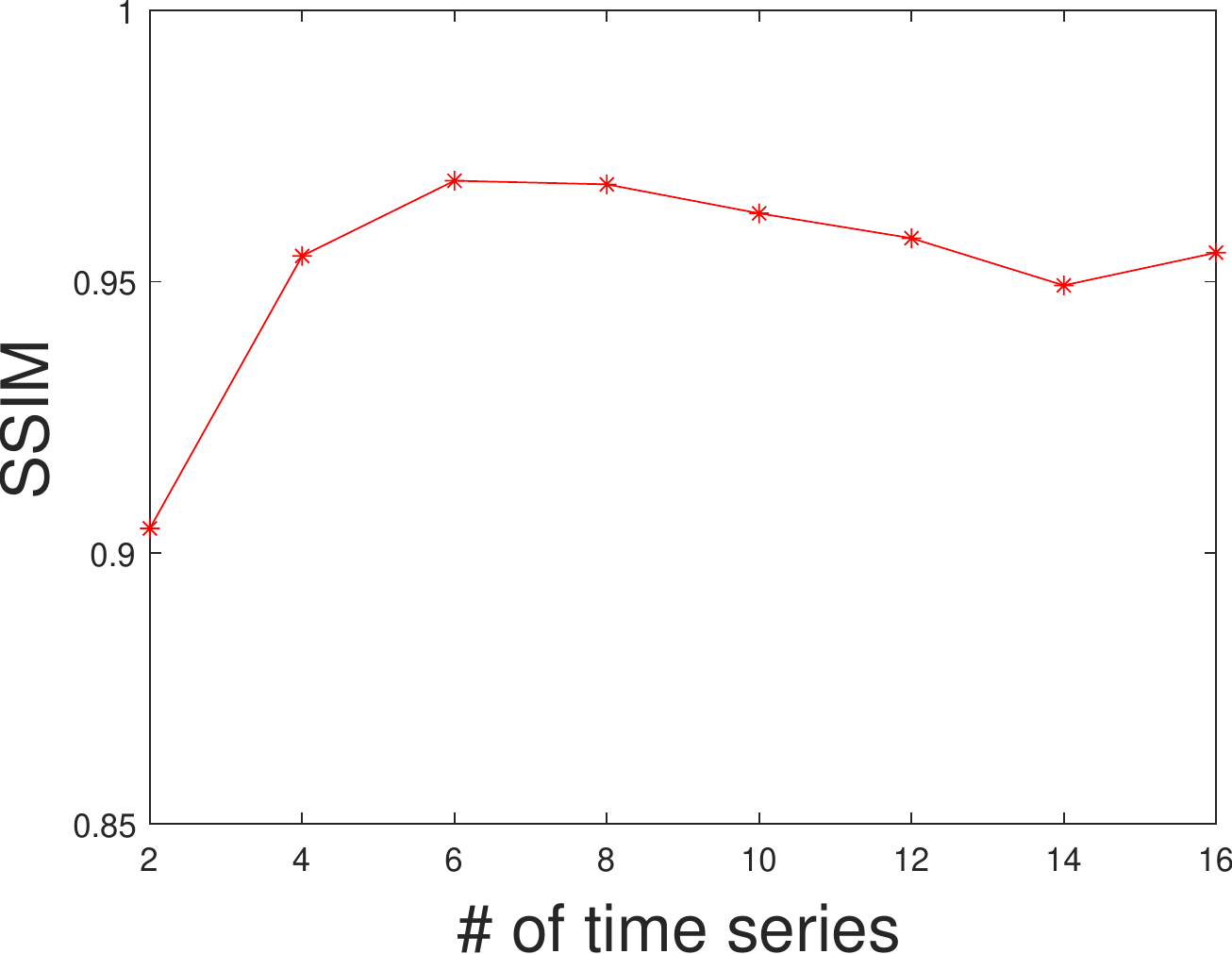}
		\includegraphics[width=0.24\textwidth]{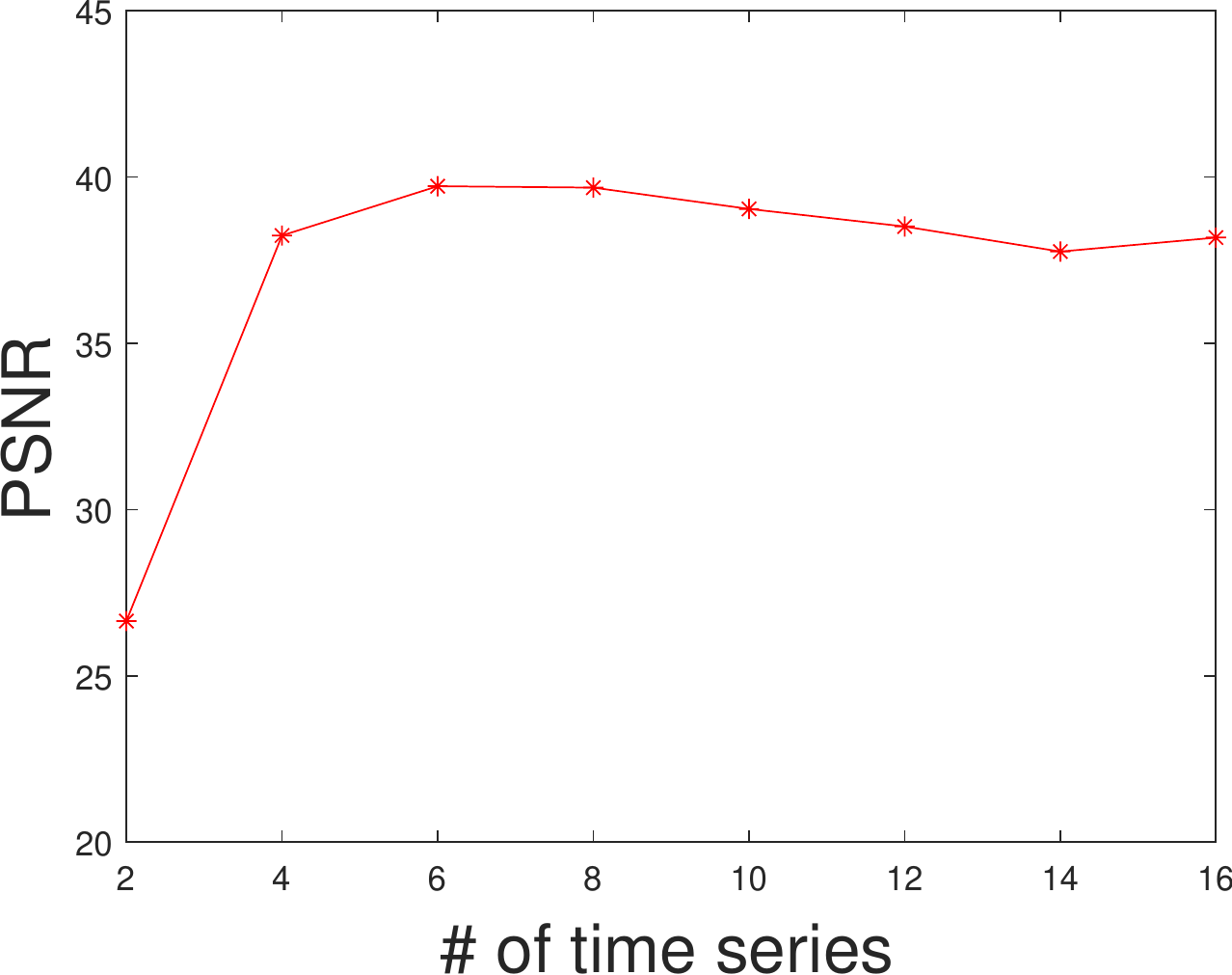}
		\caption{\footnotesize SSIM and PSNR values with respect to the numbers of time series. }
		\label{Fig:timeAnalysis}
	\end{figure}

	At last, we analyze the impact of the number of time series on the reconstruction performance by changing the number ($t=2, 4, \cdots, 16$). The test data were taken over Munich by Landsat-8 on between December, 2014 and April, 2017\footnote{The sixteen test data were taken during four years: 2014 (December), 2015 (January-April, June, August, October), 2016 (April-September), and 2017 (January, February).}. The SSIM and PSNR values with respect to the number of time series are displayed in Fig. \ref{Fig:timeAnalysis}.
		This figure shows that reconstruction results are becoming better with increasing of the number of time series. When the number of time series reach to a large amount, the PSNR and SSIM values reach to the highest with a little fluctuation. This is because more temporal data not only provide more correlative information but also contain more interference information especially when the acquisition times of the cloud-contained and reference data are far form each other.

	\subsection{Real Experiments}
\begin{figure}[ht]
		\centering
		\includegraphics[width=0.22\textwidth]{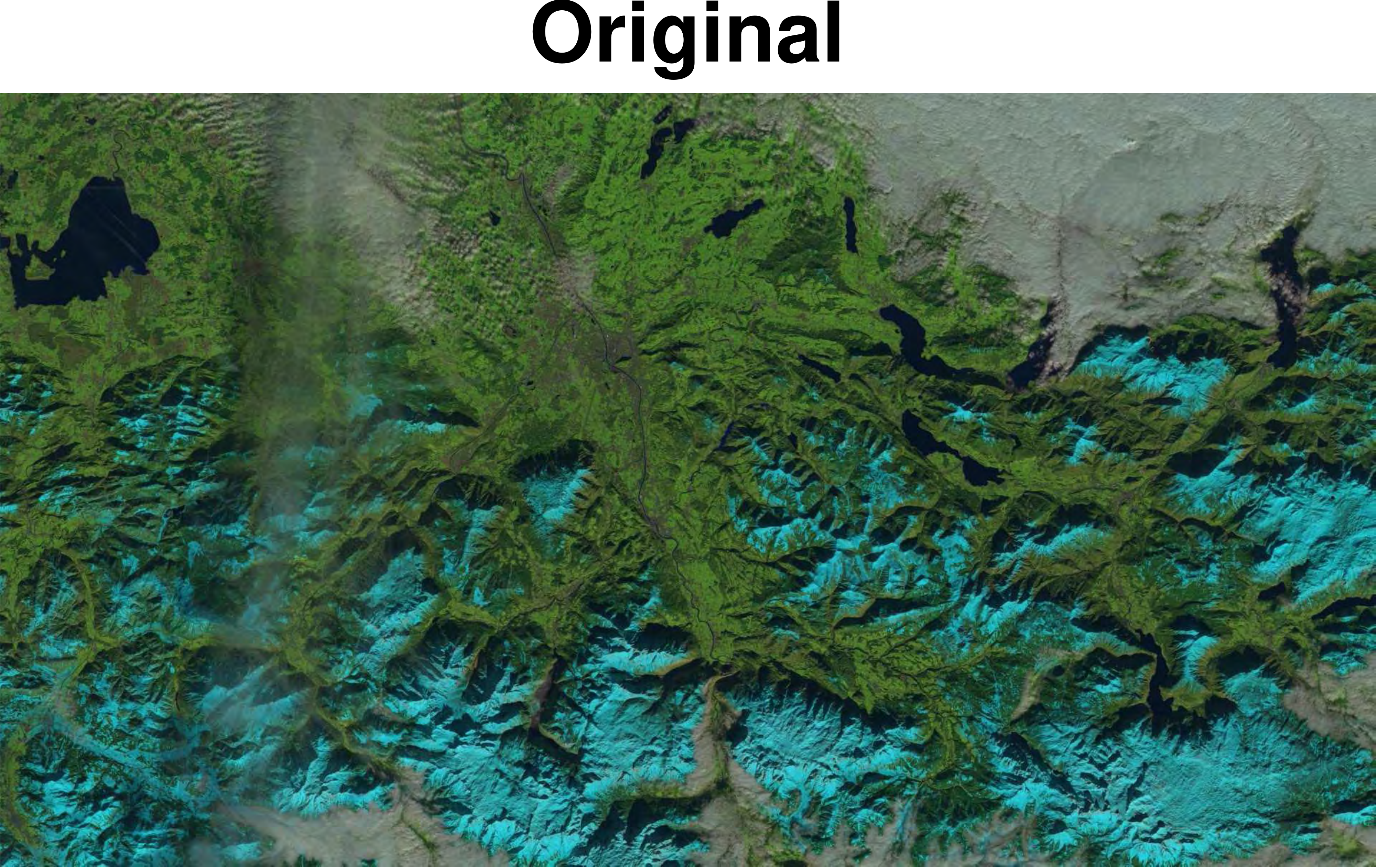}
		\includegraphics[width=0.22\textwidth]{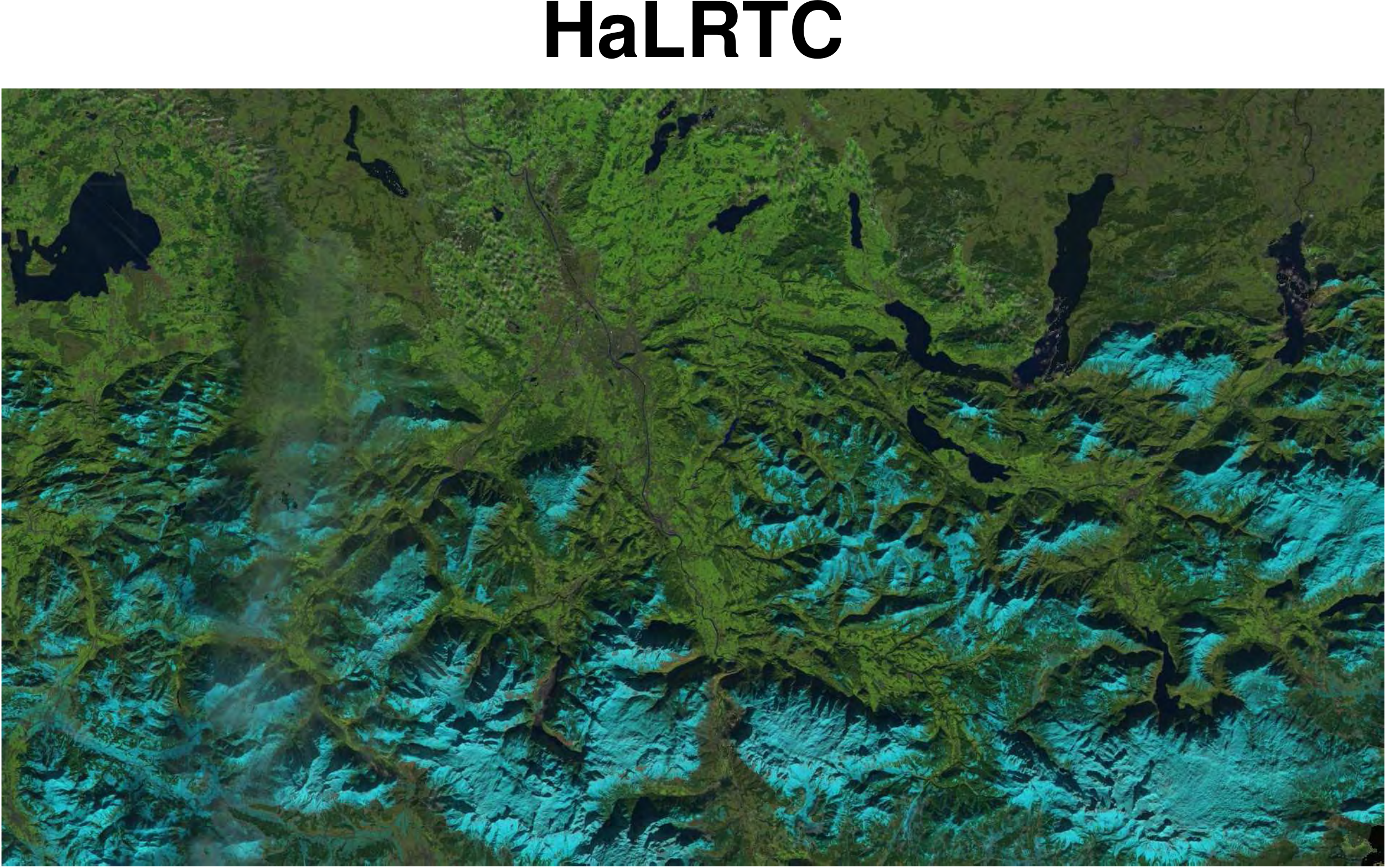} \\[0.05cm]
		\includegraphics[width=0.22\textwidth]{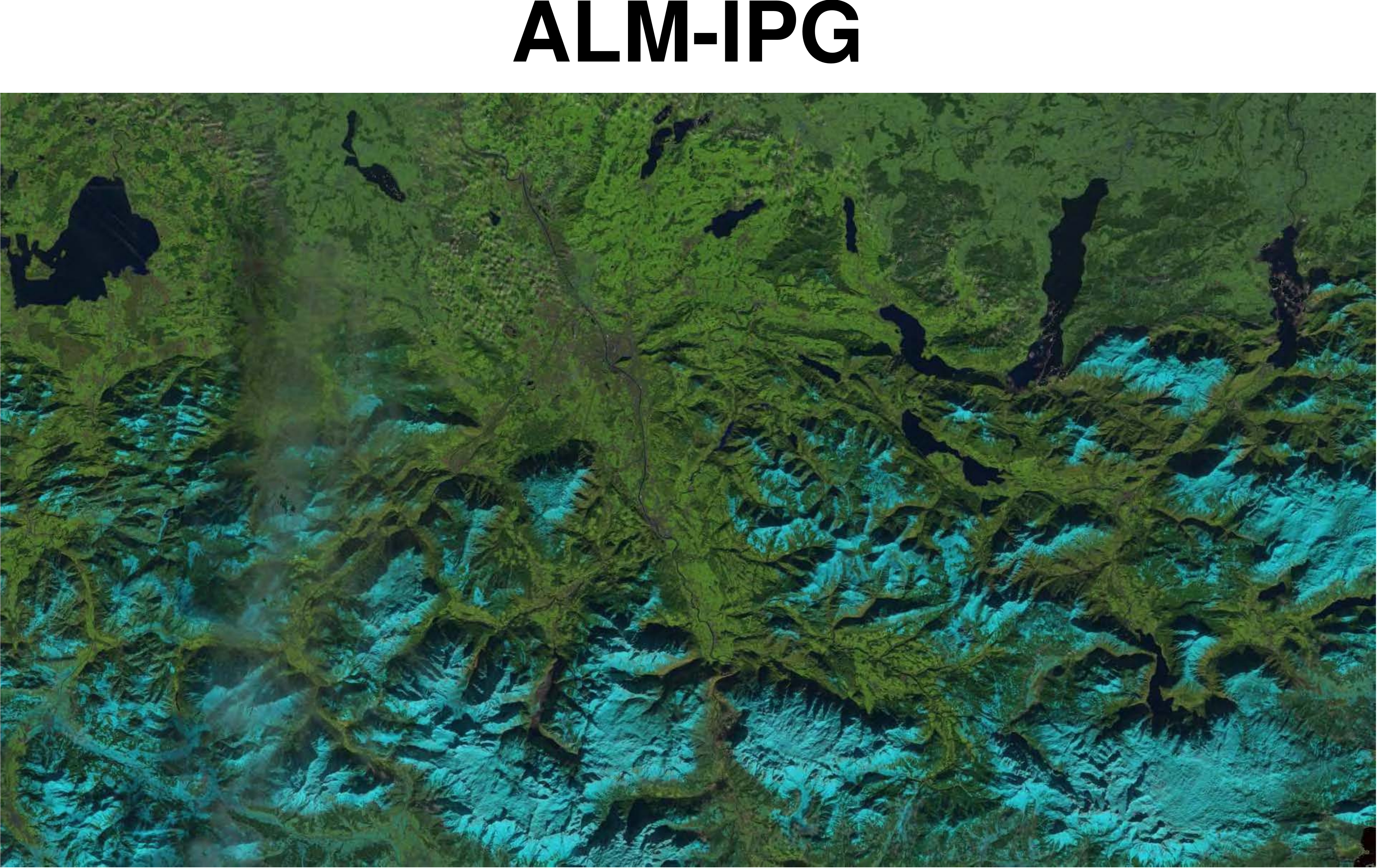}
		\includegraphics[width=0.22\textwidth]{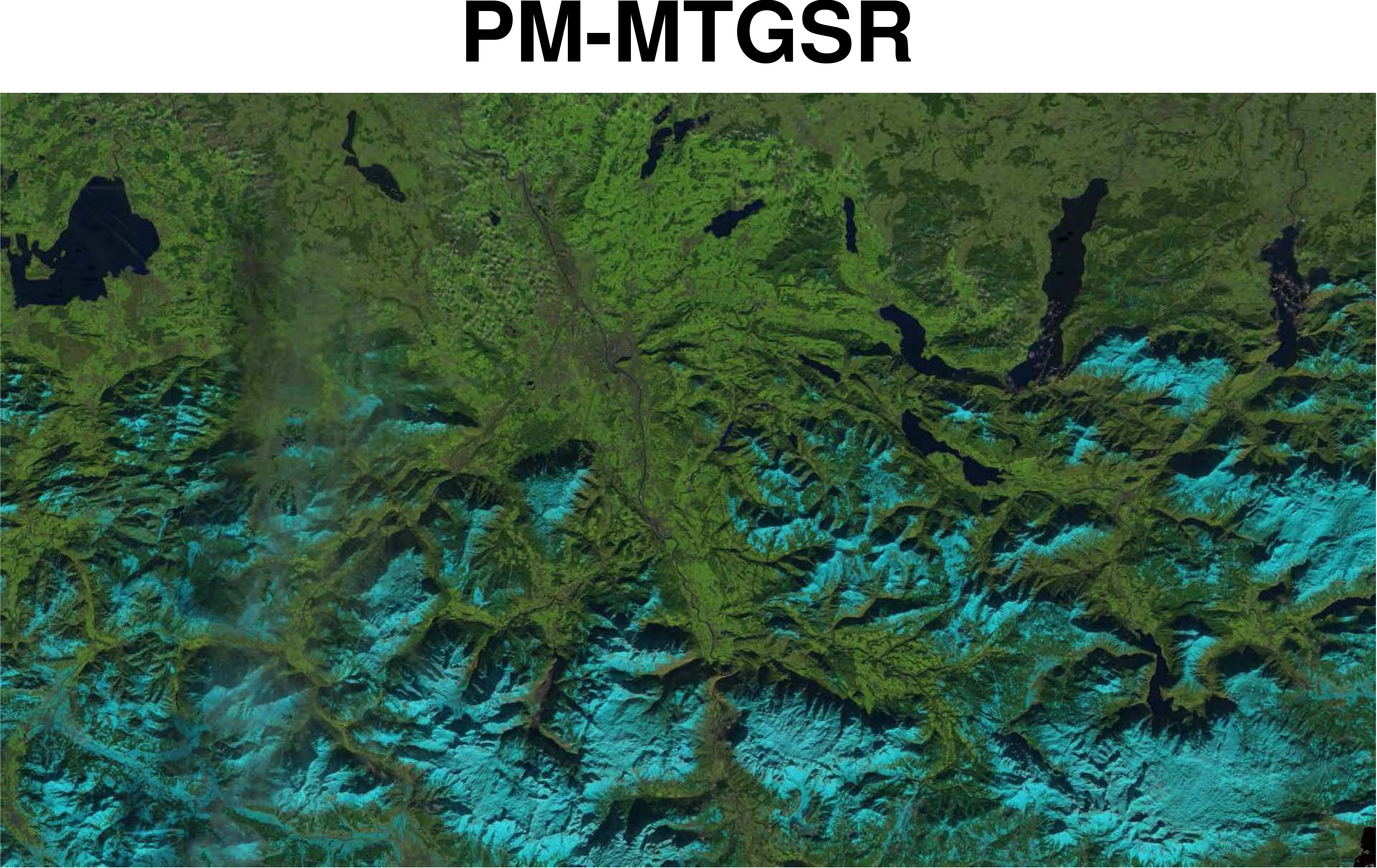} \\[0.05cm]
		\includegraphics[width=0.22\textwidth]{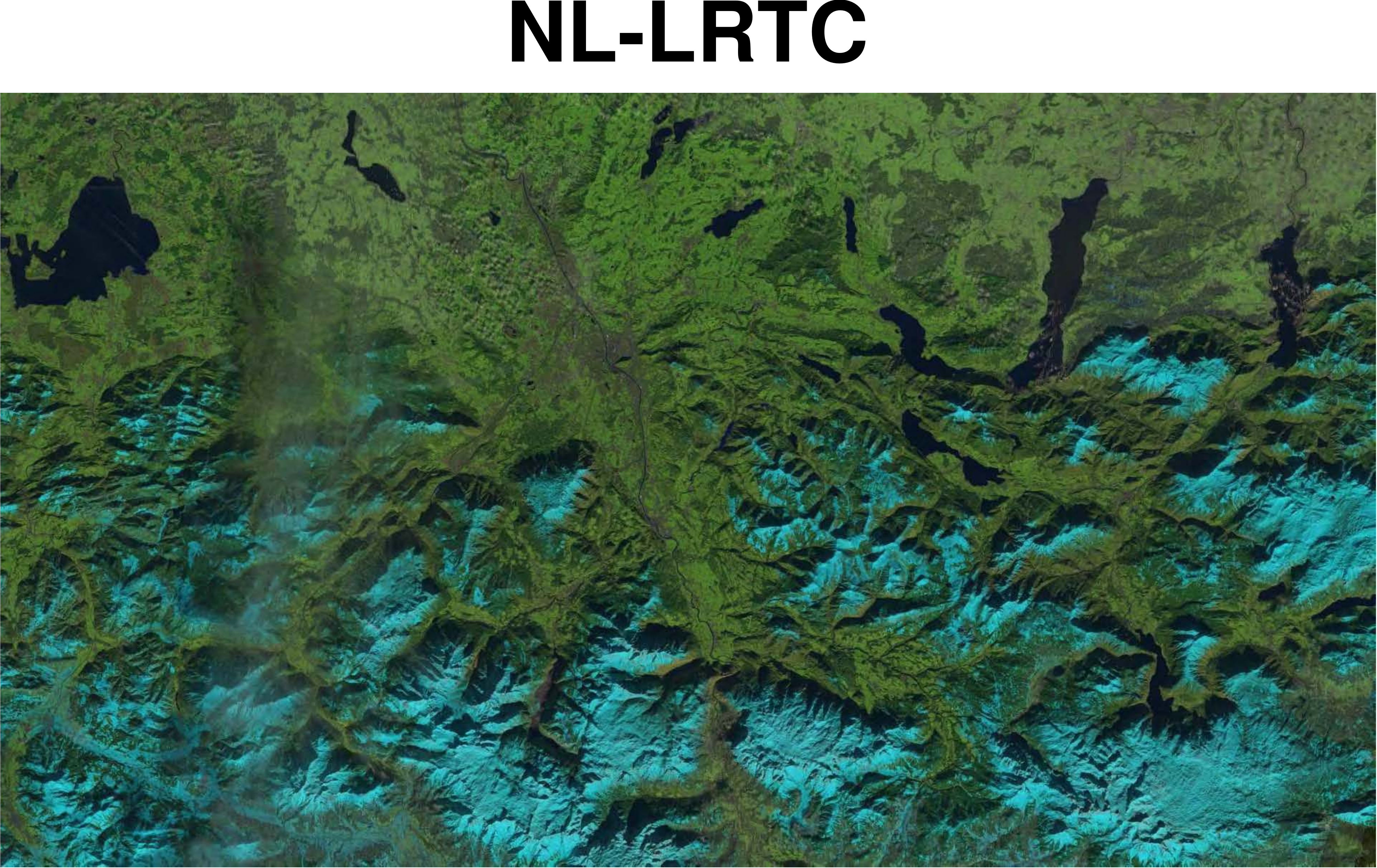}
		\caption{\footnotesize Results for real Landsat experiment. }
		\label{Fig:RealRecover}
	\end{figure}

\begin{figure}[h]
		\centering
		\includegraphics[width=0.31\linewidth]{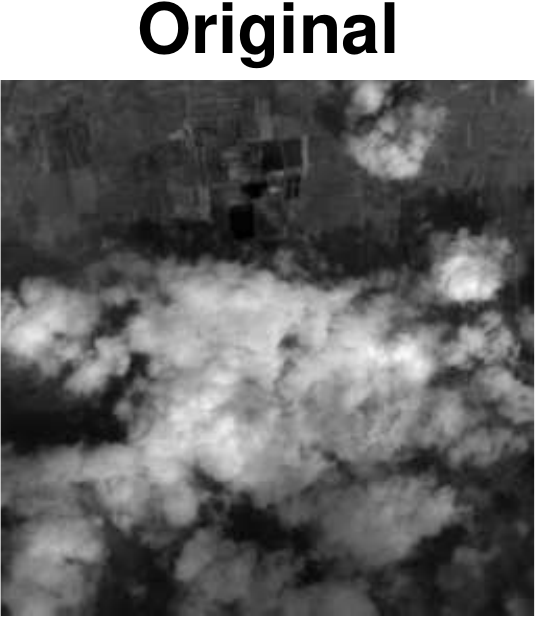}
		\includegraphics[width=0.31\linewidth]{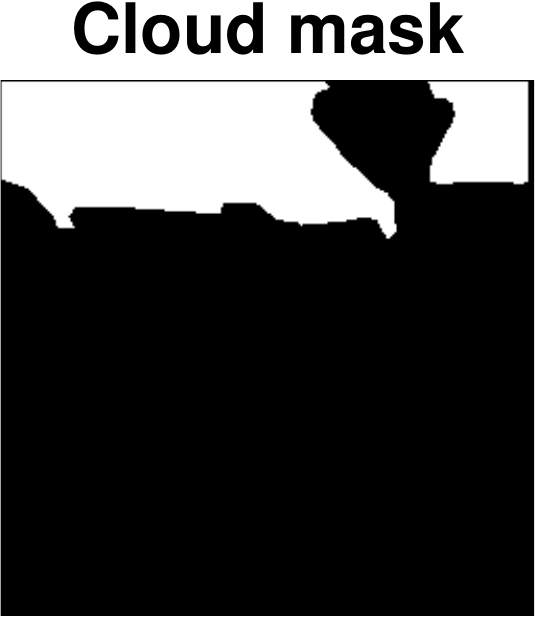}
		\includegraphics[width=0.31\linewidth]{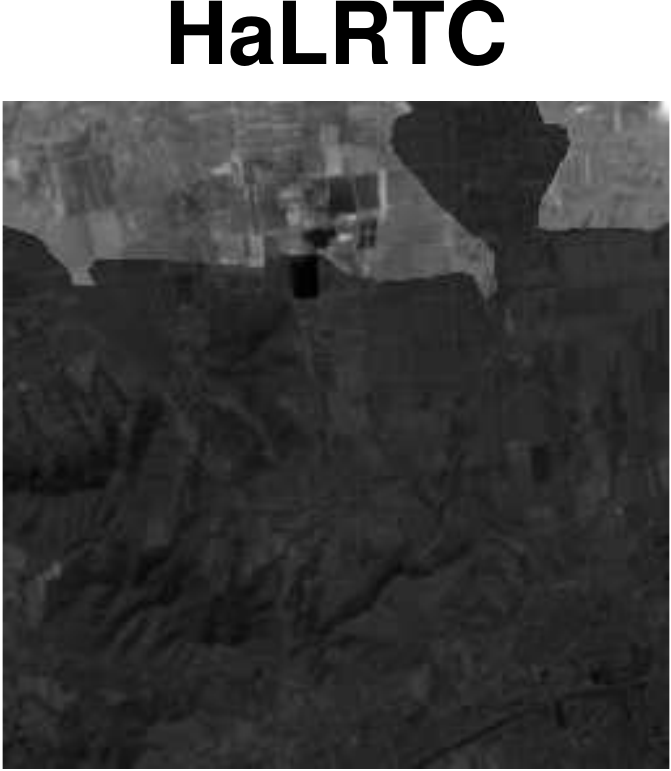} \\[0.3cm]
		\includegraphics[width=0.31\linewidth]{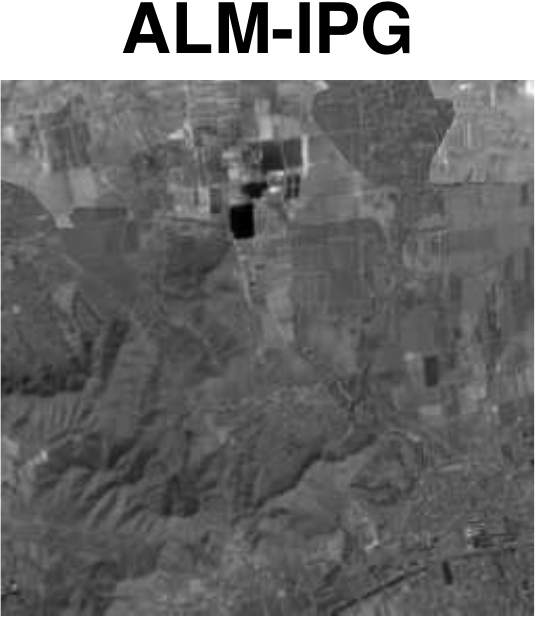}
		\includegraphics[width=0.31\linewidth]{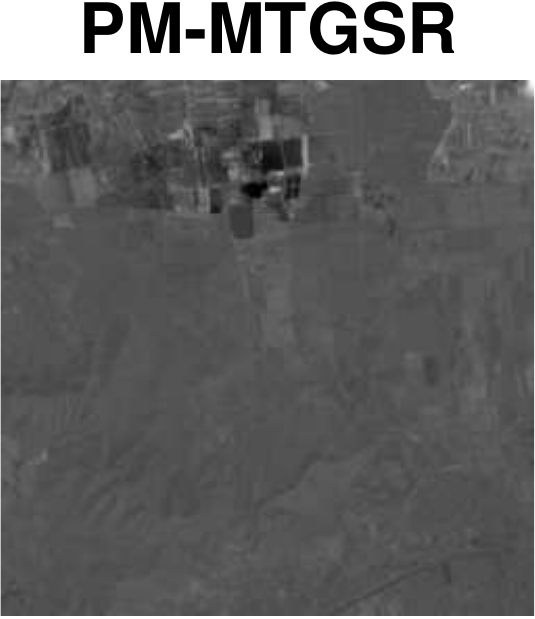}
		\includegraphics[width=0.31\linewidth]{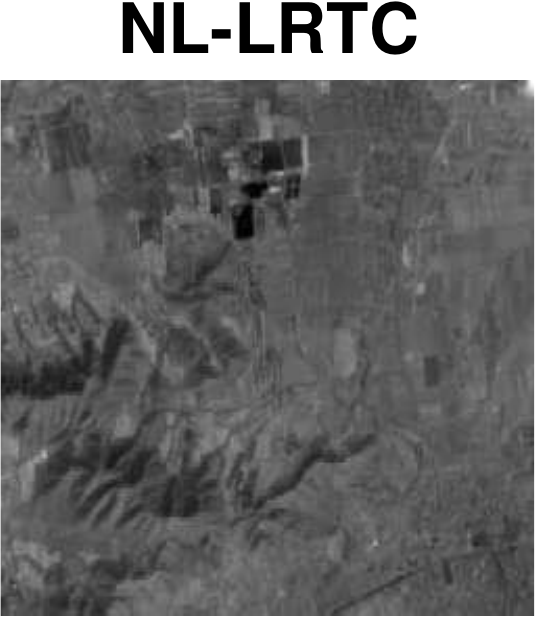}
		\caption{\footnotesize Results for real Sentinel-2 data taken over Beijing. }
		\label{Fig:RealBJ}
	\end{figure}

\begin{figure}[h]
		\centering
		\includegraphics[width=0.31\linewidth]{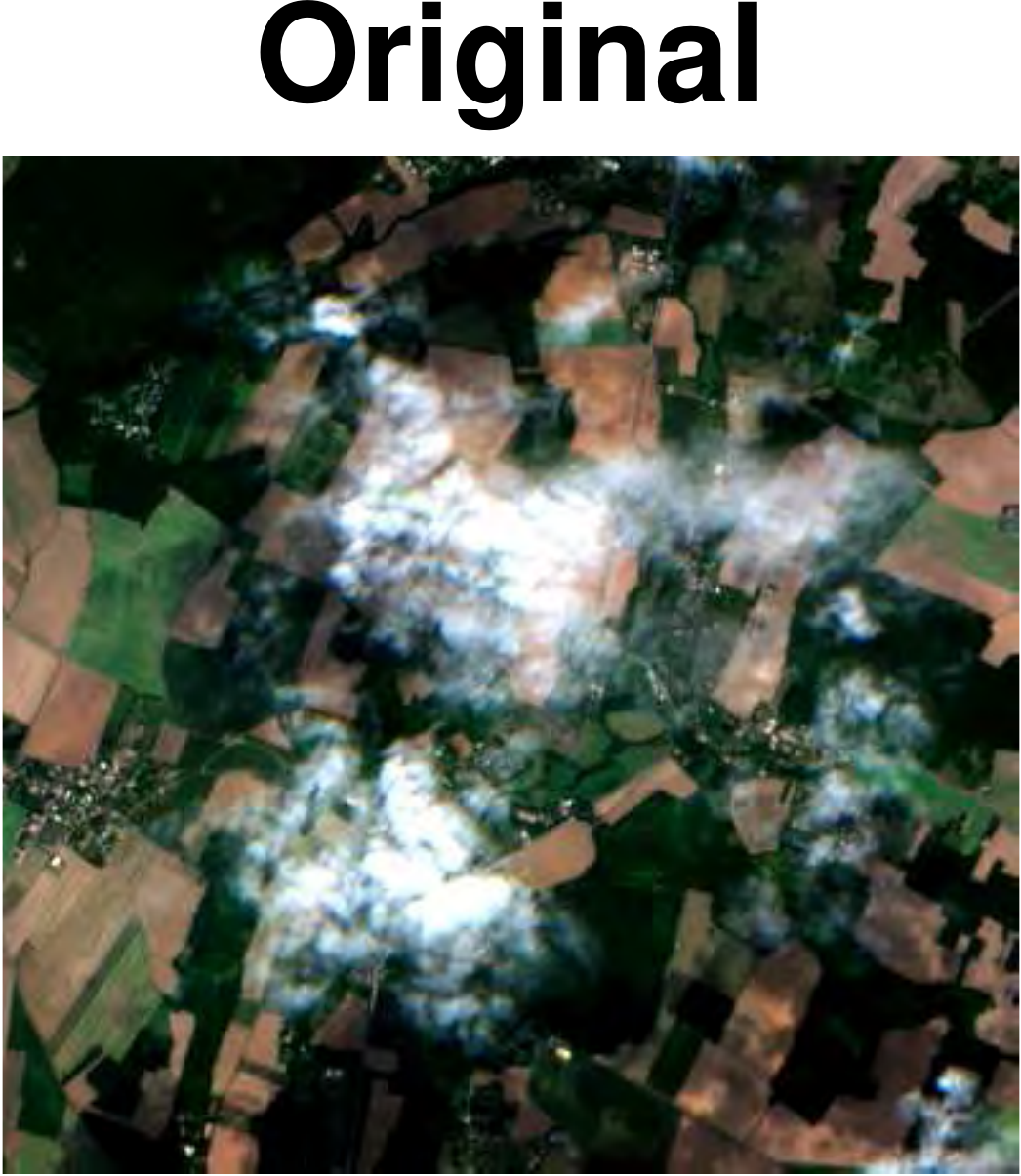}
		\includegraphics[width=0.31\linewidth]{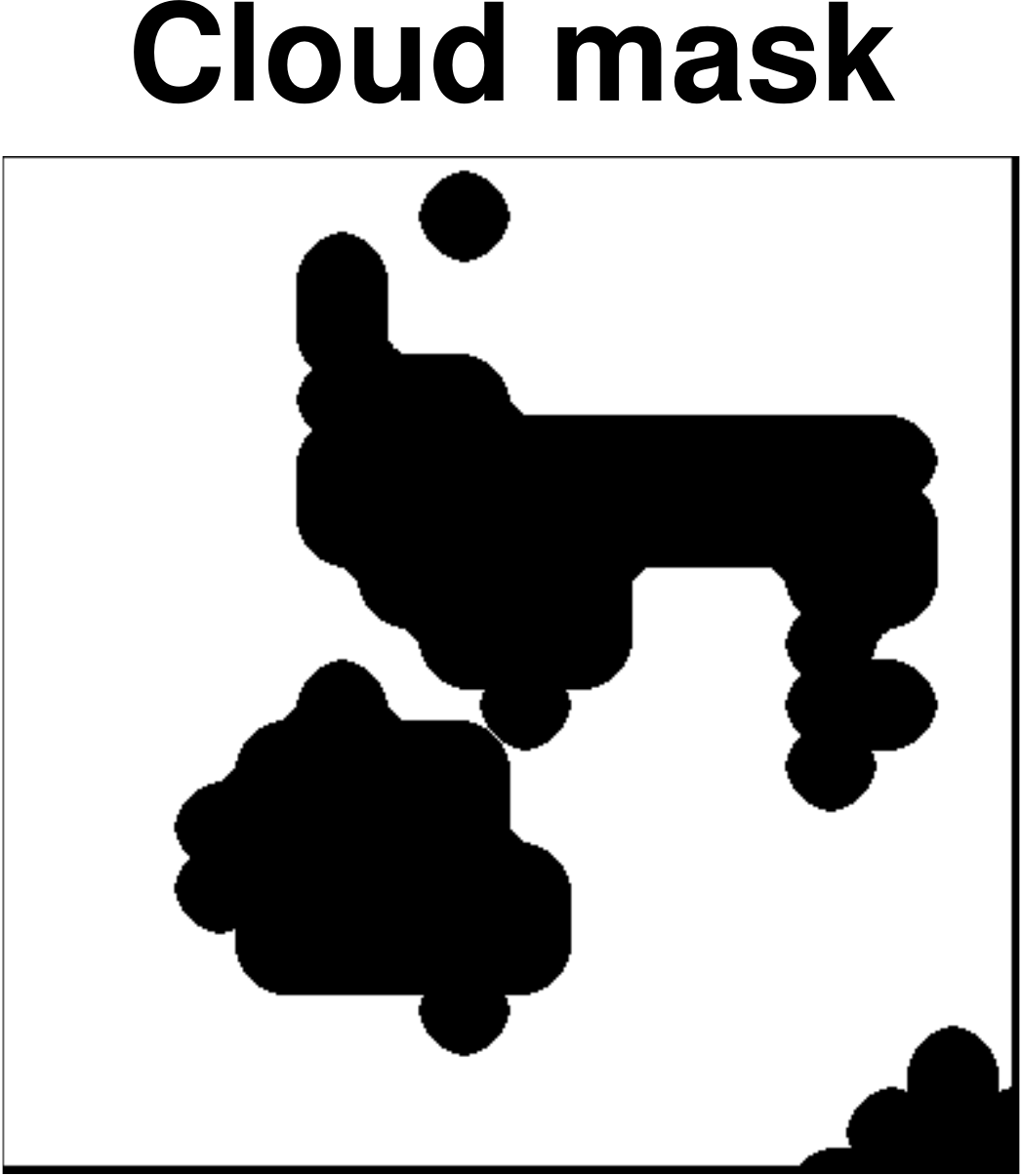}
		\includegraphics[width=0.31\linewidth]{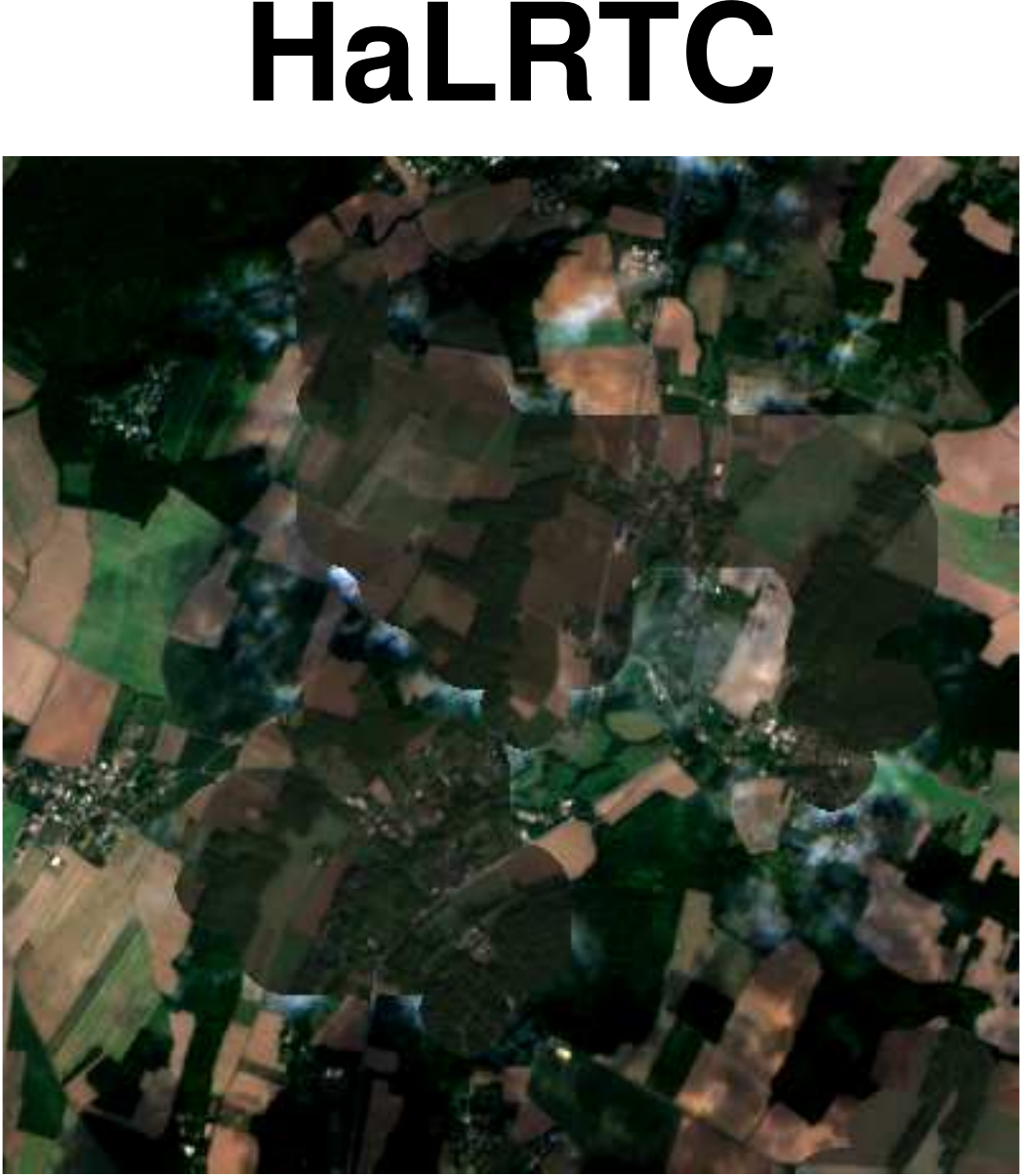}\\[0.3cm]
		\includegraphics[width=0.31\linewidth]{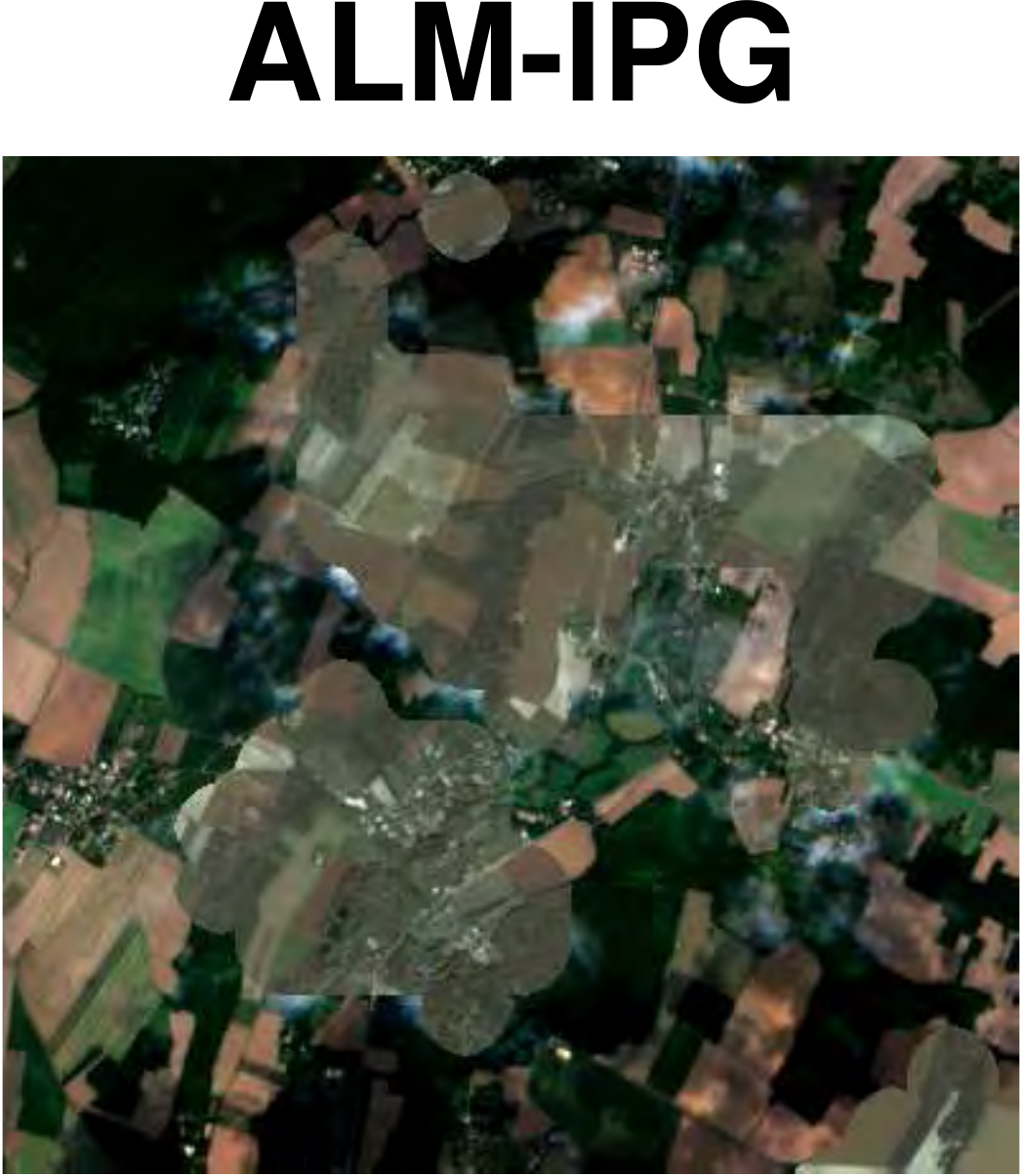}
		\includegraphics[width=0.31\linewidth]{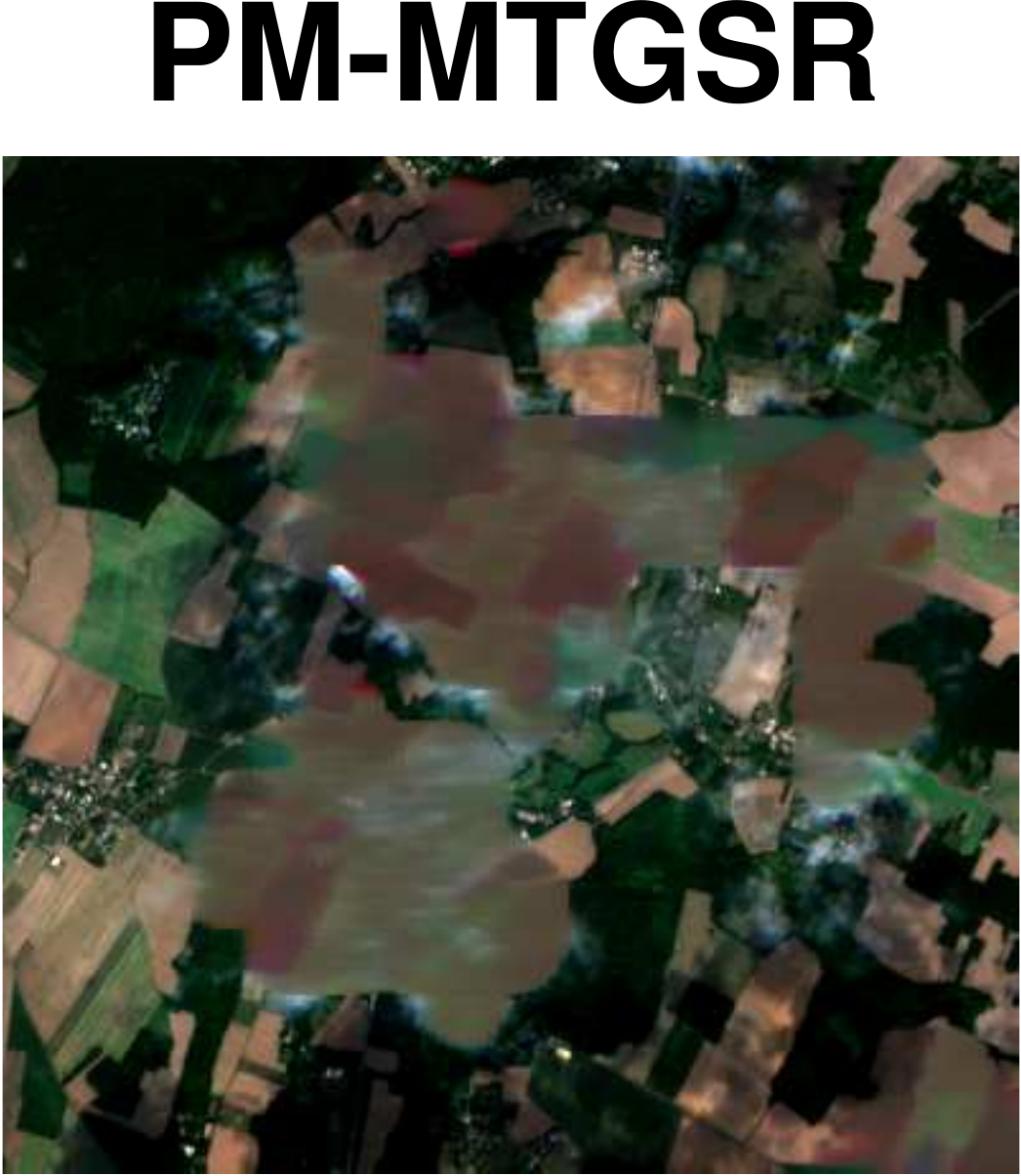}
		\includegraphics[width=0.31\linewidth]{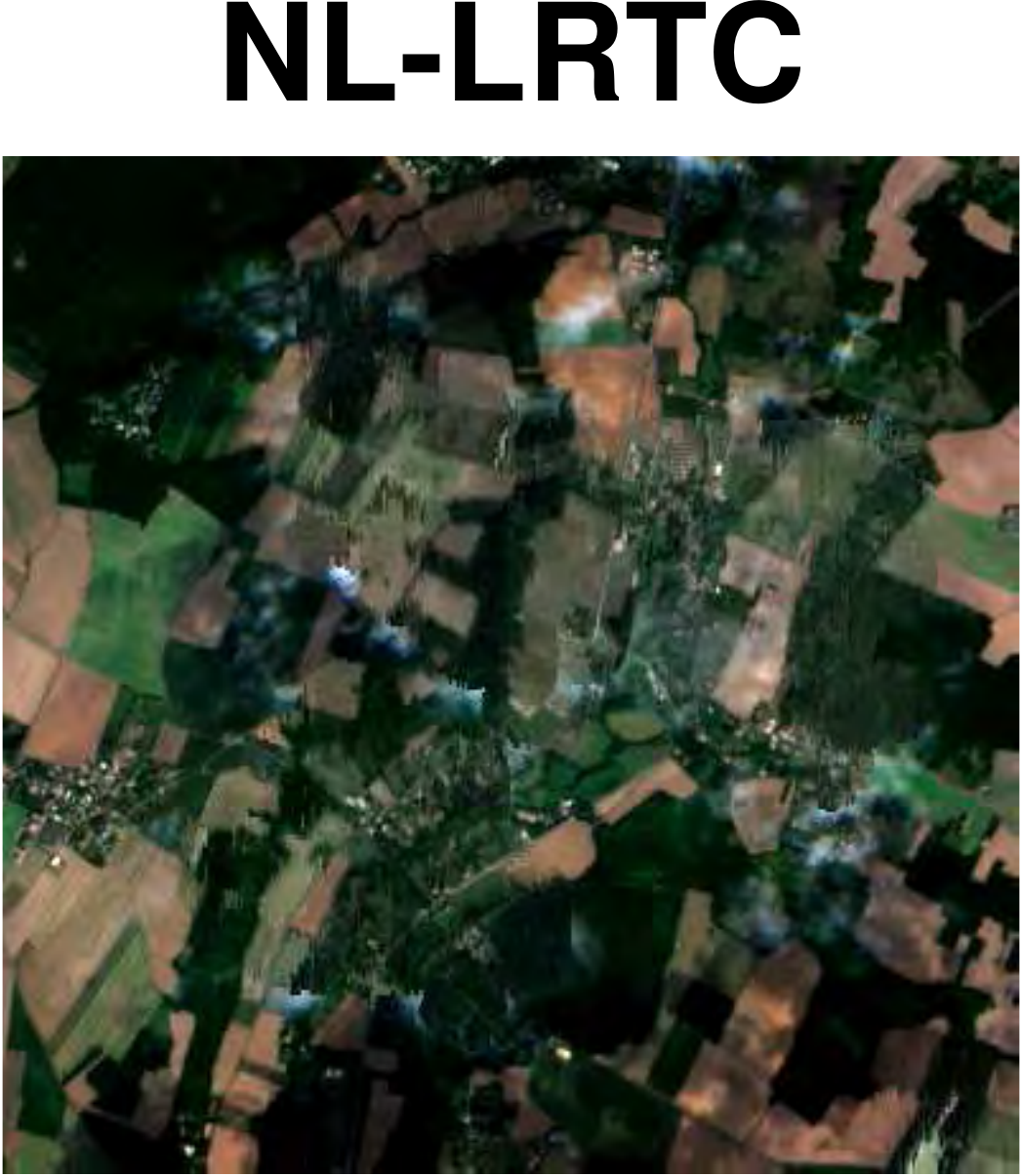}
		\caption{\footnotesize Results for real Sentinel-2 data taken over Eure, France. In this figure, the images are shown in color format using bands 4, 3, and 2.}
		\label{Fig:RealFrance}
	\end{figure}

	In this section, real-data experiments are undertaken. The experimental data are ``M102014'', ``BJ072016'', and ``EU082017''. The cloud detection is not our focus in this work and is complex for different kinds of atmospheric conditions. For the Landsat data ``M102014'', the cloud is detected via a modified version of the thresholding-based cloud detection method proposed in \cite{wang2016removing}; see Appendix \ref{Sec:CloudDetect} for more details. Beijing data ``BJ072016'' contains shadows that cannot be detected by a simple thresholding method. Thus, for ``BJ072016'', the mask for clouds and their shadows is manually drawn. For the Eure data ``EU082017'', the cloud detection is processed by MAYA \cite{Lonjou2016Data}. The corresponding recovery results are shown in Figs. \ref{Fig:RealRecover}, \ref{Fig:RealBJ}, and \ref{Fig:RealFrance}.
	For ``M102014'' (see Fig. \ref{Fig:RealRecover}), the color composite images of reconstruction areas obtained by HaLRTC, ALM-IPG, and PM-MTGSR are obviously different from the known areas. The reconstruction area of NL-LRTC shows a more natural visual effect.
	For ``BJ072016'' (see Fig. \ref{Fig:RealBJ}), the recovery results by ALM-IPG have obvious stairs in the edge of missing and known areas. HaLRTC and PM-MTGSR fail in reconstructing clear details. NL-LRTC shows better results containing more clear details and being more natural compared to the known area.
	For ``EU082017'' (see Fig. \ref{Fig:RealFrance}), the recovery results by HaLRTC, ALM-IPG, and PM-MTGSR are visually worse than that by NL-LRTC, that means the missing areas recovered by NL-LRTC are in harmony with the know areas.
	Moreover, quantitative results for Figs. \ref{Fig:RealRecover}, \ref{Fig:RealBJ}, and \ref{Fig:RealFrance} are shown in Tab. \ref{Tab:Real}. 
		For the Beijing and Eure real data, all the three index values (Q, AG, and BIQA) for NL-LRTC are better than those for HaLRTC, ALM-IPG, and PM-MTGSR. For the Landsat-8 real data, the Q and BIQA values for NL-LRTC are worse than those for HaLRTC, ALM-IPG, and PM-MTGSR. However, the difference is slight. The AG value for NL-LRTC is better than the other three compared methods. The real-data experiments also demonstrate that the proposed method is effective.

	\begin{table}[!ht]
		\caption{Quantitative results comparison for Figs. \ref{Fig:RealRecover}, \ref{Fig:RealBJ}, and \ref{Fig:RealFrance}.  }
		\begin{center}
			\begin{tabular}{c c | c c c c}
				\hline\hline
				& & HaLRTC & ALM-IPG	 & PM-MTGSR & NL-LRTC \\ \hline 
				\multicolumn{1}{c|}{\multirow{3}{*}{\rotatebox{90}{Fig. \ref{Fig:RealRecover}}}} & Q &\textbf{0.0384} &\textbf{0.0384} &0.0366 &0.0371 \\
				\multicolumn{1}{c|}{} & AG &0.0331 &0.0335 &0.0337 &\textbf{0.0339} \\
				\multicolumn{1}{c|}{} & BIQA &\textbf{0.0058} &0.0056 & \textbf{0.0058} & 0.0056 \\ \hline\hline
				\multicolumn{1}{c|}{\multirow{3}{*}{\rotatebox{90}{Fig. \ref{Fig:RealBJ}}}} & Q &0.0132 &0.0100&0.0073 &\textbf{0.0144} \\
				\multicolumn{1}{c|}{} & AG &0.0060 &0.0069 &0.0050 & \textbf{0.0088} \\
				\multicolumn{1}{c|}{} & BIQA & \textbf{0.0012} &0.0009 &0.0005&\textbf{0.0012} \\ \hline\hline
				\multicolumn{1}{c|}{\multirow{3}{*}{\rotatebox{90}{Fig. \ref{Fig:RealFrance}}}} & Q &0.0264 &0.0354&0.0258 &\textbf{0.0404} \\
				\multicolumn{1}{c|}{} & AG &0.0185 &0.0196 &0.0158 & \textbf{0.0222} \\
				\multicolumn{1}{c|}{} & BIQA & 0.0053 &0.0068 &0.0057&\textbf{0.0070} \\ \hline
			\end{tabular}
		\end{center}
		\label{Tab:Real}
	\end{table}

	%

	\section{Conclusion}\label{Sec:Conclusion}
	In this paper, a non-local low-rank tensor completion (NL-LRTC) method has been proposed to reconstruct the missing information in the multitemporal remotely sensed images. By proposing a non-convex approximation for tensors rank, all the three domains (spatial, spectral, and temporal) relationships were considered in NL-LRTC. To take advantage of the spatial correlations, we grouped the 3-order similar patches into a 4-order tensor and considered the tensor low-rankness. Because NL-LRTC made use of the global correlations of all the three domains, it is good at processing not only the temporally contiguous data but also the data that have large differences between the adjacent temporal images regarding the characteristics and conditions of the Earth's surface. In the simulations with various image data sets, NL-LRTC showed comparable or better results than HaLRTC, ALM-IPG, and PM-MTGSR, which are three of the state-of-the-art algorithms. For the real-data experiments, our method obtained visually more natural and quantitatively better reconstruction results.


	%

	\appendices
	\section{Cloud Detection}\label{Sec:CloudDetect}
	In this section, we present an automatic thresholding method for cloud detection motivated by the algorithm of \cite{wang2016removing}. This method assumes that most cloud values in the remotely sensed images are larger than other cloud free values, i.e., clouds are predominantly white. Given a 4-order observation remotely sensed image $\mathcalbf{Y}\in\mathbb{R}^{m\times n\times b\times t}$, where $m\times n$ denotes the number of pixels of remotely sensed images, $b$ denotes the number of spectral channels of remote sensors, and $t$ is the number of time series. In this research, we talk about how to restore the image at one time according to the other times. Suppose $\mathcalbf{Y}^{t_1}$ taken at time $t_1$ is the cloud contained image, then the other images (denoted as $\mathcalbf{Y}^{\hat{t}_1}$) are the references. The cloud detector is to produce a set of indices $\bm{\Omega}\in\mathbb{R}^{m\times n\times b\times t}$, where the position $(i,j,k,l)\in \mathbb{Z}^m\times\mathbb{Z}^n\times\mathbb{Z}^b\times\mathbb{Z}^t$ is covered by cloud if $\bm{\Omega}_{i,j,k,l}=0$ and is cloud free if $\bm{\Omega}_{i,j,k,l}=1$. Note that all the spectral bands of the practical remotely sensed images taken at the same position and same period will be covered by the same clouds, i.e, $\bm{\Omega}(i,j,k, t_1)=0, \forall k\in\{1, 2, \cdots, b\}$ if exist $k_1\in\{1, \cdots, b\}$ subject to $\bm{\Omega}(i,j,k_1, t_1)=0$.

	In this research, there are some other cloud free references. The key point of the detection method is to maximize the similarity of cloud contained and free images in the existing region $\bm{\Omega}$. Given a similar function $f(\bm{x}, \bm{y})$,
	find the indices set $\bm{\Omega}$ by optimizing the following problem:
	\begin{equation}
	\tilde{\bm{\Omega}} = \max_{\bm{\Omega}} f(\mathcalbf{Y}_{\bm{\Omega}}^{t_1}, \mathcalbf{Y}_{\bm{\Omega}}^{\hat{t}_1}).
	\label{Eq:IniOmg}
	\end{equation}
	The function can be any similarity functions, such as correlation coefficients, cosine coefficients, generalized Dice coefficients, and generalized Jaccard coefficients. Here, we use the correlation coefficients. Besides the mentioned method (\ref{Eq:IniOmg}), one can also minimize the distance function such as Euclidean distance, the mean absolute error (MAE), and the mean relative error (MRE) to get the indices set $\bm{\Omega}$. The thresholding produce is detailedly summarized in Algorithm \ref{Alg:thresholding}.

	\begin{algorithm}
		\caption{Thresholding method.}
		\label{Alg:thresholding}
		\begin{algorithmic}[1]
			\Require temporal sequence of cloudy images $\mathcalbf{Y}$, parameter step for increase the thresholding value $s$.
			\State Obtain the initial indices set $\bm{\Omega}_0 = \mathcalbf{Y}^{t_1}>0$;
			\State Set $\tilde{\bm{\Omega}} = \bm{\Omega}_0$ and $\gamma_1=0$;
			\While{$f(\mathcalbf{Y}_{\bm{\Omega}}^{t_1}, \mathcalbf{Y}_{\bm{\Omega}}^{\hat{t}_1})$ increase}  
			\State Update the thresholding value $\gamma_1=\gamma_1+s$;
			\State Calculate the correlation $f(\mathcalbf{Y}_{\bm{\Omega}}^{t_1}, \mathcalbf{Y}_{\bm{\Omega}}^{\hat{t}_1})$.
			\EndWhile  
			\Ensure $\tilde{\bm{\Omega}}$: initial guess for index set.
		\end{algorithmic}
	\end{algorithm}

	The above described thresholding method regards the stationary white background as the cloud, as seen in Fig. \ref{Fig:CloudDetecIllustration}. It is better that these white objects remain in the reconstructed images. Fortunately, the clouds usually cover a big continuous area, this fact motivates us to delete the discrete points. To this end, we propose a KNN-like method. In detail, the pixel in $\tilde{\bm{\Omega}}^c$ is not regarded as cloud if most of its surrounding pixels are not cloud, i.e., most of its neighbor pixels are in $\tilde{\bm{\Omega}}$. The procedure for finding the white background is shown in Fig. \ref{Fig:CloudDetecModify}.

	The two stages of cloud detection is summarized in Algorithm \ref{Alg:cloudDetect} below.

	%

	\begin{algorithm}
		\caption{Cloud detection.}
		\label{Alg:cloudDetect}
		\begin{algorithmic}[1]
			\Require temporal sequence of cloudy images $\mathcalbf{Y}$, parameter thresholding value $\gamma_1$, and parameter $r$ in the similar k-nearest-neighbors search.
			\State Obtain the initial guess $\bm{\Omega}$ via Algorithm \ref{Alg:thresholding}: $\bm{\Omega}=\tilde{\bm{\Omega}}$;
			\For{$(m,n,:,l)\notin\tilde{\bm{\Omega}}$}  
			\State Extract the patch $\bm{P}$ with the center $(m,n,:,l)$ and a radius of $r_1$: $$\bm{P}(i,j,:,l) = \bm{\Omega}(i,j,:,l),$$ where $\|(i,j)-(m,n)\|_{\infty}<r_1$;
			\State Calculate the percent ($p$) of the cloud pixels number;
			\If{$p<0.5$}
			\State $\bm{\Omega} = \bm{\Omega} \cup \{(m, n, :, l)\}$.
			\EndIf
			\EndFor  
			\Ensure $\bm{\Omega}$: index set of non-cloudy pixels.
		\end{algorithmic}
	\end{algorithm}

	\begin{figure}[h]
		\begin{center}
			\includegraphics[width=0.35\textwidth]{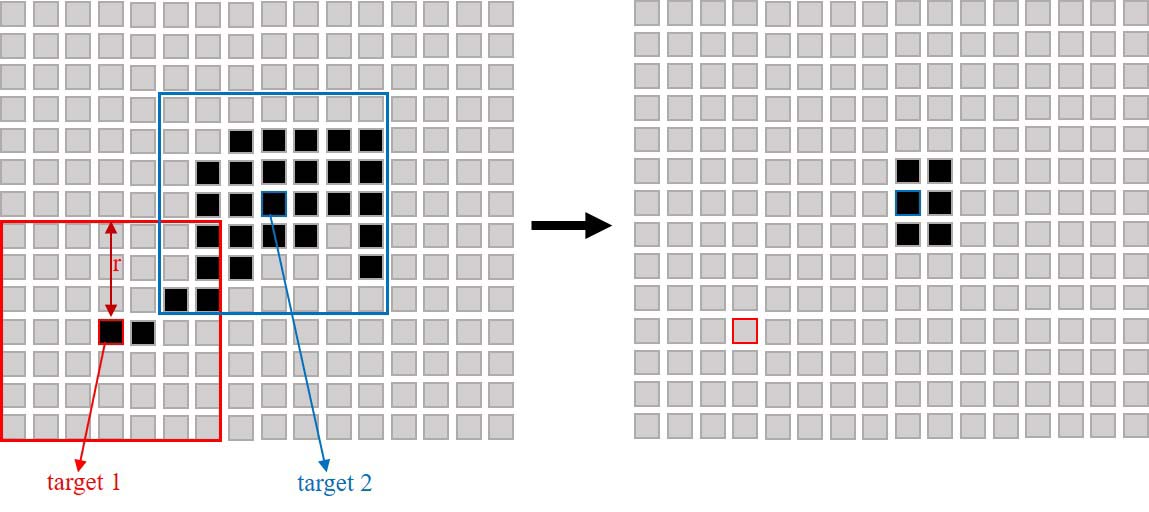}
		\end{center}
		\caption{\footnotesize Modified cloud detection procedure. The black pixels denote ��0�� (cloud), others denote ��1��. In this figure, $r$=3. For the red target pixel, it should be the white background rather than cloud, because in the search window, most pixels are ��1��.  While for the blue target pixel, it should be cloud. }
		\label{Fig:CloudDetecModify}
	\end{figure}

	\begin{figure}[h]
		\begin{center}
			\subfigure[Cloud contained]{\includegraphics[width=0.22\textwidth]{real_cloud_corrupted_1026.eps}}
			\subfigure[Cloud removal]{\includegraphics[width=0.22\textwidth]{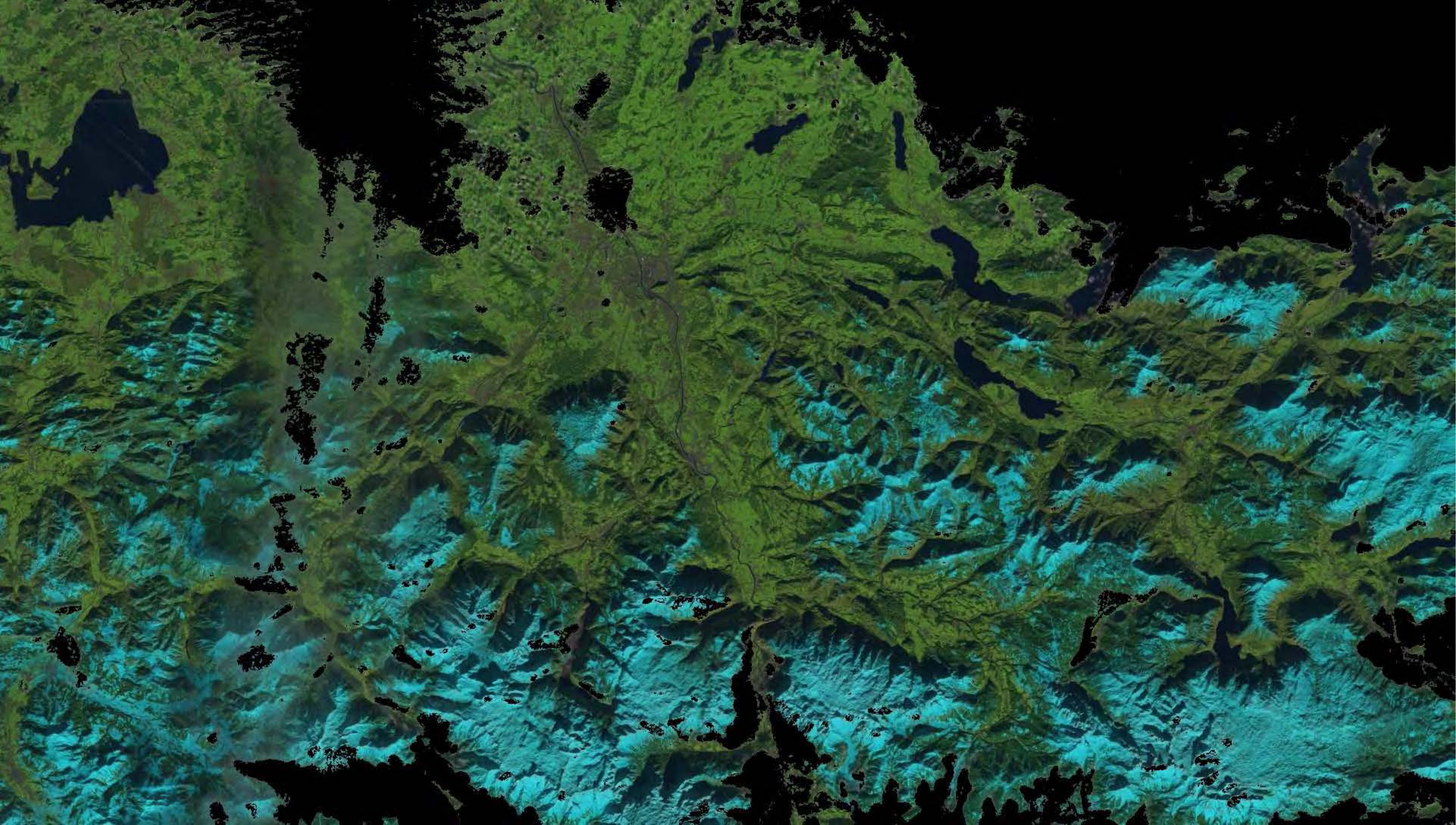}}

			\subfigure[Alg. \ref{Alg:thresholding}]{\includegraphics[width=0.22\textwidth]{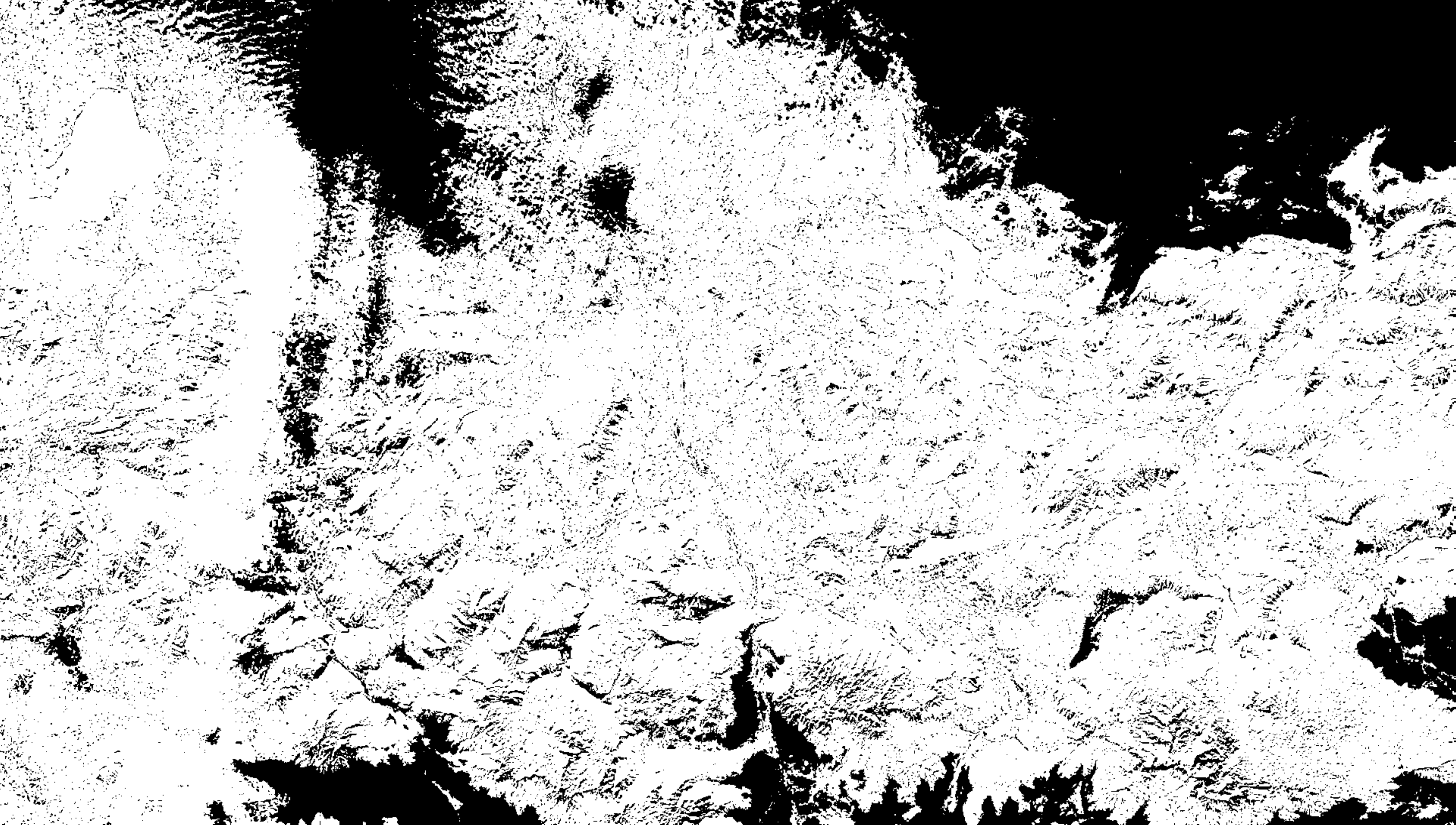}}
			\subfigure[Alg. \ref{Alg:cloudDetect}]{\includegraphics[width=0.22\textwidth]{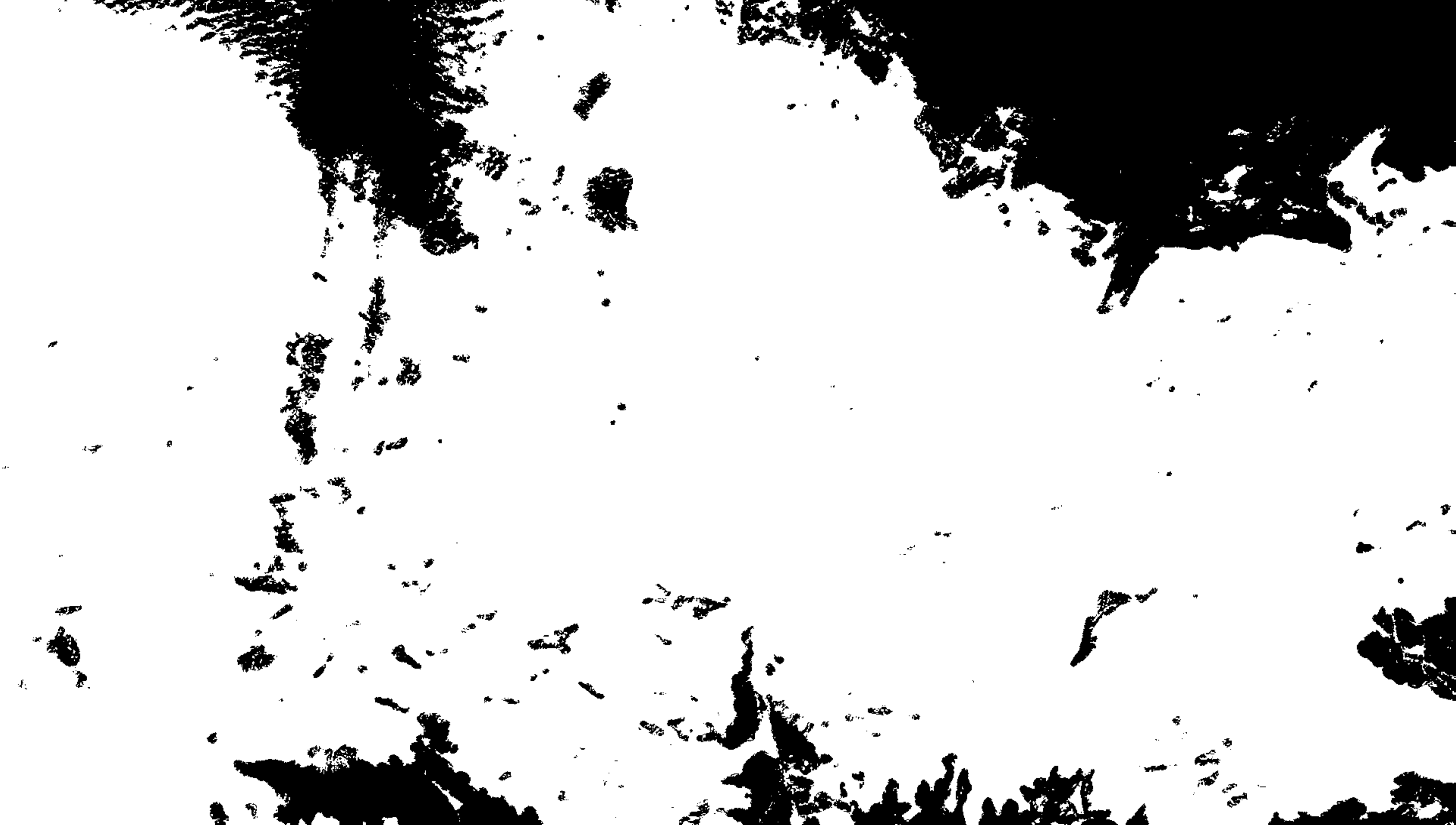}}
		\end{center}
		\caption{\footnotesize Illustration of the proposed cloud detection procedure. (a) cloud contained image, (b) removing the detected cloud, (c) cloud detected via Alg. \ref{Alg:thresholding}, (c) cloud detected via Alg. \ref{Alg:cloudDetect}. }
		\label{Fig:CloudDetecIllustration}
	\end{figure}


	\section*{Acknowledgment}

	The authors would like to thank Prof. H. Shen and X. Li from Wuhan University for sharing the codes of the PM-MTGSR method.

	\ifCLASSOPTIONcaptionsoff
	\newpage
	\fi

	\begin{IEEEbiography}[{\includegraphics[width=1in,height=1.25in,clip,keepaspectratio]{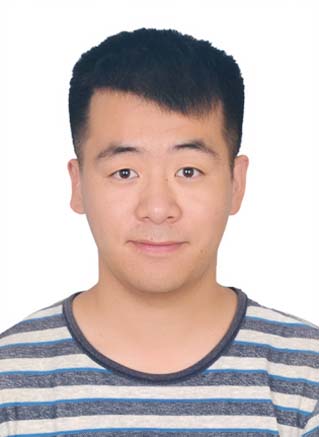}}]{Teng-Yu Ji}
	received the B.S. degree from the School of Mathematical Sciences, University of Electronic Science and Technology of China, Chengdu, China, in 2012, where he is currently pursuing the Ph.D. degree with the School of Mathematical Sciences.

	His current research interests include tensor decomposition and applications, including tensor completion and remotely sensed image reconstruction.
	\end{IEEEbiography}

	\begin{IEEEbiography}[{\includegraphics[width=1in,height=1.25in,clip,keepaspectratio]{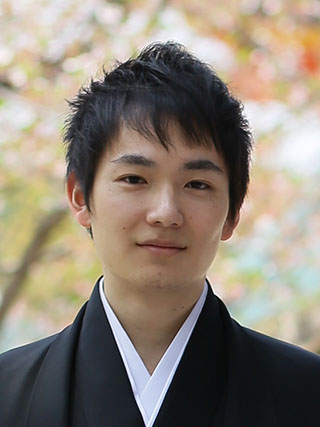}}]
	{Naoto Yokoya}
     (S'10--M'13) received the M.Sc. and Ph.D. degrees in aerospace engineering from the University of Tokyo, Tokyo, Japan, in 2010 and 2013, respectively.

From 2012 to 2013, he was a Research Fellow with Japan Society for the Promotion of Science, Tokyo, Japan. Since 2013, he is an Assistant Professor with the University of Tokyo. From 2015 to 2017, he was also an Alexander von Humboldt Research Fellow with the German Aerospace Center (DLR), Oberpfaffenhofen, and Technical University of Munich (TUM), Munich, Germany. His research interests include image analysis and data fusion in remote sensing. Since 2017, he is a Co-chair of IEEE Geoscience and Remote Sensing Image Analysis and Data Fusion Technical Committee.

	\end{IEEEbiography}

\begin{IEEEbiography}[{\includegraphics[width=1in,height=1.25in,clip,keepaspectratio]{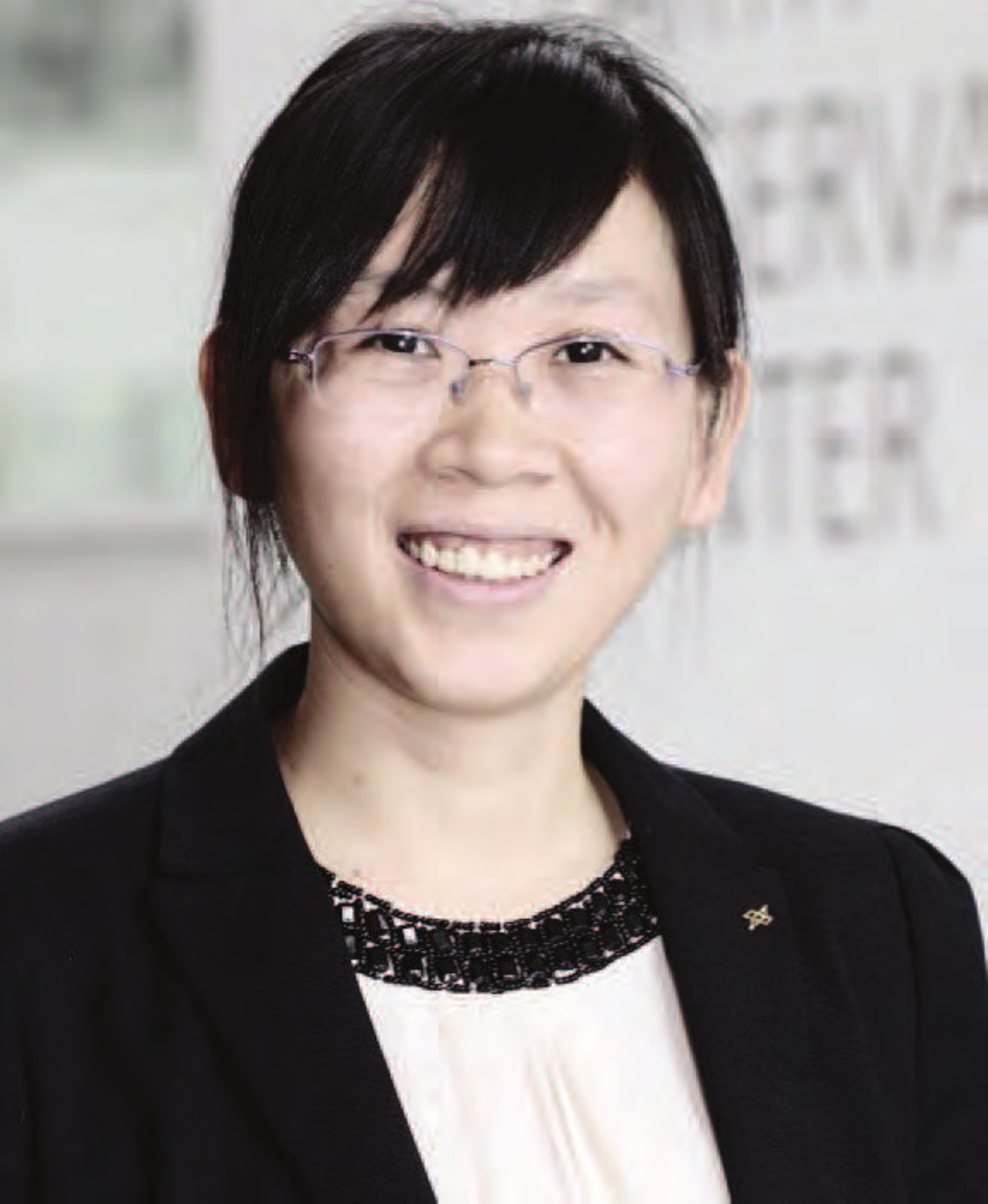}}]{Xiao Xiang Zhu}(S'10-M'12-SM'14)
received the Master (M.Sc.) degree, her doctor of engineering (Dr.-Ing.) degree and her ``Habilitation'' in the field of signal processing from Technical University of Munich (TUM), Munich, Germany, in 2008, 2011 and 2013, respectively.
\par
She is currently the Professor for Signal Processing in Earth Observation (www.sipeo.bgu.tum.de) at Technical University of Munich (TUM) and German Aerospace Center (DLR); the head of the Team Signal Analysis at DLR; and the head of the Helmholtz Young Investigator Group "SiPEO" at DLR and TUM. Prof. Zhu was a guest scientist or visiting professor at the Italian National Research Council (CNR-IREA), Naples, Italy, Fudan University, Shanghai, China, the University of Tokyo, Tokyo, Japan and University of California, Los Angeles, United States in 2009, 2014, 2015 and 2016, respectively. Her main research interests are remote sensing and Earth observation, signal processing, machine learning and data science, with a special application focus on global urban mapping.
\par
Dr. Zhu is a member of young academy (Junge Akademie/Junges Kolleg) at the Berlin-Brandenburg Academy of Sciences and Humanities and the German National Academy of Sciences Leopoldina and the Bavarian Academy of Sciences and Humanities. She is an associate Editor of IEEE Transactions on Geoscience and Remote Sensing.
\end{IEEEbiography}

	\begin{IEEEbiography}[{\includegraphics[width=1in,height=1.25in,clip,keepaspectratio]{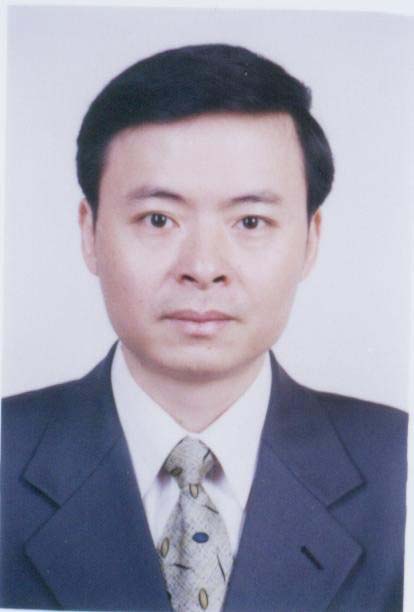}}]
	{Ting-Zhu Huang}
	received the B. S., M. S., and Ph. D. degrees in Computational Mathematics from the Department of Mathematics, Xi��an Jiaotong University, Xi��an, China. He is currently a professor in the School of Mathematical Sciences, UESTC. He is currently an editor of The Scientific World Journal, Advances in Numerical Analysis, J. Appl. Math., J. Pure and Appl. Math.: Adv. and Appl., J. Electronic Sci. and Tech. of China, etc. His current research interests include scientific computation and applications, numerical algorithms for image processing, numerical linear algebra, preconditioning technologies, and matrix analysis with applications, etc.
	\end{IEEEbiography}







\end{document}